\documentclass{aa}  
\usepackage{graphicx}
\usepackage{amsmath}	
\usepackage{amssymb}	
\usepackage{relsize}
\usepackage{txfonts}
\newcommand{\msun}{\hbox{M$_{\odot}$}}
\newcommand{\kms}{\hbox{km s$^{-1}$}}

\begin{document}

   \title{Fornax 3D project: assessing the diversity of IMF and stellar population maps within the Fornax Cluster}

   \subtitle{}

   \author{
I. Mart\'in-Navarro \inst{1,2}, F.~Pinna\inst{3}, L.~Coccato\inst{4},  J. Falc\'on-Barroso \inst{1,2}, G. van de Ven \inst{5}, M. Lyubenova \inst{4}, E.~M.~Corsini\inst{6, 7}, K. Fahrion\inst{4}, D.~A.~Gadotti\inst{4}, E.~Iodice\inst{8}, R.~M.~McDermid\inst{9, 10},  A. Poci\inst{9}, M.~Sarzi\inst{11,12}, T. W. Spriggs\inst{12}, S. Viaene\inst{12,13}, P.~T.~de~Zeeuw\inst{14,15}, \and L. Zhu\inst{16}
          }

   \institute{
Instituto de Astrof\'{\i}sica de Canarias,c/ V\'{\i}a L\'actea s/n, E38205 - La Laguna, Tenerife, Spain\\
\email{imartin@iac.es}
\and
Departamento de Astrof\'isica, Universidad de La Laguna, E-38205 La Laguna, Tenerife, Spain
\and
Max-Planck Institut f\"ur Astronomie, Konigstuhl 17, D-69117 Heidelberg, Germany
\and
European Southern Observatory, Karl-Schwarzschild-Str. 2, 85748 Garching b. M\"unchen, Germany
\and
University of Vienna, Department of Astrophysics, T\"urkenschanzstrasse 17, 1180 Vienna, Austria
\and
Dipartimento di Fisica e Astronomia `G. Galilei', Universit\`a di Padova, vicolo dell'Osservatorio 3, I-35122 Padova, Italy
\and 
INAF--Osservatorio Astronomico di Padova, vicolo dell'Osservatorio 5, I-35122 Padova, Italy
\and
INAF--Osservatorio Astronomico di Capodimonte, via Moiariello 16, I-80131 Napoli, Italy
\and
Research Centre for Astronomy, Astrophysics, and Astrophotonics, Department of Physics and Astronomy, Macquarie University, NSW 2109, Australia
\and 
ARC Centre of Excellence for All Sky Astrophysics in 3 Dimensions (ASTRO 3D), Australia
\and
Armagh Observatory and Planetarium, College Hill, Armagh, BT61 9DG, UK
\and
Centre for Astrophysics Research, University of Hertfordshire, College Lane, Hatfield AL10 9AB, UK
\and
Sterrenkundig Observatorium, Universiteit Gent, Krijgslaan 281, B-9000 Gent, Belgium
\and 
Sterrewacht Leiden, Leiden University, Postbus 9513, 2300 RA Leiden, The Netherlands
\and 
Max-Planck-Institut fuer extraterrestrische Physik, Giessenbachstrasse, 85741 Garching bei Muenchen, Germany
\and 
Shanghai Astronomical Observatory, Chinese Academy of Sciences, 80 Nandan Road, Shanghai 200030, China
}

   \date{Received; accepted}
   
   \titlerunning{A two-dimensional view of the IMF}
\authorrunning{Mart\'in-Navarro et al.}  
 
  \abstract
   {
The stellar initial mass function (IMF) is central to our interpretation of astronomical observables and to our understanding of most baryonic processes within galaxies. The universality of the IMF, suggested by observations in our own Milky Way, has been thoroughly revisited due to the apparent excess of low-mass stars in the central regions of massive quiescent galaxies. As part of the efforts within the Fornax 3D project, we aim to characterize the two-dimensional IMF variations in a sample of 23 quiescent galaxies within the Fornax cluster. For each galaxy in the sample, we measured the mean age, metallicity, [Mg/Fe], and IMF slope maps from spatially resolved integrated spectra. The IMF maps show a variety of behaviors and internal substructures, roughly following metallicity variations. However, metallicity alone is not able to fully explain the complexity exhibited by the IMF maps. In particular, for relatively metal-poor stellar populations ([M/H]$\lesssim-0.1$), the slope of the IMF seems to depend on the (specific) star formation rate at which stars were formed. Moreover, metallicity maps have systematically higher ellipticities than IMF slope ones. At the same time, both metallicity and IMF slope maps have at the same time higher ellipticities than the stellar light distribution in our sample of galaxies. In addition we find that, regardless of the stellar mass, every galaxy in our sample shows a positive radial [Mg/Fe] gradient. This results in a strong [Fe/H]--[Mg/Fe] relation, similar to what is observed in nearby, resolved galaxies. Since the formation history and chemical enrichment of galaxies are causally driven by changes in the IMF, our findings call for a physically motivated interpretation of stellar population measurements based on integrated spectra that take into account any possible time evolution of the stellar populations.
   }


   \maketitle
%

\section{Introduction} \label{sec:intro}

The observed spectrum of a galaxy is given by the sum of the fluxes of its individual stars. Since stellar evolution dictates that the radiation emitted by any given star is determined by its mass \citep[e.g.,][]{Kippenhahn}, the observed flux of a galaxy can then be understood as a summation of fluxes coming from stars with different masses

\begin{equation}
   \label{eq:flux}
   F_\mathrm{obs} = \sum_{i} F_i\,(m_\star) = M_\star \sum_{m_\star} \Phi_{m_\star} F_{m_\star} ,
\end{equation}

\noindent 
where $F_\mathrm{obs}$ is the flux of the galaxy, $M_\star$ its stellar mass and $F_i\,(m_\star)$ the fluxes of its individual stars. The quantity $\Phi_{m_\star}$ indicates the mass fraction of stars with mass $m_\star$ and radiating a flux $F_{m_\star}$ present in the galaxy, and it is a simplified approximation to the so-called stellar initial mass function (IMF), which formally describes the mass spectrum of stars at birth \citep[e.g.,][]{bastian}.

Equation~\ref{eq:flux} highlights two important features of the IMF. First, by construction, the IMF controls the fraction of light emitted by stars with different masses and therefore it is a key ingredient in modeling the integrated flux of distant galaxies \citep[e.g.,][]{Vazdekis96,TMJ,Conroy13,Villaume17}. Second, the IMF links the observed fluxes of galaxies to their stellar masses, allowing astronomical measurements to be interpreted in terms of astrophysical models. Moreover, the IMF is also ultimately responsible for modulating the chemical enrichment and the baryonic cycle within galaxies, as it determines both the number of massive stars that explode as supernovae and the amount of gas locked in long-lived, low-mass stars \citep{Ferreras15,Clauwens16,Thales18,Barber18}.

With a pivotal role in our empirical understanding of galaxies, much has been debated about the origin and properties of the IMF. \citet{Salp:55} first characterized its behavior in the solar neighborhood, finding a scale-free distribution of stars approximately given by 
\begin{equation*} 
   \Phi(\log M) = d N / d \log M \propto M^{-\Gamma} ,
   \end{equation*}
with a logarithmic slope $\Gamma=1.35$ for stars more massive than one solar mass. These findings were later expanded by \citet{Miller79} who found that, for stellar masses below the range initially explored by Edwin Salpeter, the IMF flattens out. There is now a widely adopted consensus that, at least in the Milky Way, the IMF appears to be independent of location \citep{mw,Kroupa,Chabrier}. However, with no established theoretical understanding of how star formation happens from first principles, arguments that the IMF is truly universal (that is, a fixed property in all environments at all times) remain inconclusive. Probing the IMF beyond the Milky Way has therefore been an open challenge for observational astronomy for decades.

Three main approaches have been used to measure the IMF outside the solar neighborhood, based on the principles described above. First, and given the direct connection between mass and light suggested by Equation~\ref{eq:flux}, the mass-to-light ($M/L$) ratio of a given population can be used as a proxy for the shape of its IMF. In particular, variations in the IMF can be explored by independently measuring the stellar mass and the luminosity of a galaxy and comparing this observed $M/L$ with the expectation from a Milky Way-like IMF \citep[see e.g.][\S\,4]{Smith20}. However, different IMF variations could lead to similar $M/L$ variations and thus IMF constraints based on gravitational measurements are somewhat degenerate. For example, for old stellar populations, a $M/L$ higher than expected from a Milky Way-like IMF could be equally consistent with an enhanced fraction of low-mass stars or with an enhanced fraction of stellar remnants, that is, of very massive stars \citep[e.g.,][]{cappellari}. 

Alternatively, the IMF can also be studied by analyzing the absorption spectra of galaxies, as variations in the IMF imply a change in the mixture of stars needed to reproduce the observed spectra \citep{Conroy13}. Unfortunately, changes in the spectra of unresolved galaxies induced by IMF variations are subtle and often degenerate with confounding variables such as age, metallicity, and abundance ratios \citep[e.g.,][]{conroy}. Therefore, reliable IMF measurements from integrated spectra require high quality data and careful modeling techniques. Measurements of the high-mass end of the IMF based on emission spectra are also possible \citep[e.g.,][]{Hoversten08,Meurer09,Lee09,Nanayakkara17}, but, as any other observational approach, they are sensitive to systematics and uncertainties as they probe rapidly-changing stellar evolutionary stages which are very challenging from a modeling point of view.

Finally, variations in the IMF would also lead to distinct chemical compositions in galaxies \citep{Arrigoni10,Yan19}. In particular, thanks to recent developments in millimetric and submillimetric facilities \citep{Romano17}, it has been suggested that the IMF in star-forming systems exhibits an enhanced fraction of high-mass stars, both locally \citep{Sliwa17,Brown19} and at high redshift \citep{Zhang18}. These IMF estimates, based on particular isotopic ratios, are not free from degeneracies and systematics, relying on strong assumptions about supernova yields, stellar rotation, and star formation histories \citep{Romano19}.

Beyond the Milky Way, early-type galaxies (ETGs) are arguably the systems where the (non)universality of the IMF has been most thoroughly tested \citep[see][for a recent review]{Smith20}. The observed properties of ETGs allow for assumptions regarding their internal structure and their stellar populations that greatly simplify an analysis that otherwise would be virtually impossible. In particular, stellar populations in ETGs are generally old and well-represented by single stellar population models \citep[SSP, e.g.][]{trager,Trager00,Thomas05,Thomas10,Kuntschner10,McDermid15,MN18,Parikh,Bernardi19,Lacerna20}. However, the fact that primarily old stars are found in ETGs also means that the mass range explored by IMF studies in ETGs is effectively limited to stars less massive than $\lesssim1$ \msun, meaning those with lifetimes long enough to be observed after several gigayears. Thus, it is worth emphasizing that only a small portion of the IMF is directly constrained by these studies. As noted above, the high-mass end of the IMF can be also indirectly probed through the effect of stellar remnants on the observed $M/L$, but this approach is highly degenerate with the shape of the low-mass end  \citep[e.g.,][]{Lyubenova16}.

Yet, despite these technical difficulties and the inherent limitations, there is now compelling evidence suggesting that the low-mass end slope of the IMF in massive ETGs is steeper than what it is found in the Milky Way. Dynamical measurements have long shown that galaxies with higher stellar velocity dispersion, and in general more massive galaxies, exhibit $M/L$s that are too high to be consistent with a Milky Way-like IMF \citep{Treu,thomas11,auger,cappellari,Dutton12,wegner12,Tortora13b,Lasker13,Tortora14,Posacki15,Li17,Corsini17,Sonnenfeld19}. Moreover, these enhanced $M/L$s in ETGs seem to be already in place at relatively high redshifts \citep[$z\sim$1, e.g.][]{Shetty14,Sonnenfeld15,Tortora18,Mendel20}.

Almost completely independently, detailed studies of the absorption features of massive ETGs have also suggested that the low-mass end of the IMF becomes more bottom-heavy (with an enhanced fraction of low-mass stars) with increasing galaxy mass \citep{vandokkum,spiniello12,Spiniello2013,Spiniello15,conroy12,Smith12,ferreras,labarbera,Tang17,Lagattuta17,Rosani18}. This excess of low-mass stars in massive ETGs would naturally explain the abnormally high $M/L$ derived from dynamical analysis, providing a coherent picture between dynamics- and stellar population-based studies \citep{Conroy13b,Lyubenova16}. The consistency between these two approaches is not trivial and stands as one of the strongest arguments supporting the nonuniversality of the IMF in massive ETGs. 

Although the fact that more massive galaxies tend to exhibit an enhanced fraction of low-mass stars is, in general, widely accepted now, the origin of these IMF variations remains highly debated. In this context, spatially resolved IMF studies are a natural way forward to further characterize the behavior of the IMF in ETGs. \citet{MN15a} showed that the apparent excess of low-mass stars in massive ETGs is restricted to their central regions, as the IMF slope tends to flatten toward the outskirts. This result, and in general the presence of IMF gradients in ETGs, has been independently confirmed by both stellar population \citep{LB16,LB17,LB19,vdk17,Vaughan18a,Sarzi18,Helena19} and dynamical studies \citep{Oldham,Davis17,Sonnenfeld18}, only when aperture effects are taken into account \citep{McDermid14, Lyubenova16}. 

The advent of efficient integral field units has recently revolutionized IMF studies in ETGs, allowing one to better explore the parameter space where IMF variations become relevant. The CALIFA survey \citep{califa} was the first one to provide deep-enough spectroscopic data of key spectral features to measure IMF gradients in a rather large and complete sample of ETGs, finding a tight correlation between local metallicity and the IMF low-mass end slope \citep{MN15c}. Similarly, using data from the MANGA survey \citep{manga}, \citet{Parikh} characterized the mass dependence of IMF gradients in ETGs, suggesting also a local correlation between IMF and metallicity. This correlation seems to be robust and decoupled from other quantities such as age, dust extinction, or changes in the abundance pattern \citep{Zhou19}. IMF variations in the nearby massive ETG M\,87 also track the observed metallicity gradient \citep{Sarzi18}. 

However, the observed correlation between local metallicity and IMF slope is far from giving a satisfactory explanation, as changes in metallicity alone are not able to fully explain the complexity of IMF variations in ETGs \citep{McConnell16,Alexa17,Zieleniewski17,Davis17,Alton18,LB19,Barbosa}. The need for a more complete characterization of the IMF behavior is exemplified by our findings presented in \citet{MN19}. Making use of the Multi Unit Spectroscopic Explorer (MUSE) \citep{Bacon10} data from the Fornax 3D project \citep{f3d} we showed, with exquisite spatial resolution, the first two-dimensional IMF map of the nearby ETG FCC\,167. The unprecedented quality of the Fornax 3D data allowed us to observe a suggestive connection between the internal orbital distribution of FCC\,167 and the measured IMF variations, beyond the expected IMF--metallicity trend.

The present work expands the stellar population analysis presented in \citet{MN19} to the complete sample of ETGs observed within the Fornax 3D survey, providing a complete census of the IMF variations within 24 individual galaxies in the Fornax cluster. The layout of the paper is as follows: in \S~\ref{sec:data} we present the properties of the sample and data. The analysis is described in \S~\ref{sec:analysis} and the results are presented in \S~\ref{sec:resu}. We discuss the implications of our findings in \S~\ref{sec:discu}, which are finally summarized in \S~\ref{sec:sum}.

\section{Data and sample selection} \label{sec:data}

This paper is part of the efforts within the Fornax 3D survey \citep{f3d} to provide a comprehensive picture of the formation and evolution of galaxies in the Fornax cluster \citep{Iodice19}. In short, the Fornax 3D project is a magnitude-limited survey \citep[$m_B < 15$, ][]{Ferguson89} of galaxies within the virial radius of the cluster \citep{Drinkwater01} observed with the  MUSE integral-field spectrograph \citep{Bacon10}. The unique capabilities of MUSE power the ambitious goals of the Fornax 3D project, which range from characterizing compact, unresolved systems \citep{Fahrion20,Fahrion20b,Spriggs20}, to studying the orbital heating mechanisms \citep{Pinna19,Pinna19b,Adriano21} and the dust and gas properties within the cluster \citep{Viaene19,Zabel20}.

The survey was carried out using the MUSE Wide Field Mode (without adaptive optics), with a 1$\times$1 arcmin$^2$ field-of-view and a 0.2 arcsec pixel$^{-1}$ spatial scale. This setup provides a wavelength range covering fom 4650 \AA \ to 9300 \AA, with a spectral sampling of 1.25 \AA \ pixel$^{-1}$, at a resolution of FWHM$_{7000 \AA}$ = 2.5 \AA. More details on the properties of the Fornax 3D data are given in the survey presentation paper \citep{f3d}.

As explained above, IMF measurements usually rely on the assumption that the stars contributing to the observed spectra can be modeled by a SSP model and this assumption is (usually) only valid for old stellar populations. Thus, from the full sample of 33 galaxies observed by the Fornax 3D project, we focused on the 22 classified as ETGs. However, the central regions of one of FCC\,90 are dominated by a relatively recent star formation episode \citep{Iodice19} and therefore we did not include this object in our final sample. Conversely, two objects (FCC\,176 and FCC\,179) labeled as  late-type galaxies showed stellar populations old-enough to be consistent with the SSP assumption. With all this, our final sample consists of 23 objects whose most relevant properties for the purposes of this work are listed in Table~\ref{tab:1}. 

\begin{table}
   \caption{\label{tab:1} Main sample properties}
   \centering
   \begin{tabular}{cccccc}
   \hline\hline
   Object & Alt. & Type & Distance & M$_\star$ & N$_\mathrm{bins}$ \\
       & & &[Mpc] & [10$^{10}$\msun] & \\
     (1)  &  (2) & (3) & (4) & (5) & (6) \\
   \hline
   FCC\,083 & NGC\,1351 & E5  & 19.2   & 2.27  & 1475 \\
   FCC\,119 &   & S0  & 20.9   & 0.05  & 32 \\
   FCC\,143 & NGC\,1373 & E3  & 19.3   & 0.28  & 252 \\
   FCC\,147 & NGC\,1374 & E0  & 19.6   & 2.40  & 3515 \\
   FCC\,148 & NGC\,1375 & S0  & 19.9   & 0.58  & 625 \\
   FCC\,153 & IC\,1963 & S0  & 20.8   & 0.76  & 804 \\
   FCC\,161 & NGC\,1379 & E0  & 19.9   & 2.63  & 2069 \\
   FCC\,167 & NGC\,1380 & S0/a & 21.2 & 9.85  & 6154 \\
   FCC\,170 & NGC\,1381 & S0 & 21.9   & 2.25  & 2013 \\
   FCC\,176 & NGC\,1369 & SBa & 20.9  & 0.68  & 114 \\
   FCC\,177 &   & S0 & 20.0   & 0.85  & 799 \\
   FCC\,179 & NGC\,1386 & Sa  & 20.9   & 1.58  & 1804 \\
   FCC\,182 &   & SB0  & 19.6  & 0.15  & 71 \\
   FCC\,184 & NGC\,1387 & SB0  & 19.3  & 4.70  & 8862 \\
   FCC\,190 & NGC\,1382 & SB0  & 20.3  & 0.54  & 900 \\
   FCC\,193 & NGC\,1389 & SB0  & 21.2  & 3.32  & 2804 \\
   FCC\,219 & NGC\,1404 & E2  & 20.2   & 12.7  & 5997 \\
   FCC\,249 & NGC\,1419 & E0  & 20.9   & 0.98  & 654 \\
   FCC\,255 &   & S0  & 20.9   & 0.31  & 324 \\
   FCC\,276 & NGC\,1427 & E4  & 19.6   & 1.81  & 2748 \\
   FCC\,277 & NGC\,1428 & E5  & 20.7   & 0.34  & 478 \\
   FCC\,301 &   & E4  & 19.7   & 0.20  & 238 \\
   FCC\,310 & NGC\,1460 & SB0 & 20.9  & 0.54  & 607 \\
   \hline
   \end{tabular}
   \tablefoot{(1), (2) and (3) come from \citet{Ferguson89}. (4) When available, distances come from \citet{Enrica18}. Otherwise, a fixed distance from the Fornax cluster of 20.9 Mpc is assumed. (5) Stellar masses as listed in \citet{Iodice19} (6) Number of independent Voronoi bins used in our analysis (see \S~\ref{sec:analysis}). \\
   }
   \end{table}

\section{Analysis} \label{sec:analysis}

The analysis of the Fornax 3D data was done following the procedure described in \citet{MN19} and consists of three basic steps described below. 

\subsection{Binning and photometric measurements}

In order to robustly measure IMF variations, we spatially binned the MUSE data imposing a minimum signal-to-noise ratio (SN) of 100. In practice this was done using the Voronoi tessellation routine of \citet{voronoi}. We did not include in the binning process those spaxels with a SN below 5. Given the depth of the Fornax 3D data, these SN constraints translated into a large number of independent data points per galaxy, as listed in Table~\ref{tab:1}, column (6). Even after this rather conservative approach, FCC\,083, FCC\,147 and FCC\,161 exhibited systematic offsets in their stellar population properties between different pointings \citep[see][]{f3d}. Thus, for these objects only the central pointing was finally included in the analysis. We then used the MUSE cubes to calculate the average surface brightness within each individual Voronoi bin. Assuming the distances detailed in Table~\ref{tab:1}, surface brightnesses were finally translated into average luminosities per bin. 

\begin{figure*}
   \centering
   \includegraphics[width=13cm]{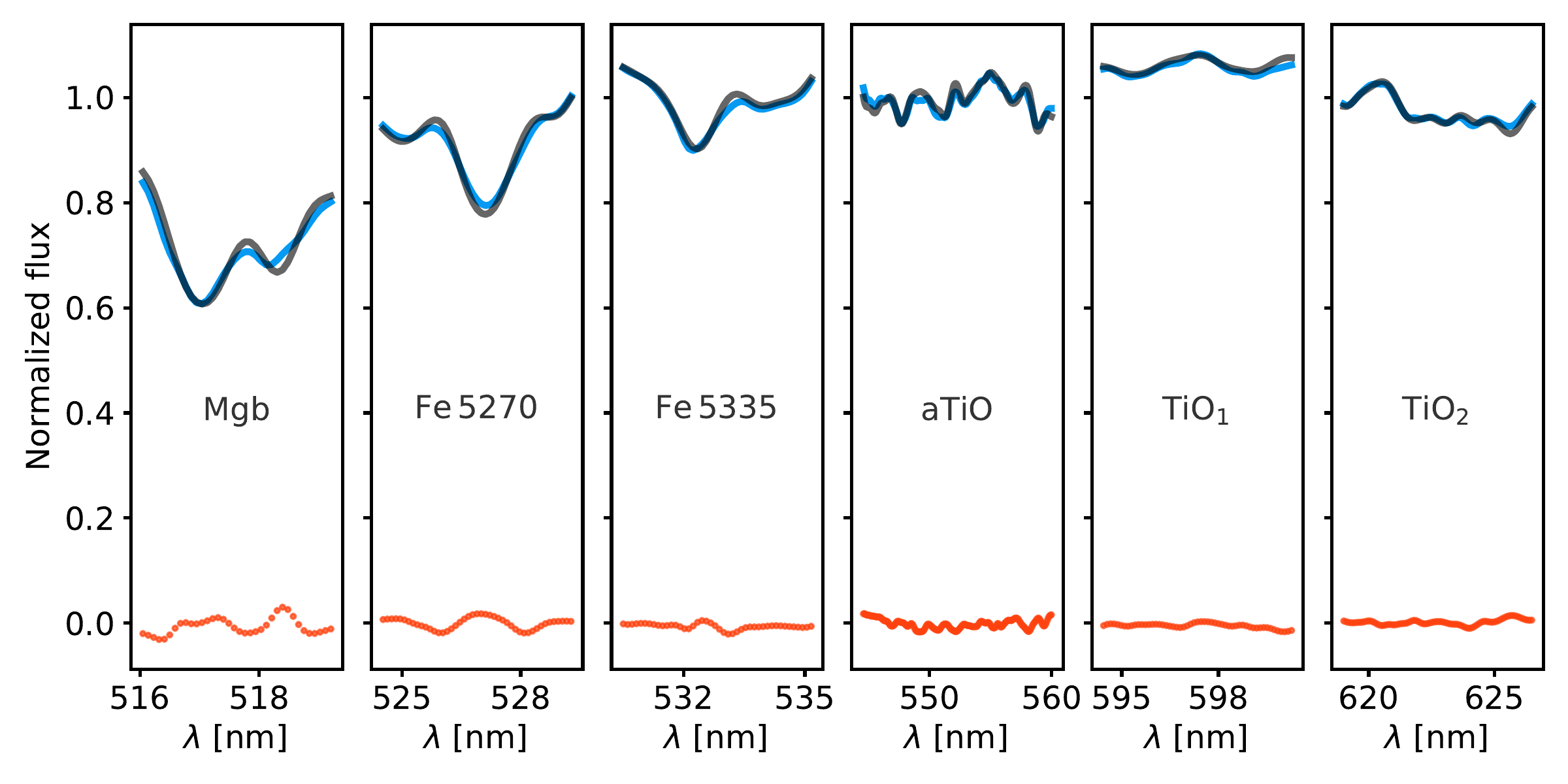}
   \caption{Full index fitting of the F3D data. The blue lines show the six (continuum-corrected) spectral features used in our analysis, for a typical F3D spectrum. The best-fitting MILES model (black solid line) was calculated by fitting the flux of each of these pixels, following Eq.~\ref{eq:fit}. Red dots at the bottom indicate the difference between observed and predicted fluxes. Each panel in this figure covers a different wavelength span.}
   \label{fig:fit}
\end{figure*}

\subsection{Kinematics and mean ages}

The kinematical properties (V and $\sigma$) of each bin were measured using pPXF \citep{ppxf,Cappellari17}. No higher moments of the velocity distribution were measured as they have a subtle effect on the stellar population estimates \citep[e.g.,][]{Kuntschner04}. We fed pPXF with the MILES stellar population synthesis models \citep[][]{miles}, based on the stellar library of \citet{Pat06}, and at a constant spectral resolution of 2.51 \AA \ \citep[FWHM, ][]{Jesus11}. Since we did not correct for the line spread function of the instrument, the measured $\sigma$ accounts for both the intrinsic velocity dispersion of the stars and the instrumental resolution. 

As discussed in \citet{MN19}, we adapted our stellar population analysis to the properties of the MUSE data. In particular, H$_\beta$ is the most reliable age sensitive feature in the observed wavelength range. However, the strength of the H$_\beta$ feature also depends on the [C/Fe] abundance ratio \citep{conroy,LB16} but no other C-sensitive spectral features are covered by MUSE to actually measure [C/Fe]. Therefore, since could not break the age-[C/Fe] degeneracy affecting the H$_\beta$ feature, we did not only use pPXF to measure the kinematics of each bin, but also to estimate the (luminosity-weighted) mean ages. In practice, we proceeded as follows.

First, we run pPXF with an un-regularized combination of MILES SSPs in order to measure V and $\sigma$. Second, we fixed the kinematics to that measured in the first step \citep[to avoid degeneracies between $\sigma$ and stellar population properties, see e.g. ][]{Pat11} and we run pPXF again, this time regularizing the age-metallicity-IMF slope plane. The imposed regularized solution ensures that the SSP weight distribution is as smooth as allowed by the data \cite[see e.g. Section 3.5 in][]{Cappellari17}. The mean luminosity-weighted age of the spectra was measured using this regularized second pPXF run. We note that we intentionally included templates with a variable IMF to account for the possible effect of the IMF in the age determination \citep{FM13,LB16}. We did not fit for [$\alpha$/Fe] as this was done later in the analysis (see \S~\ref{sec:stelpop}). Instead, we used the so-called base MILES models which inherit the [$\alpha$/Fe]--[M/H] relation of the solar neighborhood \citep{Vazdekis15}. This choice avoids nonlocal equilibrium uncertainties introduced by the theoretical response functions needed to compute stellar population models with variable abundance ratios, uncertainties that can be particularly problematic for Balmer lines \citep{Martins07,Coelho14} and thus for the age determination. To be consistent with our stellar population analysis (see details on the section below), all pPXF fits were limited to a wavelength range between 4800 \AA \ and 6400 \AA.

In summary, after measuring the kinematics, we estimated the luminosity-weighted age of each spectrum by using the best-linear combination of SSP models in the age-metallicity-IMF slope plane. This also allowed us to further correct for subtle flux calibration issues, assuming that the linear combination of templated provided by pPXF is a reliable representation of the emitted spectra. Furthermore, we also used pPXF to mask out any pixel contaminated by either telluric features or gas emission, although the latter is expected to be minimal since our sample was selected to avoid contribution from young stars.

\subsection{Stellar populations} \label{sec:stelpop}

\subsubsection{Model ingredients}
We based our stellar population analysis on the variable [$\alpha$/Fe] MILES models presented in \citet{Vazdekis15}. These models use the BaSTI set of isochrones \citep{basti1,basti2}, calculated at [$\alpha$/Fe]=0.4 and at the solar scale ([$\alpha$/Fe] = 0.0). The adopted models cover a range in ages from 0.03 to 14 Gyr and from $-2.27$ to $+0.26$ dex in total metallicity [M/H]. We note that, although models at higher metallicities are available, we did not include them in the analysis as the scarcity of solar neighborhood stars with such high [M/H] values makes these models more prone to systematics. This is a safe assumption given the observed metallicity range in our sample (see for example Fig.~\ref{fig:met}). Since our IMF inference relies on measuring the strength of titanium-sensitive features (see details below), we used the [Ti/Fe] response functions of \citet{conroy} to account for possible variations in the titanium abundance across our sample. 

For the IMF, we assumed a broken power-law parametrization. Under this functional form, changes in the IMF are controlled by varying the high-mass end slope $\Gamma_\mathrm{b}$. However, in \citet{MN19} we introduced a new parameter $\xi$ defined as 

\begin{equation*}
   \xi \equiv \frac{\int_{m=0.2}^{m=0.5} \Phi(\log m) \ dm}{\int_{m=0.2}^{m=1} \Phi(\log m) \ dm} = 
   \frac{\int_{m=0.2}^{m=0.5} m \cdot X(m) \ dm}{\int_{m=0.2}^{m=1} m \cdot X(m) \ dm} ,
   \end{equation*}

\noindent
In essence, the $\xi$ parameter describes the slope of the IMF (in terms of a mass fraction) in the stellar mass range relevant for the analysis of old stellar populations as measured from optical features  ($\sim 0.2$--1 \msun). Using the equation above, the quantity $\xi$ can be calculated for any given IMF parametrization and, in particular, for the broken power-law IMF implemented in the MILES models. Hence, throughout this paper, IMF variations will be always discussed in terms of $\xi$, whose properties are extensively detailed in \citet{MN19}. The value of the $\xi$ parameter becomes larger for steeper IMF slopes, that is, for populations with an enhanced fraction of low-mass stars. For reference, it takes a value of $\xi=0.5194$ for a \citet{Kroupa} IMF, $\xi=0.4607$ for a \citet{Chabrier} one, and $\xi=0.6370$ for that originally measured by \citet{Salp:55}.

\subsubsection{Bayesian full index fitting}

\begin{figure*}
   \centering
   \includegraphics[width=9cm]{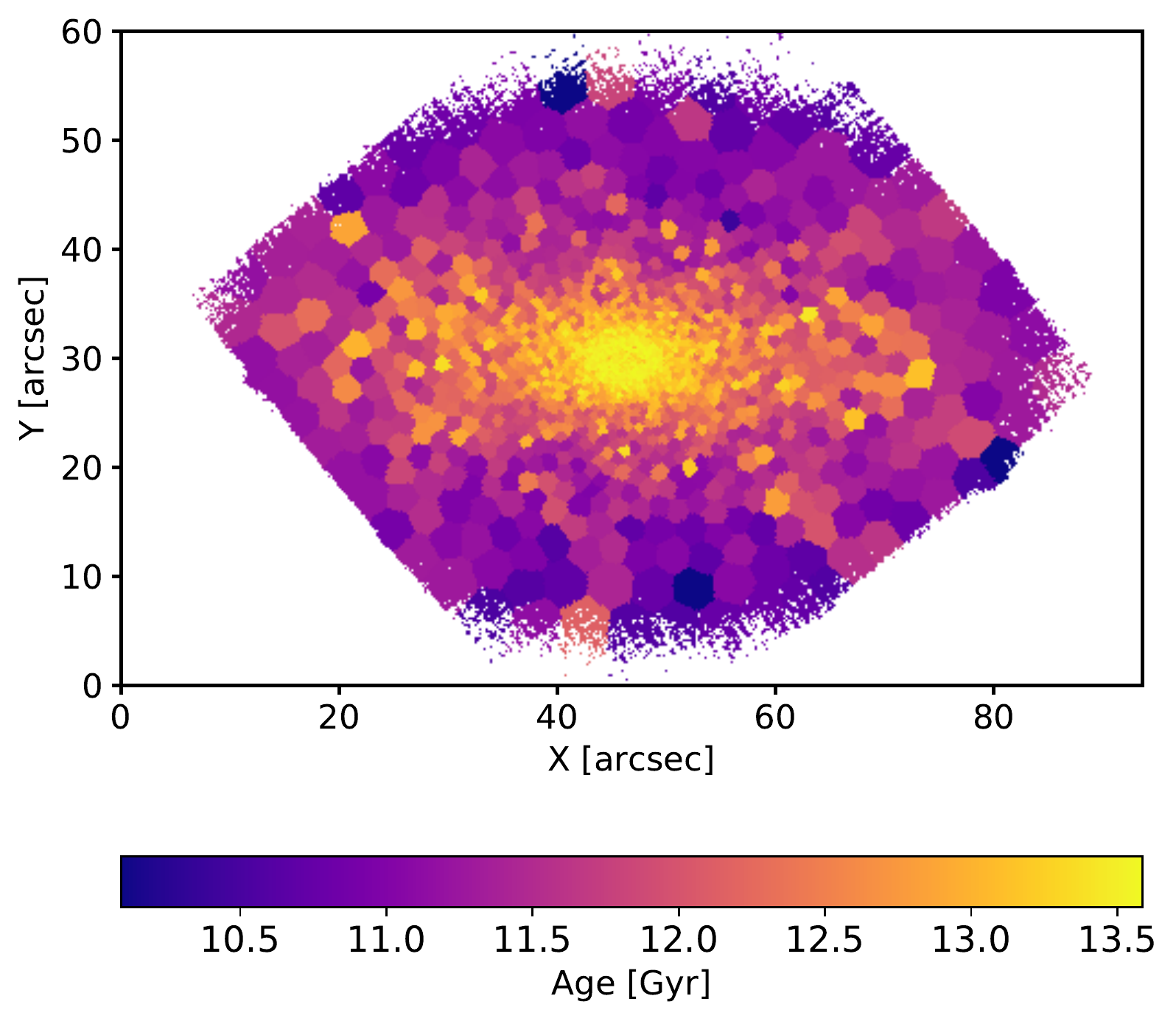}
   \includegraphics[width=9cm]{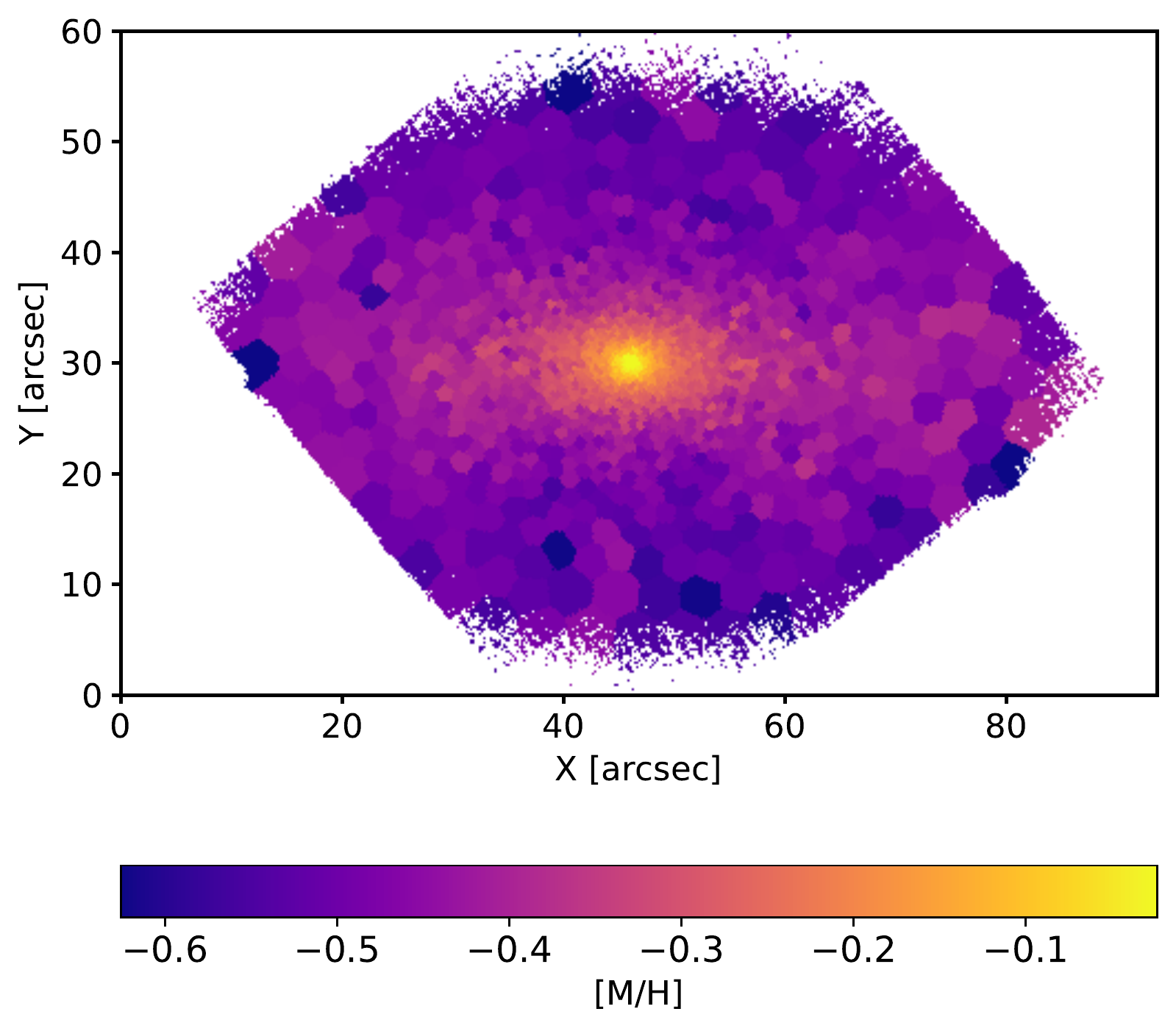}
   \includegraphics[width=9cm]{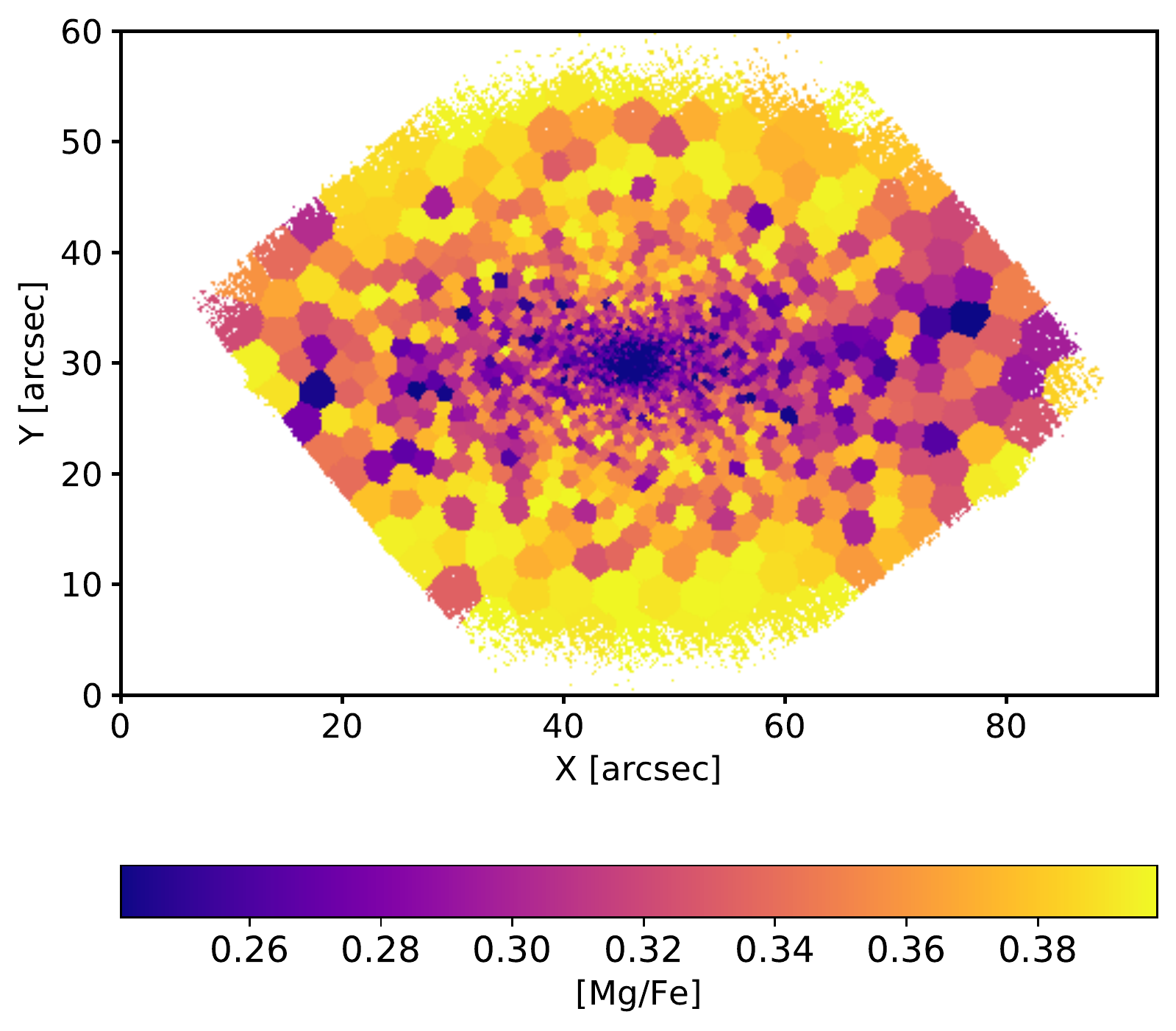}
   \includegraphics[width=9cm]{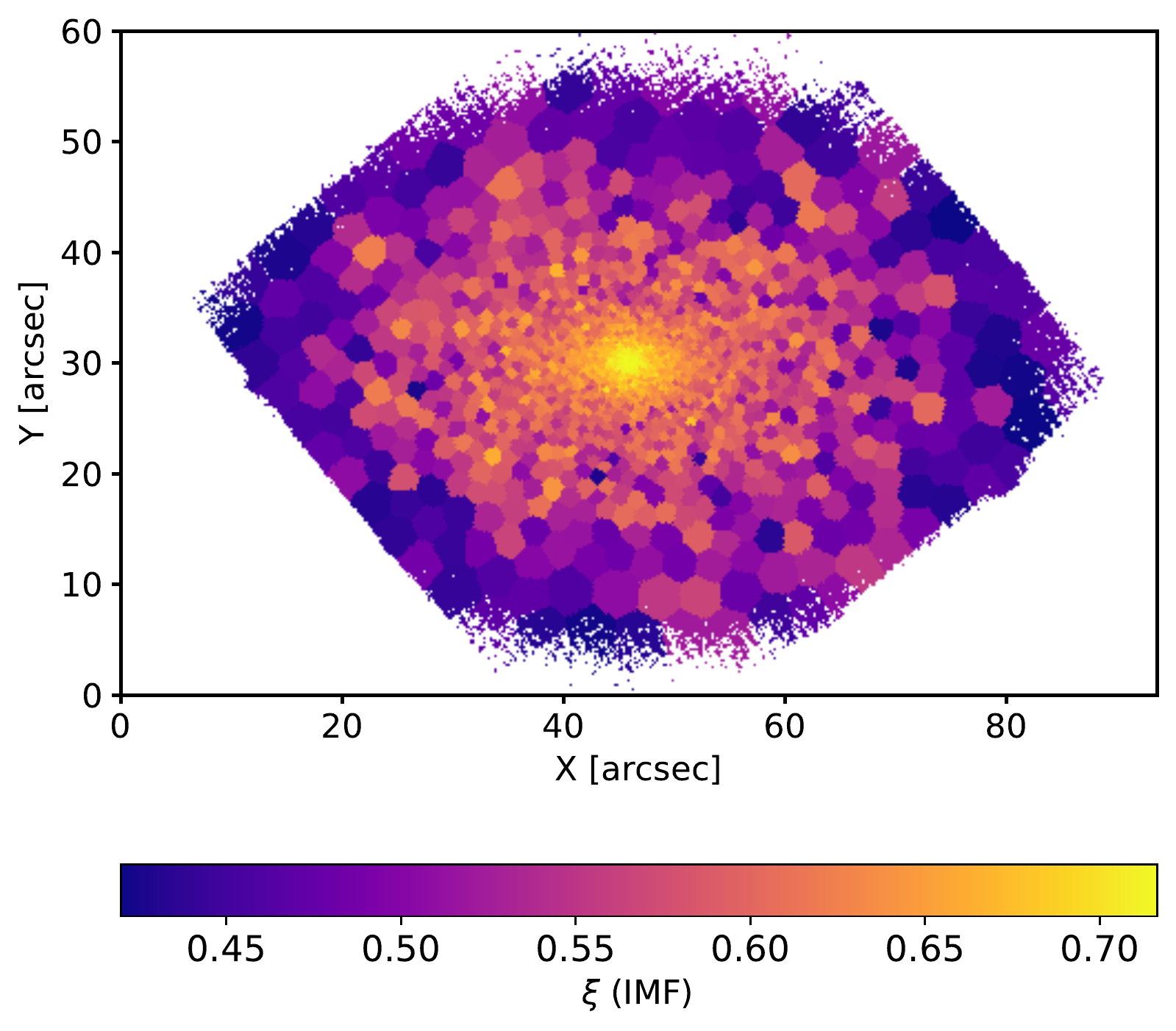}
   \caption{F3D stellar population maps. From left to right and top to bottom we show the age, metallicity, [Mg/Fe], and IMF slope maps of FCC\,083, a typical ETGs in our sample. Age and metallicity exhibit clear negative gradients, while [Mg/Fe] tends to increase with radius. There is also a clear gradient in the IMF map, becoming steeper in the central regions.}
   \label{fig:maps}
\end{figure*}

In \citet{MN19} we presented a hybrid approach to measure stellar population properties and, in particular, the slope of the IMF from integrated spectra. This approach combines the idea of the standard line-strength analysis, focusing on those wavelengths where the information about the stellar population properties is concentrated \citep[e.g.,][]{Worthey94,cat,TMB:03,Schiavon07}, with the robustness resulting from full spectral fitting techniques , where every pixel in a spectrum is compared to a model prediction \citep[e.g.,][]{CF05,Ocvirk06,Conroy09,Cappellari17,Wilkinson17}. Details on the fitting process are given in Section~4 of \citet{MN19}

In short, our approach fits every pixel within the band-pass definition of a given spectral feature, after removing the shape of the continuum by linearly fitting the slope of the spectra across the blue and red pseudo-continua. For this work, as we did in \citet{MN19}, we selected six spectral indices (Fe\,5270, Fe\,5335, Mgb5177, aTiO, TiO$_1$, and TiO$_2$), following their standard definitions \citep{trager,serven}. Then, we used the emcee Bayesian Markov chain Monte Carlo sampler \citep{emcee}, to maximize the following likelihood function

\begin{equation}\label{eq:fit}
 \ln ({\bf O} \,  | \, {\bf S} ) = -\frac{1}{2}  \mathlarger{\sum}_n \bigg[ \frac{(\mathrm{O}_n - \mathrm{M}_n)^2}{\sigma_n^2}-\ln \frac{1}{\sigma_n^2}\bigg] ,
\end{equation}

\noindent The summation extends over all the pixels within the band-passes of the six spectral features above. O$_n$ and M$_n$ are the observed and the model flux of the $n$th-pixel and $\sigma_n$ the measured uncertainty (the average residuals of the pPXF fit). MILES model predictions M$_n$ were calculated assuming the luminosity-weighted ages derived from pPXF. In addition, for each spectrum, models were convolved with a Gaussian kernel to match the stellar velocity dispersion. In practice, by minimizing Eq.~\ref{eq:fit} we found the best-fitting SSP model characterizing the observed spectra. This model is defined by five stellar population properties: mean age, metallicity [M/H], IMF slope $\xi$, [Mg/Fe], and [Ti/Fe] abundance ratios. While the age was fixed to that measured in the previous step, the other four parameters are simultaneously measured using Eq.~\ref{eq:fit}. An example of a F3D spectra and its best-fitting model is shown in Fig.~\ref{fig:fit} and a corner plot showing the expected degeneracies is shown in Figure~2 of \citet{MN19}.

It is worth noting as well that, given our choice of indices, we are only sensitive to one $\alpha$ element, namely magnesium, through the Mgb5177 spectral feature. Thus, although in principle all $\alpha$ elements are varied in the MILES models, effectively, with our approach we are only measuring [Mg/Fe]. This is a subtle yet important remark  since different $\alpha$ elements do not necessarily track each other \citep{Cenarro04,worthey14,Parikh21}. Therefore, the conclusions derived from our analysis only concern the [Mg/Fe] abundance ratio and should not be assumed to apply to all $\alpha$ elements equally.
 
\section{Results} \label{sec:resu}

\begin{figure*}
   \centering
   \includegraphics[width=15.5cm]{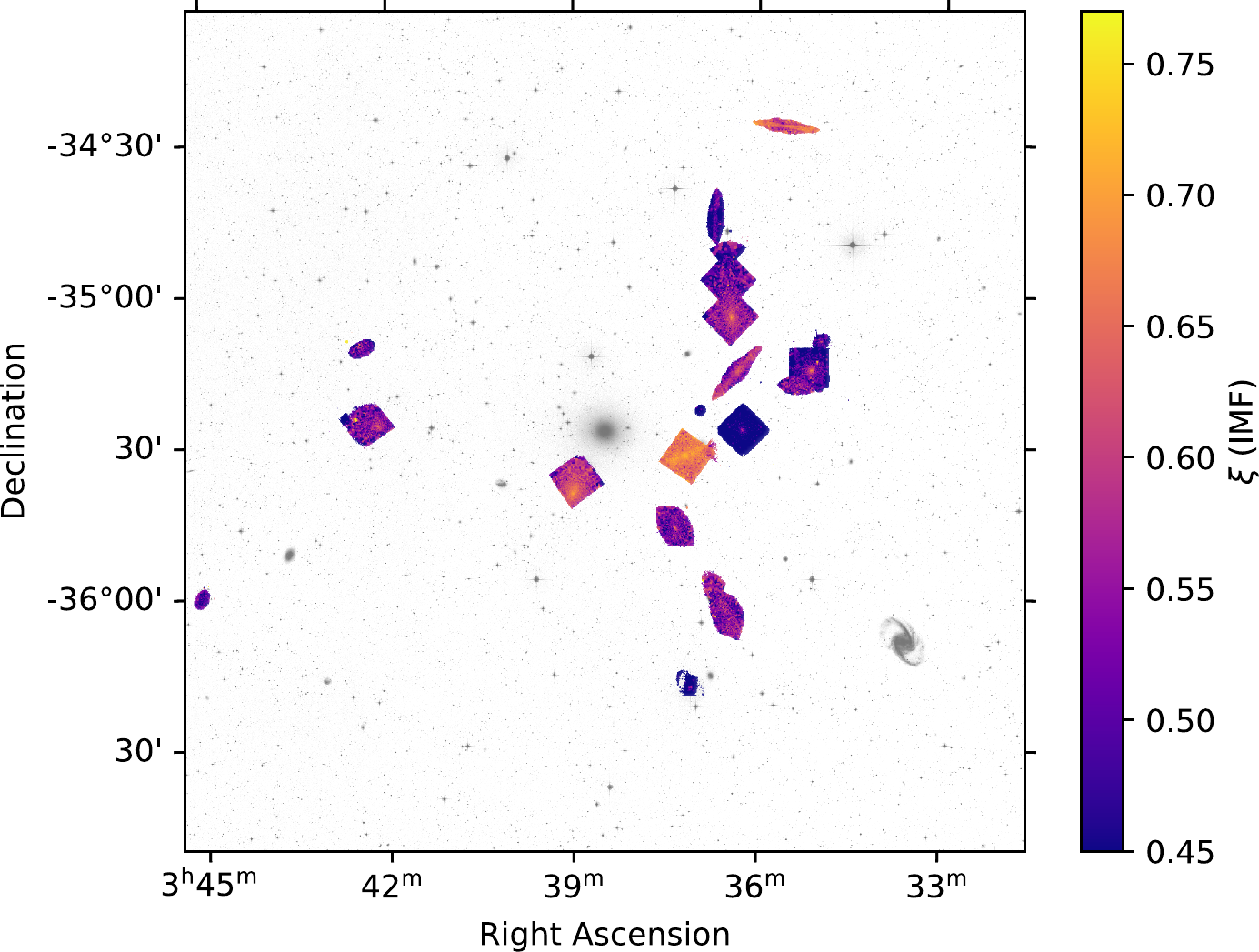}
   \includegraphics[width=15.5cm]{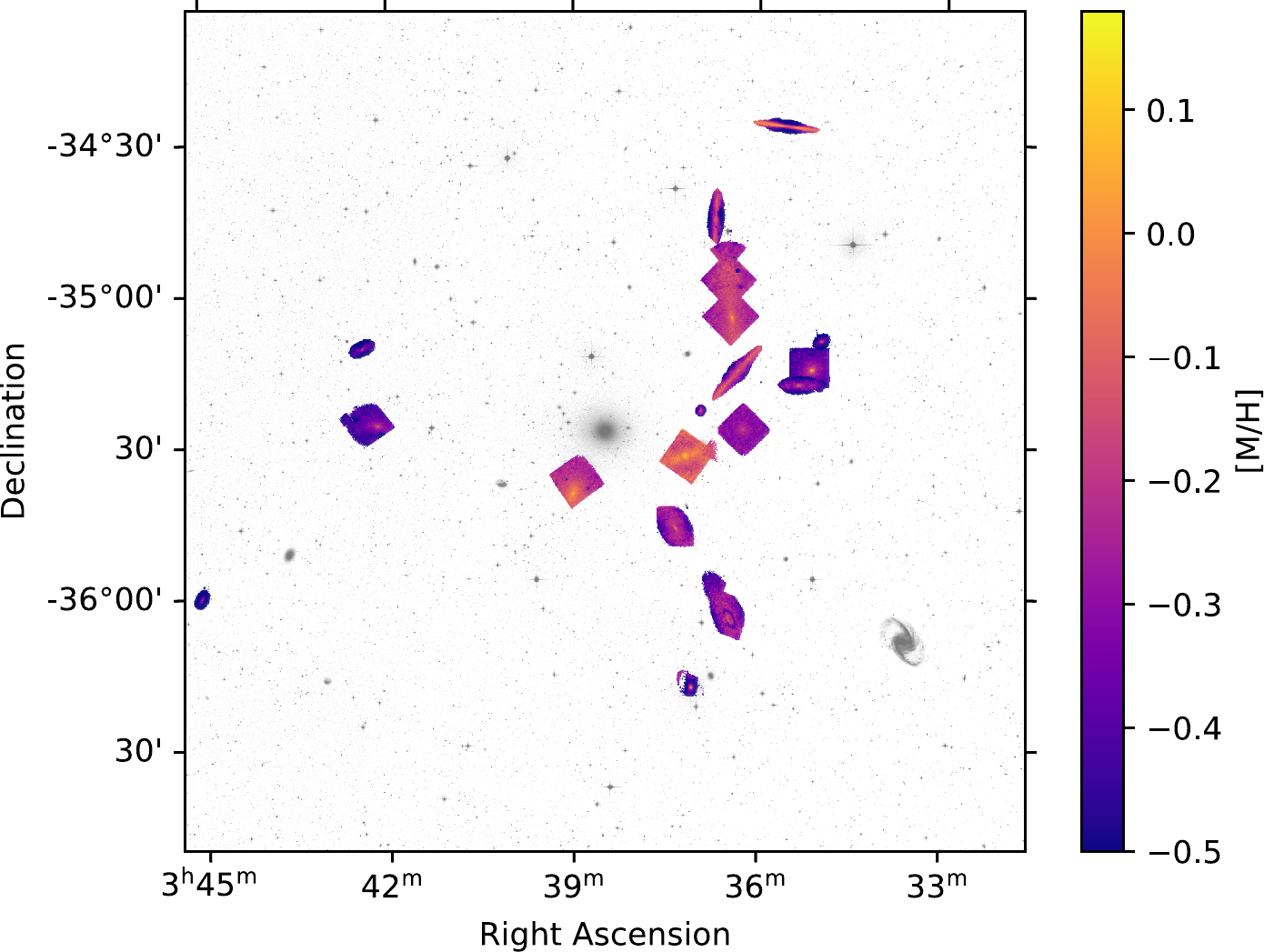}
   \caption{F3D stellar population maps projected in the sky. The IMF slope (top) and metallicity (bottom) maps of F3D galaxies near the core of the Fornax cluster. The size of each map has been amplified to reveal their internal structure. As expected, IMF and metallicity maps are rather similar, although noticeable differences emerge at low metallicities (see \S~\ref{sec:vary}). A larger version of all the derived stellar population maps can be found in Appendix~\ref{sec:maps}.}
   \label{fig:cluster_imf}
\end{figure*}

The stellar population analysis described above was designed and optimized to measure, for each F3D spectrum, four main quantities: its (luminosity-weighted) mean age, metallicity, IMF slope, and [Mg/Fe] abundance ratio. To exemplify the richness of the F3D MUSE data, Fig.~\ref{fig:maps} showcases the full two-dimensional maps of these four parameters for one of the galaxies in our sample, FCC\,083.

The upper panels in Fig.~\ref{fig:maps} show the 2D age (left) and metallicity (right) structure of FCC\,083. These two panels exhibit a rather typical behavior of ETGs, with old stellar populations that tend to be more metal-rich in the center. The metallicity distribution is particularly elongated and resembles the chemically-enriched disk structures shown by other galaxies in our sample \citep{Pinna19,Pinna19b,MN19}. Interestingly, this chemically evolved, disk-like structure is also reflected in the [Mg/Fe] map, which shows a clearly positive [Mg/Fe] gradient. Positive [Mg/Fe] gradients, very rare among relatively massive ETGs \citep{Greene15,MN18}, are ubiquitous across our sample. Finally, the bottom right panel of Fig.~\ref{fig:maps} shows the IMF map of FCC\,083. As expected, the enhanced fraction of low-mass stars (high $\xi$  values) is limited to the central (metal-rich) regions, reaching Milky Way-like values ($\xi \approx 0.5$) in the outer parts. The shape of the IMF map seems to be, however, less elongated than the metallicity and the [Mg/Fe] maps, a behavior similar to that reported in \citet{MN19} when comparing the iso-metallicity and iso-IMF contours of FCC\,167, also part of the F3D sample.

\subsection{Common features across the stellar population maps}

To provide an indication for the behavior of the IMF variation in the Fornax cluster, Fig.~\ref{fig:cluster_imf} shows the IMF (top) and metallicity (bottom) maps of our sample, as they would appear projected across the cluster. The spatial scale of the maps has been amplified to reveal the internal substructure of each individual map. It becomes obvious from this figure that local IMF and metallicity measurements behave very similarly although, as we discuss in the following subsection, with some scatter, in particular in the low-metallicity regime. All stellar population maps are shown in more detail in Appendix~\ref{sec:maps}.

When looking more in detail at the stellar population maps of individual galaxies, some features of their internal substructure are shared among different galaxies. For example, elliptical galaxies (see Table~\ref{tab:1}) show rather simple and symmetric maps with a clear coupling between metallicity and IMF slope gradients as their central regions tend to host relatively metal-rich stellar populations with an enhanced fraction of low-mass stars (bottom-heavier IMF). However, IMF slope maps do not exactly follow metallicity maps, as they tend to exhibit less elongated shapes (see for instance Fig.~\ref{fig:maps}). These differences in the internal structure of IMF slope and metallicity maps are quantified and further discussed in the following subsection. 

Moreover, the presence of a bar is imprinted in the stellar population maps of some of our galaxies. For example, FCC\,184 (Fig.~\ref{fig:fcc184}) has a photometric face-on bar that appears as an elongated high-metallicity, steep IMF slope, and low [Mg/Fe] structure in the maps. Such correspondence between photometric bar and a metal-rich and low [Mg/Fe] structure is in line with the recent results of \citet{Justus2020}, and we now show that stellar populations within bars also exhibit distinct IMF slope values, pointing toward a connection between the orbital structures and stellar population properties in galaxies. This is further exemplified by FCC\,170, where the central X-shaped component reveals the presence of an edge-on bar, which is also visible in the metallicity and IMF slope maps. 

In addition to FCC\,170, the two other edge-on S0's in our sample (FCC\,177 and FCC\,153) also exhibit a distinct behavior. A metal-rich and low [Mg/Fe] thin disk is visible in all three galaxies, characterized by relatively steep IMF slope values. Interestingly, IMF slope values within the thin disk of FCC\,153 decrease with increasing metallicity (Fig.~\ref{fig:fcc153}). This is surprising given that, in general, IMF slope values tend to increase with metallicity. However, in the case of the thin disk of FCC\,153, the anticorrelation between IMF slope and metallicity is modulated by a change in age and [Mg/Fe], demonstrating that metallicity alone does not drive IMF variations.

\subsection{Characterizing IMF variations}  \label{sec:vary}

\begin{figure}
   \centering
   \includegraphics[width=9cm]{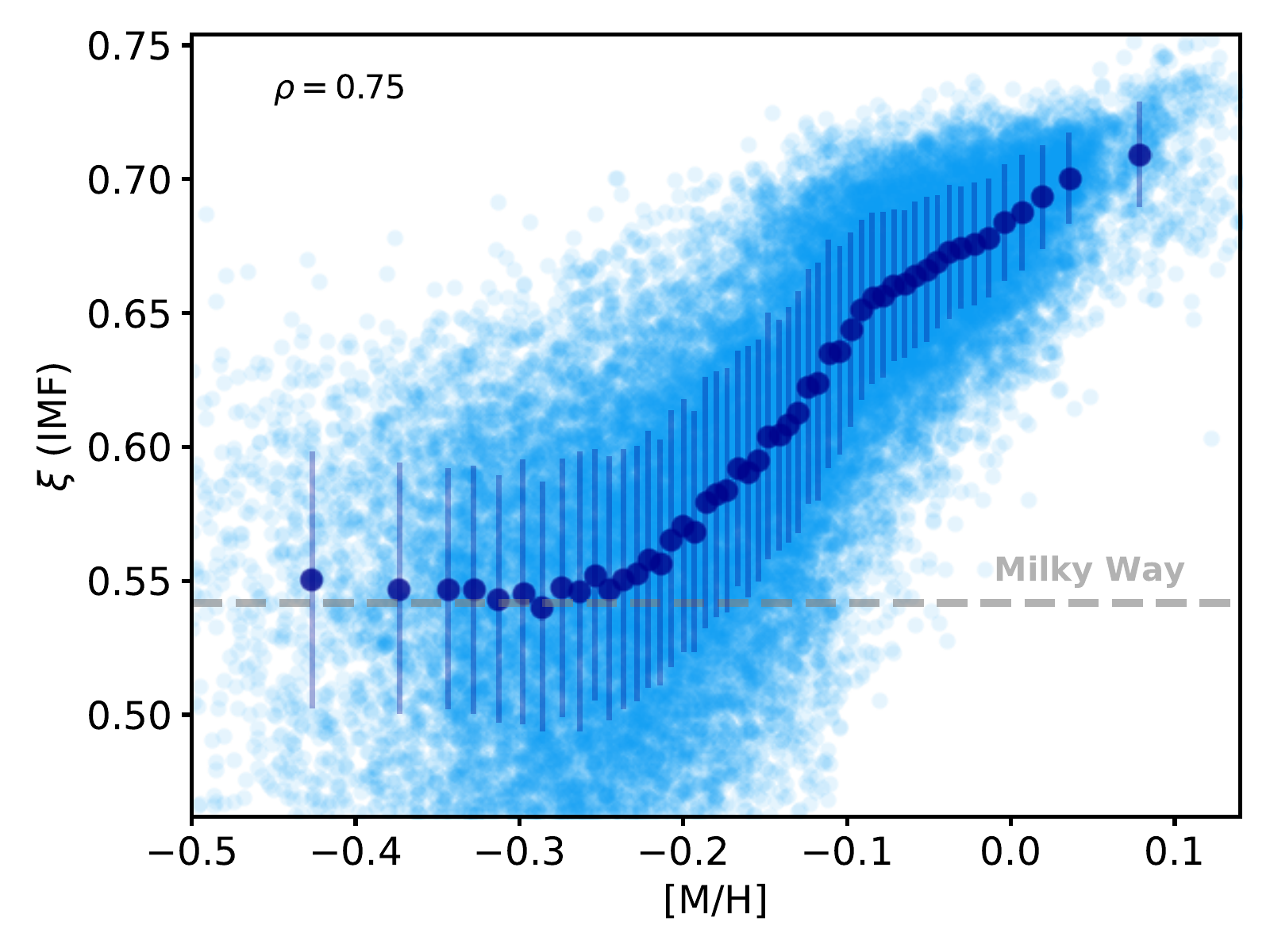}
   \caption{IMF slope -- metallicity relation. All independent IMF slope and metallicity measurements within our sample are shown in light blue, while dark blue symbols indicate the running median. Error bars indicate the (bi-weight) standard deviation. As in previous studies, we find a significant correlation between the local slope of the IMF and the metallicity, as indicated by the Pearson correlation coefficient on the top left corner of the figure. The $\xi$ value for a Milky Way-like IMF is marked with the dashed horizontal line.}
   \label{fig:met}
\end{figure}

First revealed using IFU data from the CALIFA survey \citep{MN15c}, the apparent relation between metallicity and the low-mass end slope of the IMF is becoming a well-established observational result, at least, for measurements based on absorption features \citep{vdk17,Parikh,Sarzi18,Zhou19}. However, it has also become evident that the complex behavior exhibited by the IMF slope, in particular when measured from integrated spectra, is not fully explained by changes in the local metallicity \cite[e.g.,][]{Alexa17,LB19} (and  see also e.g. \citealt{Davis17}). 

Fig.~\ref{fig:met} shows the IMF slope-metallicity relation that results from analyzing our full sample of galaxies. Each light blue point in this figure corresponds to one independent Voronoi bin from one of the 23 galaxies listed in Table~\ref{tab:1}. Dark blue symbols indicate the running median of the distribution and the horizontal dashed line indicates the expected $\xi$ value for the Milky Way. As already hinted by Fig.~\ref{fig:cluster_imf}, changes in the IMF within the Fornax cluster do follow metallicity variations, with more metal-rich bins showing steeper IMF slope values, that is, an enhanced fraction of low-mass stars. For lower metallicities the IMF slope tends to approach the Milky Way value.

\begin{figure}
   \centering
   \includegraphics[width=9cm]{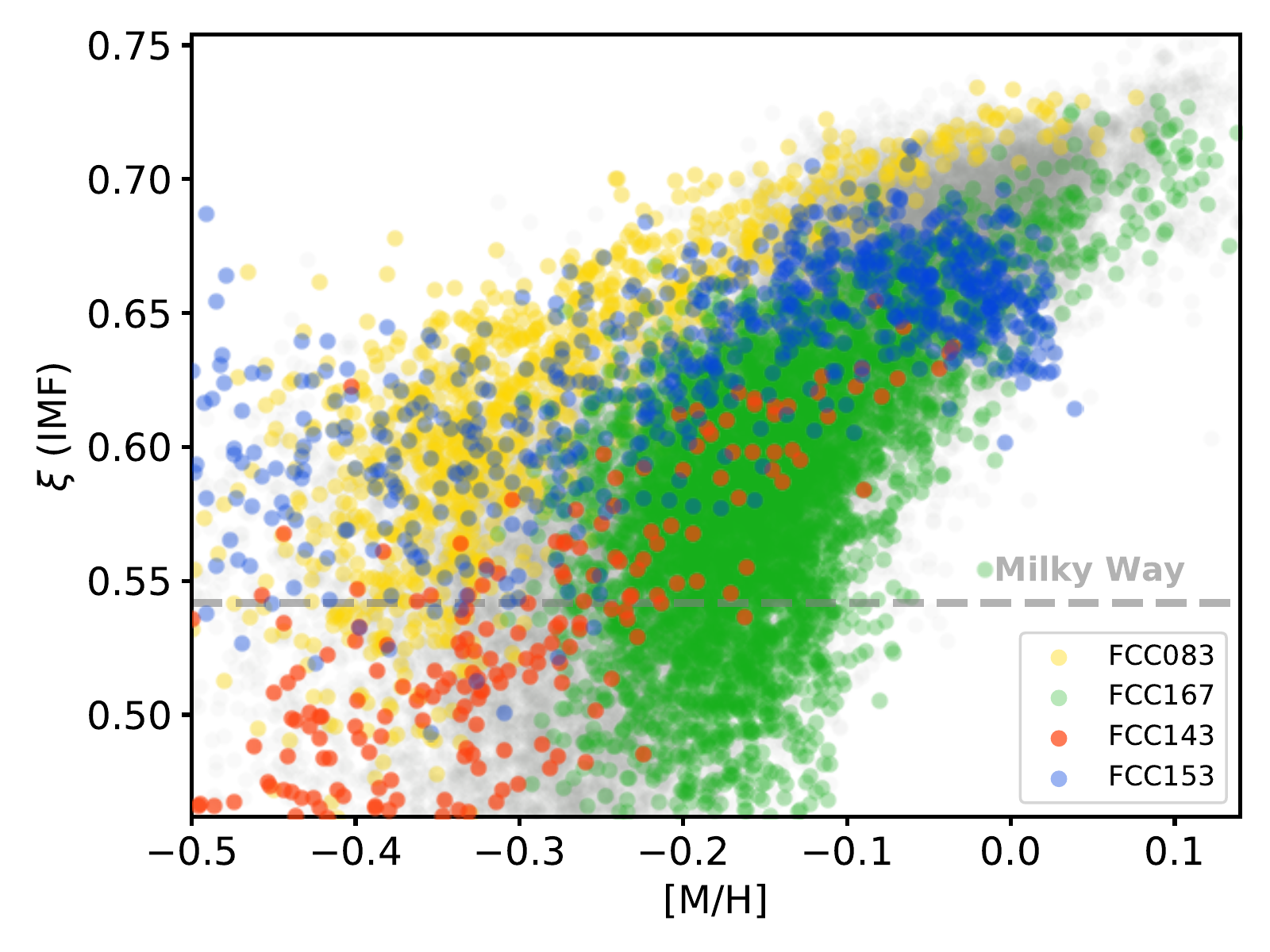}
   \caption{Individual galaxies in the IMF slope -- metallicity plane. Yellow, green, blue, and red symbols correspond to four individual F3D galaxies, namely, FCC\,083, FCC\,167, FCC\,153, and FCC\,143, respectively, showcasing the variety of behaviors observed in our sample. Although each galaxy follows a rather well-defined IMF slope metallicity relation, there are clear differences among them, revealing that the observed scatter is due to internal variation and to changes from galaxy to galaxy. As in Fig.~\ref{fig:met}, the $\xi$ value for a Milky Way-like IMF is marked with the dashed horizontal line.}
   \label{fig:indi_met}
\end{figure}

Complementary, Fig.~\ref{fig:indi_met} shows the behavior of four individual galaxies in the IMF slope - metallicity plane. Yellow points in this figure correspond to the massive E5 galaxy FCC\,083, that follows a tight IMF slope-metallicity relation and populates the upper envelope of the distribution. In green, measurements from FCC\,167 indicate that its stellar populations have similar metallicities as those in FCC\,083 but with lower IMF slope values. For metallicities below [M/H]$\lesssim-0.1$, changes in the IMF slope of FCC\,167 only weakly correlate with changes in metallicity. Thus, the scatter in Figs.~\ref{fig:met} and~\ref{fig:indi_met} is driven by both galaxy-to-galaxy and internal variations. Two more galaxies are highlighted in Fig.~\ref{fig:indi_met}. In blue, IMF slope values in FCC\,153 tend to increase with increasing metallicity, but this relation changes at high metallicities. As noted above, this happens in the thin disk of this galaxy and it is related to the presence of younger stellar populations. Finally, red points in Fig.~\ref{fig:indi_met} show the behavior of FCC\,143, a relatively low-mass galaxy that also follows a clear IMF slope-metallicity relation, in particular those regions with higher metallicities. This is a common feature of low-mass galaxies in our sample.

Metallicity is, however, not the only interesting quantity derived from our stellar population analysis that could be potentially coupled to the IMF variations. As detailed above, for each spatial bin we also measured its age and [Mg/Fe] abundance ratio. Moreover, given a best-fitting stellar population model, an expected $M/L$ ratio can also be estimated for each bin by linearly interpolating the discrete predictions from the MILES models. In our case, we naturally take into account the possible effect of the IMF in the derived $M/L$, which is usually one of the main sources of systematic uncertainties when deriving stellar masses from spectro-photometric measurements \citep[e.g.,][]{Mitchell13}. With a proper $M/L$ estimation for each bin and its luminosity as measured from the MUSE data cubes (see \S~\ref{sec:analysis}), we derived the stellar surface mass density\footnote{Both luminosities and $M/L$s were calculated in the $r$-band.}. 

\begin{figure}
   \centering
   \includegraphics[width=9cm]{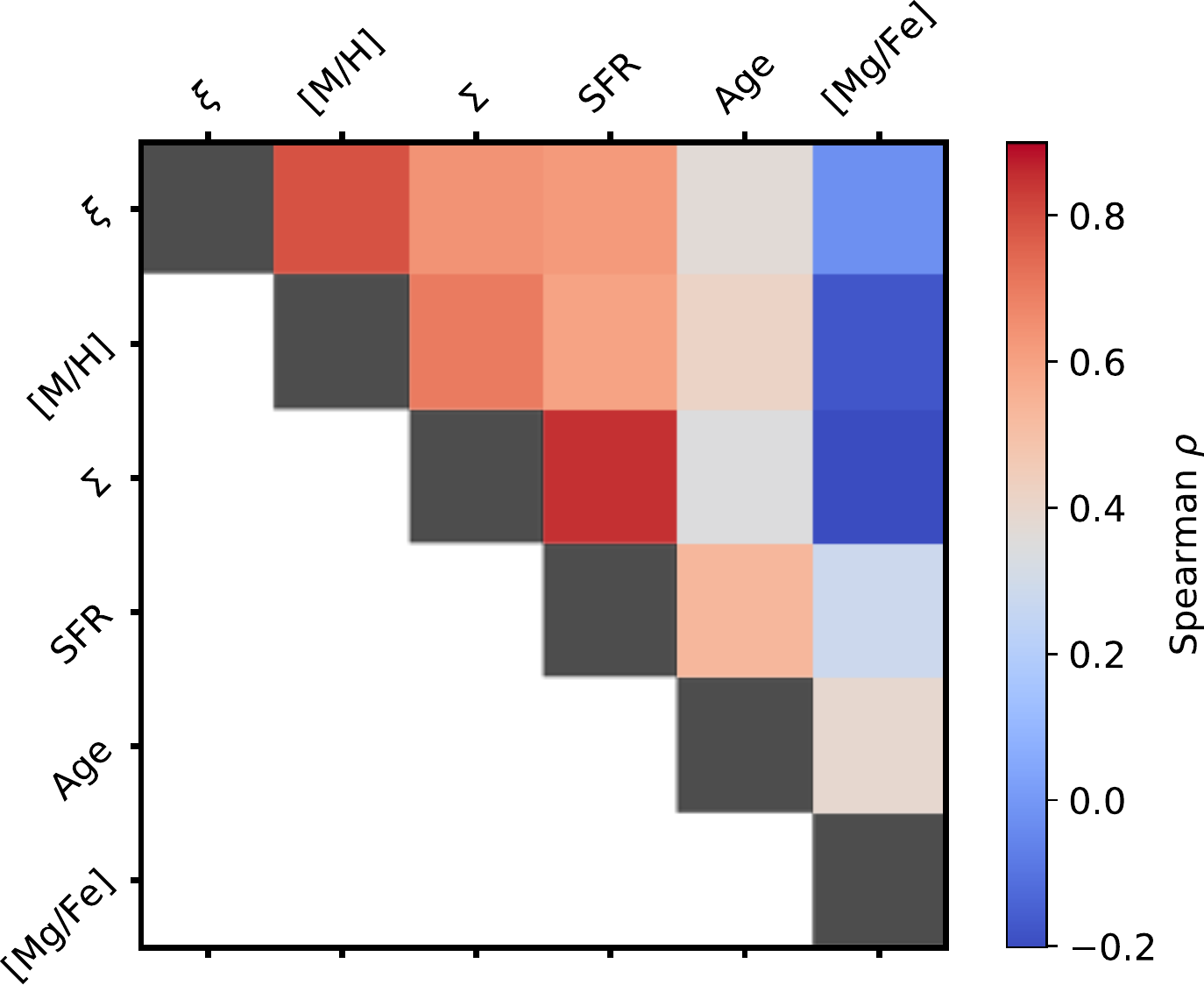}
   \caption{ Spearman correlation matrix. Each element in the matrix is color-coded by the value of the Spearman rank correlation coefficient between the two corresponding quantities. Metallicity, surface stellar density $\Sigma$, and star formation rate are the parameters more strongly correlated with the IMF slope $\xi$ variations. Conversely, [Mg/Fe] shows a significant anticorrelation with these three quantities due to the ubiquity of positive [Mg/Fe] gradients in our sample.} 
   \label{fig:corre}
\end{figure}

Among all the quantities measured on the F3D data, metallicity is the one that shows a stronger correlation with the IMF slope, with a Spearman correlation coefficient of $\rho=0.75$. Local stellar surface density also exhibits a noticeable yet weaker correlation ($\rho=0.64$). The strength of the correlation between all these derived quantities is summarized in Fig.~\ref{fig:corre}, which shows the Spearman correlation matrix for the IMF slope, metallicity, surface stellar density, surface star formation rate, age, and [Mg/Fe].

\begin{figure*}
   \centering
   \includegraphics[width=9cm]{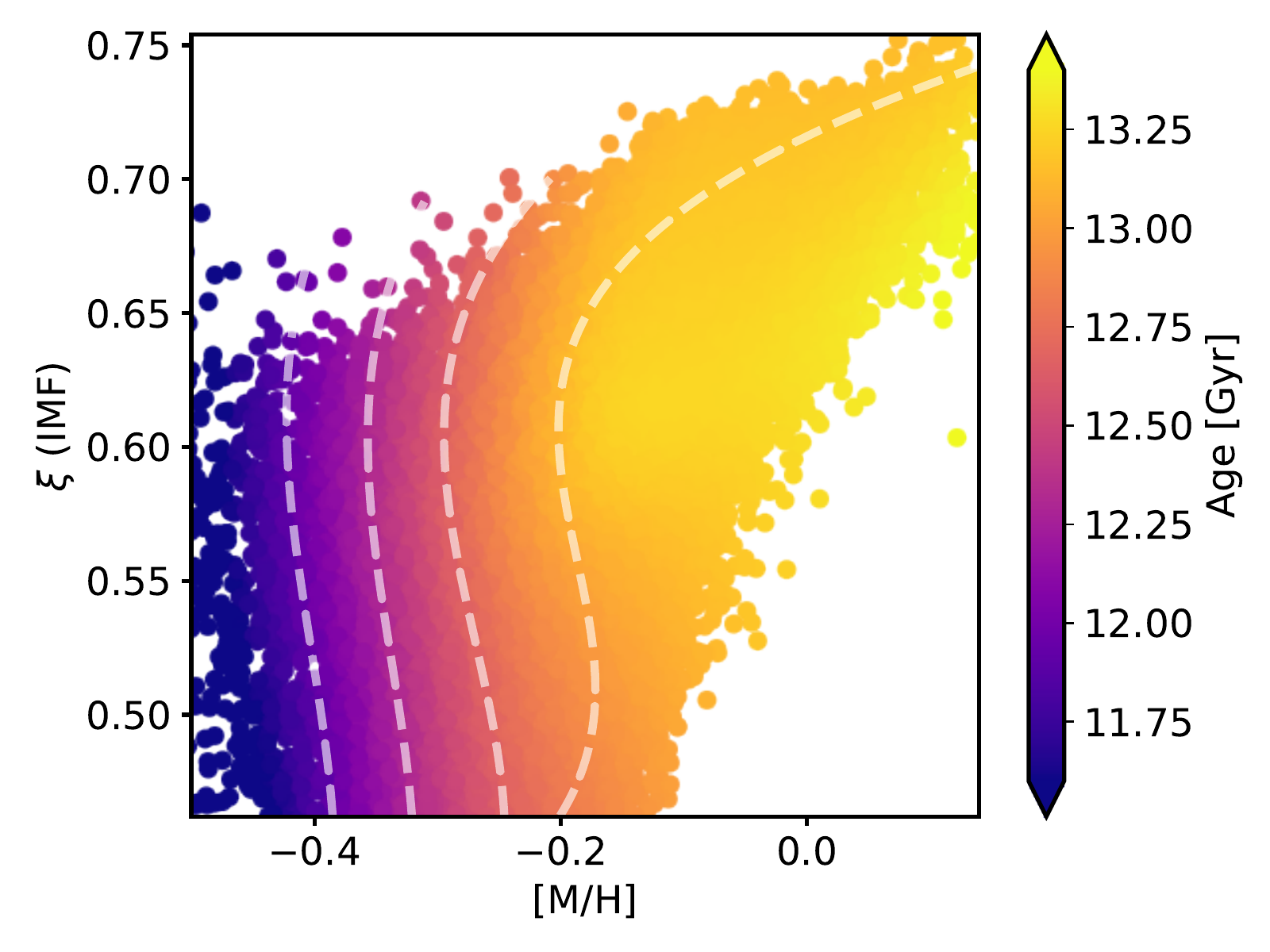}
   \includegraphics[width=9cm]{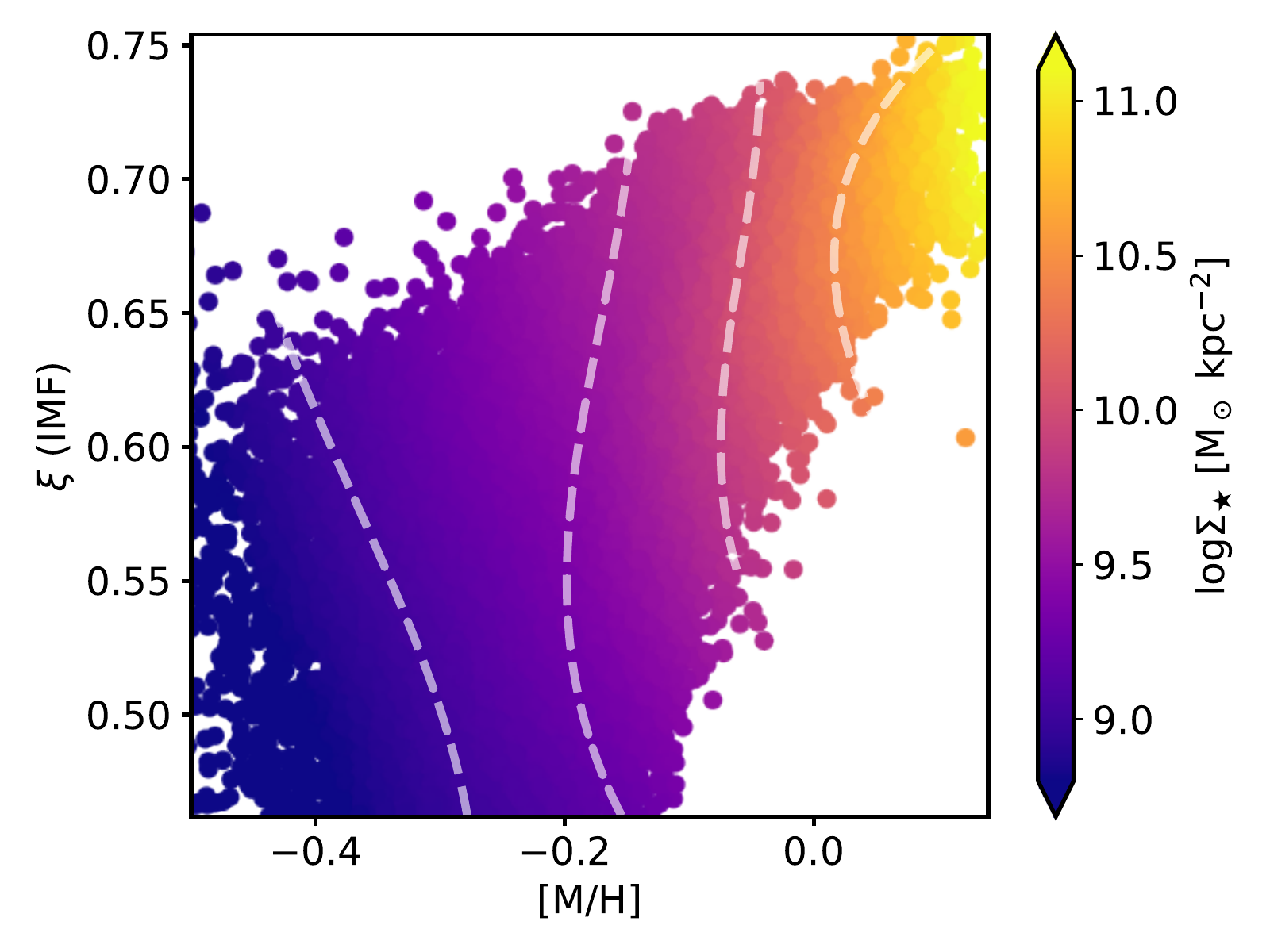}
   \includegraphics[width=9cm]{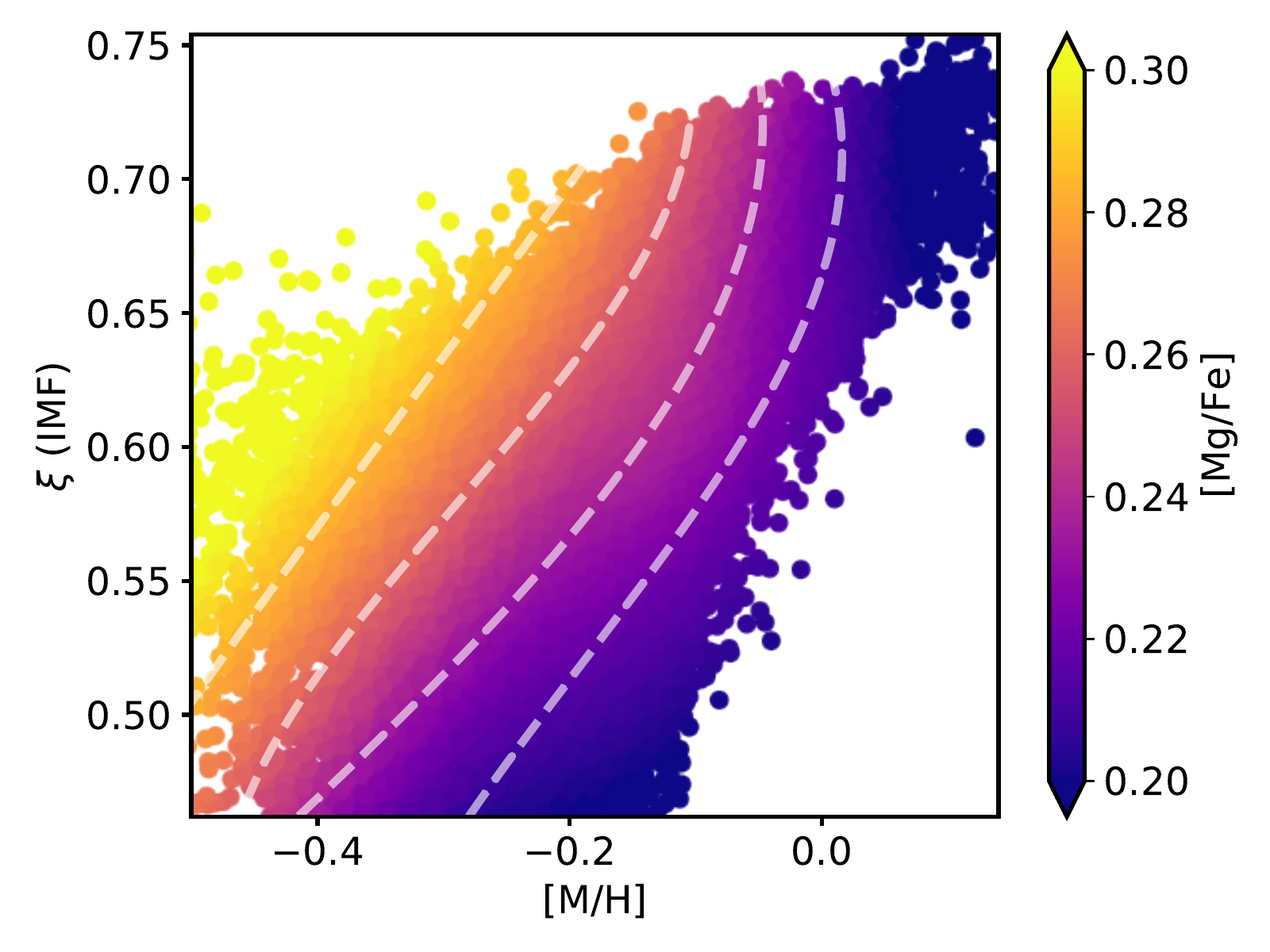}
   \includegraphics[width=9cm]{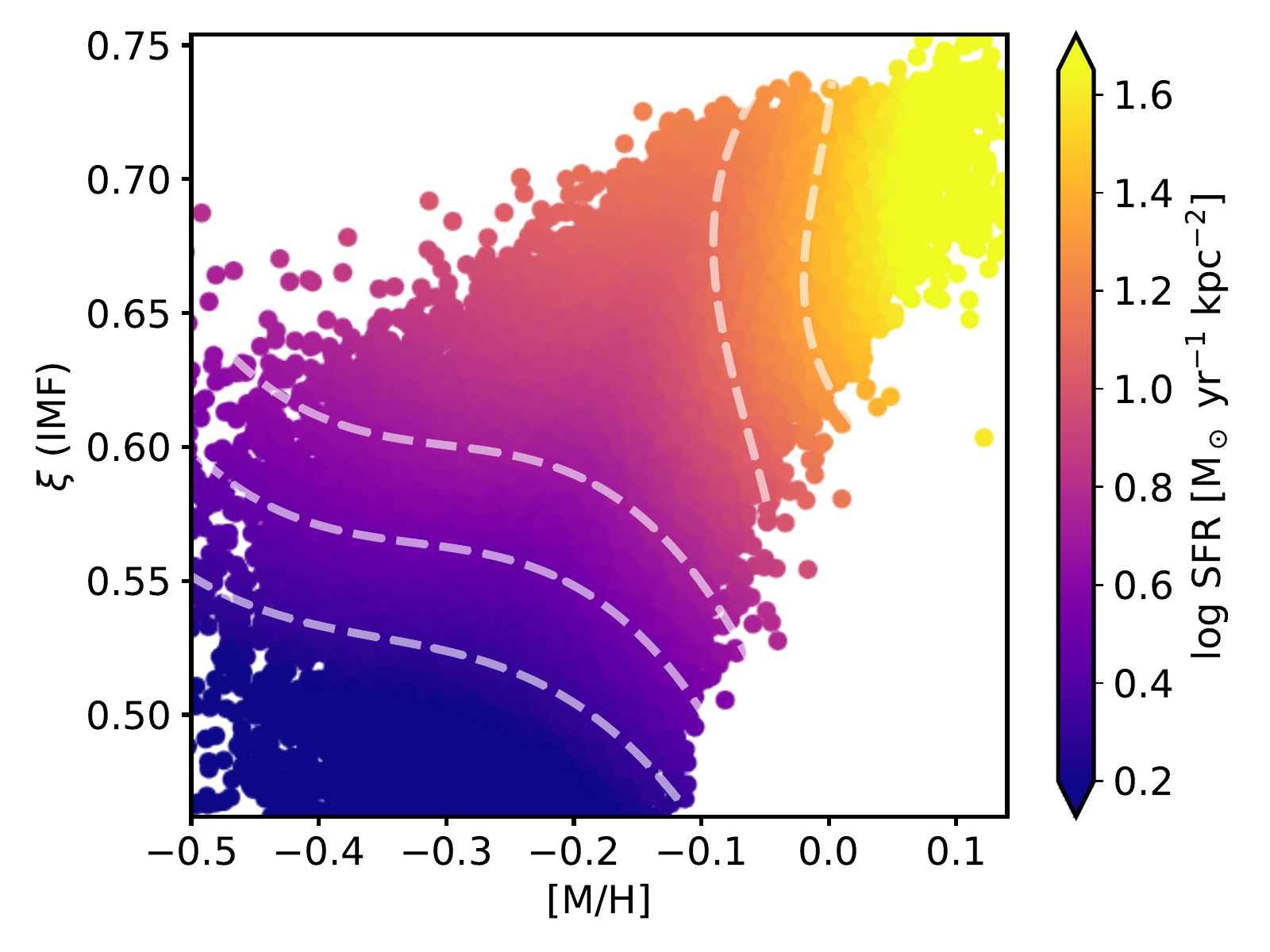}
   \caption{Secondary correlations. Each panel in this figure shows the metallicity -- IMF slope plane, color-coded by a different quantity derived from our stellar population analysis.  White dashed lines indicate the shape of the iso-contours. The color-coding of the top panels corresponds to the age (left) and the stellar surface density (right) of each Voronoi bin within our sample of F3D galaxies. No noticeable correlation is found between these two quantities and the scatter in the metallicity-IMF slope plane. However, both the [Mg/Fe], and particularly the effective star formation rate surface density, seem to correlate with the scatter in IMF slope values in the low metallicity ([M/H]$\lesssim -0.1$) regime. This may indicate a secondary dependence of IMF variations on star formation rate-related quantities, on top of the main trend with metallicity shown in Fig.~\ref{fig:met}.}
   \label{fig:secondary}
\end{figure*}

The large number of independent data points provided by the F3D data offers a unique opportunity to go beyond this zero-order IMF-metallicity relation, looking for secondary correlations. This is particularly interesting for metallicities below [M/H]$\sim-0.1$ where there is a significant scatter around the mean IMF slope value, as shown in Fig.~\ref{fig:met}. Fig.~\ref{fig:secondary} shows the metallicity--IMF slope plane, color-coded by four quantities directly derived from our stellar population analysis. A Locally Weighted Regression \citep[LOESS, ][]{Cappellari13} smoothing has been applied to highlight any underlying trends. The top left and right panels show the dependence on the local age and stellar surface density, respectively. In both panels, the color-coding changes rather independently of the IMF value, suggesting that these two quantities are not contributing to the (vertical) scatter in the metallicity--IMF slope plane.

In contrast, the bottom left panel of Fig.~\ref{fig:secondary} reveals that the scatter around the mean IMF slope value, in particular at low metallicities ([M/H]$\lesssim -0.1$), is partially coupled to the local [Mg/Fe] values. At these relatively low metallicities, regions with higher [Mg/Fe] values tend to exhibit steeper IMF slope values. The possible meaning of this secondary correlation is discussed in \S~\ref{sec:discu}, but the [Mg/Fe] ratio is usually interpreted as proxy for the formation time-scale $\tau_\star$ of a given stellar population \citep{Worthey92,Thomas99}, or equivalently, as the inverse of its effective specific star formation rate ($\log$ sSFR [yr$^{-1}$] $\propto 1 / \log \tau_\star \propto \mathrm{[Mg/Fe]}$). 

We further explored this tentative connection between IMF slope and star formation rate-related quantities by actually translating our [Mg/Fe] values into time-scale units following Equation 4 in \citet{Thomas05}. Then, for each Voronoi bin in our sample, we divided the best-fitting stellar surface density by its corresponding $\tau_\star$, resulting in a measurement of the effective star formation rate of that particular stellar population. We note that this quantity does not reflect the current star formation rate value, but it approximates the typical star formation rate at which a given population was formed. In our sample, with ages of 10 Gyr or more, we are estimating the star formation rate values that one would observe at redshift $z\sim2$ or above. As with the other quantities, the bottom right panel in Fig.~\ref{fig:secondary} shows the metallicity-IMF slope plane, this time color-coded by this effective star formation rate surface density. In the low-metallicity regime, the correlation between IMF slope and effective star formation rate surface density is clear, even more decoupled from the metallicity variations than the [Mg/Fe] shown on the bottom left panel.

\begin{figure}
   \centering
   \includegraphics[width=9cm]{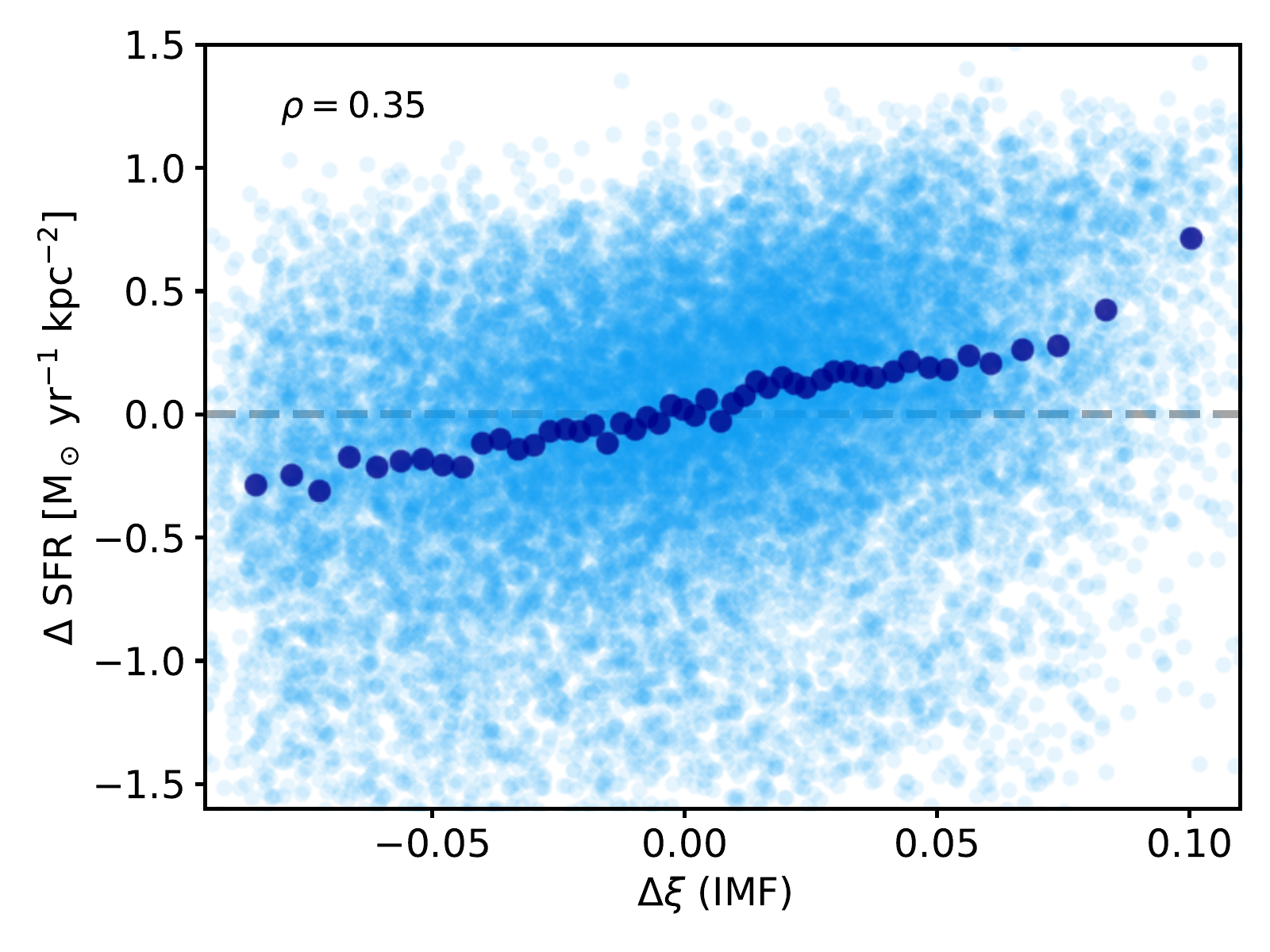}
   \caption{IMF dependence on the star formation rate. Light blue symbols show how the scatter in the metallicity-IMF relation (horizontal axis) correlate with the scatter in the metallicity-effective star formation rate density plane (vertical axis) for low-metallicty F3D bins ([M/H]$\lesssim -0.1$). Dark blue symbols indicate the running median of the distribution, revealing a significant correlation between IMF slope and star formation rate at fixed metallicity.}
   \label{fig:sfr}
\end{figure}

The additional correlation between IMF slope and effective star formation rate surface density is explicitly shown in Fig.~\ref{fig:sfr} where the scatter in the metallicity-IMF slope plane is plotted against the scatter in the metallicity-star formation rate surface density relation (light blue symbols), for Voronoi bins with relatively low metallicity ([M/H]$\lesssim -0.1$). The running median trend is shown with dark blue symbols, revealing the underlying correlation between the two quantities ($\rho=0.35$). It is evident from Fig.~\ref{fig:sfr} that, at fixed metallicity, the observed IMF slope values depend on the effective star formation rate. We would like to note that translating [Mg/Fe] values into formation time-scales and thus sSFR is subject to strong assumptions on the formation histories of galaxies and it is highly coupled to the IMF slope itself \citep[see e.g.][]{Thomas99,MN16}. Therefore, our estimations of the sSFR should be carefully  interpreted. 

\subsection{The geometry of IMF slope and metallicity variations} 

In the sections above we have discussed the similarities and differences between IMF slope and metallicity variations. In particular, we noted above that the two dimensional structure of the metallicity map of FCC\,083 seems to be more elongated than the IMF slope one (right panels in Fig.~\ref{fig:maps}). In order to properly quantify these differences between IMF slope and metallicity maps across the whole sample, we followed the same approach as in \citet{MN19} measuring the ellipticities of the metallicity and IMF slope maps for each galaxy. In addition, we also measured the ellipticity of the [Mg/Fe] maps, as well as that of the $r$-band light distribution of the galaxies as measured from the MUSE data cubes. Ellipticities were measured using the Multi-Gaussian Expansion routine described in \citet{Emsellem94} and  \citet{Cappellari02}. 

\begin{figure}
   \centering
   \includegraphics[width=9cm]{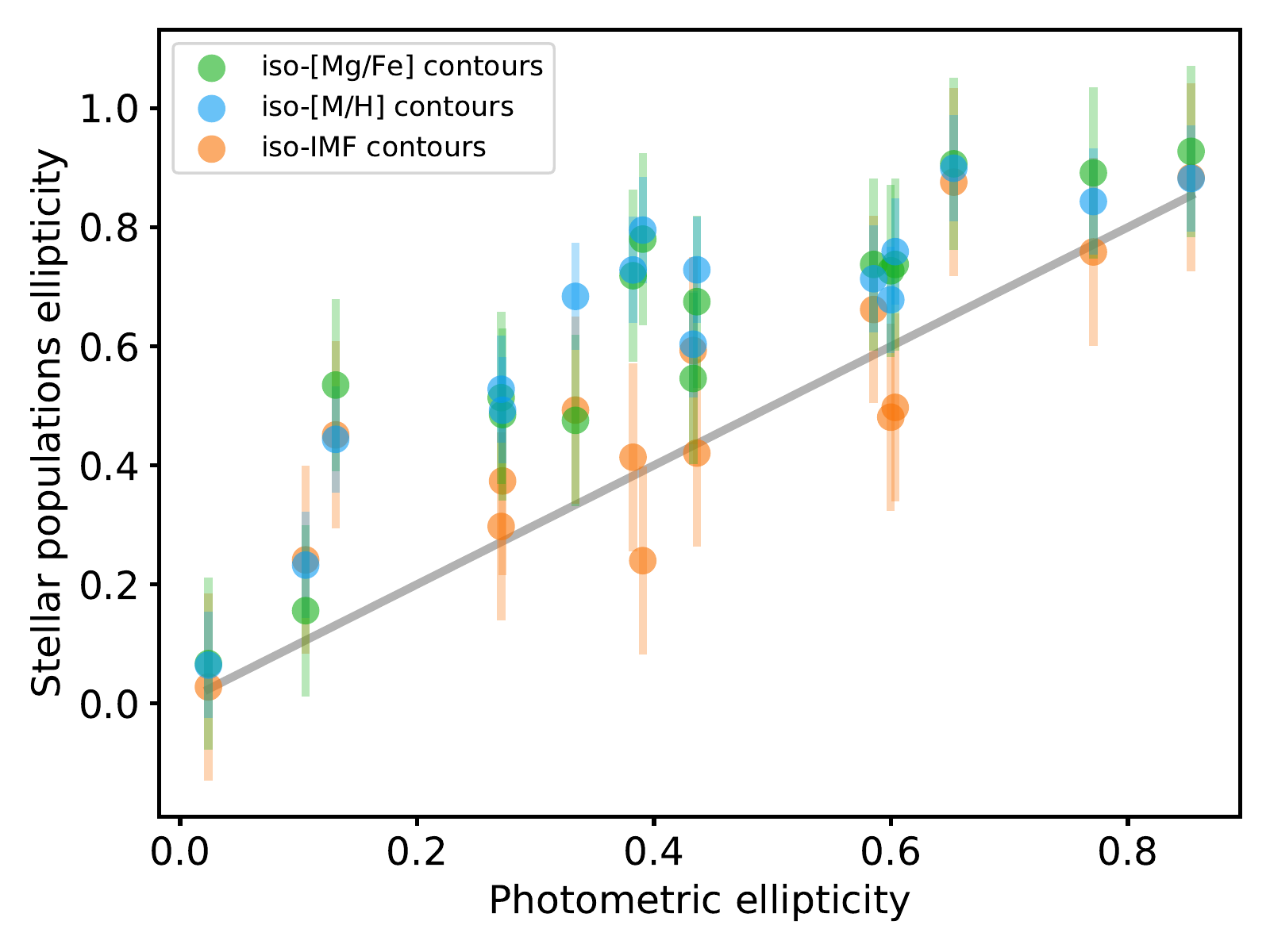}
   \caption{Photometric vs stellar population-based ellipticities. The mean ellipticity of the metallicity (orange),  [Mg/Fe] (green), and IMF slope maps (blue) is shown as a function of the photometric ellipticity. Stellar population-based ellipticities are higher than the photometric ones and the difference is larger for the metallicity and the [Mg/Fe] maps than for the IMF slope ones, hinting toward a different internal distribution of IMF slope and chemical enrichment values.}
   \label{fig:elli}
\end{figure}

Fig.~\ref{fig:elli} shows how the four ellipticities (from the light, metallicity, [Mg/Fe], and IMF slope distributions) compare for those galaxies in our sample with high enough spatial resolution to carry out the analysis. Blue, green, and orange symbols in Fig.~\ref{fig:elli} represent the average ellipticity of the metallicity, [Mg/Fe], and IMF slope maps as a function of the ellipticity of the luminous component, respectively. The solid gray line indicates the one-to-one relation. 

The most evident feature revealed by Fig.~\ref{fig:elli} is the fact that stellar population-based ellipticities are systematically higher than those measured in the light distribution. Moreover, these differences depend on the ellipticity, being maximal for intermediate values but negligible for both high and low ellipticity values.

The trends in Fig.~\ref{fig:elli} can be naturally explained as projection effect if stellar population properties are arranged in an elongated disk-like structure embedded in a more spherically symmetric representing the light distribution. If such a disk-like structure is observed in a face-on projection it would exhibit a low ellipticity, similarly to luminous spheroidal component. In the edge-on case, integration along the line-of-sight will make the disk-like structure overshine the spheroidal one, leading to a high averaged ellipticity. It would be only at intermediate inclinations that the differences between the two structures become more evident. Alternatively, galaxies properties may actually depend on ellipticity beyond any orientation effect. Although in principle this could also explain the differences observed in Fig.~\ref{fig:elli}, changes in the properties of ETGs appear to be more related to their internal kinematics and sizes rather than to their ellipticities \citep{Emsellem07,kine,McDermid15,FB19}. 

Using data from the SAURON project, \citet{Kuntschner06} already showed how the mean ellipticity of their Mgb maps was higher than that measured over the light distribution. Since the Mgb feature is mostly sensitive to metallicity and [Mg/Fe], their results are likely driven by the change in ellipticity revealed by our metallicity maps, rather than the actual [Mg/Fe] ones. Moreover, it is also clear from Fig.~\ref{fig:elli} that the mean ellipticity of the IMF slope maps is still higher than the photometric ones, yet, lower than those measured in the metallicity and [Mg/Fe] maps. This is in agreement with our analysis of FCC\,167 where we found that the structure of the IMF slope variations was partially decoupled from the metallicity one, resembling the distribution of those stars in warm orbits, while the metallicity map was likely tracking a colder dynamical structure \citep{MN19}.

\section{Discussion} \label{sec:discu}

\subsection{On the IMF variations}
Our analysis of the F3D data has revealed the rich and complex two-dimensional structure of the stellar population properties of galaxies within the Fornax cluster. This is true, not only for ages and metallicities (the main parameters regulating the properties of the observed spectra), but also for [Mg/Fe] and IMF slope values, whose variations have a much more subtle effect on the data. Galaxies like FCC\,153 show clearly separated thin and thick disks \citep{Pinna19} in the IMF map, with spatially coherent substructures in the central regions. Moreover, bar-like features can be seen in the IMF slope maps of galaxies like FCC\,170 and FCC\,184. As shown in \citet{MN19}, in those galaxies where the Voronoi-binned maps have particularly high spatial resolution (for example FCC\,167 or FCC\,083 shown in Fig.~\ref{fig:maps}), metallicity and IMF two-dimensional distributions seem to deviate from each other in a spatially coherent manner. This may reflect a difference in the underlying orbital distribution \citep{Ling20,Adriano21}, which can only be captured by two-dimensional integral field spectroscopic data. Measurements based on long-slit data or on elliptically averaged maps will, by construction, wash out any of these features revealed by our analysis. 

The strength of our two-dimensional analysis is exemplified by Fig.~\ref{fig:elli}, showing the comparison between stellar population-based and photometric ellipticities. This figure demonstrates that metallicity alone can not be responsible for determining the low-mass end slope of the IMF, since internal distribution of both stellar population properties is different. In \citet{MN19} we showed that, while the metallicity distribution of FCC\,167 mimics that of a cold dynamical component in the galaxy, IMF slope values seem to follow the distribution of warm orbits, with lower ellipticities as shown in Fig.~\ref{fig:elli}. The differences between the metallicity, [Mg/Fe], IMF slope, and light distributions in our sample can be explained if the observed IMF slope values are set at an early evolutionary stage of galaxies, but the chemical enrichment can continue if there is enough gas, leading to a dynamically colder, elongated and chemically evolved (high metallicity and low [Mg/Fe]) structure. The late accretion of satellites can build up the outskirts of ETGs, shaping their observed photometric structure.

Regarding the origin of the observed IMF slope variations, the relation between metallicity and IMF shown in Fig.~\ref{fig:met} is in agreement with previous studies \citep{MN15c,vdk17,Parikh,Sarzi18,Zhou19,MN19,Barbosa}. Fig.~\ref{fig:met} also reveals a clear increase in the scatter in the low-metallicity ([M/H]$\lesssim -0.1$), which corresponds to IMF slopes relatively close to the Milky Way value. This scatter is partially due to the fact that the effect of the IMF in the integrated spectra becomes weaker for flatter IMF slopes \citep[see e.g.][]{labarbera}. However, we find that the observed scatter correlates with the [Mg/Fe] values and even more strongly with the effective star formation rate surface density, as described in \S~\ref{sec:resu}. Since the [Mg/Fe] ratio can be understood as a star formation time-scale \citep[e.g.,][]{Thomas05,dlr11,McDermid15}, and therefore as the inverse of an averaged specific star formation rate, our observations seem to hint toward a connection between the star formation rate when the bulk of stars were formed (at redshifts $z\gtrsim2$) and the slope of the IMF.  Interestingly, the ellipticities of the iso-[Mg/Fe] contours are similar to the iso-[M/H] ones, suggesting that this secondary correlation between IMF slope and star formation rate-related quantities is induced by galaxy wide, mass-dependent offset in the [Mg/Fe]-[M/H] relation (see discussion below on the chemical enrichment).

Theoretical arguments support a possible connection between star formation rate and IMF slope as revealed by our analysis. For example it has been long argued that high star formation rate events could lead to a top-heavy IMF, that is, an IMF biased toward massive stars \citep{Papadopoulos10,Papadopoulos11,Jerabkova18,Fontanot18,Fontanot18b}. Observations of star-forming systems seem to support this idea, as an enhanced fraction of massive stars have been recently found in both distant \citep{Zhang18} and nearby \citep{Sliwa17,Brown19} star-forming galaxies. We note that in these studies the effect of the star formation rate is limited to the high mass end of the IMF. These massive stars, however, are no longer present in the old stellar populations of the F3D sample and could in fact explode as supernovae even before the low-mass stars that now dominate the observed spectra reached the main sequence. Therefore, our low-mass end IMF measurements are, in principle, decoupled from changes in the relative number of massive stars. The time-scale differences between low- and high-mass end slopes could be particularly relevant if the IMF, as suggested, also evolves in time \citep{Pieter08,Dave08,weidner:13,Ferreras15,Jerabkova18,DM18}.

It is also worth considering the possibility that the observed IMF variations might not be causally driven by changes in metallicity or star formation rate. In particular, although stability arguments can be used to link metallicity and IMF variations via cooling efficiency \citep[e.g.,][]{Dopcke13,MN15c,Guszejnov17,Chon21}, detailed numerical simulations struggle to reproduce significant metallicity-driven IMF variations \citep[e.g.,][]{Myers11,Guszejnov19}. Different prescriptions for IMF variations have been also implemented in cosmological numerical simulations, showing that different subgrid physics implementations of IMF variations can lead to results in reasonable agreement with observations \citep{Thales18,Barber18,Barber18b,Barber19}. In short, in the absence of a comprehensive theoretical framework that explains the complex behavior exhibited by the IMF in nearby quiescent galaxies, one has to be careful when considering the possible physical origin of a variable IMF, in particular given that it is strongly coupled to many observables as described in \S~\ref{sec:intro}.

\subsection{On the chemical enrichment}

One of the most noticeable features in the F3D stellar population maps is the ubiquity of positive [Mg/Fe] radial gradients across the whole mass range covered by our sample. Positive [Mg/Fe] radial gradients ([Mg/Fe] increasing with radius) are a natural prediction of outside-in monolithic collapse models \citep[e.g.,][]{Pipino04,Pipino06}, which can successfully explain the prevalence of negative metallicity gradients in ETGs. These models predict prolonged star formation in the central regions of ETGs as the potential well of the galaxy prevents stellar ejecta to efficiently escape, leading to a more developed chemical enrichment in the inner regions, with high metallicity values but low [Mg/Fe]. However, most observational studies of ETGs have reported rather flat or slightly decreasing [Mg/Fe] gradients, in particular for the most massive galaxies \citep[][]{Davies93,Trager00,Mehlert03,SB07,Greene15,MN18,Corsini18,Parikh21}.

This apparent inconsistency between predicted and observed [Mg/Fe] and [M/H] radial gradients has been traditionally interpreted as a consequence of the accretion-driven late evolution of massive ETGs. The cores of present-day ETGs are supposedly formed in rapid in situ-dominated star formation episodes \citep[e.g.,][]{oser}. These cores would exhibit steep negative metallicity gradients, while at the same time [Mg/Fe] would increase with radius as expected from chemical evolution models. However, as dry mergers start to populate the outskirts of ETGs late in their evolution, these pristine chemical gradients are washed out. This interpretation is supported by the fact that nearby naked red nuggets, arguably the unaltered progenitors of present-day massive galaxies, do show indeed positive [Mg/Fe] radial gradients and strongly negative metallicity ones \citep{akin17,MN18}.

In this context, the environment of the Fornax cluster may be responsible for preserving the pristine chemical enrichment of its members. With an estimated halo mass of $7\times10^{13}$ \msun \ and an intrinsic velocity dispersion of $\sim 380$ \kms \citep{Drinkwater01}, galaxy-galaxy interactions in such an environment are less frequent than in the field or in less massive (and thus lower velocity dispersion) halos \citep{Merritt83,Binney87,Gnedin03}. We note, however, that the internal velocity dispersion of the Fornax cluster is not sufficiently high to completely prevent galaxy-galaxy interactions, which may still play a role in the evolution of galaxies within the cluster. In addition to the internal dynamics of galaxies within Fornax, the gas in the cluster is found to be shock heated at a temperature of $\kappa T = 1.48$ keV \citep{Frank13}, which would also prevent fueling of cool gas and thus star formation. This interpretation is in agreement with the recent measurements of similarly positive [Mg/Fe] gradients in some nearby cluster galaxies \citep{Ferreras19,Barbosa20}.

\begin{figure}
   \centering
   \includegraphics[width=9cm]{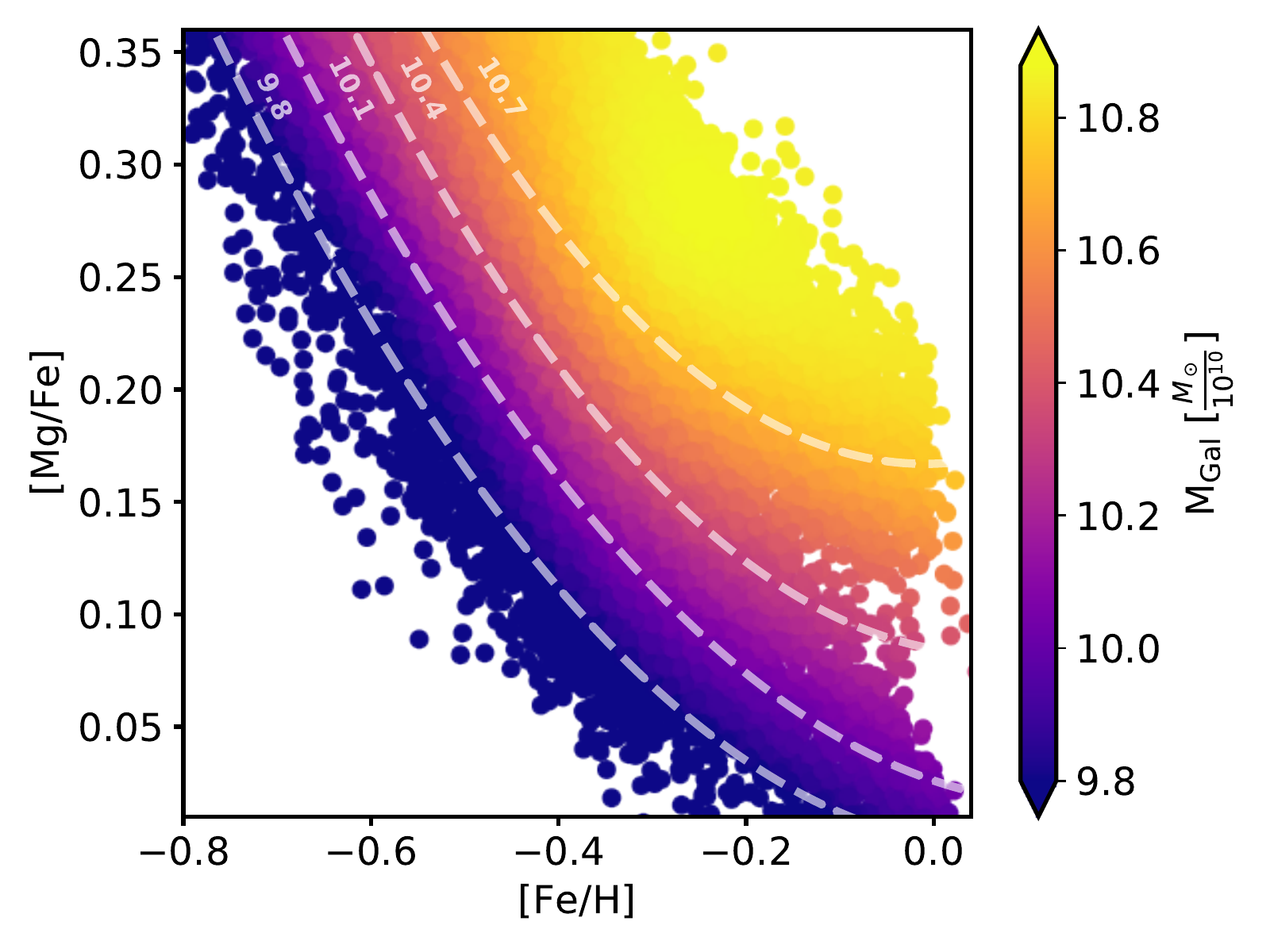}
   \caption{Iron metallicity [Fe/H] -- [Mg/Fe] plane from integrated F3D spectra. Each symbol indicates the [Fe/H] and [Mg/Fe] values of one Voronoi bin within our F3D sample, color-coded by the mass of the galaxy. At fixed stellar mass, there is a clear trend where more iron-rich bins tend to be less [Mg/Fe]-enhanced. The normalization of the [Fe/H]--[Mg/Fe] relation depends on galaxy mass, with more massive galaxies showing higher [Mg/Fe] values. Light shaded lines indicate the average distribution of points for galaxies with different masses.}
   \label{fig:chemi}
\end{figure}

Moreover, it is worth noting that metallicity and [Mg/Fe] measurements of F3D galaxies share interesting similarities with the chemical enrichment of nearby resolved galaxies \citet{Wallerstein62,Tinsley79}. In particular, Fig.~\ref{fig:chemi} shows the [Fe/H]-[Mg/Fe] plane derived from our stellar population analysis, color-coded by the mass of the host galaxy. Iron metallicities were estimated following Equation~4 in \citet{Vazdekis15}:
\begin{equation*}
\mathrm{[Fe/H]}=\mathrm{[M/H]}-0.75\times\mathrm{[Mg/Fe]} ,
\end{equation*}

\noindent 
At fixed galaxy stellar mass, F3D data points show a clear pattern, where low-metallicity spatial bins tend to be more [Mg/Fe] enhanced. For iron metallicities above [Fe/H] $\sim-0.4$, the [Fe/H]--[Mg/Fe] relation tends to flatten out. The normalization of this relation does, however, clearly depend on galaxy mass, with more massive galaxies being more [Mg/Fe]-enhanced at fixed iron metallicity. We note that the equation above leads to an subtle correlation between [Fe/H] and [Mg/Fe] since with our approach based on the MILES models does not measure iron but total metallicities. However, the main driver of the trends shown in Fig.~\ref{fig:chemi} is the observed relation between [Mg/Fe] and [M/H], which is not imposed by the models.

The behavior exhibited by our sample of F3D galaxies in Fig.~\ref{fig:chemi} is suggestively similar to the trends shown by nearby galaxies, where the properties of individual stars can be measured \citep[e.g.,][]{Tolstoy09,Sven21}. These mass-dependent tracks are usually interpreted as a chemical evolution record history: the oldest and most chemically pristine stars have been mainly polluted by the $\alpha$-enhanced stellar ejecta of core-collapse supernovae. However, after the onset of Type Ia supernovae, late generations of stars form from iron-rich (and thus relatively low [Mg/Fe]) gas, leading to the observed [Fe/H]--[Mg/Fe] trend. More massive galaxies are more efficient at retaining the stellar ejecta and turning them into stars and therefore exhibit a higher normalization.

The consistency between Fig.~\ref{fig:chemi} and studies based on resolved stars suggests that the [Fe/H]--[Mg/Fe] plane is indeed sensitive to the details of the chemical evolution of galaxies, even when measured from their integrated spectra \citep{Vincenzo18}. Our results are in agreement with those presented by \citet{Sybilska18}, who found similar trends in a sample of nearby Virgo and field galaxies. However, a major difference should be noted between integrated and resolved studies. Although individual stars contributing to an integrated spectrum do follow an intrinsic [Fe/H]--[Mg/Fe] relation, they are all assigned the same metallicity and [Mg/Fe] abundance in our stellar population analysis. Hence, the observed trends in Fig.~\ref{fig:chemi} should be interpreted as a weighted average. For the particular case of metallicities and abundance ratios in (old) composite stellar populations as it is the case of the F3D sample, \citet{Serra07} showed that single-stellar-population equivalent measurements are close to the expected mass-weighted values.

This difference between integrated and resolved studies becomes more obvious when looking at the age distributions. In resolved studies, the [Fe/H]--[Mg/Fe] relation can be understood as a time-evolution, as young stars form out of more chemically enriched material. However, this is the opposite to what we see in our sample of F3D galaxies (see for example Fig.~\ref{fig:maps}), where metal-poor and Mg-enhanced values tend to correspond to younger stellar populations. This results from the fact that we are not directly measuring the intrinsic age--[Fe/H]--[Mg/Fe] distribution of the stars within our sample \citep{Walcher15}, but making use of the gradients in [M/H] and [Mg/Fe] shown by our data to probe differences in the chemical enrichment. 

Finally, while the [Fe/H]--[Mg/Fe] distribution of resolved stars shows a clear plateau at low metallicities ([Fe/H]$\lesssim-0.5$), we do not find evidence of such saturation in the recovered [Mg/Fe] values. As discussed above, the averaged nature of our measurements is likely responsible for the absence of a plateau in the recovered [Fe/H]--[Mg/Fe] distribution, as we are not probing the very low surface brightness regime dominated in light by stars metal-poor enough to be in the [Mg/Fe] plateau. In the future, population-orbital modeling \citep{Ling20} will hopefully help to recover the underlying [Mg/Fe]--[Fe/H] distributions. In addition, from a purely technical point of view, the plateau occurs at [Mg/Fe]$\sim$0.4 which (by construction) is the maximum value allowed by the MILES models. In order to safely study the knee in the [Fe/H]--[Mg/Fe] distribution, models with predictions at higher [Mg/Fe] are required.

\section{Summary and conclusions} \label{sec:sum}

We have analyzed a sample of 23 galaxies within the virial radius of the Fornax cluster. These galaxies were observed as part of the Fornax 3D project with the MUSE spectrograph, resulting in an unprecedented spatial resolution that has allowed us to explore the richness and complexity of the IMF and stellar population maps over a range of galaxy masses. In particular, for each galaxy in the sample we measured the full two-dimensional age, metallicity, [Mg/Fe], and low-mass end IMF slope maps. The analysis of these maps leads to the following main conclusions:

\begin{itemize}
   \item Stellar population maps show a variety of features and behaviors, from disks to bar-like structures. These features are not completely symmetric and thus they are only accessible to integral field spectrographs. 
	\item Steep IMF slope values (populations with an enhanced fraction of low-mass stars) are predominantly metal-rich and usually concentrated in the innermost regions of galaxies. Within our sample, local metallicity is the best predictor for the observed IMF slope values.
	\item For metallicities below $\sim-0.1$[M/H] the scatter around the main [M/H]--IMF slope relation increases. The observed scatter correlates with quantities related to the star formation rate at which stars were formed, with regions with higher star formation rates showing steeper IMF slope values. 
	\item The ellipticities of the stellar population maps are higher than the photometric ones. Moreover, the metallicity maps are systematically more elongated than the IMF slope ones, revealing that their internal structures are intrinsically different.
	\item Every galaxy in our sample, even the more massive ones, exhibits a clearly positive [Mg/Fe] gradient. We hypothesize that this is a signature of the early chemical evolution of galaxies preserved by the particular environment of the cluster.
	\item The chemical properties of our galaxies, in particular the [Fe/H]--[Mg/Fe] plane, share suggestive similarities with measurements based on nearby resolved stars. This demonstrates the potential of studying the [Fe/H]--[Mg/Fe] relation from integrated spectra to understand the formation history of galaxies, but physically motivated models and predictions are needed to fully explain the observed trends. These predictions need to take into account how the mass growth and the chemical enrichment of galaxies are affected by (time-evolving) variations in the IMF slope.
\end{itemize}

In summary, our analysis has revealed systematic differences between IMF slope and metallicity variations in our sample of quiescent galaxies. These differences are imprinted in the orbital structure of galaxies and suggest a link between the early assembly of these galaxies and their observed properties in the local Universe. In particular, we hypothesize that the low-mass end IMF slope values that we measure today were set before the cold gas reservoir was exhausted and thus while the chemical enrichment and star formation were still taking place. This interpretation could explain the structural differences in the IMF slope and metallicity distributions, as well why galaxies and regions of galaxies with the same metallicity exhibit different IMF slope values that seem to correlate with star formation rate-related quantities. 

Moreover, our results emphasize the need for a comprehensive approach in order to measure and understand the observed properties of galaxies. The slope of the IMF, the chemical enrichment, and the star formation rate are all highly coupled quantities, both from a physical and an observational perspective. 
The mass growth of a galaxy gets encoded on its optical spectrum, but star formation histories derived from absorption spectra need to be consistent with the measured chemical properties and IMF slope values. Thus, physically motivated models that can reproduce the observed spectra of galaxies would be greatly beneficial to break the intrinsic degeneracies of such a complex inversion problem. Complementary, with the arrival of new facilities and a new class of ground-based and space telescopes, high quality spectra of distant (and therefore younger) massive galaxies will become the norm, hopefully unveiling the early stages of massive galaxy formation through detailed stellar population analyses.

\begin{acknowledgement}

We would like to thank the anonymous referee for providing constructive and detailed comments on the manuscript. IMN would like to thank F. La Barbera for his help during the early stages of the project. IMN, FP, and JFB acknowledge support from grant PID2019-107427GB-C32 from the Spanish Ministry of Science and Innovation. GvdV acknowledges funding from the  European  Research  Council  (ERC)  under  the  European  Union’s  Horizon2020  research  and  innovation  programme  under  grant  agreement  No  724857(Consolidator  Grant  ArcheoDyn).  RMcD is the recipient of an Australian Research Council Future Fellowship (project number FT150100333).  EMC is supported by MIUR grant PRIN 2017 20173ML3WW001 and Padua University grants DOR1885254/18, DOR1935272/19, and DOR2013080/20.  This research made use of Astropy,\footnote{http://www.astropy.org} a community-developed core Python package for Astronomy \citep{astropya,astropyb}, and of the Numpy \citep{harris2020array} and Matplotlib \citep{Hunter:2007} libraries. 
\end{acknowledgement}

\nopagebreak
\bibliographystyle{aa}  
\bibliography{IMF} 

\begin{thebibliography}{188}
\expandafter\ifx\csname natexlab\endcsname\relax\def\natexlab#1{#1}\fi

\bibitem[{{Alton} {et~al.}(2018){Alton}, {Smith}, \& {Lucey}}]{Alton18}
{Alton}, P.~D., {Smith}, R.~J., \& {Lucey}, J.~R. 2018, \mnras, 478, 4464

\bibitem[{{Arrigoni} {et~al.}(2010){Arrigoni}, {Trager}, {Somerville}, \&
  {Gibson}}]{Arrigoni10}
{Arrigoni}, M., {Trager}, S.~C., {Somerville}, R.~S., \& {Gibson}, B.~K. 2010,
  \mnras, 402, 173

\bibitem[{{Astropy Collaboration} {et~al.}(2018){Astropy Collaboration},
  {Price-Whelan}, {Sip{\H o}cz}, {G{\"u}nther}, {Lim}, {Crawford}, {Conseil},
  {Shupe}, {Craig}, {Dencheva}, {Ginsburg}, {VanderPlas}, {Bradley},
  {P{\'e}rez-Su{\'a}rez}, {de Val-Borro}, {Aldcroft}, {Cruz}, {Robitaille},
  {Tollerud}, {Ardelean}, {Babej}, {Bach}, {Bachetti}, {Bakanov}, {Bamford},
  {Barentsen}, {Barmby}, {Baumbach}, {Berry}, {Biscani}, {Boquien}, {Bostroem},
  {Bouma}, {Brammer}, {Bray}, {Breytenbach}, {Buddelmeijer}, {Burke},
  {Calderone}, {Cano Rodr{\'{\i}}guez}, {Cara}, {Cardoso}, {Cheedella},
  {Copin}, {Corrales}, {Crichton}, {D'Avella}, {Deil}, {Depagne}, {Dietrich},
  {Donath}, {Droettboom}, {Earl}, {Erben}, {Fabbro}, {Ferreira}, {Finethy},
  {Fox}, {Garrison}, {Gibbons}, {Goldstein}, {Gommers}, {Greco}, {Greenfield},
  {Groener}, {Grollier}, {Hagen}, {Hirst}, {Homeier}, {Horton}, {Hosseinzadeh},
  {Hu}, {Hunkeler}, {Ivezi{\'c}}, {Jain}, {Jenness}, {Kanarek}, {Kendrew},
  {Kern}, {Kerzendorf}, {Khvalko}, {King}, {Kirkby}, {Kulkarni}, {Kumar},
  {Lee}, {Lenz}, {Littlefair}, {Ma}, {Macleod}, {Mastropietro}, {McCully},
  {Montagnac}, {Morris}, {Mueller}, {Mumford}, {Muna}, {Murphy}, {Nelson},
  {Nguyen}, {Ninan}, {N{\"o}the}, {Ogaz}, {Oh}, {Parejko}, {Parley}, {Pascual},
  {Patil}, {Patil}, {Plunkett}, {Prochaska}, {Rastogi}, {Reddy Janga},
  {Sabater}, {Sakurikar}, {Seifert}, {Sherbert}, {Sherwood-Taylor}, {Shih},
  {Sick}, {Silbiger}, {Singanamalla}, {Singer}, {Sladen}, {Sooley},
  {Sornarajah}, {Streicher}, {Teuben}, {Thomas}, {Tremblay}, {Turner},
  {Terr{\'o}n}, {van Kerkwijk}, {de la Vega}, {Watkins}, {Weaver}, {Whitmore},
  {Woillez}, {Zabalza}, \& {Astropy Contributors}}]{astropyb}
{Astropy Collaboration}, {Price-Whelan}, A.~M., {Sip{\H o}cz}, B.~M., {et~al.}
  2018, \aj, 156, 123

\bibitem[{{Astropy Collaboration} {et~al.}(2013){Astropy Collaboration},
  {Robitaille}, {Tollerud}, {Greenfield}, {Droettboom}, {Bray}, {Aldcroft},
  {Davis}, {Ginsburg}, {Price-Whelan}, {Kerzendorf}, {Conley}, {Crighton},
  {Barbary}, {Muna}, {Ferguson}, {Grollier}, {Parikh}, {Nair}, {Unther},
  {Deil}, {Woillez}, {Conseil}, {Kramer}, {Turner}, {Singer}, {Fox}, {Weaver},
  {Zabalza}, {Edwards}, {Azalee Bostroem}, {Burke}, {Casey}, {Crawford},
  {Dencheva}, {Ely}, {Jenness}, {Labrie}, {Lim}, {Pierfederici}, {Pontzen},
  {Ptak}, {Refsdal}, {Servillat}, \& {Streicher}}]{astropya}
{Astropy Collaboration}, {Robitaille}, T.~P., {Tollerud}, E.~J., {et~al.} 2013,
  \aap, 558, A33

\bibitem[{{Auger} {et~al.}(2010){Auger}, {Treu}, {Gavazzi}, {Bolton},
  {Koopmans}, \& {Marshall}}]{auger}
{Auger}, M.~W., {Treu}, T., {Gavazzi}, R., {et~al.} 2010, \apjl, 721, L163

\bibitem[{{Bacon} {et~al.}(2010){Bacon}, {Accardo}, {Adjali}, {Anwand},
  {Bauer}, {Biswas}, {Blaizot}, {Boudon}, {Brau-Nogue}, {Brinchmann},
  {Caillier}, {Capoani}, {Carollo}, {Contini}, {Couderc}, {Daguis{\'e}},
  {Deiries}, {Delabre}, {Dreizler}, {Dubois}, {Dupieux}, {Dupuy}, {Emsellem},
  {Fechner}, {Fleischmann}, {Fran{\c c}ois}, {Gallou}, {Gharsa}, {Glindemann},
  {Gojak}, {Guiderdoni}, {Hansali}, {Hahn}, {Jarno}, {Kelz}, {Koehler},
  {Kosmalski}, {Laurent}, {Le Floch}, {Lilly}, {Lizon}, {Loupias}, {Manescau},
  {Monstein}, {Nicklas}, {Olaya}, {Pares}, {Pasquini}, {P{\'e}contal-Rousset},
  {Pell{\'o}}, {Petit}, {Popow}, {Reiss}, {Remillieux}, {Renault}, {Roth},
  {Rupprecht}, {Serre}, {Schaye}, {Soucail}, {Steinmetz}, {Streicher}, {Stuik},
  {Valentin}, {Vernet}, {Weilbacher}, {Wisotzki}, \& {Yerle}}]{Bacon10}
{Bacon}, R., {Accardo}, M., {Adjali}, L., {et~al.} 2010, in \procspie, Vol.
  7735, Ground-based and Airborne Instrumentation for Astronomy III, 773508

\bibitem[{{Barber} {et~al.}(2018){Barber}, {Crain}, \& {Schaye}}]{Barber18}
{Barber}, C., {Crain}, R.~A., \& {Schaye}, J. 2018, \mnras, 479, 5448

\bibitem[{{Barber} {et~al.}(2019{\natexlab{a}}){Barber}, {Schaye}, \&
  {Crain}}]{Barber19}
{Barber}, C., {Schaye}, J., \& {Crain}, R.~A. 2019{\natexlab{a}}, \mnras, 482,
  2515

\bibitem[{{Barber} {et~al.}(2019{\natexlab{b}}){Barber}, {Schaye}, \&
  {Crain}}]{Barber18b}
{Barber}, C., {Schaye}, J., \& {Crain}, R.~A. 2019{\natexlab{b}}, \mnras, 483,
  985

\bibitem[{{Barbosa} {et~al.}(2021{\natexlab{a}}){Barbosa}, {Spiniello},
  {Arnaboldi}, {Coccato}, {Hilker}, \& {Richtler}}]{Barbosa20}
{Barbosa}, C.~E., {Spiniello}, C., {Arnaboldi}, M., {et~al.}
  2021{\natexlab{a}}, \aap, 649, A93

\bibitem[{{Barbosa} {et~al.}(2021{\natexlab{b}}){Barbosa}, {Spiniello},
  {Arnaboldi}, {Coccato}, {Hilker}, \& {Richtler}}]{Barbosa}
{Barbosa}, C.~E., {Spiniello}, C., {Arnaboldi}, M., {et~al.}
  2021{\natexlab{b}}, \aap, 645, L1

\bibitem[{{Bastian} {et~al.}(2010){Bastian}, {Covey}, \& {Meyer}}]{bastian}
{Bastian}, N., {Covey}, K.~R., \& {Meyer}, M.~R. 2010, \araa, 48, 339

\bibitem[{{Bernardi} {et~al.}(2019){Bernardi}, {Dom{\'\i}nguez S{\'a}nchez},
  {Brownstein}, {Drory}, \& {Sheth}}]{Bernardi19}
{Bernardi}, M., {Dom{\'\i}nguez S{\'a}nchez}, H., {Brownstein}, J.~R., {Drory},
  N., \& {Sheth}, R.~K. 2019, \mnras, 489, 5633

\bibitem[{{Binney} \& {Tremaine}(1987)}]{Binney87}
{Binney}, J. \& {Tremaine}, S. 1987, {Galactic dynamics}

\bibitem[{{Brown} \& {Wilson}(2019)}]{Brown19}
{Brown}, T. \& {Wilson}, C.~D. 2019, \apj, 879, 17

\bibitem[{{Buder} {et~al.}(2021){Buder}, {Sharma}, {Kos}, {Amarsi},
  {Nordlander}, {Lind}, {Martell}, {Asplund}, {Bland-Hawthorn}, {Casey}, {de
  Silva}, {D'Orazi}, {Freeman}, {Hayden}, {Lewis}, {Lin}, {Schlesinger},
  {Simpson}, {Stello}, {Zucker}, {Zwitter}, {Beeson}, {Buck}, {Casagrande},
  {Clark}, {{\v{C}}otar}, {da Costa}, {de Grijs}, {Feuillet}, {Horner},
  {Kafle}, {Khanna}, {Kobayashi}, {Liu}, {Montet}, {Nandakumar}, {Nataf},
  {Ness}, {Spina}, {Tepper-Garc{\'\i}a}, {Ting}, {Traven},
  {Vogrin{\v{c}}i{\v{c}}}, {Wittenmyer}, {Wyse}, {{\v{Z}}erjal},
  {{\v{Z}}erjal}, \& {Galah Collaboration}}]{Sven21}
{Buder}, S., {Sharma}, S., {Kos}, J., {et~al.} 2021, \mnras, 506, 150

\bibitem[{{Bundy} {et~al.}(2015){Bundy}, {Bershady}, {Law}, {Yan}, {Drory},
  {MacDonald}, {Wake}, {Cherinka}, {S{\'a}nchez-Gallego}, {Weijmans}, {Thomas},
  {Tremonti}, {Masters}, {Coccato}, {Diamond-Stanic}, {Arag{\'o}n-Salamanca},
  {Avila-Reese}, {Badenes}, {Falc{\'o}n-Barroso}, {Belfiore}, {Bizyaev},
  {Blanc}, {Bland-Hawthorn}, {Blanton}, {Brownstein}, {Byler}, {Cappellari},
  {Conroy}, {Dutton}, {Emsellem}, {Etherington}, {Frinchaboy}, {Fu}, {Gunn},
  {Harding}, {Johnston}, {Kauffmann}, {Kinemuchi}, {Klaene}, {Knapen},
  {Leauthaud}, {Li}, {Lin}, {Maiolino}, {Malanushenko}, {Malanushenko}, {Mao},
  {Maraston}, {McDermid}, {Merrifield}, {Nichol}, {Oravetz}, {Pan}, {Parejko},
  {Sanchez}, {Schlegel}, {Simmons}, {Steele}, {Steinmetz}, {Thanjavur},
  {Thompson}, {Tinker}, {van den Bosch}, {Westfall}, {Wilkinson}, {Wright},
  {Xiao}, \& {Zhang}}]{manga}
{Bundy}, K., {Bershady}, M.~A., {Law}, D.~R., {et~al.} 2015, \apj, 798, 7

\bibitem[{{Cappellari}(2002)}]{Cappellari02}
{Cappellari}, M. 2002, \mnras, 333, 400

\bibitem[{{Cappellari}(2017)}]{Cappellari17}
{Cappellari}, M. 2017, \mnras, 466, 798

\bibitem[{{Cappellari} \& {Copin}(2003)}]{voronoi}
{Cappellari}, M. \& {Copin}, Y. 2003, \mnras, 342, 345

\bibitem[{{Cappellari} \& {Emsellem}(2004)}]{ppxf}
{Cappellari}, M. \& {Emsellem}, E. 2004, \pasp, 116, 138

\bibitem[{{Cappellari} {et~al.}(2012){Cappellari}, {McDermid}, {Alatalo},
  {Blitz}, {Bois}, {Bournaud}, {Bureau}, {Crocker}, {Davies}, {Davis}, {de
  Zeeuw}, {Duc}, {Emsellem}, {Khochfar}, {Krajnovi{\'c}}, {Kuntschner},
  {Lablanche}, {Morganti}, {Naab}, {Oosterloo}, {Sarzi}, {Scott}, {Serra},
  {Weijmans}, \& {Young}}]{cappellari}
{Cappellari}, M., {McDermid}, R.~M., {Alatalo}, K., {et~al.} 2012, \nat, 484,
  485

\bibitem[{{Cappellari} {et~al.}(2013){Cappellari}, {McDermid}, {Alatalo},
  {Blitz}, {Bois}, {Bournaud}, {Bureau}, {Crocker}, {Davies}, {Davis}, {de
  Zeeuw}, {Duc}, {Emsellem}, {Khochfar}, {Krajnovi{\'c}}, {Kuntschner},
  {Morganti}, {Naab}, {Oosterloo}, {Sarzi}, {Scott}, {Serra}, {Weijmans}, \&
  {Young}}]{Cappellari13}
{Cappellari}, M., {McDermid}, R.~M., {Alatalo}, K., {et~al.} 2013, \mnras, 432,
  1862

\bibitem[{{Cenarro} {et~al.}(2001){Cenarro}, {Cardiel}, {Gorgas}, {Peletier},
  {Vazdekis}, \& {Prada}}]{cat}
{Cenarro}, A.~J., {Cardiel}, N., {Gorgas}, J., {et~al.} 2001, \mnras, 326, 959

\bibitem[{{Cenarro} {et~al.}(2004){Cenarro}, {S{\'a}nchez-Bl{\'a}zquez},
  {Cardiel}, \& {Gorgas}}]{Cenarro04}
{Cenarro}, A.~J., {S{\'a}nchez-Bl{\'a}zquez}, P., {Cardiel}, N., \& {Gorgas},
  J. 2004, \apjl, 614, L101

\bibitem[{{Chabrier}(2003)}]{Chabrier}
{Chabrier}, G. 2003, \pasp, 115, 763

\bibitem[{{Chon} {et~al.}(2021){Chon}, {Omukai}, \& {Schneider}}]{Chon21}
{Chon}, S., {Omukai}, K., \& {Schneider}, R. 2021, arXiv e-prints,
  arXiv:2103.04997

\bibitem[{{Cid Fernandes} {et~al.}(2005){Cid Fernandes}, {Mateus}, {Sodr{\'e}},
  {Stasi{\'n}ska}, \& {Gomes}}]{CF05}
{Cid Fernandes}, R., {Mateus}, A., {Sodr{\'e}}, L., {Stasi{\'n}ska}, G., \&
  {Gomes}, J.~M. 2005, \mnras, 358, 363

\bibitem[{{Clauwens} {et~al.}(2016){Clauwens}, {Schaye}, \&
  {Franx}}]{Clauwens16}
{Clauwens}, B., {Schaye}, J., \& {Franx}, M. 2016, \mnras, 462, 2832

\bibitem[{{Coelho}(2014)}]{Coelho14}
{Coelho}, P.~R.~T. 2014, \mnras, 440, 1027

\bibitem[{{Conroy}(2013)}]{Conroy13}
{Conroy}, C. 2013, \araa, 51, 393

\bibitem[{{Conroy} {et~al.}(2013){Conroy}, {Dutton}, {Graves}, {Mendel}, \&
  {van Dokkum}}]{Conroy13b}
{Conroy}, C., {Dutton}, A.~A., {Graves}, G.~J., {Mendel}, J.~T., \& {van
  Dokkum}, P.~G. 2013, \apjl, 776, L26

\bibitem[{{Conroy} {et~al.}(2009){Conroy}, {Gunn}, \& {White}}]{Conroy09}
{Conroy}, C., {Gunn}, J.~E., \& {White}, M. 2009, \apj, 699, 486

\bibitem[{{Conroy} \& {van Dokkum}(2012{\natexlab{a}})}]{conroy}
{Conroy}, C. \& {van Dokkum}, P. 2012{\natexlab{a}}, \apj, 747, 69

\bibitem[{{Conroy} \& {van Dokkum}(2012{\natexlab{b}})}]{conroy12}
{Conroy}, C. \& {van Dokkum}, P.~G. 2012{\natexlab{b}}, \apj, 760, 71

\bibitem[{{Corsini} {et~al.}(2018){Corsini}, {Morelli}, {Zarattini}, {Aguerri},
  {Costantin}, {D'Onghia}, {Girardi}, {Kundert}, {M{\'e}ndez-Abreu}, \&
  {Thomas}}]{Corsini18}
{Corsini}, E.~M., {Morelli}, L., {Zarattini}, S., {et~al.} 2018, \aap, 618,
  A172

\bibitem[{{Corsini} {et~al.}(2017){Corsini}, {Wegner}, {Thomas}, {Saglia}, \&
  {Bender}}]{Corsini17}
{Corsini}, E.~M., {Wegner}, G.~A., {Thomas}, J., {Saglia}, R.~P., \& {Bender},
  R. 2017, \mnras, 466, 974

\bibitem[{{Dav{\'e}}(2008)}]{Dave08}
{Dav{\'e}}, R. 2008, \mnras, 385, 147

\bibitem[{{Davies} {et~al.}(1993){Davies}, {Sadler}, \& {Peletier}}]{Davies93}
{Davies}, R.~L., {Sadler}, E.~M., \& {Peletier}, R.~F. 1993, \mnras, 262, 650

\bibitem[{{Davis} \& {McDermid}(2017)}]{Davis17}
{Davis}, T.~A. \& {McDermid}, R.~M. 2017, \mnras, 464, 453

\bibitem[{{de La Rosa} {et~al.}(2011){de La Rosa}, {La Barbera}, {Ferreras}, \&
  {de Carvalho}}]{dlr11}
{de La Rosa}, I.~G., {La Barbera}, F., {Ferreras}, I., \& {de Carvalho}, R.~R.
  2011, \mnras, 418, L74

\bibitem[{{De Masi} {et~al.}(2019){De Masi}, {Vincenzo}, {Matteucci}, {Rosani},
  {La Barbera}, {Pasquali}, \& {Spitoni}}]{DM18}
{De Masi}, C., {Vincenzo}, F., {Matteucci}, F., {et~al.} 2019, \mnras, 483,
  2217

\bibitem[{{Dom{\'\i}nguez S{\'a}nchez} {et~al.}(2019){Dom{\'\i}nguez
  S{\'a}nchez}, {Bernardi}, {Brownstein}, {Drory}, \& {Sheth}}]{Helena19}
{Dom{\'\i}nguez S{\'a}nchez}, H., {Bernardi}, M., {Brownstein}, J.~R., {Drory},
  N., \& {Sheth}, R.~K. 2019, \mnras, 489, 5612

\bibitem[{{Dopcke} {et~al.}(2013){Dopcke}, {Glover}, {Clark}, \&
  {Klessen}}]{Dopcke13}
{Dopcke}, G., {Glover}, S. C.~O., {Clark}, P.~C., \& {Klessen}, R.~S. 2013,
  \apj, 766, 103

\bibitem[{{Drinkwater} {et~al.}(2001){Drinkwater}, {Gregg}, \&
  {Colless}}]{Drinkwater01}
{Drinkwater}, M.~J., {Gregg}, M.~D., \& {Colless}, M. 2001, \apjl, 548, L139

\bibitem[{{Dutton} {et~al.}(2012){Dutton}, {Mendel}, \& {Simard}}]{Dutton12}
{Dutton}, A.~A., {Mendel}, J.~T., \& {Simard}, L. 2012, \mnras, 422, L33

\bibitem[{{Emsellem} {et~al.}(2011){Emsellem}, {Cappellari}, {Krajnovi{\'c}},
  {Alatalo}, {Blitz}, {Bois}, {Bournaud}, {Bureau}, {Davies}, {Davis}, {de
  Zeeuw}, {Khochfar}, {Kuntschner}, {Lablanche}, {McDermid}, {Morganti},
  {Naab}, {Oosterloo}, {Sarzi}, {Scott}, {Serra}, {van de Ven}, {Weijmans}, \&
  {Young}}]{kine}
{Emsellem}, E., {Cappellari}, M., {Krajnovi{\'c}}, D., {et~al.} 2011, \mnras,
  414, 888

\bibitem[{{Emsellem} {et~al.}(2007){Emsellem}, {Cappellari}, {Krajnovi{\'c}},
  {van de Ven}, {Bacon}, {Bureau}, {Davies}, {de Zeeuw}, {Falc{\'o}n-Barroso},
  {Kuntschner}, {McDermid}, {Peletier}, \& {Sarzi}}]{Emsellem07}
{Emsellem}, E., {Cappellari}, M., {Krajnovi{\'c}}, D., {et~al.} 2007, \mnras,
  379, 401

\bibitem[{{Emsellem} {et~al.}(1994){Emsellem}, {Monnet}, \&
  {Bacon}}]{Emsellem94}
{Emsellem}, E., {Monnet}, G., \& {Bacon}, R. 1994, \aap, 285, 723

\bibitem[{{Fahrion} {et~al.}(2020{\natexlab{a}}){Fahrion}, {Lyubenova},
  {Hilker}, {van de Ven}, {Falc{\'o}n-Barroso}, {Leaman},
  {Mart{\'\i}n-Navarro}, {Bittner}, {Coccato}, {Corsini}, {Gadotti}, {Iodice},
  {McDermid}, {Pinna}, {Sarzi}, {Viaene}, {de Zeeuw}, \& {Zhu}}]{Fahrion20b}
{Fahrion}, K., {Lyubenova}, M., {Hilker}, M., {et~al.} 2020{\natexlab{a}},
  \aap, 637, A26

\bibitem[{{Fahrion} {et~al.}(2020{\natexlab{b}}){Fahrion}, {Lyubenova},
  {Hilker}, {van de Ven}, {Falc{\'o}n-Barroso}, {Leaman},
  {Mart{\'\i}n-Navarro}, {Bittner}, {Coccato}, {Corsini}, {Gadotti}, {Iodice},
  {McDermid}, {Pinna}, {Sarzi}, {Viaene}, {de Zeeuw}, \& {Zhu}}]{Fahrion20}
{Fahrion}, K., {Lyubenova}, M., {Hilker}, M., {et~al.} 2020{\natexlab{b}},
  \aap, 637, A27

\bibitem[{{Falc{\'o}n-Barroso} {et~al.}(2011){Falc{\'o}n-Barroso},
  {S{\'a}nchez-Bl{\'a}zquez}, {Vazdekis}, {Ricciardelli}, {Cardiel}, {Cenarro},
  {Gorgas}, \& {Peletier}}]{Jesus11}
{Falc{\'o}n-Barroso}, J., {S{\'a}nchez-Bl{\'a}zquez}, P., {Vazdekis}, A.,
  {et~al.} 2011, \aap, 532, A95

\bibitem[{{Falc{\'o}n-Barroso} {et~al.}(2019){Falc{\'o}n-Barroso}, {van de
  Ven}, {Lyubenova}, {Mendez-Abreu}, {Aguerri}, {Garc{\'\i}a-Lorenzo},
  {Bekerait{\'e}}, {S{\'a}nchez}, {Husemann}, {Garc{\'\i}a-Benito},
  {Gonz{\'a}lez Delgado}, {Mast}, {Walcher}, {Zibetti}, {Zhu},
  {Barrera-Ballesteros}, {Galbany}, {S{\'a}nchez-Bl{\'a}zquez}, {Singh}, {van
  den Bosch}, {Wild}, {Bland-Hawthorn}, {Cid Fernandes}, {de
  Lorenzo-C{\'a}ceres}, {Gallazzi}, {Marino}, {M{\'a}rquez}, {Peletier},
  {P{\'e}rez}, {P{\'e}rez}, {Roth}, {Rosales-Ortega}, {Ruiz-Lara}, {Wisotzki},
  \& {Ziegler}}]{FB19}
{Falc{\'o}n-Barroso}, J., {van de Ven}, G., {Lyubenova}, M., {et~al.} 2019,
  \aap, 632, A59

\bibitem[{{Ferguson}(1989)}]{Ferguson89}
{Ferguson}, H.~C. 1989, \aj, 98, 367

\bibitem[{{Ferr{\'e}-Mateu} {et~al.}(2013){Ferr{\'e}-Mateu}, {Vazdekis}, \& {de
  la Rosa}}]{FM13}
{Ferr{\'e}-Mateu}, A., {Vazdekis}, A., \& {de la Rosa}, I.~G. 2013, \mnras,
  431, 440

\bibitem[{{Ferreras} {et~al.}(2013){Ferreras}, {La Barbera}, {de la Rosa},
  {Vazdekis}, {de Carvalho}, {Falc{\'o}n-Barroso}, \&
  {Ricciardelli}}]{ferreras}
{Ferreras}, I., {La Barbera}, F., {de la Rosa}, I.~G., {et~al.} 2013, \mnras,
  429, L15

\bibitem[{{Ferreras} {et~al.}(2019){Ferreras}, {Scott}, {La Barbera}, {Croom},
  {van de Sande}, {Hopkins}, {Colless}, {Barone}, {d'Eugenio},
  {Bland-Hawthorn}, {Brough}, {Bryant}, {Konstantopoulos}, {Lagos}, {Lawrence},
  {L{\'o}pez-S{\'a}nchez}, {Medling}, {Owers}, \& {Richards}}]{Ferreras19}
{Ferreras}, I., {Scott}, N., {La Barbera}, F., {et~al.} 2019, \mnras, 489, 608

\bibitem[{{Ferreras} {et~al.}(2015){Ferreras}, {Weidner}, {Vazdekis}, \& {La
  Barbera}}]{Ferreras15}
{Ferreras}, I., {Weidner}, C., {Vazdekis}, A., \& {La Barbera}, F. 2015,
  \mnras, 448, L82

\bibitem[{{Fontanot} {et~al.}(2018{\natexlab{a}}){Fontanot}, {De Lucia}, {Xie},
  {Hirschmann}, {Bruzual}, \& {Charlot}}]{Fontanot18b}
{Fontanot}, F., {De Lucia}, G., {Xie}, L., {et~al.} 2018{\natexlab{a}}, \mnras,
  475, 2467

\bibitem[{{Fontanot} {et~al.}(2018{\natexlab{b}}){Fontanot}, {La Barbera}, {De
  Lucia}, {Pasquali}, \& {Vazdekis}}]{Fontanot18}
{Fontanot}, F., {La Barbera}, F., {De Lucia}, G., {Pasquali}, A., \&
  {Vazdekis}, A. 2018{\natexlab{b}}, \mnras, 479, 5678

\bibitem[{{Foreman-Mackey} {et~al.}(2013){Foreman-Mackey}, {Hogg}, {Lang}, \&
  {Goodman}}]{emcee}
{Foreman-Mackey}, D., {Hogg}, D.~W., {Lang}, D., \& {Goodman}, J. 2013, \pasp,
  125, 306

\bibitem[{{Frank} {et~al.}(2013){Frank}, {Peterson}, {Andersson}, {Fabian}, \&
  {Sanders}}]{Frank13}
{Frank}, K.~A., {Peterson}, J.~R., {Andersson}, K., {Fabian}, A.~C., \&
  {Sanders}, J.~S. 2013, \apj, 764, 46

\bibitem[{{Gnedin}(2003)}]{Gnedin03}
{Gnedin}, O.~Y. 2003, \apj, 582, 141

\bibitem[{{Greene} {et~al.}(2015){Greene}, {Janish}, {Ma}, {McConnell},
  {Blakeslee}, {Thomas}, \& {Murphy}}]{Greene15}
{Greene}, J.~E., {Janish}, R., {Ma}, C.-P., {et~al.} 2015, \apj, 807, 11

\bibitem[{{Guszejnov} {et~al.}(2019){Guszejnov}, {Hopkins}, \&
  {Graus}}]{Guszejnov19}
{Guszejnov}, D., {Hopkins}, P.~F., \& {Graus}, A.~S. 2019, \mnras, 485, 4852

\bibitem[{{Guszejnov} {et~al.}(2017){Guszejnov}, {Hopkins}, \&
  {Ma}}]{Guszejnov17}
{Guszejnov}, D., {Hopkins}, P.~F., \& {Ma}, X. 2017, \mnras, 472, 2107

\bibitem[{{Gutcke} \& {Springel}(2019)}]{Thales18}
{Gutcke}, T.~A. \& {Springel}, V. 2019, \mnras, 482, 118

\bibitem[{Harris {et~al.}(2020)Harris, Millman, van~der Walt, Gommers,
  Virtanen, Cournapeau, Wieser, Taylor, Berg, Smith, Kern, Picus, Hoyer, van
  Kerkwijk, Brett, Haldane, del R{\'{i}}o, Wiebe, Peterson,
  G{\'{e}}rard-Marchant, Sheppard, Reddy, Weckesser, Abbasi, Gohlke, \&
  Oliphant}]{harris2020array}
Harris, C.~R., Millman, K.~J., van~der Walt, S.~J., {et~al.} 2020, Nature, 585,
  357

\bibitem[{{Hoversten} \& {Glazebrook}(2008)}]{Hoversten08}
{Hoversten}, E.~A. \& {Glazebrook}, K. 2008, \apj, 675, 163

\bibitem[{Hunter(2007)}]{Hunter:2007}
Hunter, J.~D. 2007, Computing in Science \& Engineering, 9, 90

\bibitem[{{Iodice} {et~al.}(2019{\natexlab{a}}){Iodice}, {Sarzi}, {Bittner},
  {Coccato}, {Costantin}, {Corsini}, {van de Ven}, {de Zeeuw},
  {Falc{\'o}n-Barroso}, {Gadotti}, {Lyubenova}, {Mart{\'\i}n-Navarro},
  {McDermid}, {Nedelchev}, {Pinna}, {Pizzella}, {Spavone}, \&
  {Viaene}}]{Iodice19}
{Iodice}, E., {Sarzi}, M., {Bittner}, A., {et~al.} 2019{\natexlab{a}}, \aap,
  627, A136

\bibitem[{{Iodice} {et~al.}(2019{\natexlab{b}}){Iodice}, {Spavone},
  {Capaccioli}, {Peletier}, {van de Ven}, {Napolitano}, {Hilker}, {Mieske},
  {Smith}, {Pasquali}, {Limatola}, {Grado}, {Venhola}, {Cantiello}, {Paolillo},
  {Falcon-Barroso}, {D'Abrusco}, \& {Schipani}}]{Enrica18}
{Iodice}, E., {Spavone}, M., {Capaccioli}, M., {et~al.} 2019{\natexlab{b}},
  \aap, 623, A1

\bibitem[{{Je{\v{r}}{\'a}bkov{\'a}} {et~al.}(2018){Je{\v{r}}{\'a}bkov{\'a}},
  {Hasani Zonoozi}, {Kroupa}, {Beccari}, {Yan}, {Vazdekis}, \&
  {Zhang}}]{Jerabkova18}
{Je{\v{r}}{\'a}bkov{\'a}}, T., {Hasani Zonoozi}, A., {Kroupa}, P., {et~al.}
  2018, \aap, 620, A39

\bibitem[{{Kippenhahn} {et~al.}(2012){Kippenhahn}, {Weigert}, \&
  {Weiss}}]{Kippenhahn}
{Kippenhahn}, R., {Weigert}, A., \& {Weiss}, A. 2012, {Stellar Structure and
  Evolution}

\bibitem[{{Kroupa}(2001)}]{mw}
{Kroupa}, P. 2001, \mnras, 322, 231

\bibitem[{{Kroupa}(2002)}]{Kroupa}
{Kroupa}, P. 2002, Science, 295, 82

\bibitem[{{Kuntschner}(2004)}]{Kuntschner04}
{Kuntschner}, H. 2004, \aap, 426, 737

\bibitem[{{Kuntschner} {et~al.}(2006){Kuntschner}, {Emsellem}, {Bacon},
  {Bureau}, {Cappellari}, {Davies}, {de Zeeuw}, {Falc{\'o}n-Barroso},
  {Krajnovi{\'c}}, {McDermid}, {Peletier}, \& {Sarzi}}]{Kuntschner06}
{Kuntschner}, H., {Emsellem}, E., {Bacon}, R., {et~al.} 2006, \mnras, 369, 497

\bibitem[{{Kuntschner} {et~al.}(2010){Kuntschner}, {Emsellem}, {Bacon},
  {Cappellari}, {Davies}, {de Zeeuw}, {Falc{\'o}n-Barroso}, {Krajnovi{\'c}},
  {McDermid}, {Peletier}, {Sarzi}, {Shapiro}, {van den Bosch}, \& {van de
  Ven}}]{Kuntschner10}
{Kuntschner}, H., {Emsellem}, E., {Bacon}, R., {et~al.} 2010, \mnras, 408, 97

\bibitem[{{La Barbera} {et~al.}(2013){La Barbera}, {Ferreras}, {Vazdekis}, {de
  la Rosa}, {de Carvalho}, {Trevisan}, {Falc{\'o}n-Barroso}, \&
  {Ricciardelli}}]{labarbera}
{La Barbera}, F., {Ferreras}, I., {Vazdekis}, A., {et~al.} 2013, \mnras, 433,
  3017

\bibitem[{{La Barbera} {et~al.}(2019){La Barbera}, {Vazdekis}, {Ferreras},
  {Pasquali}, {Allende Prieto}, {Mart{\'\i}n-Navarro}, {Aguado}, {de Carvalho},
  {Rembold}, {Falc{\'o}n-Barroso}, \& {van de Ven}}]{LB19}
{La Barbera}, F., {Vazdekis}, A., {Ferreras}, I., {et~al.} 2019, \mnras, 489,
  4090

\bibitem[{{La Barbera} {et~al.}(2017){La Barbera}, {Vazdekis}, {Ferreras},
  {Pasquali}, {Allende Prieto}, {R{\"o}ck}, {Aguado}, \& {Peletier}}]{LB17}
{La Barbera}, F., {Vazdekis}, A., {Ferreras}, I., {et~al.} 2017, \mnras, 464,
  3597

\bibitem[{{La Barbera} {et~al.}(2016){La Barbera}, {Vazdekis}, {Ferreras},
  {Pasquali}, {Cappellari}, {Mart{\'{\i}}n-Navarro}, {Sch{\"o}nebeck}, \&
  {Falc{\'o}n-Barroso}}]{LB16}
{La Barbera}, F., {Vazdekis}, A., {Ferreras}, I., {et~al.} 2016, \mnras, 457,
  1468

\bibitem[{{Lacerna} {et~al.}(2020){Lacerna}, {Ibarra-Medel}, {Avila-Reese},
  {Hern{\'a}ndez-Toledo}, {V{\'a}zquez-Mata}, \& {S{\'a}nchez}}]{Lacerna20}
{Lacerna}, I., {Ibarra-Medel}, H., {Avila-Reese}, V., {et~al.} 2020, \aap, 644,
  A117

\bibitem[{{Lagattuta} {et~al.}(2017){Lagattuta}, {Mould}, {Forbes}, {Monson},
  {Pastorello}, \& {Persson}}]{Lagattuta17}
{Lagattuta}, D.~J., {Mould}, J.~R., {Forbes}, D.~A., {et~al.} 2017, \apj, 846,
  166

\bibitem[{{L{\"a}sker} {et~al.}(2013){L{\"a}sker}, {van den Bosch}, {van de
  Ven}, {Ferreras}, {La Barbera}, {Vazdekis}, \&
  {Falc{\'o}n-Barroso}}]{Lasker13}
{L{\"a}sker}, R., {van den Bosch}, R.~C.~E., {van de Ven}, G., {et~al.} 2013,
  \mnras, 434, L31

\bibitem[{{Lee} {et~al.}(2009){Lee}, {Gil de Paz}, {Tremonti}, {Kennicutt},
  {Salim}, {Bothwell}, {Calzetti}, {Dalcanton}, {Dale}, {Engelbracht}, {Funes},
  {Johnson}, {Sakai}, {Skillman}, {van Zee}, {Walter}, \& {Weisz}}]{Lee09}
{Lee}, J.~C., {Gil de Paz}, A., {Tremonti}, C., {et~al.} 2009, \apj, 706, 599

\bibitem[{{Li} {et~al.}(2017){Li}, {Ge}, {Mao}, {Cappellari}, {Long}, {Li},
  {Emsellem}, {Dutton}, {Li}, {Bundy}, {Thomas}, {Drory}, \& {Lopes}}]{Li17}
{Li}, H., {Ge}, J., {Mao}, S., {et~al.} 2017, \apj, 838, 77

\bibitem[{{Lyubenova} {et~al.}(2016){Lyubenova}, {Mart{\'{\i}}n-Navarro}, {van
  de Ven}, {Falc{\'o}n-Barroso}, {Galbany}, {Gallazzi}, {Garc{\'{\i}}a-Benito},
  {Gonz{\'a}lez Delgado}, {Husemann}, {La Barbera}, {Marino}, {Mast},
  {Mendez-Abreu}, {Peletier}, {S{\'a}nchez-Bl{\'a}zquez}, {S{\'a}nchez},
  {Trager}, {van den Bosch}, {Vazdekis}, {Walcher}, {Zhu}, {Zibetti},
  {Ziegler}, {Bland-Hawthorn}, \& {CALIFA Collaboration}}]{Lyubenova16}
{Lyubenova}, M., {Mart{\'{\i}}n-Navarro}, I., {van de Ven}, G., {et~al.} 2016,
  \mnras, 463, 3220

\bibitem[{{Mart{\'{\i}}n-Navarro}(2016)}]{MN16}
{Mart{\'{\i}}n-Navarro}, I. 2016, \mnras, 456, L104

\bibitem[{{Mart{\'{\i}}n-Navarro}
  {et~al.}(2015{\natexlab{a}}){Mart{\'{\i}}n-Navarro}, {La Barbera},
  {Vazdekis}, {Falc{\'o}n-Barroso}, \& {Ferreras}}]{MN15a}
{Mart{\'{\i}}n-Navarro}, I., {La Barbera}, F., {Vazdekis}, A.,
  {Falc{\'o}n-Barroso}, J., \& {Ferreras}, I. 2015{\natexlab{a}}, \mnras, 447,
  1033

\bibitem[{{Mart{\'\i}n-Navarro} {et~al.}(2019){Mart{\'\i}n-Navarro},
  {Lyubenova}, {van de Ven}, {Falc{\'o}n-Barroso}, {Coccato}, {Corsini},
  {Gadotti}, {Iodice}, {La Barbera}, {McDermid}, {Pinna}, {Sarzi}, {Viaene},
  {de Zeeuw}, \& {Zhu}}]{MN19}
{Mart{\'\i}n-Navarro}, I., {Lyubenova}, M., {van de Ven}, G., {et~al.} 2019,
  \aap, 626, A124

\bibitem[{{Mart{\'{\i}}n-Navarro} {et~al.}(2018){Mart{\'{\i}}n-Navarro},
  {Vazdekis}, {Falc{\'o}n-Barroso}, {La Barbera}, {Y{\i}ld{\i}r{\i}m}, \& {van
  de Ven}}]{MN18}
{Mart{\'{\i}}n-Navarro}, I., {Vazdekis}, A., {Falc{\'o}n-Barroso}, J., {et~al.}
  2018, \mnras, 475, 3700

\bibitem[{{Mart{\'{\i}}n-Navarro}
  {et~al.}(2015{\natexlab{b}}){Mart{\'{\i}}n-Navarro}, {Vazdekis}, {La
  Barbera}, {Falc{\'o}n-Barroso}, {Lyubenova}, {van de Ven}, {Ferreras},
  {S{\'a}nchez}, {Trager}, {Garc{\'{\i}}a-Benito}, {Mast}, {Mendoza},
  {S{\'a}nchez-Bl{\'a}zquez}, {Gonz{\'a}lez Delgado}, {Walcher}, \& {The CALIFA
  Team}}]{MN15c}
{Mart{\'{\i}}n-Navarro}, I., {Vazdekis}, A., {La Barbera}, F., {et~al.}
  2015{\natexlab{b}}, \apjl, 806, L31

\bibitem[{{Martins} \& {Coelho}(2007)}]{Martins07}
{Martins}, L.~P. \& {Coelho}, P. 2007, \mnras, 381, 1329

\bibitem[{{McConnell} {et~al.}(2016){McConnell}, {Lu}, \& {Mann}}]{McConnell16}
{McConnell}, N.~J., {Lu}, J.~R., \& {Mann}, A.~W. 2016, \apj, 821, 39

\bibitem[{{McDermid} {et~al.}(2015){McDermid}, {Alatalo}, {Blitz}, {Bournaud},
  {Bureau}, {Cappellari}, {Crocker}, {Davies}, {Davis}, {de Zeeuw}, {Duc},
  {Emsellem}, {Khochfar}, {Krajnovi{\'c}}, {Kuntschner}, {Morganti}, {Naab},
  {Oosterloo}, {Sarzi}, {Scott}, {Serra}, {Weijmans}, \& {Young}}]{McDermid15}
{McDermid}, R.~M., {Alatalo}, K., {Blitz}, L., {et~al.} 2015, \mnras, 448, 3484

\bibitem[{{McDermid} {et~al.}(2014){McDermid}, {Cappellari}, {Alatalo},
  {Bayet}, {Blitz}, {Bois}, {Bournaud}, {Bureau}, {Crocker}, {Davies}, {Davis},
  {de Zeeuw}, {Duc}, {Emsellem}, {Khochfar}, {Krajnovi{\'c}}, {Kuntschner},
  {Morganti}, {Naab}, {Oosterloo}, {Sarzi}, {Scott}, {Serra}, {Weijmans}, \&
  {Young}}]{McDermid14}
{McDermid}, R.~M., {Cappellari}, M., {Alatalo}, K., {et~al.} 2014, \apjl, 792,
  L37

\bibitem[{{Mehlert} {et~al.}(2003){Mehlert}, {Thomas}, {Saglia}, {Bender}, \&
  {Wegner}}]{Mehlert03}
{Mehlert}, D., {Thomas}, D., {Saglia}, R.~P., {Bender}, R., \& {Wegner}, G.
  2003, \aap, 407, 423

\bibitem[{{Mendel} {et~al.}(2020){Mendel}, {Beifiori}, {Saglia}, {Bender},
  {Brammer}, {Chan}, {F{\"o}rster Schreiber}, {Fossati}, {Galametz},
  {Momcheva}, {Nelson}, {Wilman}, \& {Wuyts}}]{Mendel20}
{Mendel}, J.~T., {Beifiori}, A., {Saglia}, R.~P., {et~al.} 2020, \apj, 899, 87

\bibitem[{{Merritt}(1983)}]{Merritt83}
{Merritt}, D. 1983, \apj, 264, 24

\bibitem[{{Meurer} {et~al.}(2009){Meurer}, {Wong}, {Kim}, {Hanish}, {Heckman},
  {Werk}, {Bland-Hawthorn}, {Dopita}, {Zwaan}, {Koribalski}, {Seibert},
  {Thilker}, {Ferguson}, {Webster}, {Putman}, {Knezek}, {Doyle}, {Drinkwater},
  {Hoopes}, {Kilborn}, {Meyer}, {Ryan-Weber}, {Smith}, \&
  {Staveley-Smith}}]{Meurer09}
{Meurer}, G.~R., {Wong}, O.~I., {Kim}, J.~H., {et~al.} 2009, \apj, 695, 765

\bibitem[{{Miller} \& {Scalo}(1979)}]{Miller79}
{Miller}, G.~E. \& {Scalo}, J.~M. 1979, \apjs, 41, 513

\bibitem[{{Mitchell} {et~al.}(2013){Mitchell}, {Lacey}, {Baugh}, \&
  {Cole}}]{Mitchell13}
{Mitchell}, P.~D., {Lacey}, C.~G., {Baugh}, C.~M., \& {Cole}, S. 2013, \mnras,
  435, 87

\bibitem[{{Myers} {et~al.}(2011){Myers}, {Krumholz}, {Klein}, \&
  {McKee}}]{Myers11}
{Myers}, A.~T., {Krumholz}, M.~R., {Klein}, R.~I., \& {McKee}, C.~F. 2011,
  \apj, 735, 49

\bibitem[{{Nanayakkara} {et~al.}(2017){Nanayakkara}, {Glazebrook}, {Kacprzak},
  {Yuan}, {Fisher}, {Tran}, {Kewley}, {Spitler}, {Alcorn}, {Cowley}, {Labbe},
  {Straatman}, \& {Tomczak}}]{Nanayakkara17}
{Nanayakkara}, T., {Glazebrook}, K., {Kacprzak}, G.~G., {et~al.} 2017, \mnras,
  468, 3071

\bibitem[{{Neumann} {et~al.}(2020){Neumann}, {Fragkoudi}, {P{\'e}rez},
  {Gadotti}, {Falc{\'o}n-Barroso}, {S{\'a}nchez-Bl{\'a}zquez}, {Bittner},
  {Husemann}, {G{\'o}mez}, {Grand}, {Donohoe-Keyes}, {Kim}, {de
  Lorenzo-C{\'a}ceres}, {Martig}, {M{\'e}ndez-Abreu}, {Pakmor}, {Seidel}, \&
  {van de Ven}}]{Justus2020}
{Neumann}, J., {Fragkoudi}, F., {P{\'e}rez}, I., {et~al.} 2020, \aap, 637, A56

\bibitem[{{Ocvirk} {et~al.}(2006){Ocvirk}, {Pichon}, {Lan{\c c}on}, \&
  {Thi{\'e}baut}}]{Ocvirk06}
{Ocvirk}, P., {Pichon}, C., {Lan{\c c}on}, A., \& {Thi{\'e}baut}, E. 2006,
  \mnras, 365, 74

\bibitem[{{Oldham} \& {Auger}(2018)}]{Oldham}
{Oldham}, L. \& {Auger}, M. 2018, \mnras, 474, 4169

\bibitem[{{Oser} {et~al.}(2010){Oser}, {Ostriker}, {Naab}, {Johansson}, \&
  {Burkert}}]{oser}
{Oser}, L., {Ostriker}, J.~P., {Naab}, T., {Johansson}, P.~H., \& {Burkert}, A.
  2010, \apj, 725, 2312

\bibitem[{{Papadopoulos}(2010)}]{Papadopoulos10}
{Papadopoulos}, P.~P. 2010, \apj, 720, 226

\bibitem[{{Papadopoulos} {et~al.}(2011){Papadopoulos}, {Thi}, {Miniati}, \&
  {Viti}}]{Papadopoulos11}
{Papadopoulos}, P.~P., {Thi}, W.-F., {Miniati}, F., \& {Viti}, S. 2011, \mnras,
  414, 1705

\bibitem[{{Parikh} {et~al.}(2021){Parikh}, {Thomas}, {Maraston}, {Westfall},
  {Andrews}, {Boardman}, {Drory}, \& {Oyarzun}}]{Parikh21}
{Parikh}, T., {Thomas}, D., {Maraston}, C., {et~al.} 2021, \mnras, 502, 5508

\bibitem[{{Parikh} {et~al.}(2018){Parikh}, {Thomas}, {Maraston}, {Westfall},
  {Goddard}, {Lian}, {Meneses-Goytia}, {Jones}, {Vaughan}, {Andrews},
  {Bershady}, {Bizyaev}, {Brinkmann}, {Brownstein}, {Bundy}, {Drory},
  {Emsellem}, {Law}, {Newman}, {Roman-Lopes}, {Wake}, {Yan}, \&
  {Zheng}}]{Parikh}
{Parikh}, T., {Thomas}, D., {Maraston}, C., {et~al.} 2018, \mnras, 477, 3954

\bibitem[{{Pietrinferni} {et~al.}(2004){Pietrinferni}, {Cassisi}, {Salaris}, \&
  {Castelli}}]{basti1}
{Pietrinferni}, A., {Cassisi}, S., {Salaris}, M., \& {Castelli}, F. 2004, \apj,
  612, 168

\bibitem[{{Pietrinferni} {et~al.}(2006){Pietrinferni}, {Cassisi}, {Salaris}, \&
  {Castelli}}]{basti2}
{Pietrinferni}, A., {Cassisi}, S., {Salaris}, M., \& {Castelli}, F. 2006, \apj,
  642, 797

\bibitem[{{Pinna} {et~al.}(2019{\natexlab{a}}){Pinna}, {Falc{\'o}n-Barroso},
  {Martig}, {Coccato}, {Corsini}, {de Zeeuw}, {Gadotti}, {Iodice}, {Leaman},
  {Lyubenova}, {Mart{\'\i}n-Navarro}, {Morelli}, {Sarzi}, {van de Ven},
  {Viaene}, \& {McDermid}}]{Pinna19}
{Pinna}, F., {Falc{\'o}n-Barroso}, J., {Martig}, M., {et~al.}
  2019{\natexlab{a}}, \aap, 625, A95

\bibitem[{{Pinna} {et~al.}(2019{\natexlab{b}}){Pinna}, {Falc{\'o}n-Barroso},
  {Martig}, {Sarzi}, {Coccato}, {Iodice}, {Corsini}, {de Zeeuw}, {Gadotti},
  {Leaman}, {Lyubenova}, {McDermid}, {Minchev}, {Morelli}, {van de Ven}, \&
  {Viaene}}]{Pinna19b}
{Pinna}, F., {Falc{\'o}n-Barroso}, J., {Martig}, M., {et~al.}
  2019{\natexlab{b}}, \aap, 623, A19

\bibitem[{{Pipino} \& {Matteucci}(2004)}]{Pipino04}
{Pipino}, A. \& {Matteucci}, F. 2004, \mnras, 347, 968

\bibitem[{{Pipino} {et~al.}(2006){Pipino}, {Matteucci}, \&
  {Chiappini}}]{Pipino06}
{Pipino}, A., {Matteucci}, F., \& {Chiappini}, C. 2006, \apj, 638, 739

\bibitem[{{Poci} {et~al.}(2021){Poci}, {McDermid}, {Lyubenova}, {Zhu}, {van de
  Ven}, {Iodice}, {Coccato}, {Pinna}, {Corsini}, {Falc{\'o}n-Barroso},
  {Gadotti}, {Grand}, {Fahrion}, {Mart{\'\i}n-Navarro}, {Sarzi}, {Viaene}, \&
  {de Zeeuw}}]{Adriano21}
{Poci}, A., {McDermid}, R.~M., {Lyubenova}, M., {et~al.} 2021, \aap, 647, A145

\bibitem[{{Posacki} {et~al.}(2015){Posacki}, {Cappellari}, {Treu},
  {Pellegrini}, \& {Ciotti}}]{Posacki15}
{Posacki}, S., {Cappellari}, M., {Treu}, T., {Pellegrini}, S., \& {Ciotti}, L.
  2015, \mnras, 446, 493

\bibitem[{{Romano} {et~al.}(2019){Romano}, {Matteucci}, {Zhang}, {Ivison}, \&
  {Ventura}}]{Romano19}
{Romano}, D., {Matteucci}, F., {Zhang}, Z.-Y., {Ivison}, R.~J., \& {Ventura},
  P. 2019, \mnras, 490, 2838

\bibitem[{{Romano} {et~al.}(2017){Romano}, {Matteucci}, {Zhang},
  {Papadopoulos}, \& {Ivison}}]{Romano17}
{Romano}, D., {Matteucci}, F., {Zhang}, Z.~Y., {Papadopoulos}, P.~P., \&
  {Ivison}, R.~J. 2017, \mnras, 470, 401

\bibitem[{{Rosani} {et~al.}(2018){Rosani}, {Pasquali}, {La Barbera},
  {Ferreras}, \& {Vazdekis}}]{Rosani18}
{Rosani}, G., {Pasquali}, A., {La Barbera}, F., {Ferreras}, I., \& {Vazdekis},
  A. 2018, \mnras, 476, 5233

\bibitem[{{Salpeter}(1955)}]{Salp:55}
{Salpeter}, E.~E. 1955, \apj, 121, 161

\bibitem[{{S{\'a}nchez} {et~al.}(2012){S{\'a}nchez}, {Kennicutt}, {Gil de Paz},
  {van de Ven}, {V{\'{\i}}lchez}, {Wisotzki}, {Walcher}, {Mast}, {Aguerri},
  {Albiol-P{\'e}rez}, {Alonso-Herrero}, {Alves}, {Bakos}, {Bart{\'a}kov{\'a}},
  {Bland-Hawthorn}, {Boselli}, {Bomans}, {Castillo-Morales}, {Cortijo-Ferrero},
  {de Lorenzo-C{\'a}ceres}, {Del Olmo}, {Dettmar}, {D{\'{\i}}az}, {Ellis},
  {Falc{\'o}n-Barroso}, {Flores}, {Gallazzi}, {Garc{\'{\i}}a-Lorenzo},
  {Gonz{\'a}lez Delgado}, {Gruel}, {Haines}, {Hao}, {Husemann},
  {Igl{\'e}sias-P{\'a}ramo}, {Jahnke}, {Johnson}, {Jungwiert}, {Kalinova},
  {Kehrig}, {Kupko}, {L{\'o}pez-S{\'a}nchez}, {Lyubenova}, {Marino},
  {M{\'a}rmol-Queralt{\'o}}, {M{\'a}rquez}, {Masegosa}, {Meidt},
  {Mendez-Abreu}, {Monreal-Ibero}, {Montijo}, {Mour{\~a}o}, {Palacios-Navarro},
  {Papaderos}, {Pasquali}, {Peletier}, {P{\'e}rez}, {P{\'e}rez}, {Quirrenbach},
  {Rela{\~n}o}, {Rosales-Ortega}, {Roth}, {Ruiz-Lara},
  {S{\'a}nchez-Bl{\'a}zquez}, {Sengupta}, {Singh}, {Stanishev}, {Trager},
  {Vazdekis}, {Viironen}, {Wild}, {Zibetti}, \& {Ziegler}}]{califa}
{S{\'a}nchez}, S.~F., {Kennicutt}, R.~C., {Gil de Paz}, A., {et~al.} 2012,
  \aap, 538, A8

\bibitem[{{S{\'a}nchez-Bl{\'a}zquez} {et~al.}(2007){S{\'a}nchez-Bl{\'a}zquez},
  {Forbes}, {Strader}, {Brodie}, \& {Proctor}}]{SB07}
{S{\'a}nchez-Bl{\'a}zquez}, P., {Forbes}, D.~A., {Strader}, J., {Brodie}, J.,
  \& {Proctor}, R. 2007, \mnras, 377, 759

\bibitem[{{S{\'a}nchez-Bl{\'a}zquez} {et~al.}(2011){S{\'a}nchez-Bl{\'a}zquez},
  {Ocvirk}, {Gibson}, {P{\'e}rez}, \& {Peletier}}]{Pat11}
{S{\'a}nchez-Bl{\'a}zquez}, P., {Ocvirk}, P., {Gibson}, B.~K., {P{\'e}rez}, I.,
  \& {Peletier}, R.~F. 2011, \mnras, 415, 709

\bibitem[{{S{\'a}nchez-Bl{\'a}zquez} {et~al.}(2006){S{\'a}nchez-Bl{\'a}zquez},
  {Peletier}, {Jim{\'e}nez-Vicente}, {Cardiel}, {Cenarro},
  {Falc{\'o}n-Barroso}, {Gorgas}, {Selam}, \& {Vazdekis}}]{Pat06}
{S{\'a}nchez-Bl{\'a}zquez}, P., {Peletier}, R.~F., {Jim{\'e}nez-Vicente}, J.,
  {et~al.} 2006, \mnras, 371, 703

\bibitem[{{Sarzi} {et~al.}(2018{\natexlab{a}}){Sarzi}, {Iodice}, {Coccato},
  {Corsini}, {de Zeeuw}, {Falc{\'o}n-Barroso}, {Gadotti}, {Lyubenova},
  {McDermid}, {van de Ven}, {Fahrion}, {Pizzella}, \& {Zhu}}]{f3d}
{Sarzi}, M., {Iodice}, E., {Coccato}, L., {et~al.} 2018{\natexlab{a}}, \aap,
  616, A121

\bibitem[{{Sarzi} {et~al.}(2018{\natexlab{b}}){Sarzi}, {Spiniello}, {La
  Barbera}, {Krajnovi{\'c}}, \& {van den Bosch}}]{Sarzi18}
{Sarzi}, M., {Spiniello}, C., {La Barbera}, F., {Krajnovi{\'c}}, D., \& {van
  den Bosch}, R. 2018{\natexlab{b}}, \mnras, 478, 4084

\bibitem[{{Schiavon}(2007)}]{Schiavon07}
{Schiavon}, R.~P. 2007, \apjs, 171, 146

\bibitem[{{Serra} \& {Trager}(2007)}]{Serra07}
{Serra}, P. \& {Trager}, S.~C. 2007, \mnras, 374, 769

\bibitem[{{Serven} {et~al.}(2005){Serven}, {Worthey}, \& {Briley}}]{serven}
{Serven}, J., {Worthey}, G., \& {Briley}, M.~M. 2005, \apj, 627, 754

\bibitem[{{Shetty} \& {Cappellari}(2014)}]{Shetty14}
{Shetty}, S. \& {Cappellari}, M. 2014, \apjl, 786, L10

\bibitem[{{Sliwa} {et~al.}(2017){Sliwa}, {Wilson}, {Aalto}, \&
  {Privon}}]{Sliwa17}
{Sliwa}, K., {Wilson}, C.~D., {Aalto}, S., \& {Privon}, G.~C. 2017, \apjl, 840,
  L11

\bibitem[{{Smith}(2020)}]{Smith20}
{Smith}, R.~J. 2020, \araa, 58, 577

\bibitem[{{Smith} {et~al.}(2012){Smith}, {Lucey}, \& {Carter}}]{Smith12}
{Smith}, R.~J., {Lucey}, J.~R., \& {Carter}, D. 2012, \mnras, 426, 2994

\bibitem[{{Sonnenfeld} {et~al.}(2019){Sonnenfeld}, {Jaelani}, {Chan}, {More},
  {Suyu}, {Wong}, {Oguri}, \& {Lee}}]{Sonnenfeld19}
{Sonnenfeld}, A., {Jaelani}, A.~T., {Chan}, J., {et~al.} 2019, \aap, 630, A71

\bibitem[{{Sonnenfeld} {et~al.}(2018){Sonnenfeld}, {Leauthaud}, {Auger},
  {Gavazzi}, {Treu}, {More}, \& {Komiyama}}]{Sonnenfeld18}
{Sonnenfeld}, A., {Leauthaud}, A., {Auger}, M.~W., {et~al.} 2018, \mnras, 481,
  164

\bibitem[{{Sonnenfeld} {et~al.}(2015){Sonnenfeld}, {Treu}, {Marshall}, {Suyu},
  {Gavazzi}, {Auger}, \& {Nipoti}}]{Sonnenfeld15}
{Sonnenfeld}, A., {Treu}, T., {Marshall}, P.~J., {et~al.} 2015, \apj, 800, 94

\bibitem[{{Spiniello} {et~al.}(2014){Spiniello}, {Trager}, {Koopmans}, \&
  {Conroy}}]{Spiniello2013}
{Spiniello}, C., {Trager}, S., {Koopmans}, L.~V.~E., \& {Conroy}, C. 2014,
  \mnras, 438, 1483

\bibitem[{{Spiniello} {et~al.}(2015){Spiniello}, {Trager}, \&
  {Koopmans}}]{Spiniello15}
{Spiniello}, C., {Trager}, S.~C., \& {Koopmans}, L.~V.~E. 2015, \apj, 803, 87

\bibitem[{{Spiniello} {et~al.}(2012){Spiniello}, {Trager}, {Koopmans}, \&
  {Chen}}]{spiniello12}
{Spiniello}, C., {Trager}, S.~C., {Koopmans}, L.~V.~E., \& {Chen}, Y.~P. 2012,
  \apjl, 753, L32

\bibitem[{{Spriggs} {et~al.}(2020){Spriggs}, {Sarzi}, {Napiwotzki},
  {Gal{\'a}n-de Anta}, {Viaene}, {Nedelchev}, {Coccato}, {Corsini}, {de Zeeuw},
  {Falc{\'o}n-Barroso}, {Gadotti}, {Iodice}, {Lyubenova},
  {Mart{\'\i}n-Navarro}, {McDermid}, {Pinna}, {van de Ven}, \&
  {Zhu}}]{Spriggs20}
{Spriggs}, T.~W., {Sarzi}, M., {Napiwotzki}, R., {et~al.} 2020, \aap, 637, A62

\bibitem[{{Sybilska} {et~al.}(2018){Sybilska}, {Kuntschner}, {van de Ven},
  {Vazdekis}, {Falc{\'o}n-Barroso}, {Peletier}, \& {Lisker}}]{Sybilska18}
{Sybilska}, A., {Kuntschner}, H., {van de Ven}, G., {et~al.} 2018, \mnras, 476,
  4501

\bibitem[{{Tang} \& {Worthey}(2017)}]{Tang17}
{Tang}, B. \& {Worthey}, G. 2017, \mnras, 467, 674

\bibitem[{{Thomas} {et~al.}(1999){Thomas}, {Greggio}, \& {Bender}}]{Thomas99}
{Thomas}, D., {Greggio}, L., \& {Bender}, R. 1999, \mnras, 302, 537

\bibitem[{{Thomas} {et~al.}(2003){Thomas}, {Maraston}, \& {Bender}}]{TMB:03}
{Thomas}, D., {Maraston}, C., \& {Bender}, R. 2003, \mnras, 339, 897

\bibitem[{{Thomas} {et~al.}(2005){Thomas}, {Maraston}, {Bender}, \& {Mendes de
  Oliveira}}]{Thomas05}
{Thomas}, D., {Maraston}, C., {Bender}, R., \& {Mendes de Oliveira}, C. 2005,
  \apj, 621, 673

\bibitem[{{Thomas} {et~al.}(2011{\natexlab{a}}){Thomas}, {Maraston}, \&
  {Johansson}}]{TMJ}
{Thomas}, D., {Maraston}, C., \& {Johansson}, J. 2011{\natexlab{a}}, \mnras,
  412, 2183

\bibitem[{{Thomas} {et~al.}(2010){Thomas}, {Maraston}, {Schawinski}, {Sarzi},
  \& {Silk}}]{Thomas10}
{Thomas}, D., {Maraston}, C., {Schawinski}, K., {Sarzi}, M., \& {Silk}, J.
  2010, \mnras, 404, 1775

\bibitem[{{Thomas} {et~al.}(2011{\natexlab{b}}){Thomas}, {Saglia}, {Bender},
  {Thomas}, {Gebhardt}, {Magorrian}, {Corsini}, {Wegner}, \&
  {Seitz}}]{thomas11}
{Thomas}, J., {Saglia}, R.~P., {Bender}, R., {et~al.} 2011{\natexlab{b}},
  \mnras, 415, 545

\bibitem[{{Tinsley}(1979)}]{Tinsley79}
{Tinsley}, B.~M. 1979, \apj, 229, 1046

\bibitem[{{Tolstoy} {et~al.}(2009){Tolstoy}, {Hill}, \& {Tosi}}]{Tolstoy09}
{Tolstoy}, E., {Hill}, V., \& {Tosi}, M. 2009, \araa, 47, 371

\bibitem[{{Tortora} {et~al.}(2014){Tortora}, {La Barbera}, {Napolitano},
  {Romanowsky}, {Ferreras}, \& {de Carvalho}}]{Tortora14}
{Tortora}, C., {La Barbera}, F., {Napolitano}, N.~R., {et~al.} 2014, \mnras,
  445, 115

\bibitem[{{Tortora} {et~al.}(2018){Tortora}, {Napolitano}, {Roy}, {Radovich},
  {Getman}, {Koopmans}, {Verdoes Kleijn}, \& {Kuijken}}]{Tortora18}
{Tortora}, C., {Napolitano}, N.~R., {Roy}, N., {et~al.} 2018, \mnras, 473, 969

\bibitem[{{Tortora} {et~al.}(2013){Tortora}, {Romanowsky}, \&
  {Napolitano}}]{Tortora13b}
{Tortora}, C., {Romanowsky}, A.~J., \& {Napolitano}, N.~R. 2013, \apj, 765, 8

\bibitem[{{Trager} {et~al.}(2000){Trager}, {Faber}, {Worthey}, \&
  {Gonz{\'a}lez}}]{Trager00}
{Trager}, S.~C., {Faber}, S.~M., {Worthey}, G., \& {Gonz{\'a}lez}, J.~J. 2000,
  \aj, 119, 1645

\bibitem[{{Trager} {et~al.}(1998){Trager}, {Worthey}, {Faber}, {Burstein}, \&
  {Gonzalez}}]{trager}
{Trager}, S.~C., {Worthey}, G., {Faber}, S.~M., {Burstein}, D., \& {Gonzalez},
  J.~J. 1998, \apjs, 116, 1

\bibitem[{{Treu} {et~al.}(2010){Treu}, {Auger}, {Koopmans}, {Gavazzi},
  {Marshall}, \& {Bolton}}]{Treu}
{Treu}, T., {Auger}, M.~W., {Koopmans}, L.~V.~E., {et~al.} 2010, \apj, 709,
  1195

\bibitem[{{van Dokkum} {et~al.}(2017){van Dokkum}, {Conroy}, {Villaume},
  {Brodie}, \& {Romanowsky}}]{vdk17}
{van Dokkum}, P., {Conroy}, C., {Villaume}, A., {Brodie}, J., \& {Romanowsky},
  A.~J. 2017, \apj, 841, 68

\bibitem[{{van Dokkum}(2008)}]{Pieter08}
{van Dokkum}, P.~G. 2008, \apj, 674, 29

\bibitem[{{van Dokkum} \& {Conroy}(2010)}]{vandokkum}
{van Dokkum}, P.~G. \& {Conroy}, C. 2010, \nat, 468, 940

\bibitem[{{Vaughan} {et~al.}(2018){Vaughan}, {Davies}, {Zieleniewski}, \&
  {Houghton}}]{Vaughan18a}
{Vaughan}, S.~P., {Davies}, R.~L., {Zieleniewski}, S., \& {Houghton}, R.~C.~W.
  2018, \mnras, 479, 2443

\bibitem[{{Vazdekis} {et~al.}(1996){Vazdekis}, {Casuso}, {Peletier}, \&
  {Beckman}}]{Vazdekis96}
{Vazdekis}, A., {Casuso}, E., {Peletier}, R.~F., \& {Beckman}, J.~E. 1996,
  \apjs, 106, 307

\bibitem[{{Vazdekis} {et~al.}(2015){Vazdekis}, {Coelho}, {Cassisi},
  {Ricciardelli}, {Falc{\'o}n-Barroso}, {S{\'a}nchez-Bl{\'a}zquez}, {La
  Barbera}, {Beasley}, \& {Pietrinferni}}]{Vazdekis15}
{Vazdekis}, A., {Coelho}, P., {Cassisi}, S., {et~al.} 2015, \mnras, 449, 1177

\bibitem[{{Vazdekis} {et~al.}(2010){Vazdekis}, {S{\'a}nchez-Bl{\'a}zquez},
  {Falc{\'o}n-Barroso}, {Cenarro}, {Beasley}, {Cardiel}, {Gorgas}, \&
  {Peletier}}]{miles}
{Vazdekis}, A., {S{\'a}nchez-Bl{\'a}zquez}, P., {Falc{\'o}n-Barroso}, J.,
  {et~al.} 2010, \mnras, 404, 1639

\bibitem[{{Viaene} {et~al.}(2019){Viaene}, {Sarzi}, {Zabel}, {Coccato},
  {Corsini}, {Davis}, {De Vis}, {de Zeeuw}, {Falc{\'o}n-Barroso}, {Gadotti},
  {Iodice}, {Lyubenova}, {McDermid}, {Morelli}, {Nedelchev}, {Pinna},
  {Spriggs}, \& {van de Ven}}]{Viaene19}
{Viaene}, S., {Sarzi}, M., {Zabel}, N., {et~al.} 2019, \aap, 622, A89

\bibitem[{{Villaume} {et~al.}(2017{\natexlab{a}}){Villaume}, {Brodie},
  {Conroy}, {Romanowsky}, \& {van Dokkum}}]{Alexa17}
{Villaume}, A., {Brodie}, J., {Conroy}, C., {Romanowsky}, A.~J., \& {van
  Dokkum}, P. 2017{\natexlab{a}}, \apjl, 850, L14

\bibitem[{{Villaume} {et~al.}(2017{\natexlab{b}}){Villaume}, {Conroy},
  {Johnson}, {Rayner}, {Mann}, \& {van Dokkum}}]{Villaume17}
{Villaume}, A., {Conroy}, C., {Johnson}, B., {et~al.} 2017{\natexlab{b}},
  \apjs, 230, 23

\bibitem[{{Vincenzo} {et~al.}(2018){Vincenzo}, {Kobayashi}, \&
  {Taylor}}]{Vincenzo18}
{Vincenzo}, F., {Kobayashi}, C., \& {Taylor}, P. 2018, \mnras, 480, L38

\bibitem[{{Walcher} {et~al.}(2015){Walcher}, {Coelho}, {Gallazzi}, {Bruzual},
  {Charlot}, \& {Chiappini}}]{Walcher15}
{Walcher}, C.~J., {Coelho}, P.~R.~T., {Gallazzi}, A., {et~al.} 2015, \aap, 582,
  A46

\bibitem[{{Wallerstein}(1962)}]{Wallerstein62}
{Wallerstein}, G. 1962, \apjs, 6, 407

\bibitem[{{Wegner} {et~al.}(2012){Wegner}, {Corsini}, {Thomas}, {Saglia},
  {Bender}, \& {Pu}}]{wegner12}
{Wegner}, G.~A., {Corsini}, E.~M., {Thomas}, J., {et~al.} 2012, \aj, 144, 78

\bibitem[{{Weidner} {et~al.}(2013){Weidner}, {Ferreras}, {Vazdekis}, \& {La
  Barbera}}]{weidner:13}
{Weidner}, C., {Ferreras}, I., {Vazdekis}, A., \& {La Barbera}, F. 2013,
  \mnras, 435, 2274

\bibitem[{{Wilkinson} {et~al.}(2017){Wilkinson}, {Maraston}, {Goddard},
  {Thomas}, \& {Parikh}}]{Wilkinson17}
{Wilkinson}, D.~M., {Maraston}, C., {Goddard}, D., {Thomas}, D., \& {Parikh},
  T. 2017, \mnras, 472, 4297

\bibitem[{{Worthey}(1994)}]{Worthey94}
{Worthey}, G. 1994, \apjs, 95, 107

\bibitem[{{Worthey} {et~al.}(1992){Worthey}, {Faber}, \&
  {Gonzalez}}]{Worthey92}
{Worthey}, G., {Faber}, S.~M., \& {Gonzalez}, J.~J. 1992, \apj, 398, 69

\bibitem[{{Worthey} {et~al.}(2014){Worthey}, {Tang}, \& {Serven}}]{worthey14}
{Worthey}, G., {Tang}, B., \& {Serven}, J. 2014, \apj, 783, 20

\bibitem[{{Yan} {et~al.}(2019){Yan}, {Jerabkova}, {Kroupa}, \&
  {Vazdekis}}]{Yan19}
{Yan}, Z., {Jerabkova}, T., {Kroupa}, P., \& {Vazdekis}, A. 2019, \aap, 629,
  A93

\bibitem[{{Y{\i}ld{\i}r{\i}m} {et~al.}(2017){Y{\i}ld{\i}r{\i}m}, {van den
  Bosch}, {van de Ven}, {Mart{\'{\i}}n-Navarro}, {Walsh}, {Husemann},
  {G{\"u}ltekin}, \& {Gebhardt}}]{akin17}
{Y{\i}ld{\i}r{\i}m}, A., {van den Bosch}, R.~C.~E., {van de Ven}, G., {et~al.}
  2017, \mnras, 468, 4216

\bibitem[{{Zabel} {et~al.}(2020){Zabel}, {Davis}, {Sarzi}, {Nedelchev},
  {Chevance}, {Kruijssen}, {Iodice}, {Baes}, {Bendo}, {Corsini}, {De Looze},
  {de Zeeuw}, {Gadotti}, {Grossi}, {Peletier}, {Pinna}, {Serra}, {van de
  Voort}, {Venhola}, {Viaene}, \& {Vlahakis}}]{Zabel20}
{Zabel}, N., {Davis}, T.~A., {Sarzi}, M., {et~al.} 2020, \mnras, 496, 2155

\bibitem[{{Zhang} {et~al.}(2018){Zhang}, {Romano}, {Ivison}, {Papadopoulos}, \&
  {Matteucci}}]{Zhang18}
{Zhang}, Z.-Y., {Romano}, D., {Ivison}, R.~J., {Papadopoulos}, P.~P., \&
  {Matteucci}, F. 2018, \nat, 558, 260

\bibitem[{{Zhou} {et~al.}(2019){Zhou}, {Mo}, {Li}, {Zheng}, {Li}, {Du}, {Mao},
  {Parikh}, {Lane}, \& {Thomas}}]{Zhou19}
{Zhou}, S., {Mo}, H.~J., {Li}, C., {et~al.} 2019, \mnras, 485, 5256

\bibitem[{{Zhu} {et~al.}(2020){Zhu}, {van de Ven}, {Leaman}, {Grand},
  {Falc{\'o}n-Barroso}, {Jethwa}, {Watkins}, {Mao}, {Poci}, {McDermid}, \&
  {Nelson}}]{Ling20}
{Zhu}, L., {van de Ven}, G., {Leaman}, R., {et~al.} 2020, \mnras, 496, 1579

\bibitem[{{Zieleniewski} {et~al.}(2017){Zieleniewski}, {Houghton}, {Thatte},
  {Davies}, \& {Vaughan}}]{Zieleniewski17}
{Zieleniewski}, S., {Houghton}, R.~C.~W., {Thatte}, N., {Davies}, R.~L., \&
  {Vaughan}, S.~P. 2017, \mnras, 465, 192

\end{thebibliography}

\begin{appendix} 
   \section{Stellar population maps}
   \label{sec:maps}
   \begin{figure*}
      \centering
      \includegraphics[width=6.1cm]{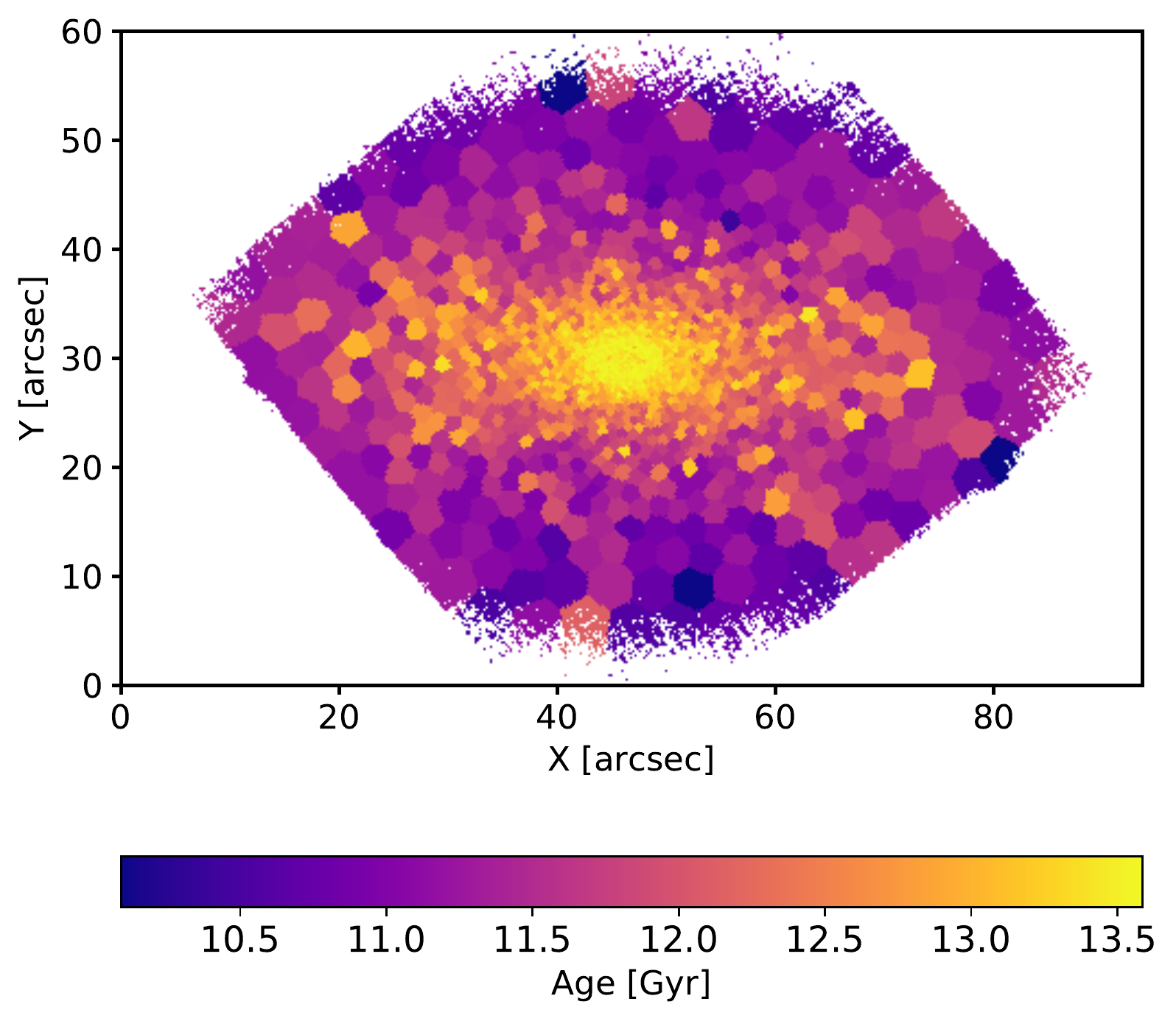}
      \includegraphics[width=6.1cm]{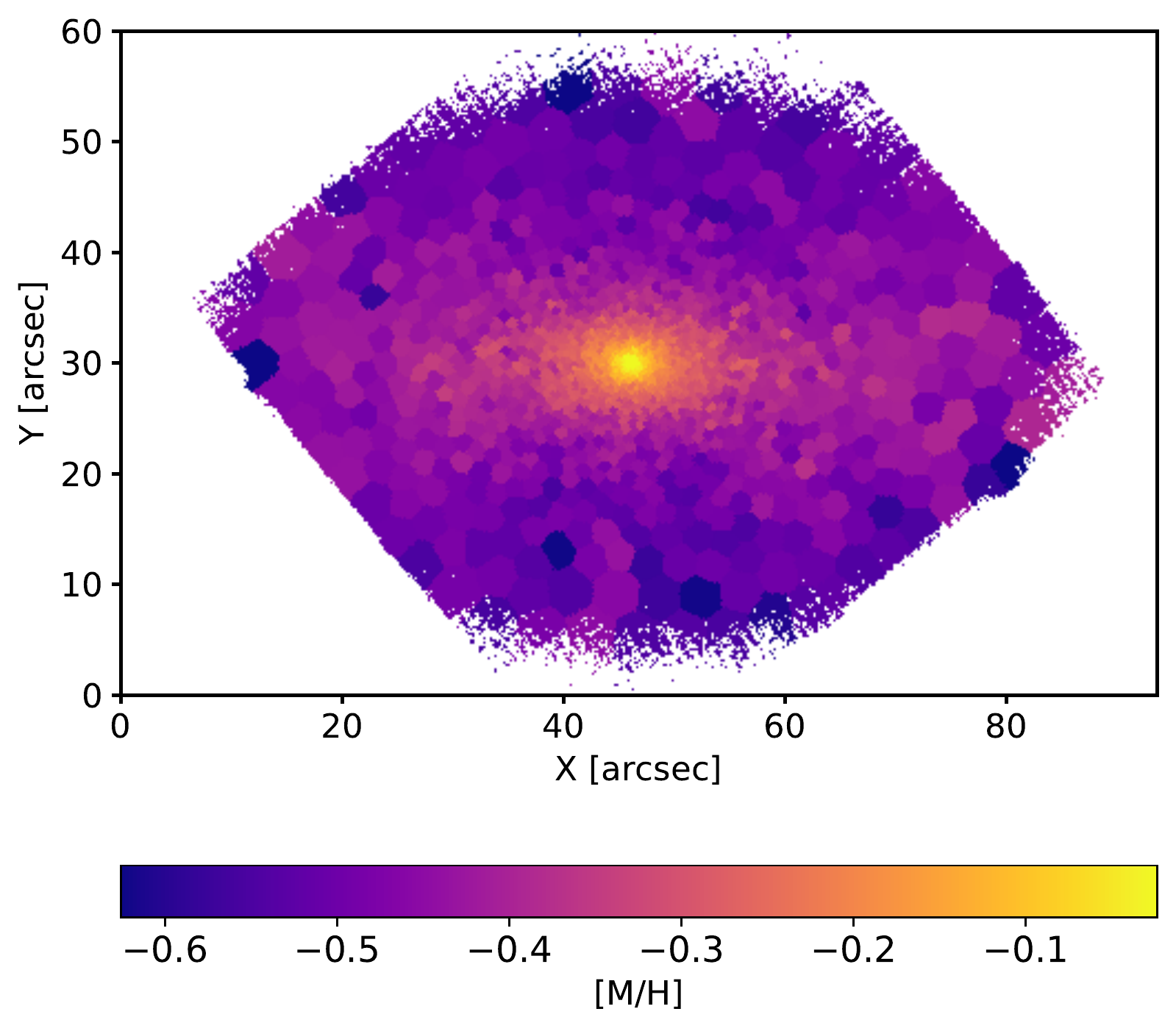}
      \includegraphics[width=6.1cm]{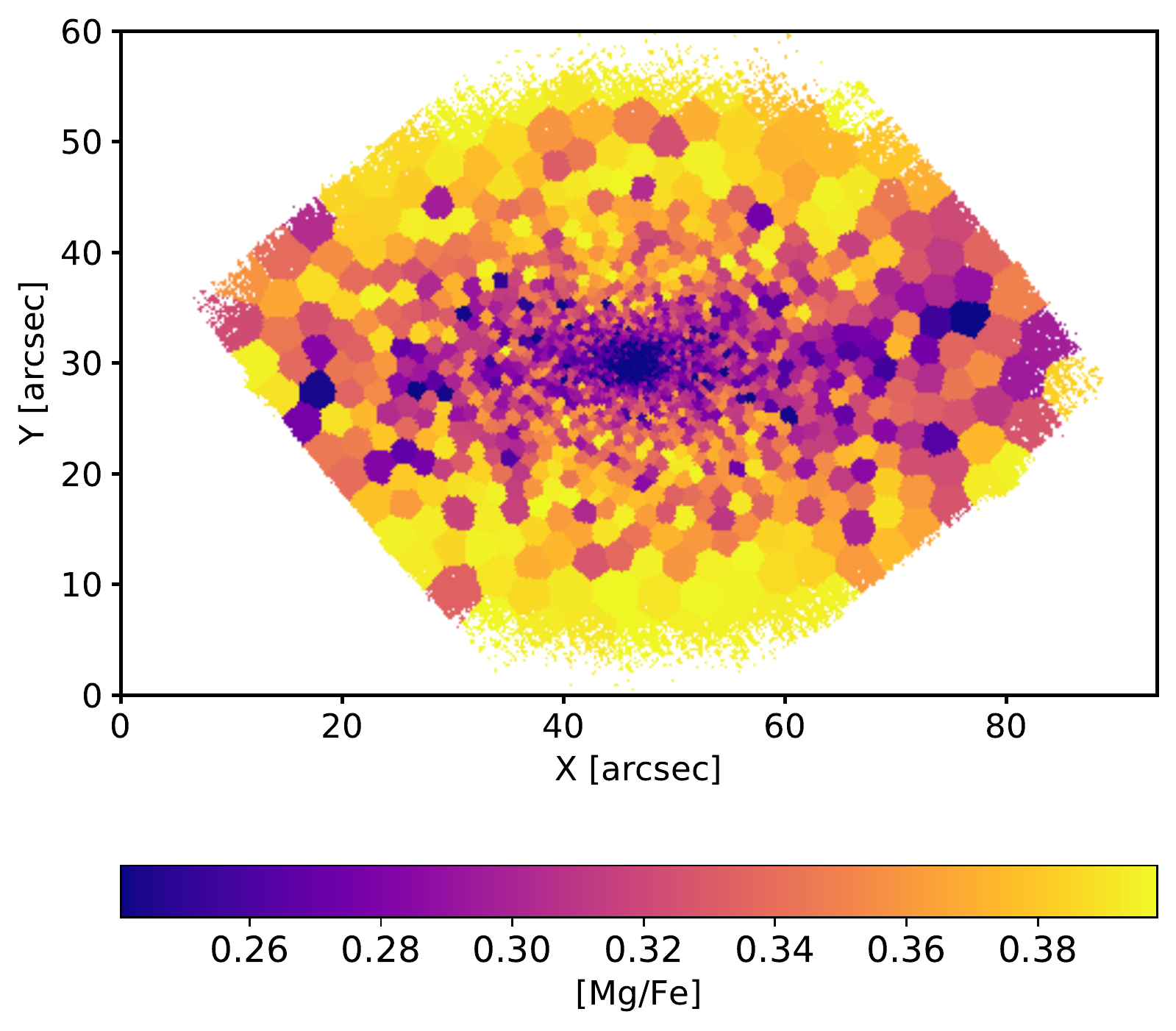}
      \includegraphics[width=6.1cm]{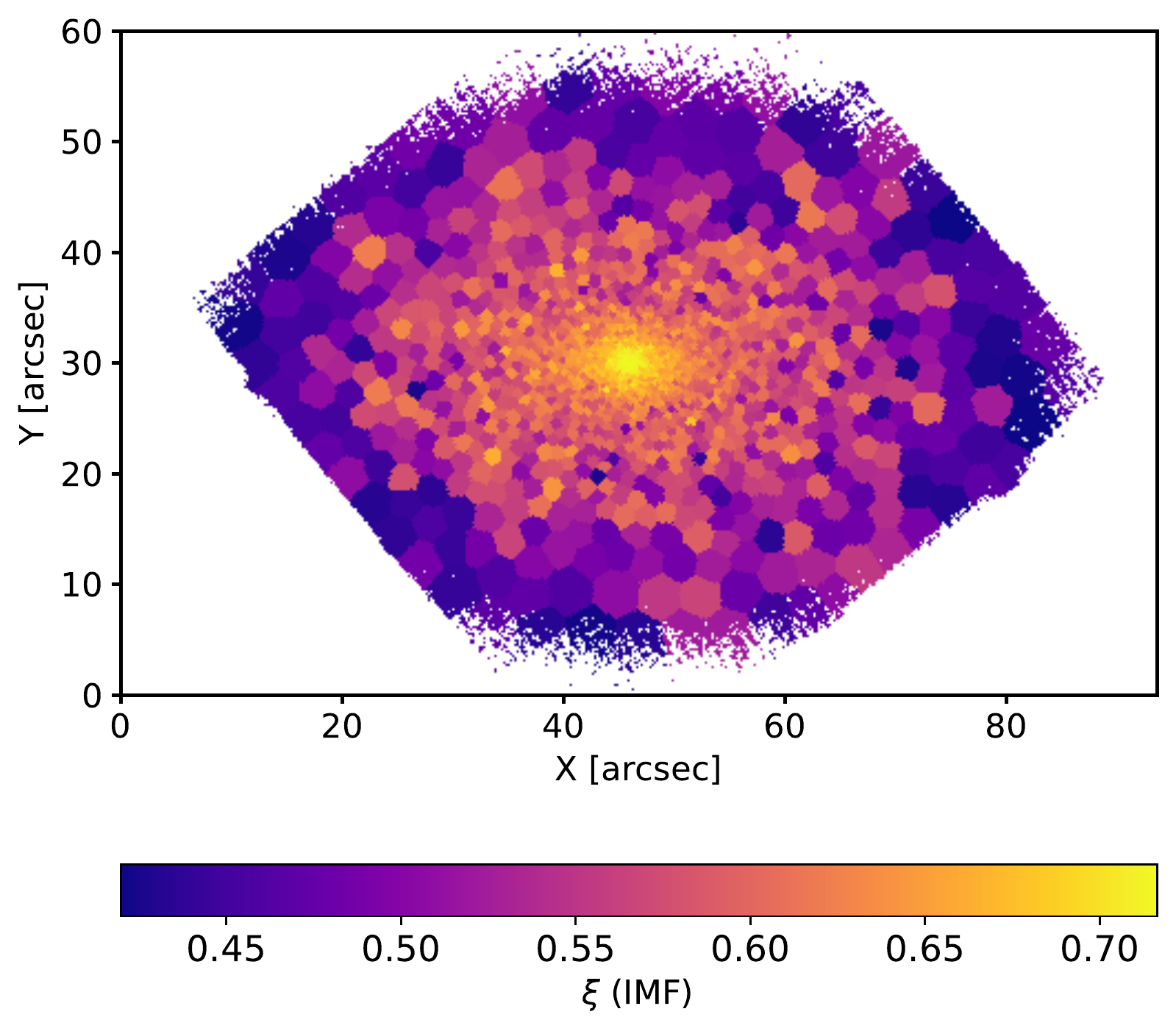}
      \caption{F3D stellar population maps of FCC\,083. From left to right and top to bottom: age, metallicity, [Mg/Fe], and IMF slope maps} 
   \end{figure*}
   \begin{figure*}
      \centering
      \includegraphics[width=6.1cm]{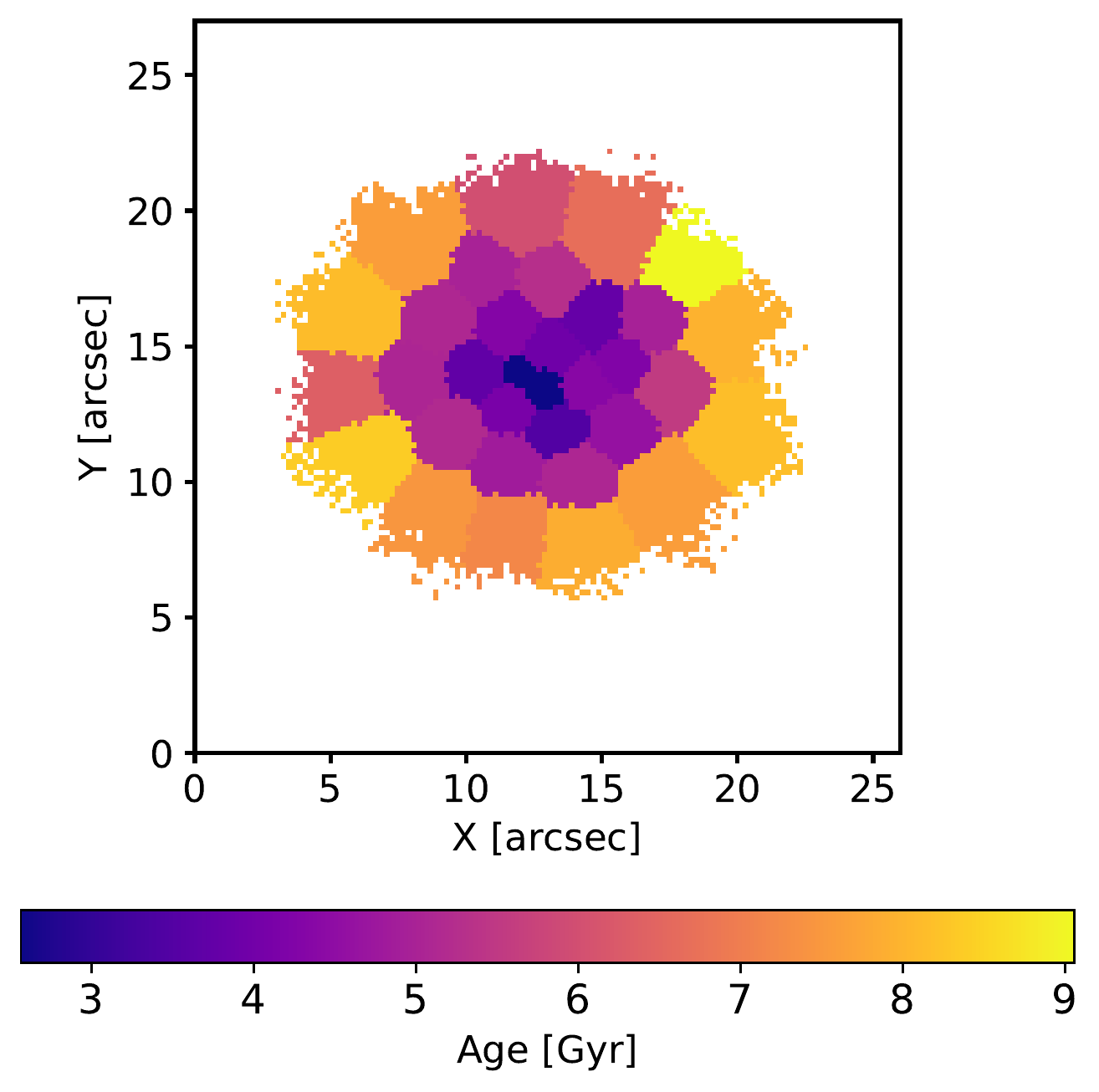}
      \includegraphics[width=6.1cm]{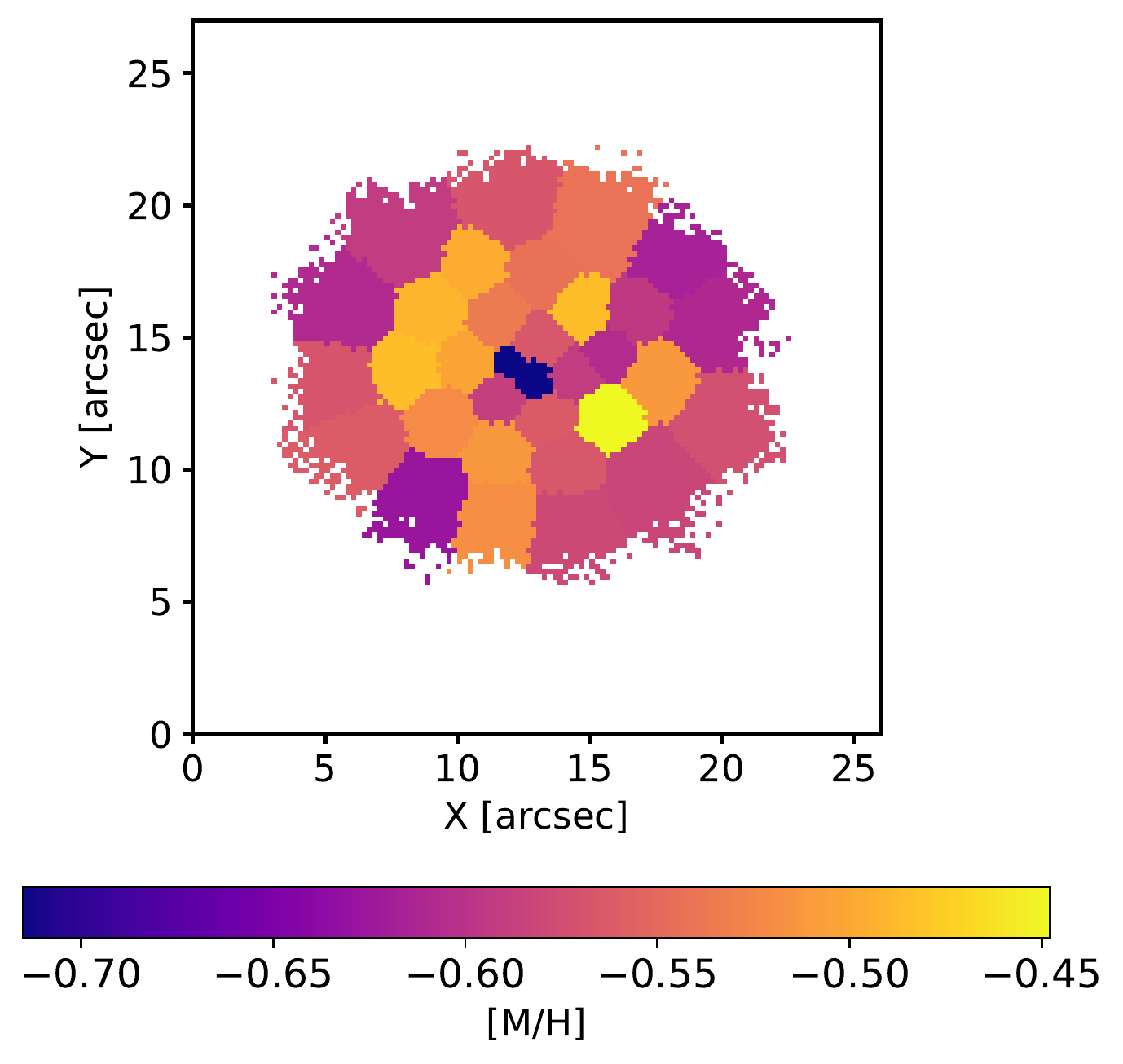}
      \includegraphics[width=6.1cm]{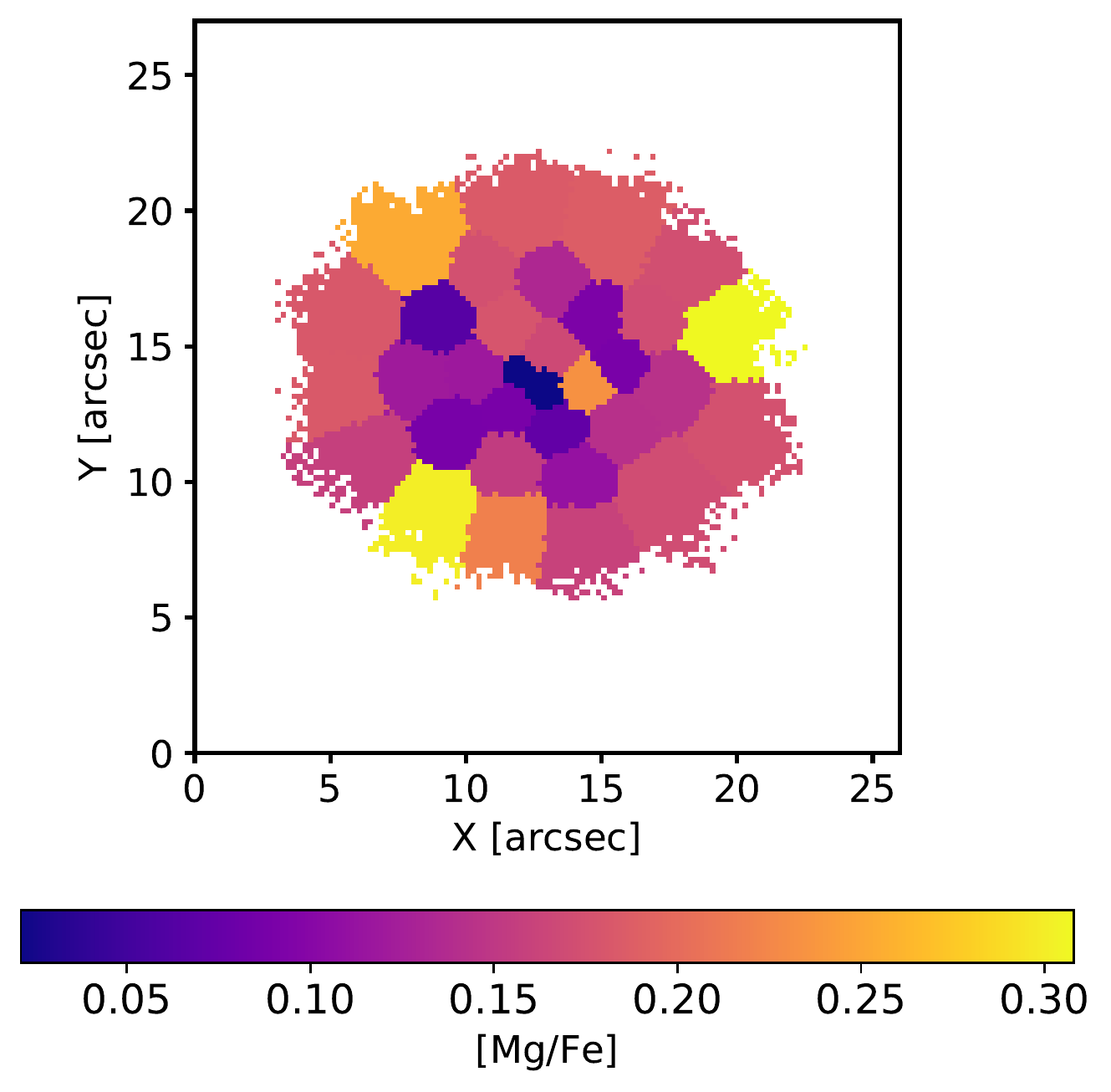}
      \includegraphics[width=6.1cm]{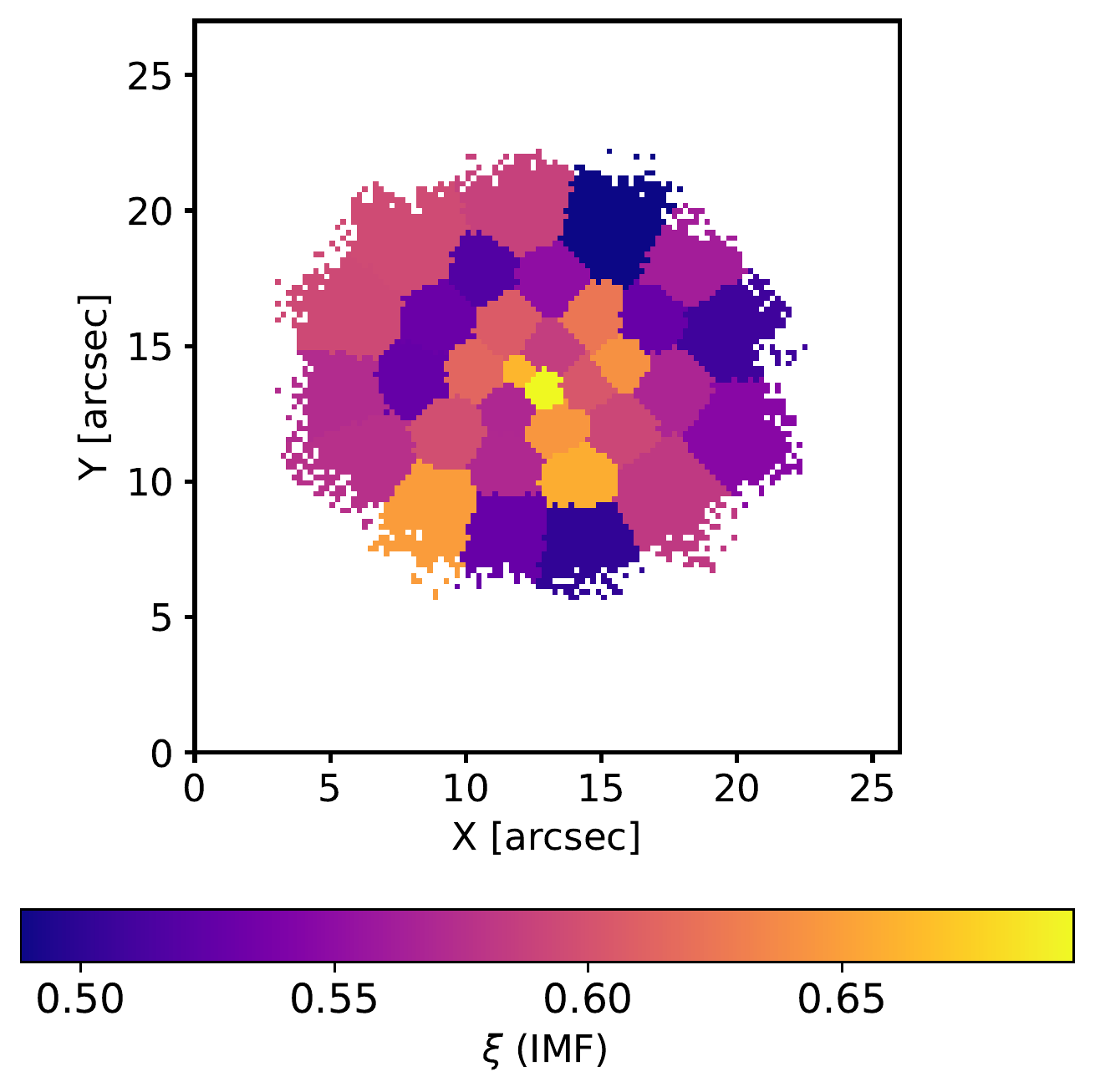}
      \caption{F3D stellar population maps of FCC\,119.} 
   \end{figure*}

   \begin{figure*}
      \centering
      \includegraphics[width=6.1cm]{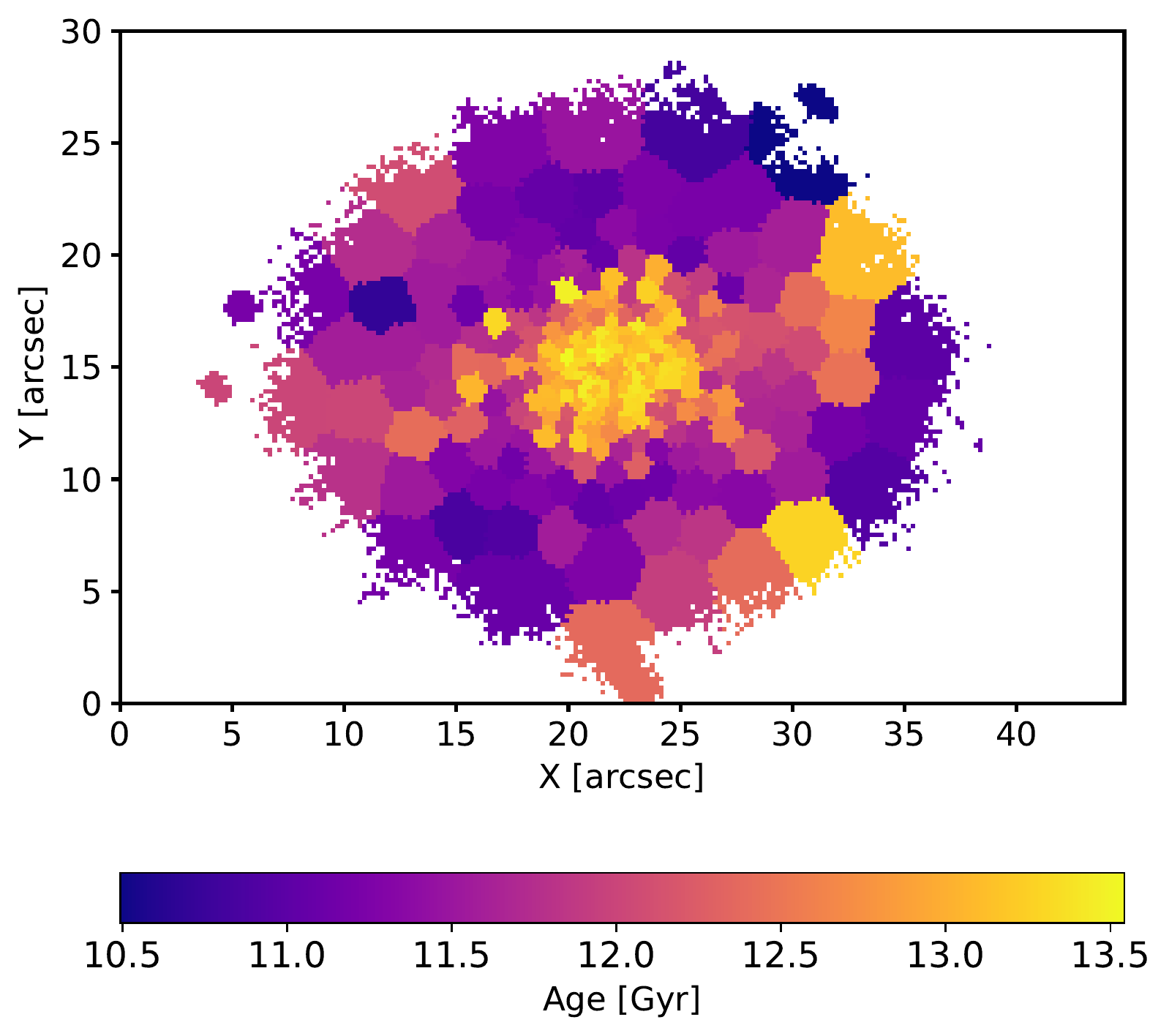}
      \includegraphics[width=6.1cm]{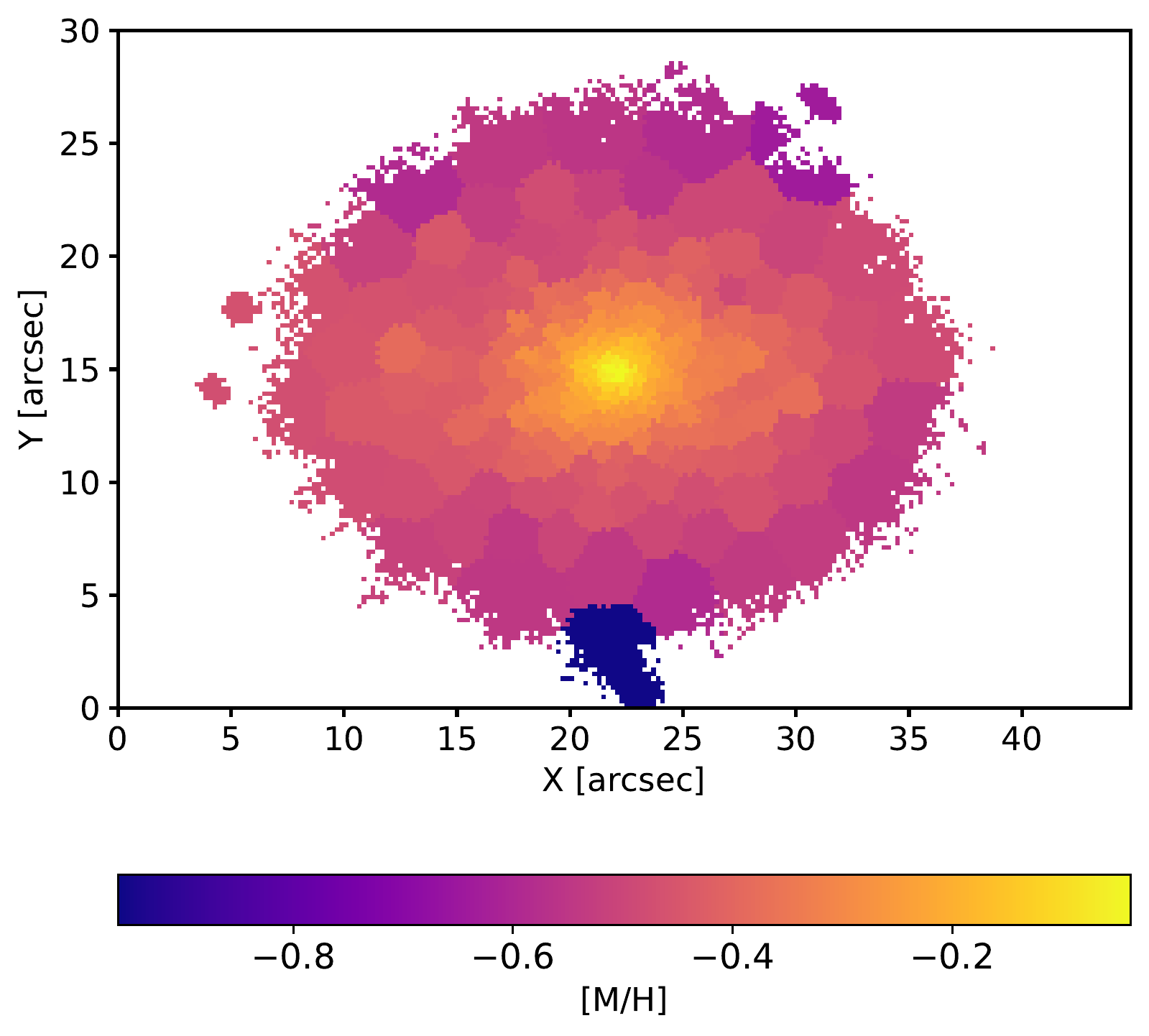}
      \includegraphics[width=6.1cm]{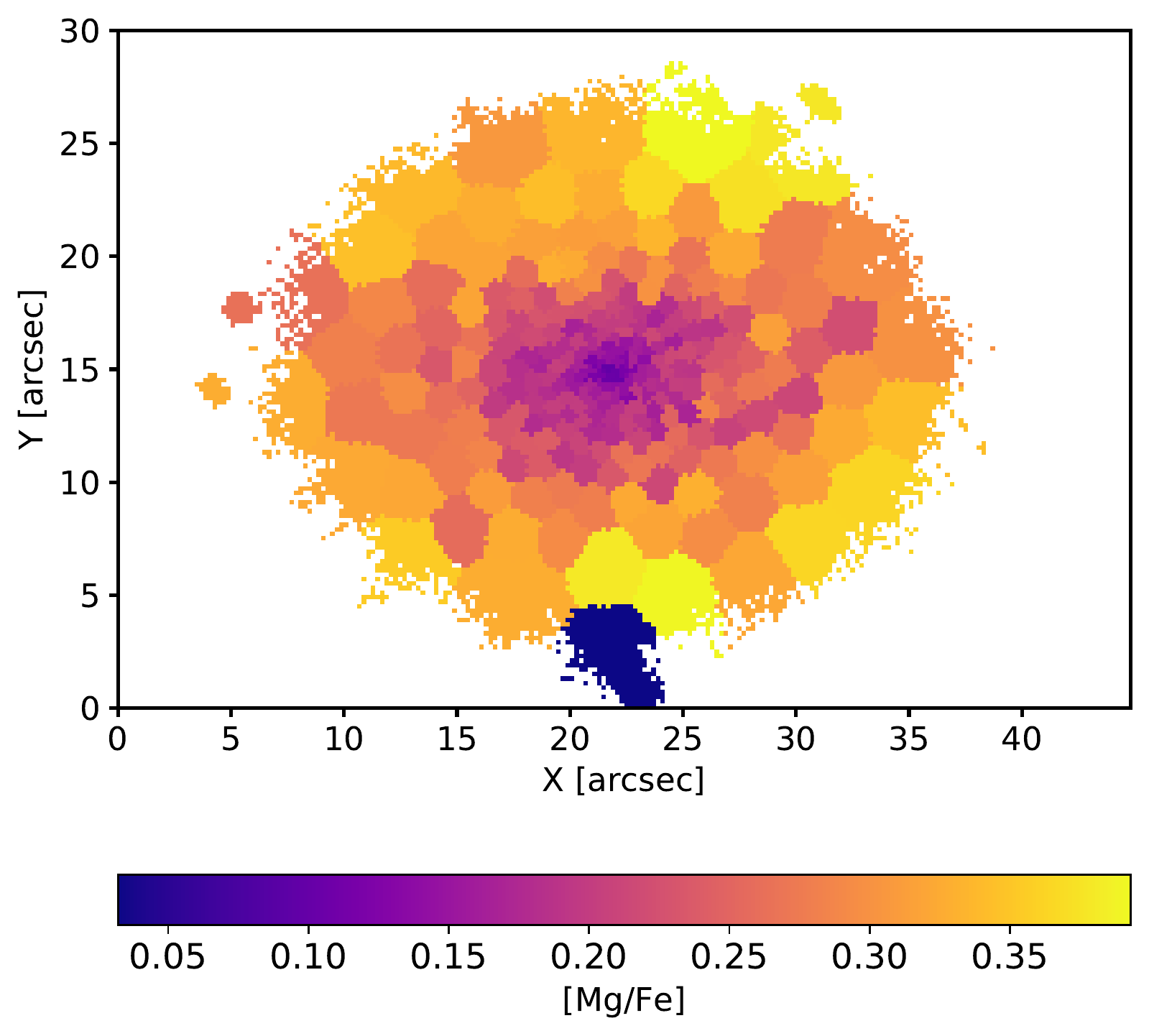}
      \includegraphics[width=6.1cm]{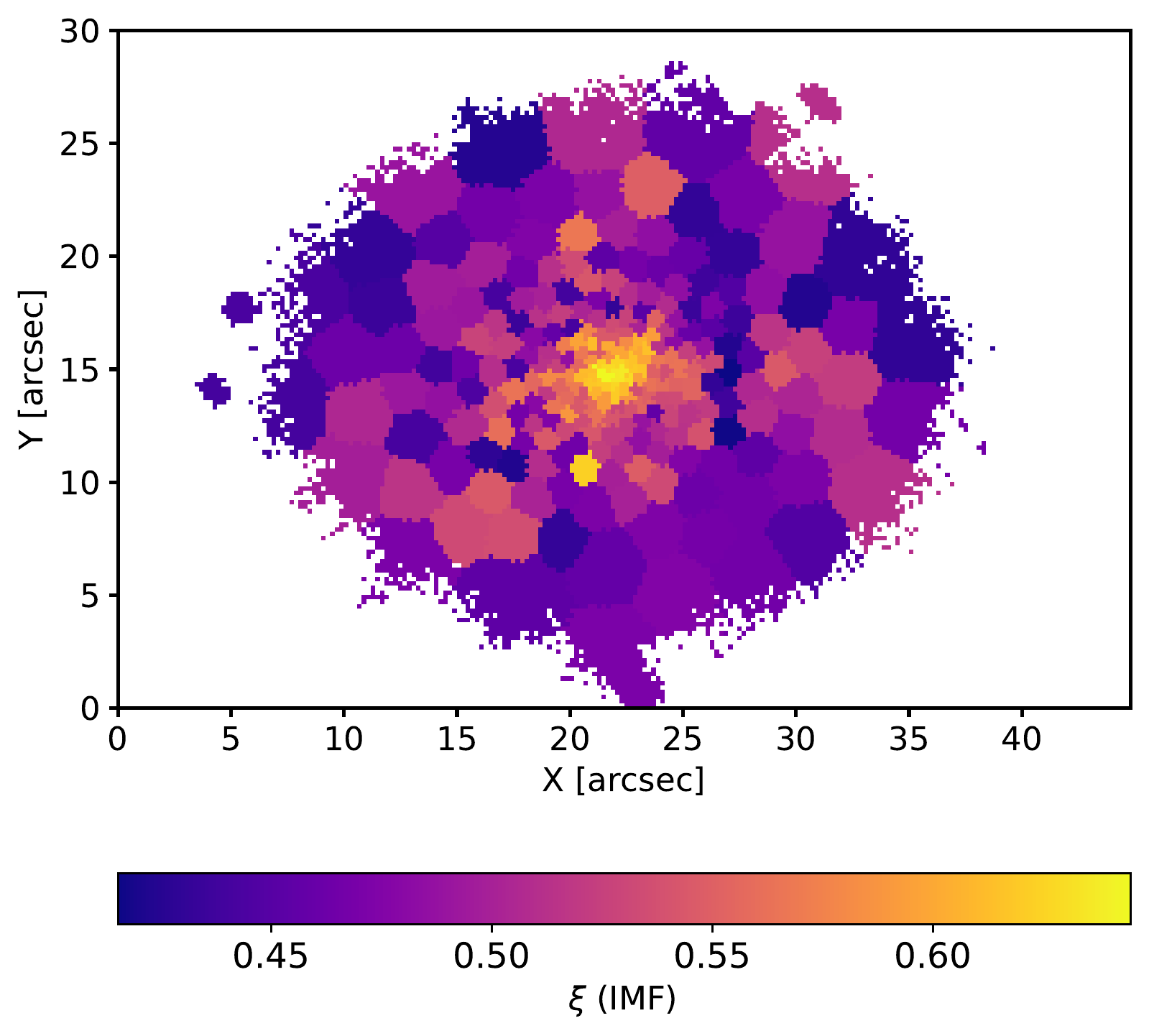}
      \caption{F3D stellar population maps of FCC\,143. From left to right and top to bottom: age, metallicity, [Mg/Fe], and IMF slope maps} 
   \end{figure*}

   \begin{figure*}
      \centering
      \includegraphics[width=6.1cm]{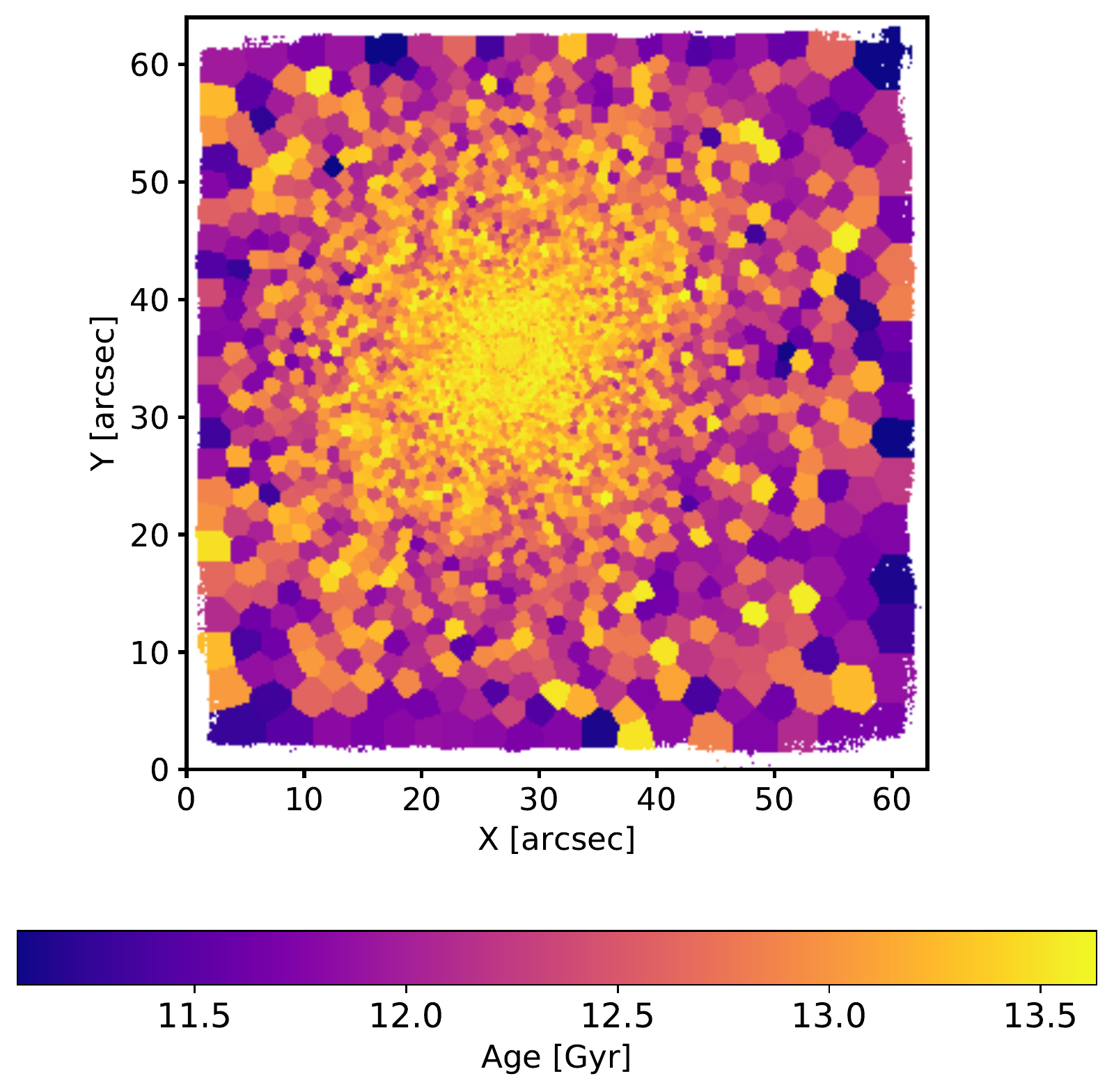}
      \includegraphics[width=6.1cm]{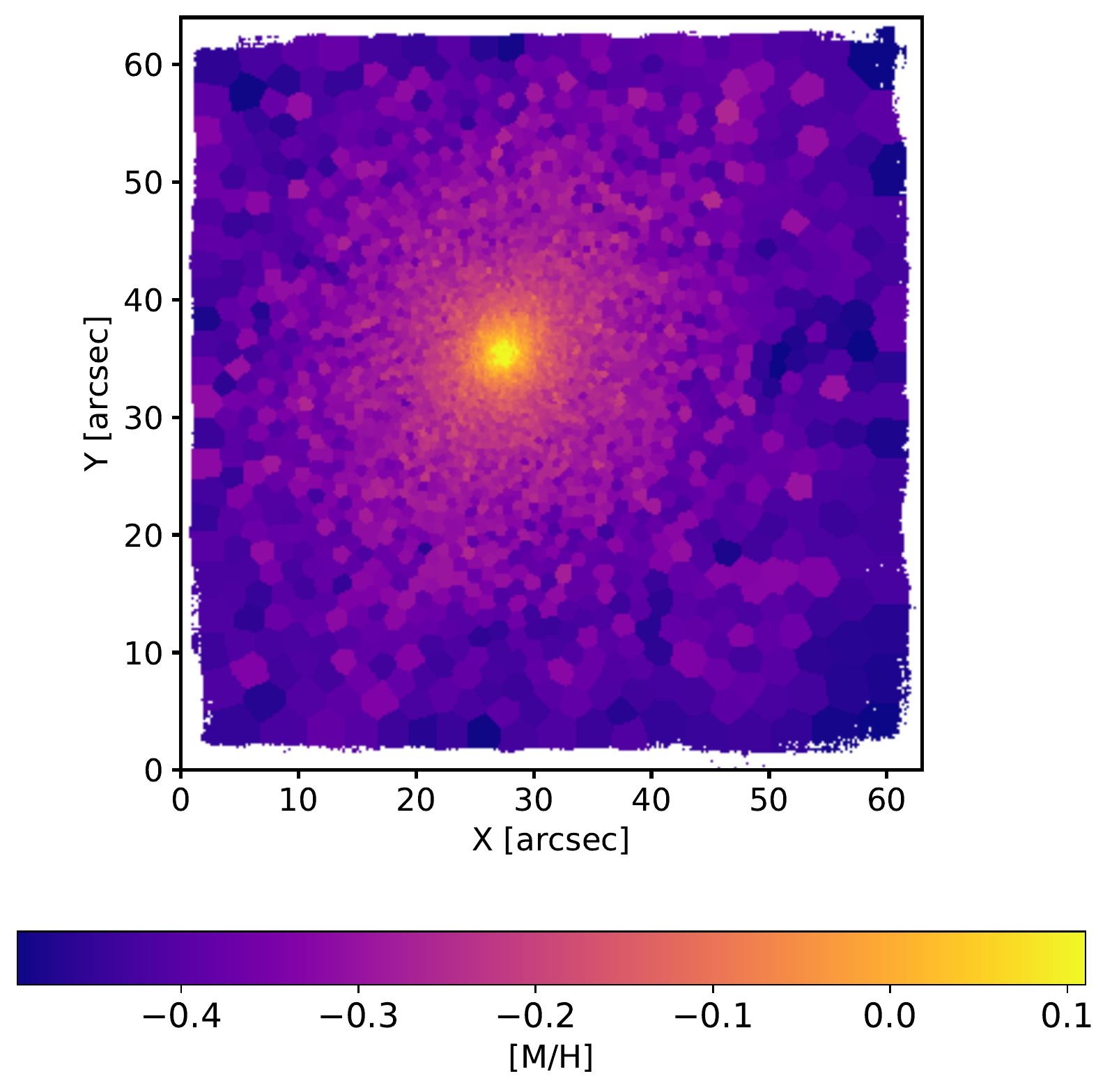}
      \includegraphics[width=6.1cm]{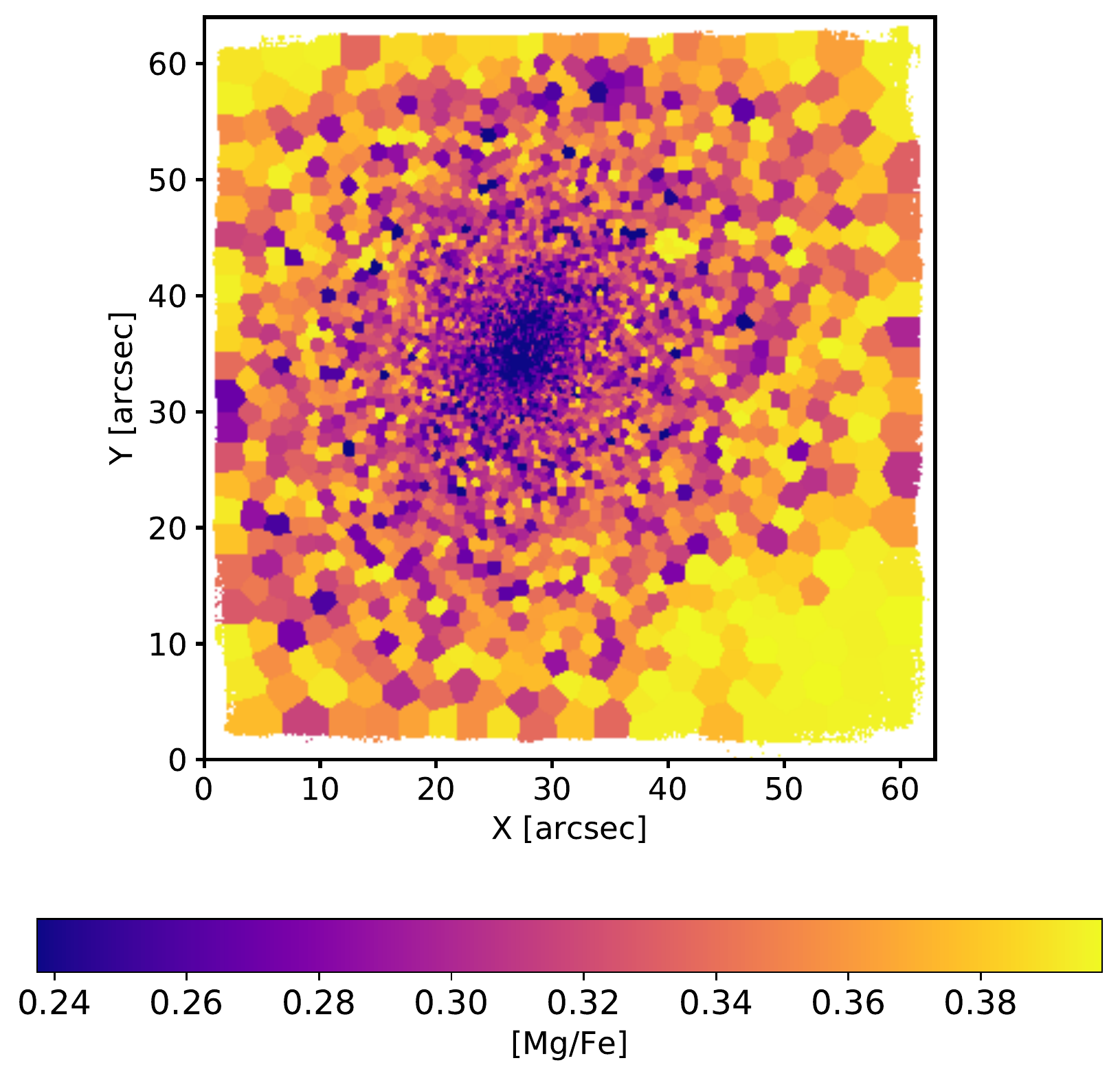}
      \includegraphics[width=6.1cm]{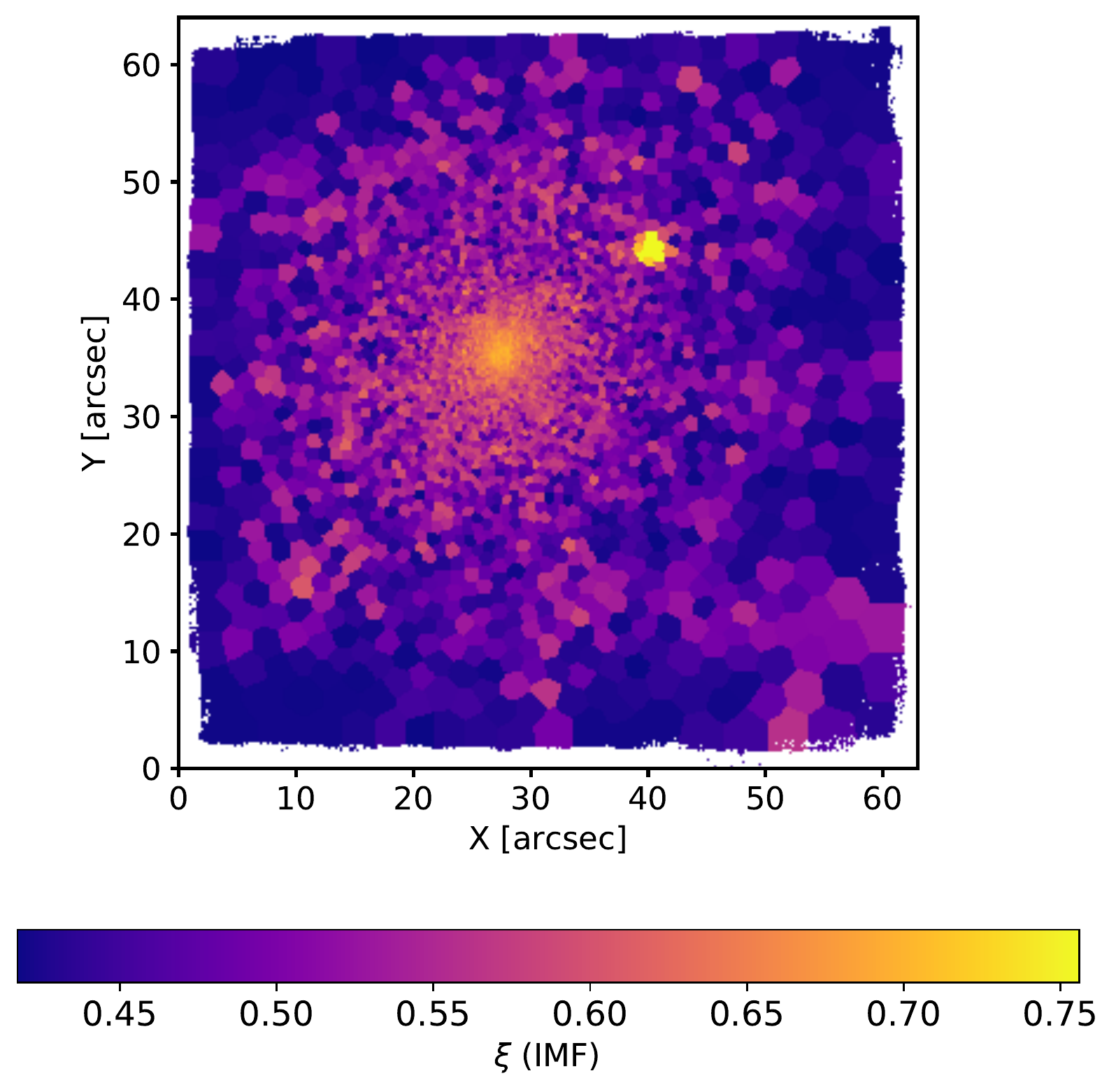}
      \caption{F3D stellar population maps of FCC\,147.} 
   \end{figure*}

   \begin{figure*}
      \centering
      \includegraphics[width=7.1cm]{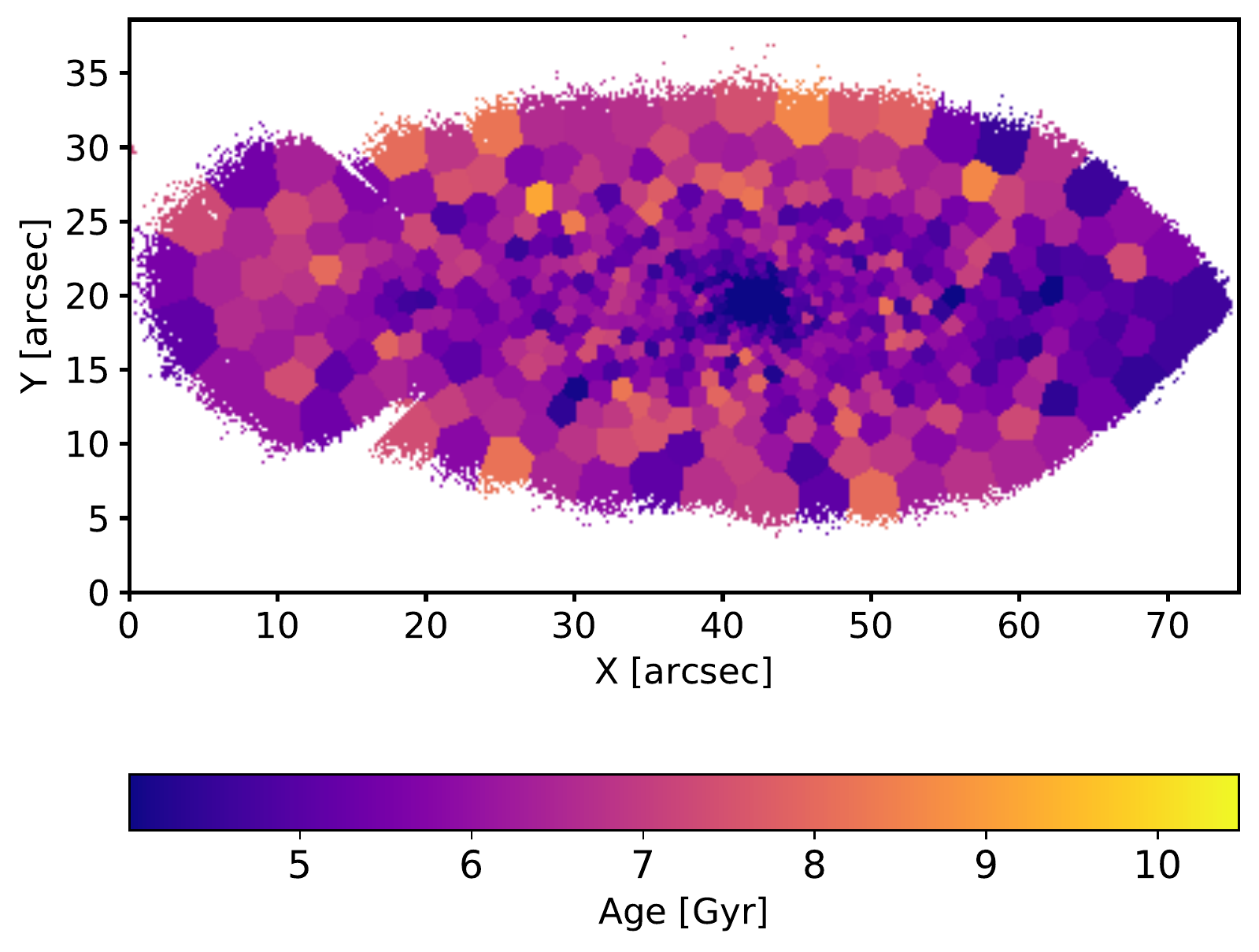}
      \includegraphics[width=7.1cm]{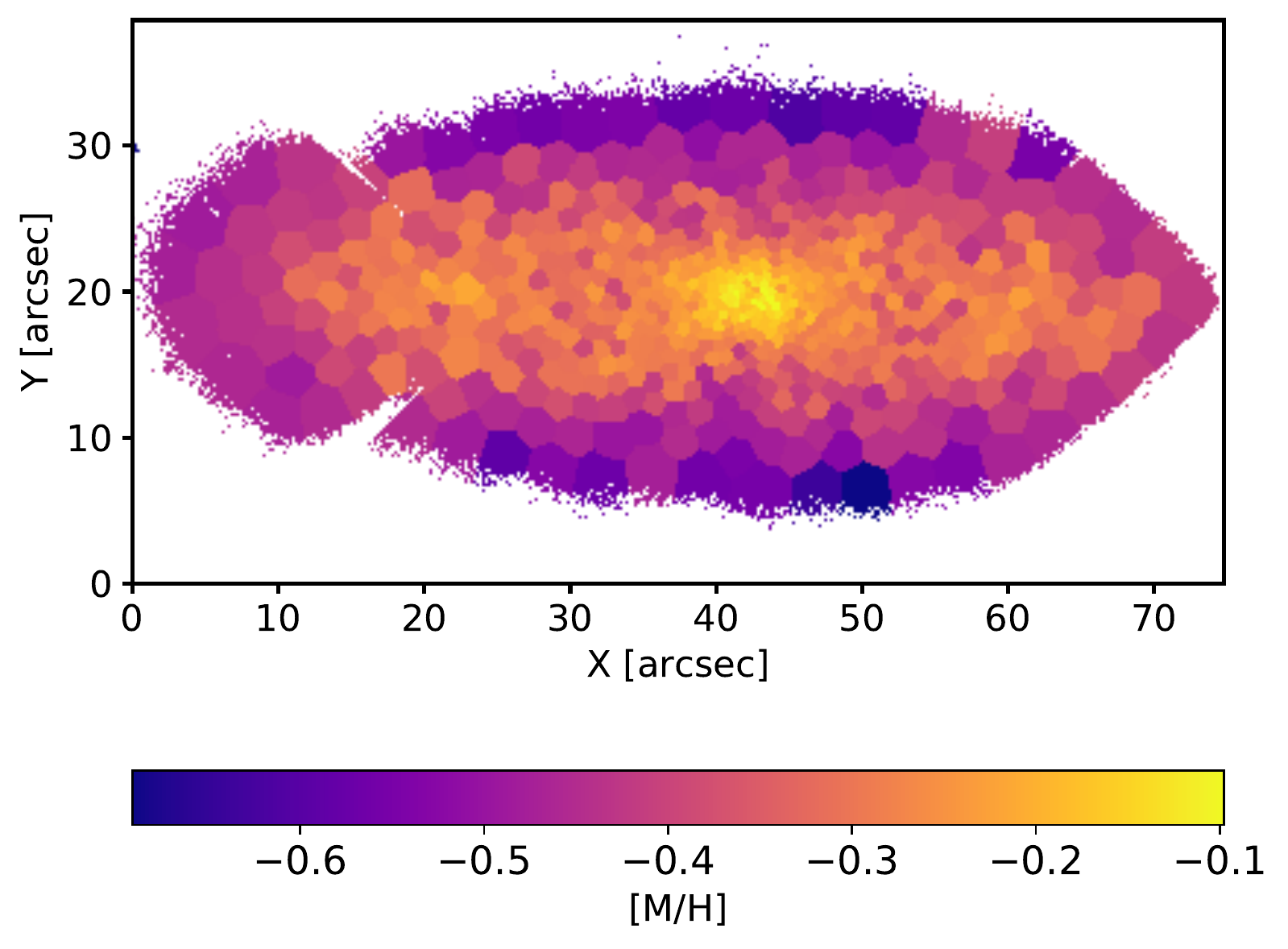}
      \includegraphics[width=7.1cm]{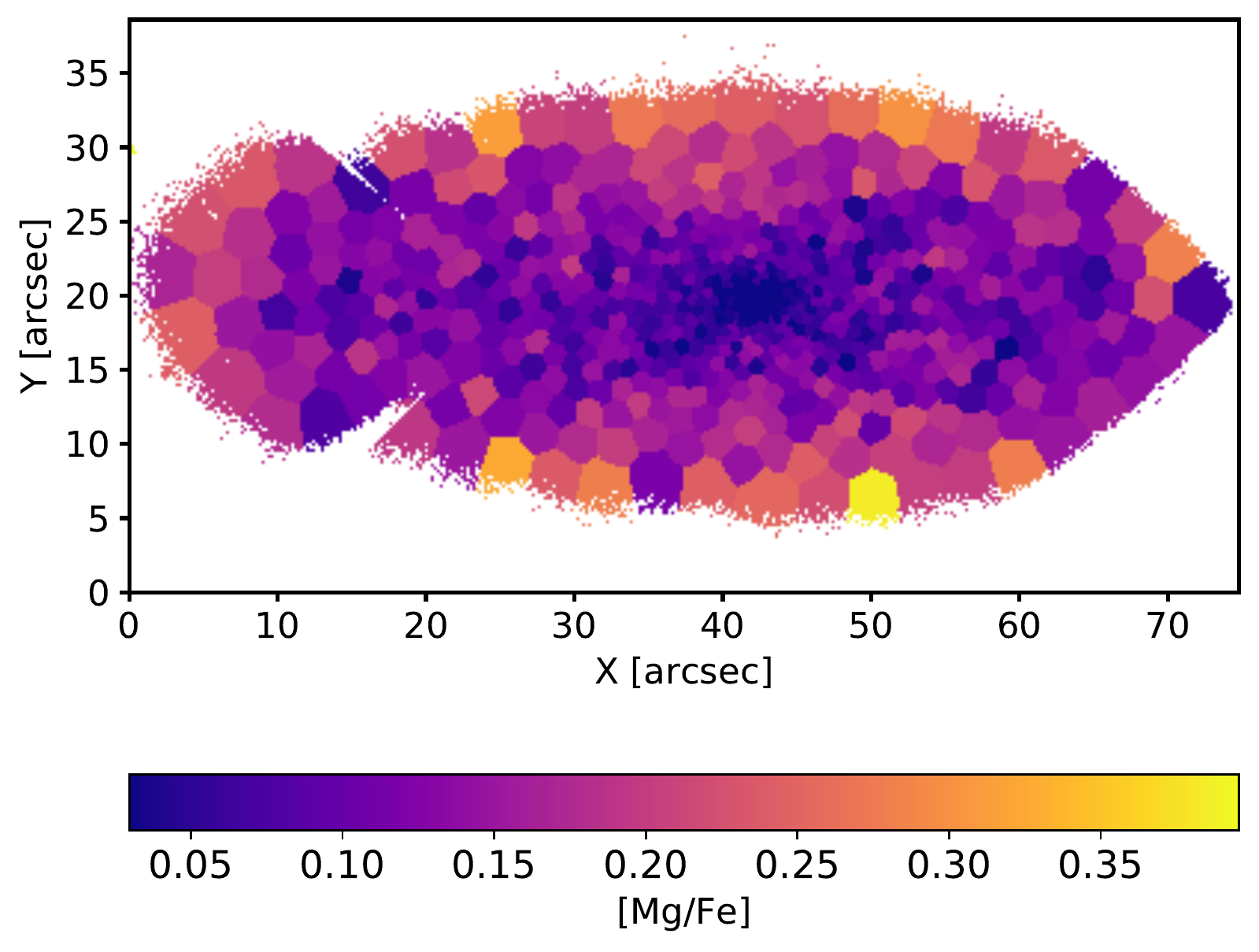}
      \includegraphics[width=7.1cm]{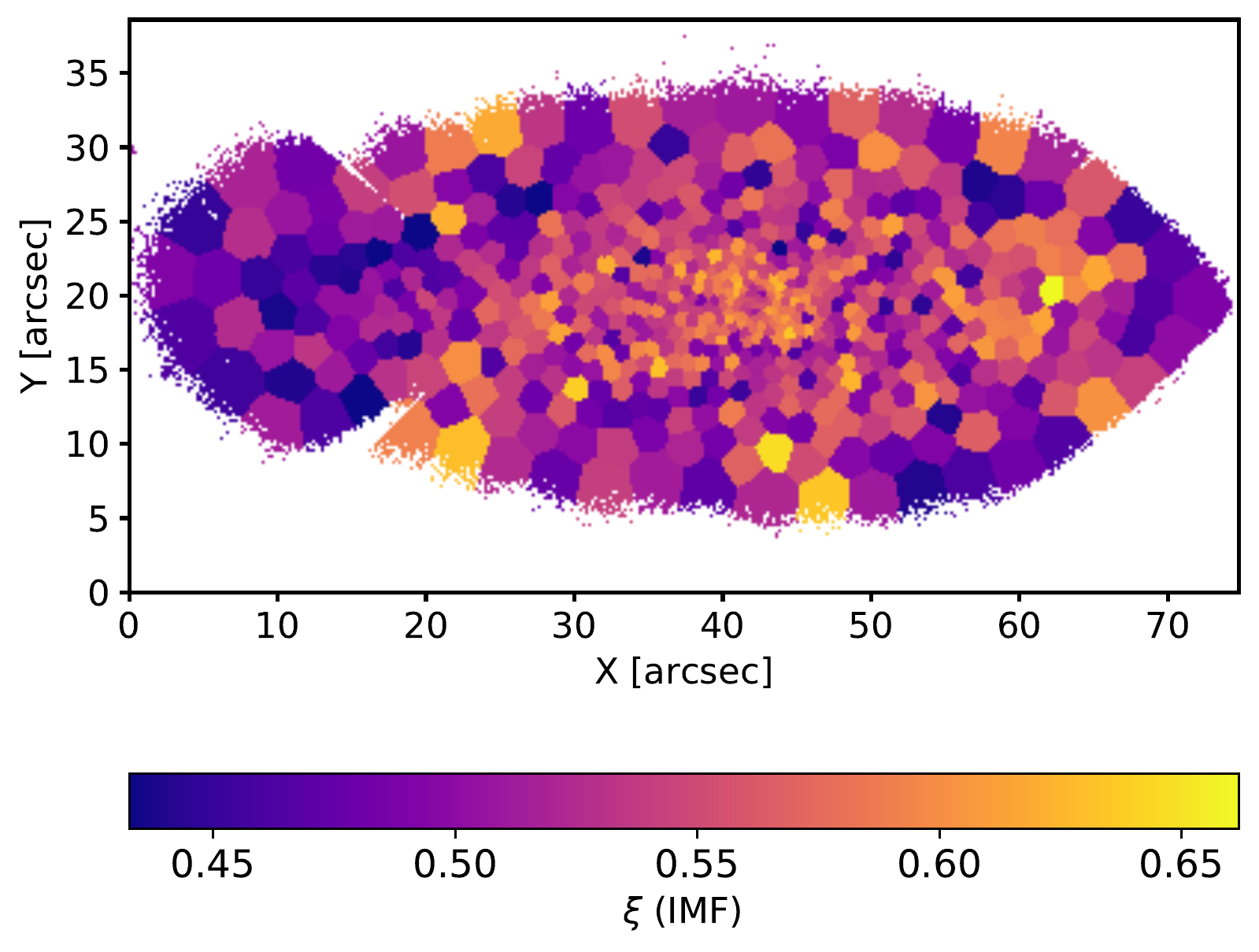}
      \caption{F3D stellar population maps of FCC\,148. From left to right and top to bottom: age, metallicity, [Mg/Fe], and IMF slope maps} 
   \end{figure*}

   \begin{figure*}
      \includegraphics[width=8.5cm]{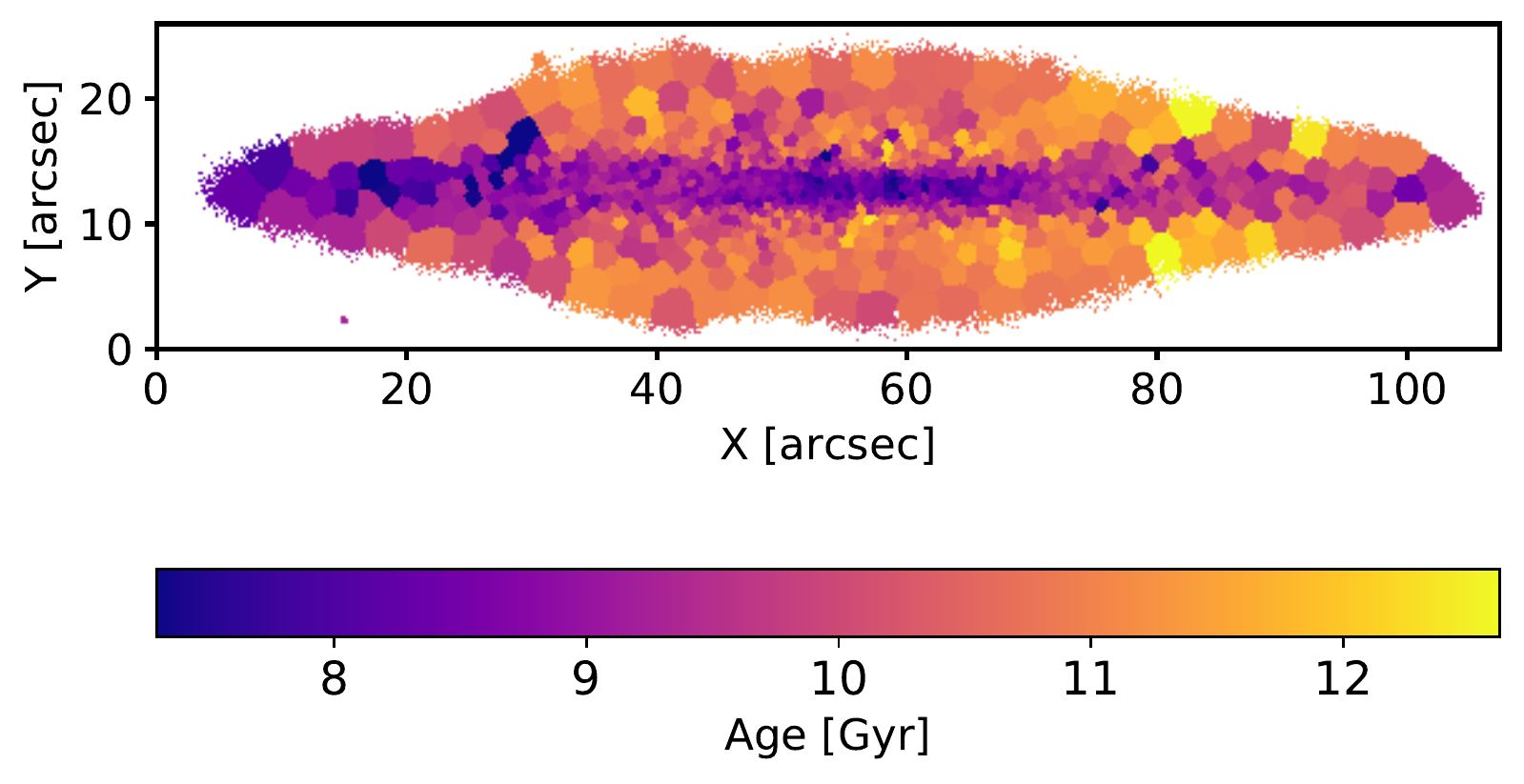}
      \includegraphics[width=8.5cm]{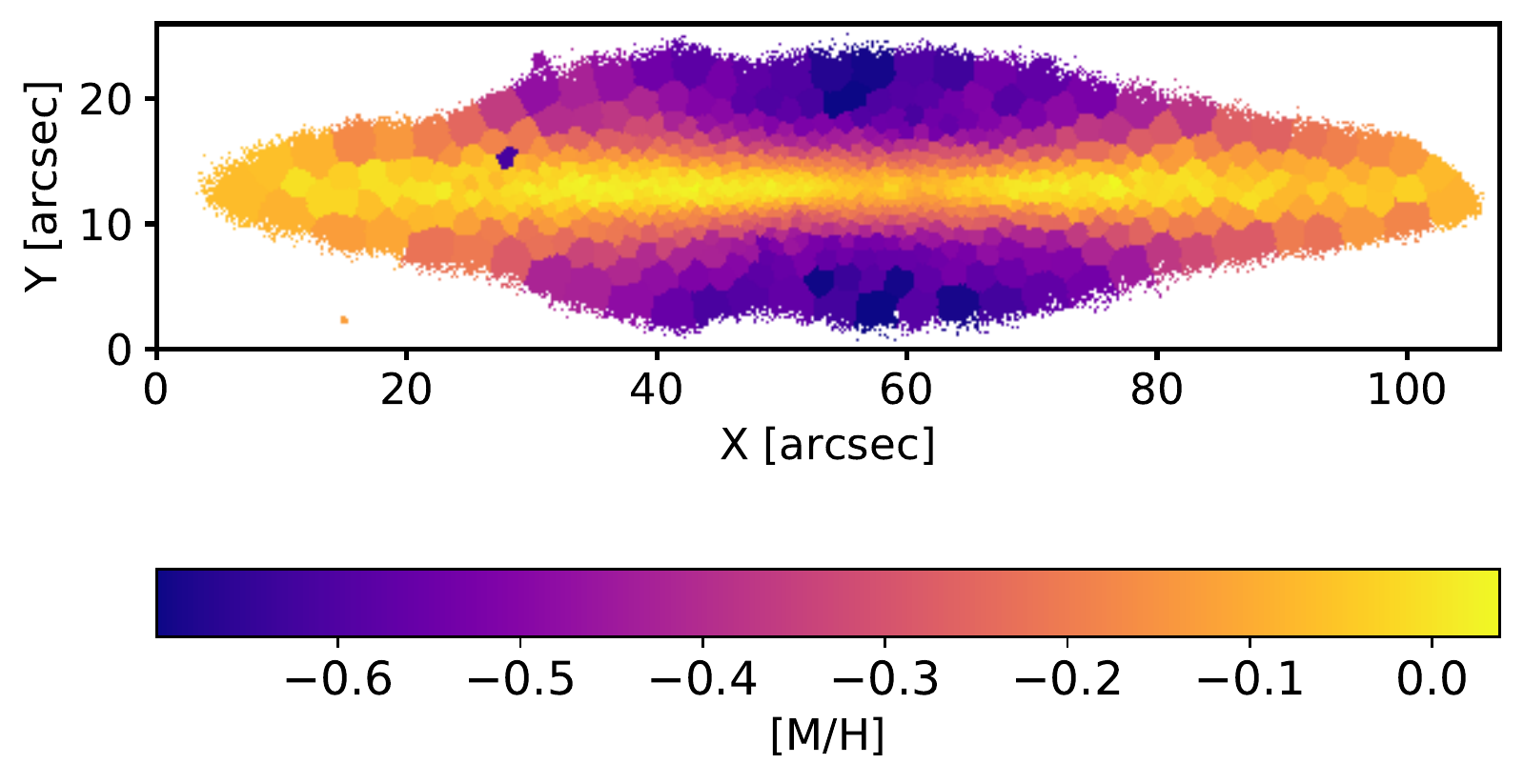}
      \includegraphics[width=8.5cm]{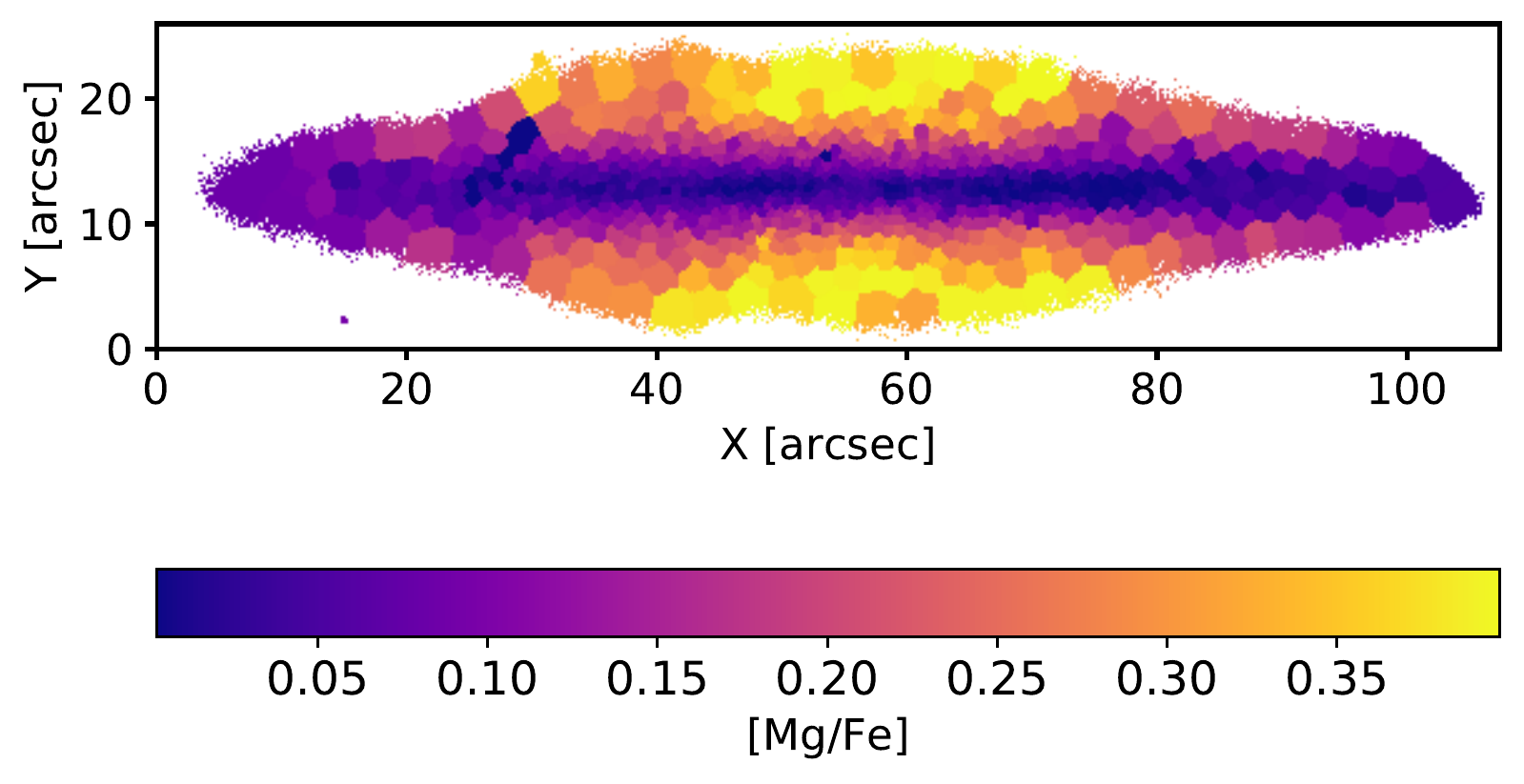}
      \includegraphics[width=8.5cm]{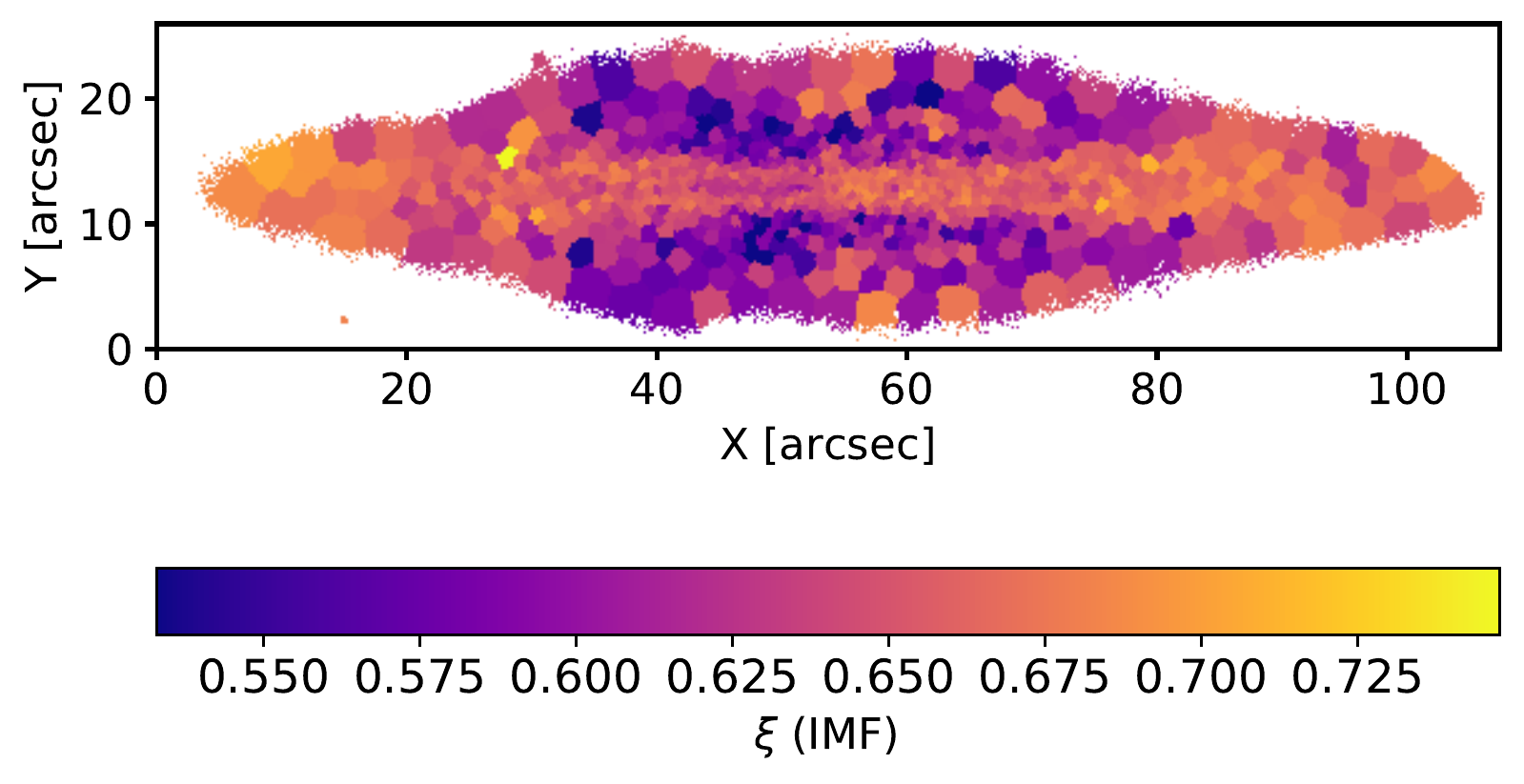}
      \caption{F3D stellar population maps of FCC\,153.} 
      \centering\label{fig:fcc153}
   \end{figure*}

   \begin{figure*}
      \centering
      \includegraphics[width=6.1cm]{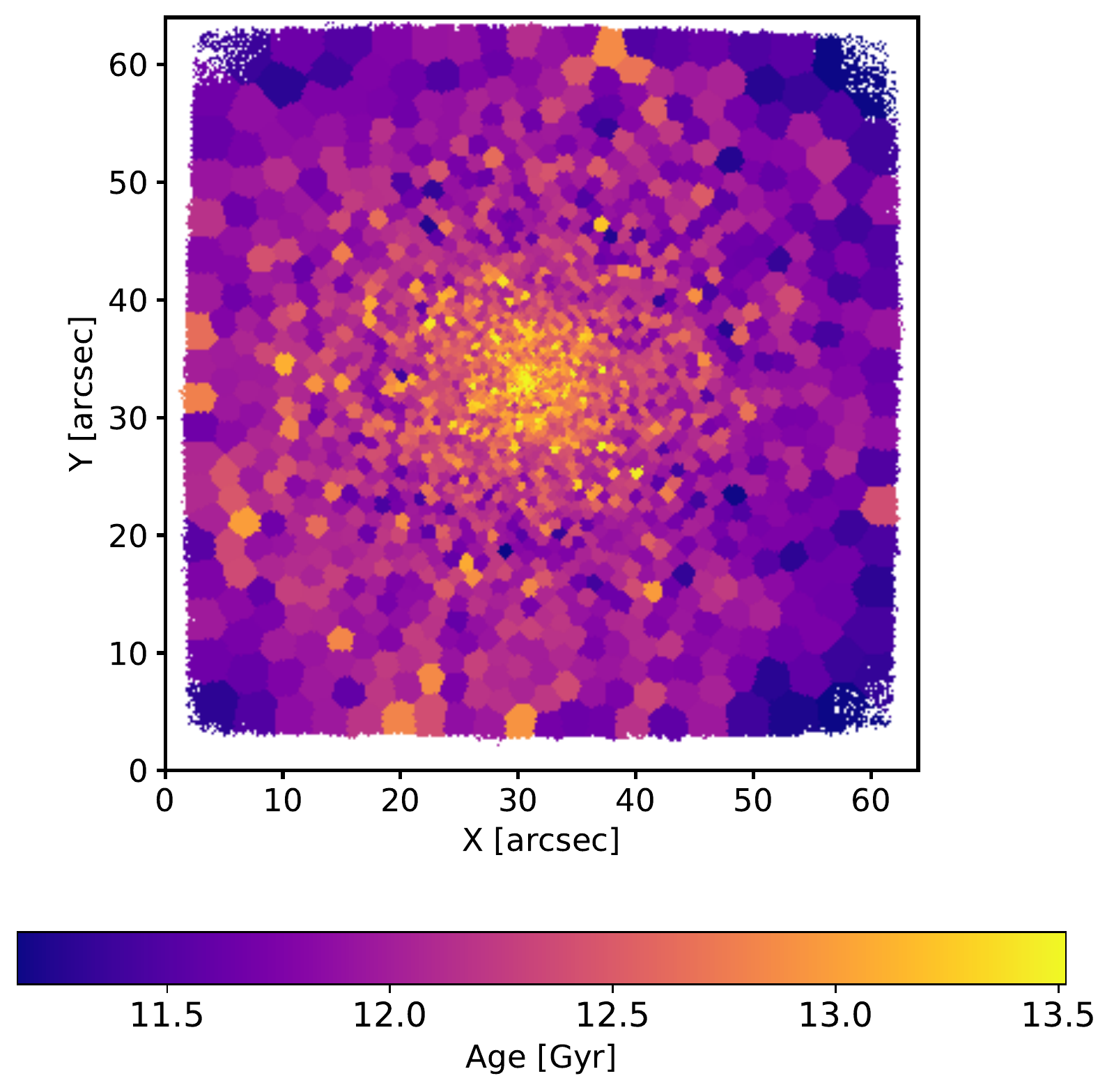}
      \includegraphics[width=6.1cm]{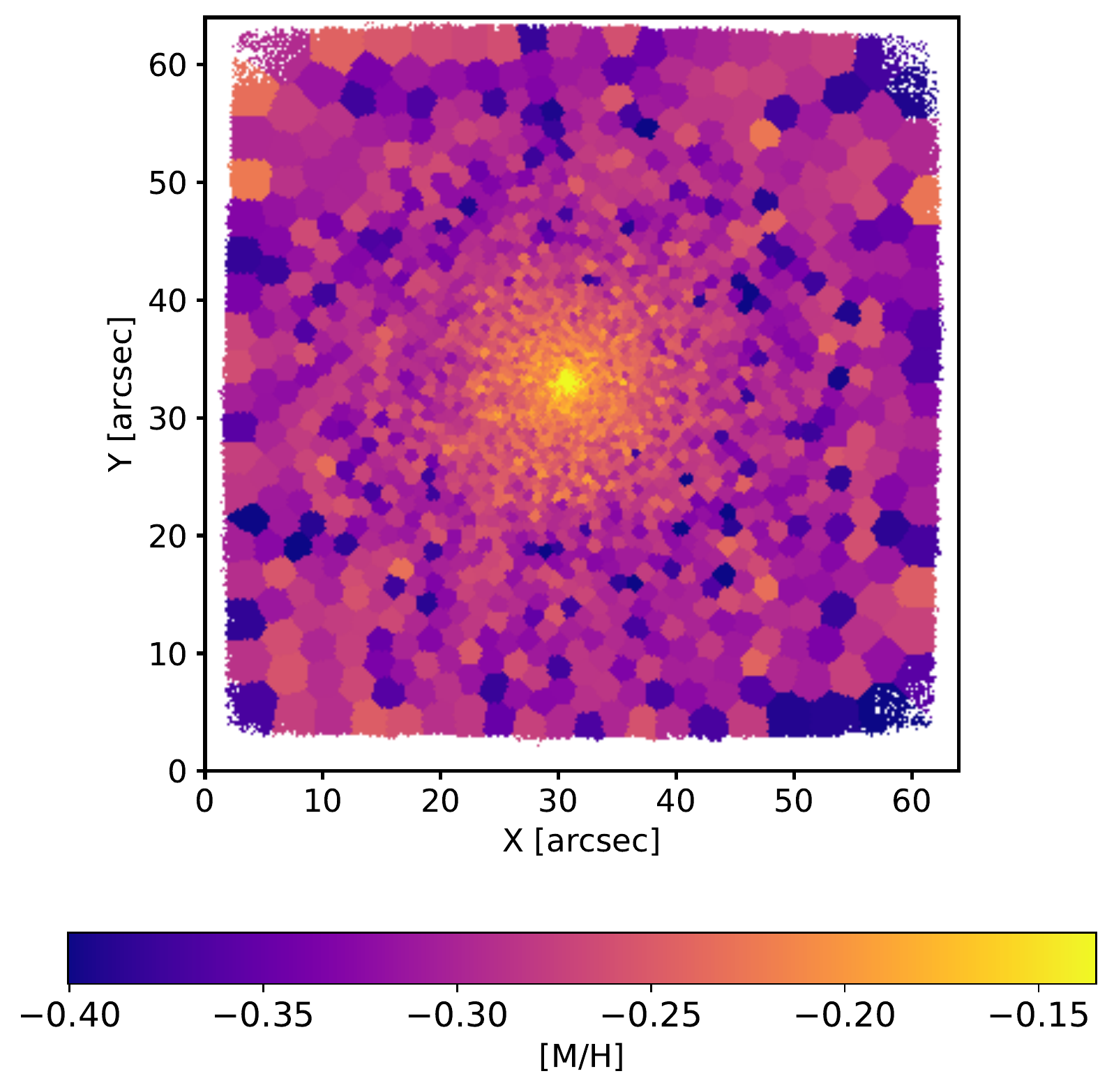}
      \includegraphics[width=6.1cm]{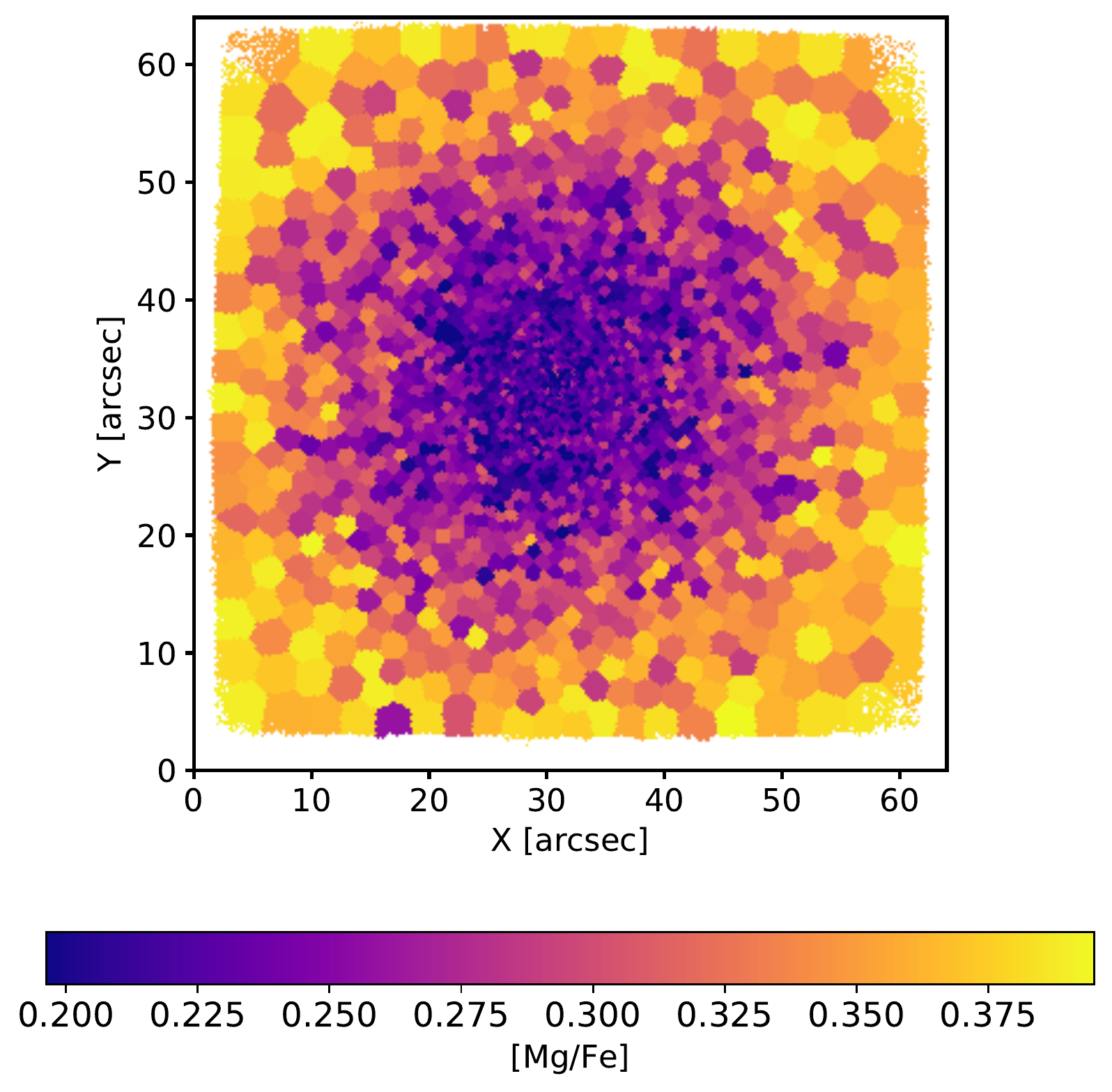}
      \includegraphics[width=6.1cm]{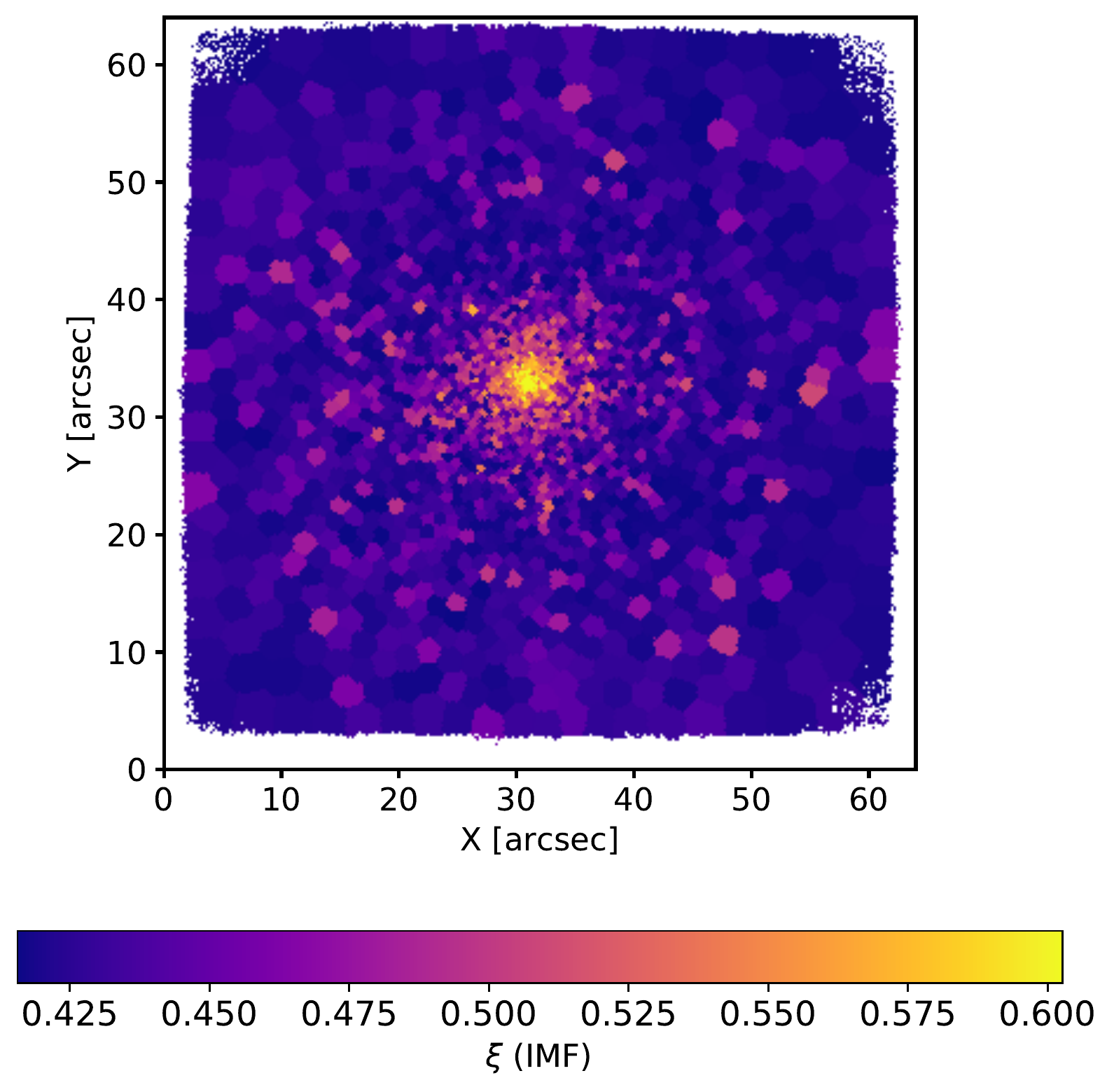}
      \caption{F3D stellar population maps of FCC\,161. From left to right and top to bottom: age, metallicity, [Mg/Fe], and IMF slope maps} 
   \end{figure*}

   \begin{figure*}
      \centering
      \includegraphics[width=6.9cm]{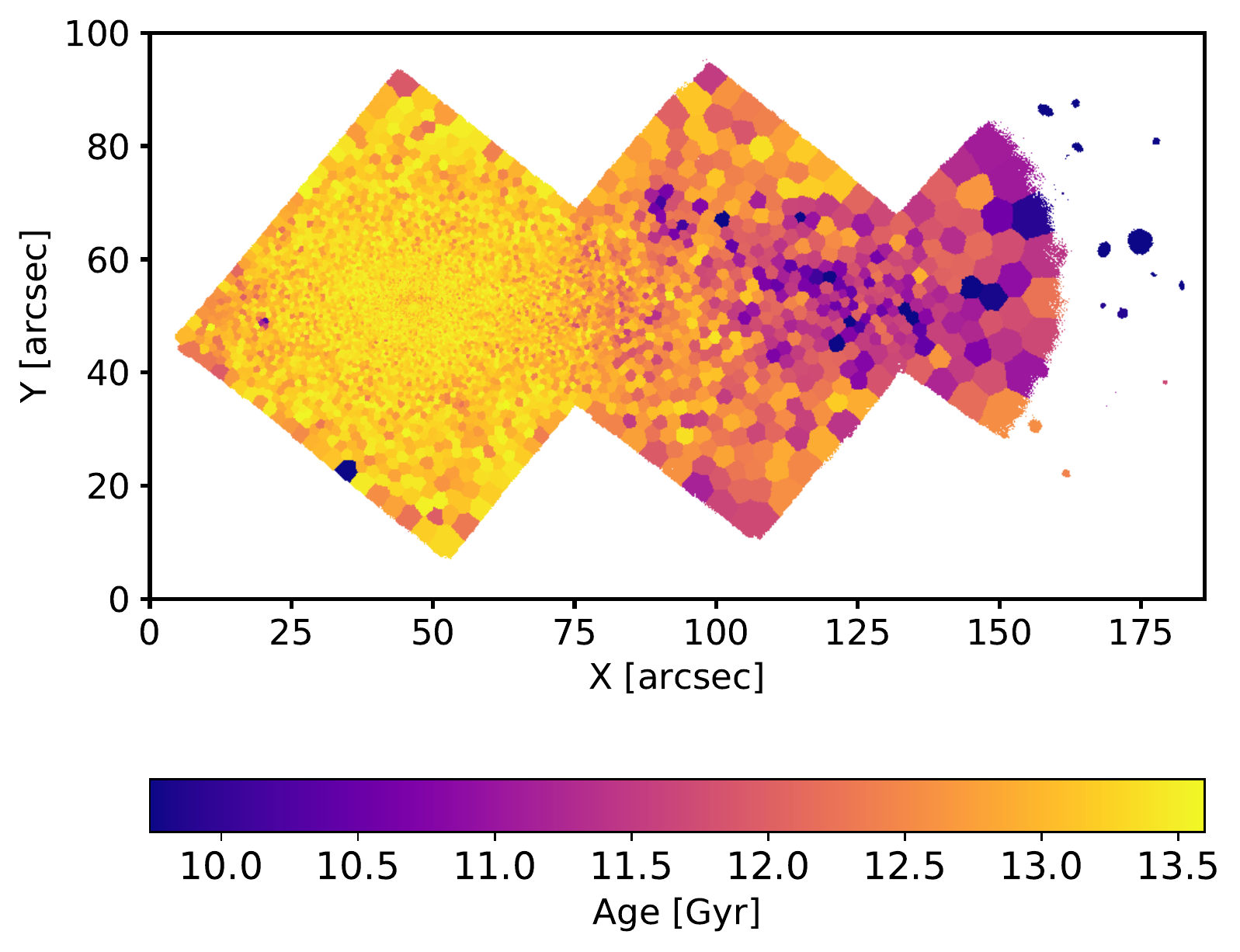}
      \includegraphics[width=6.9cm]{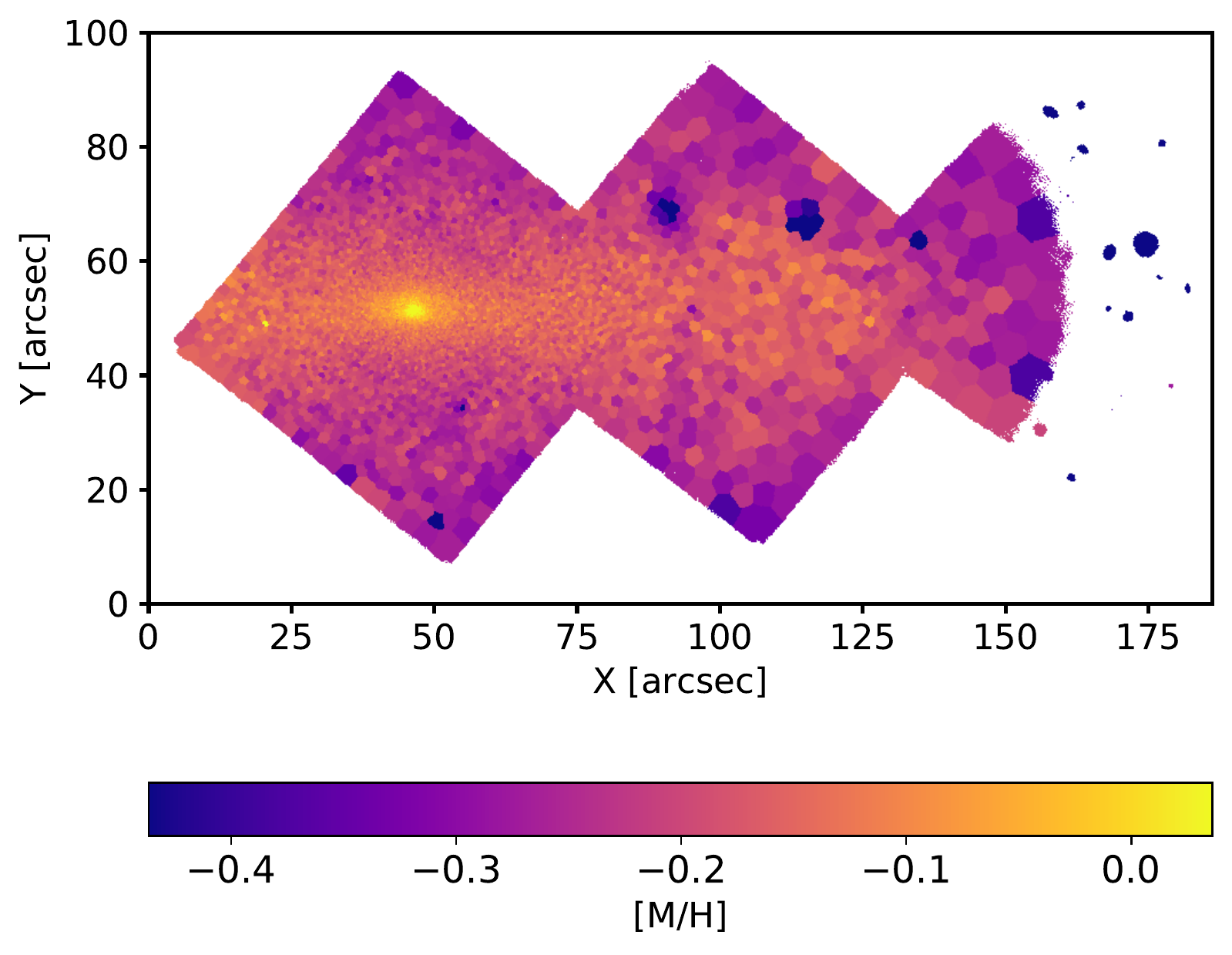}
      \includegraphics[width=6.9cm]{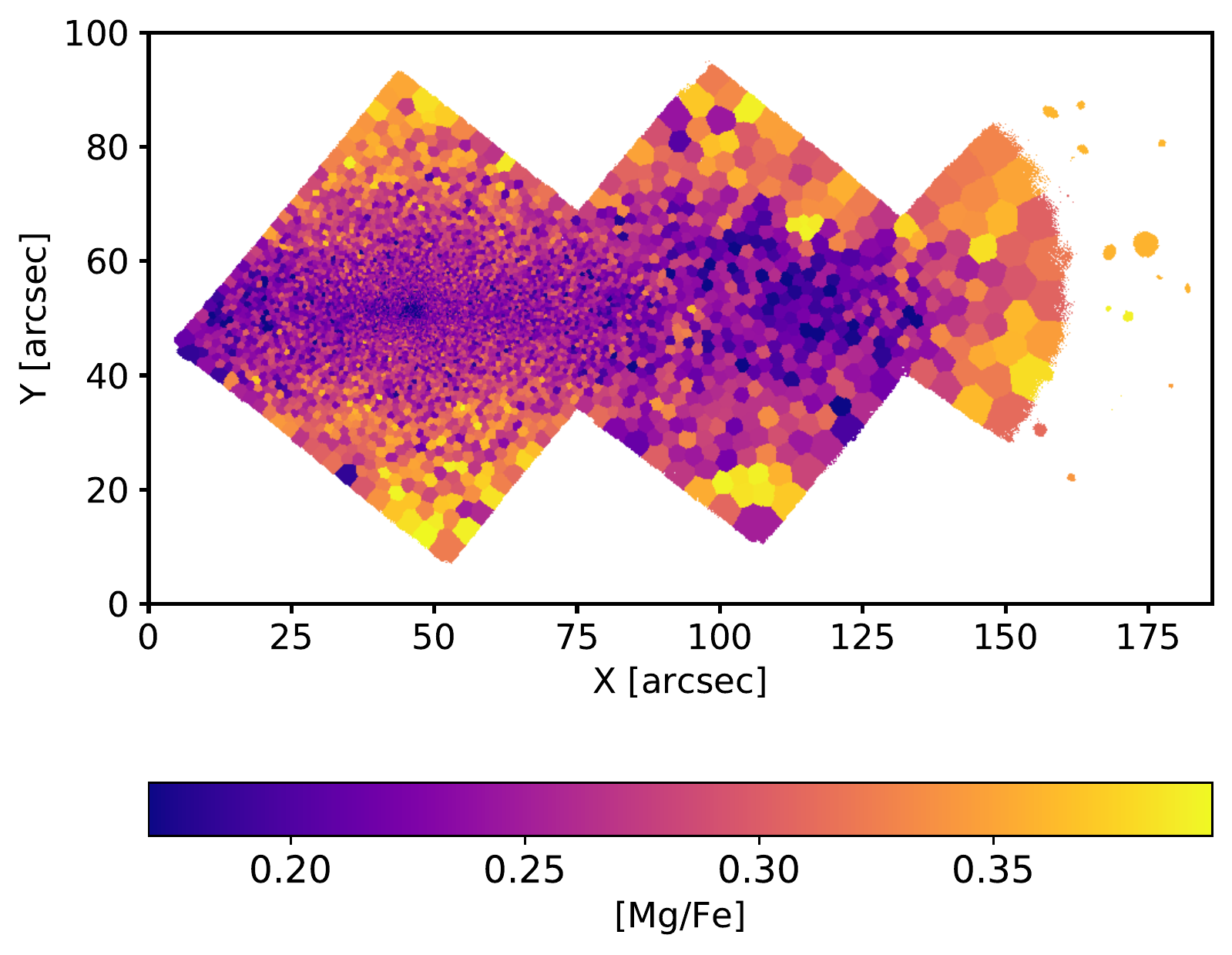}
      \includegraphics[width=6.9cm]{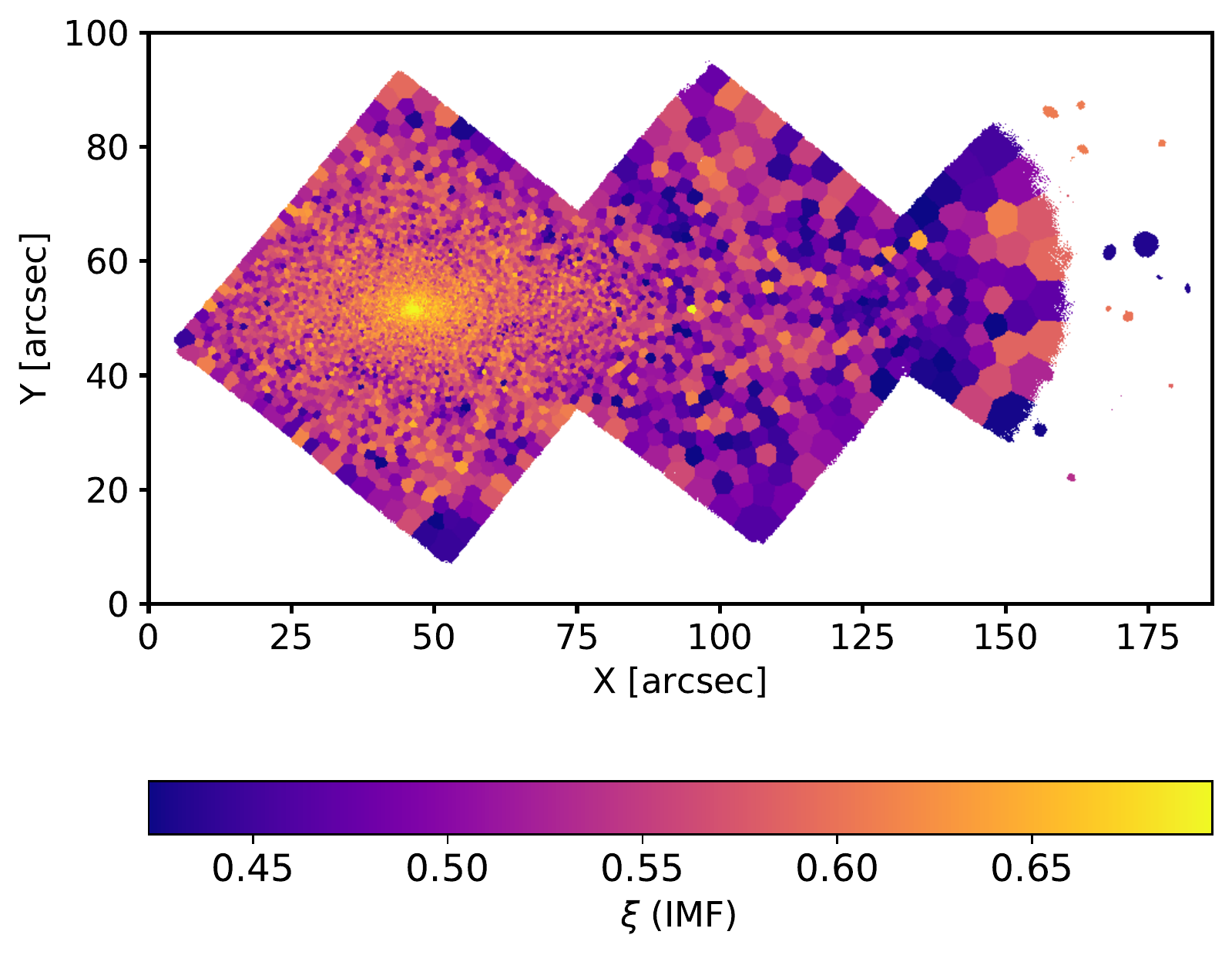}
      \caption{F3D stellar population maps of FCC\,167.} 
   \end{figure*}

   \begin{figure*}
      \centering
      \includegraphics[width=7.9cm]{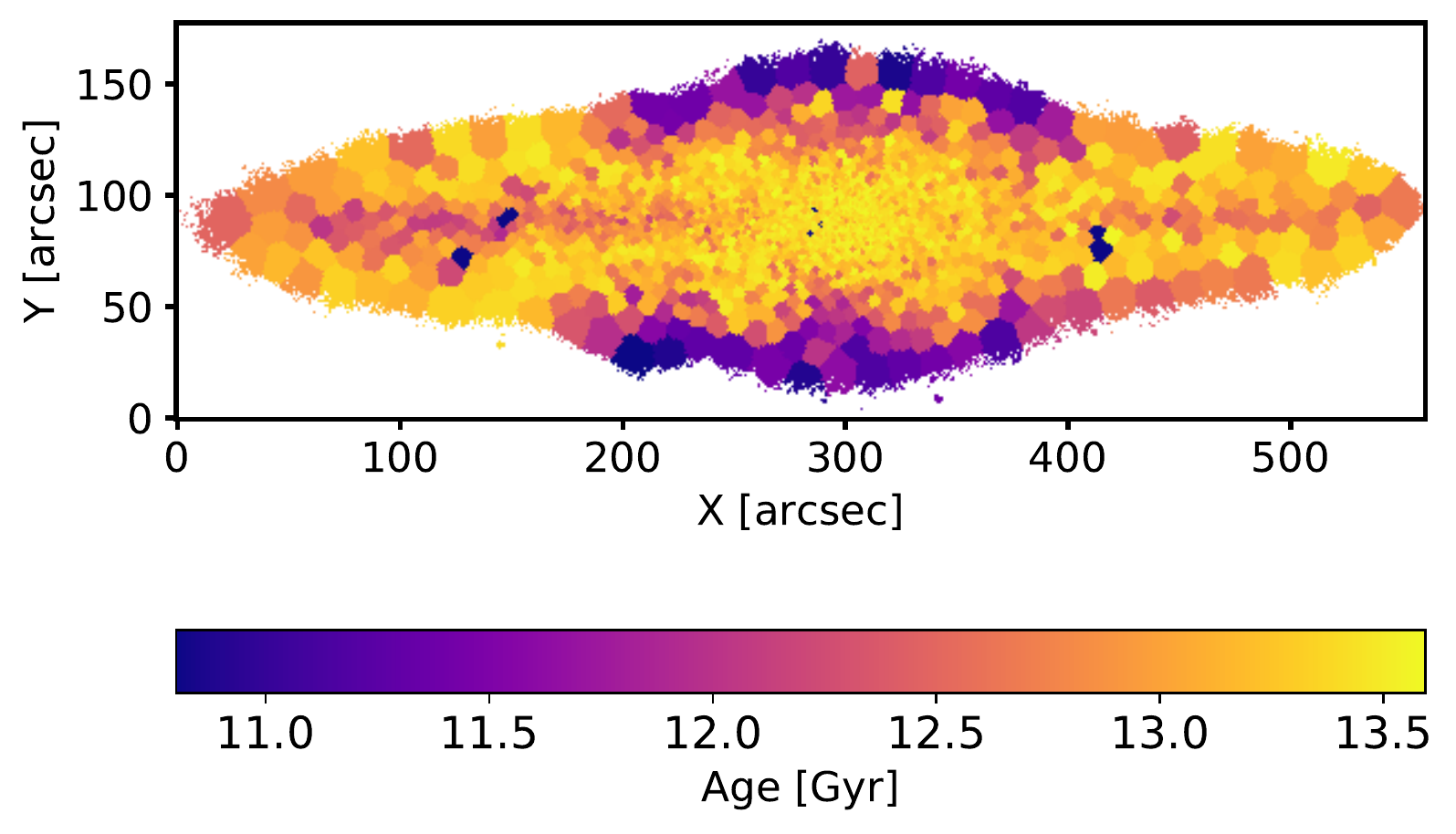}
      \includegraphics[width=7.9cm]{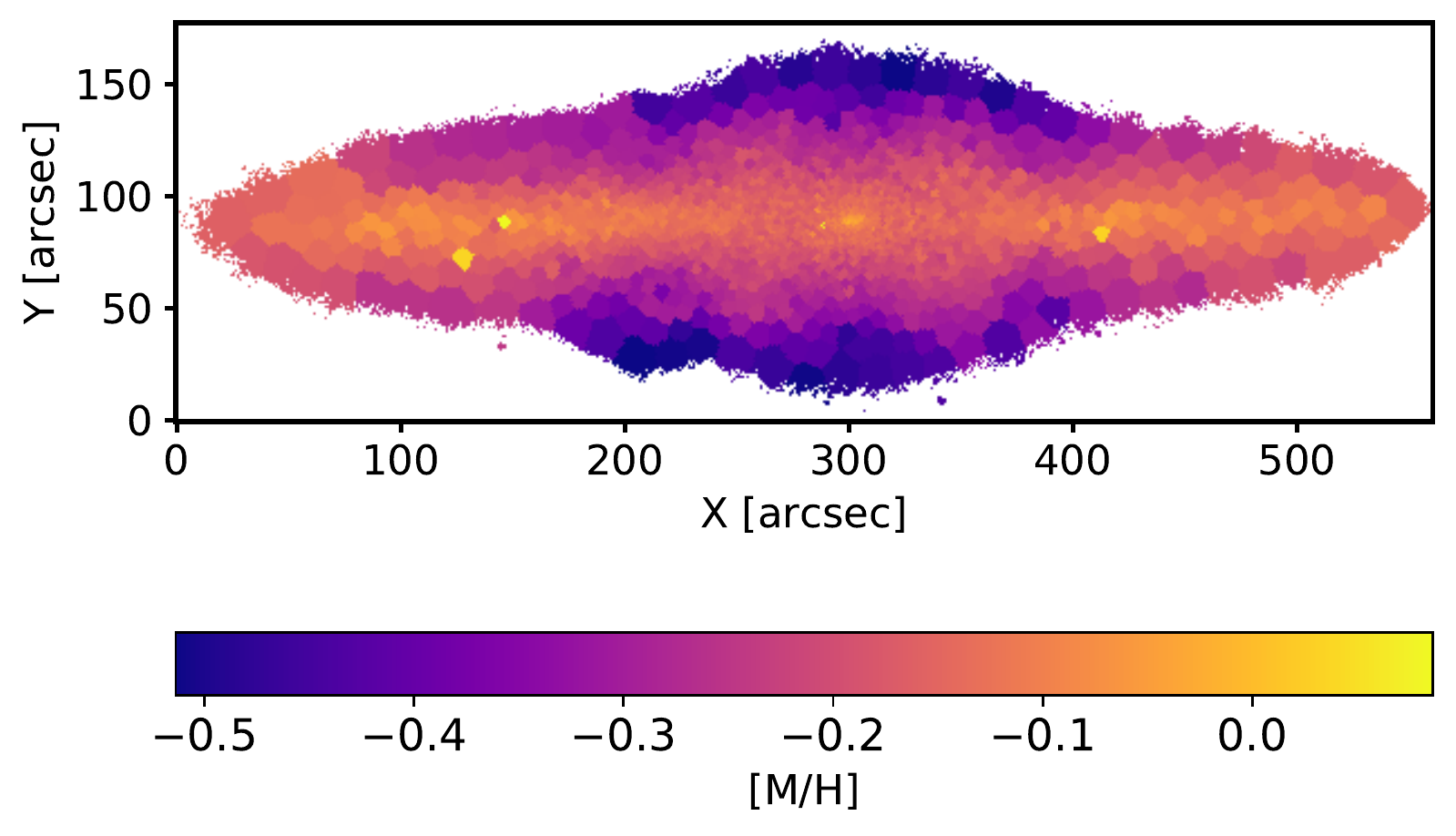}
      \includegraphics[width=7.9cm]{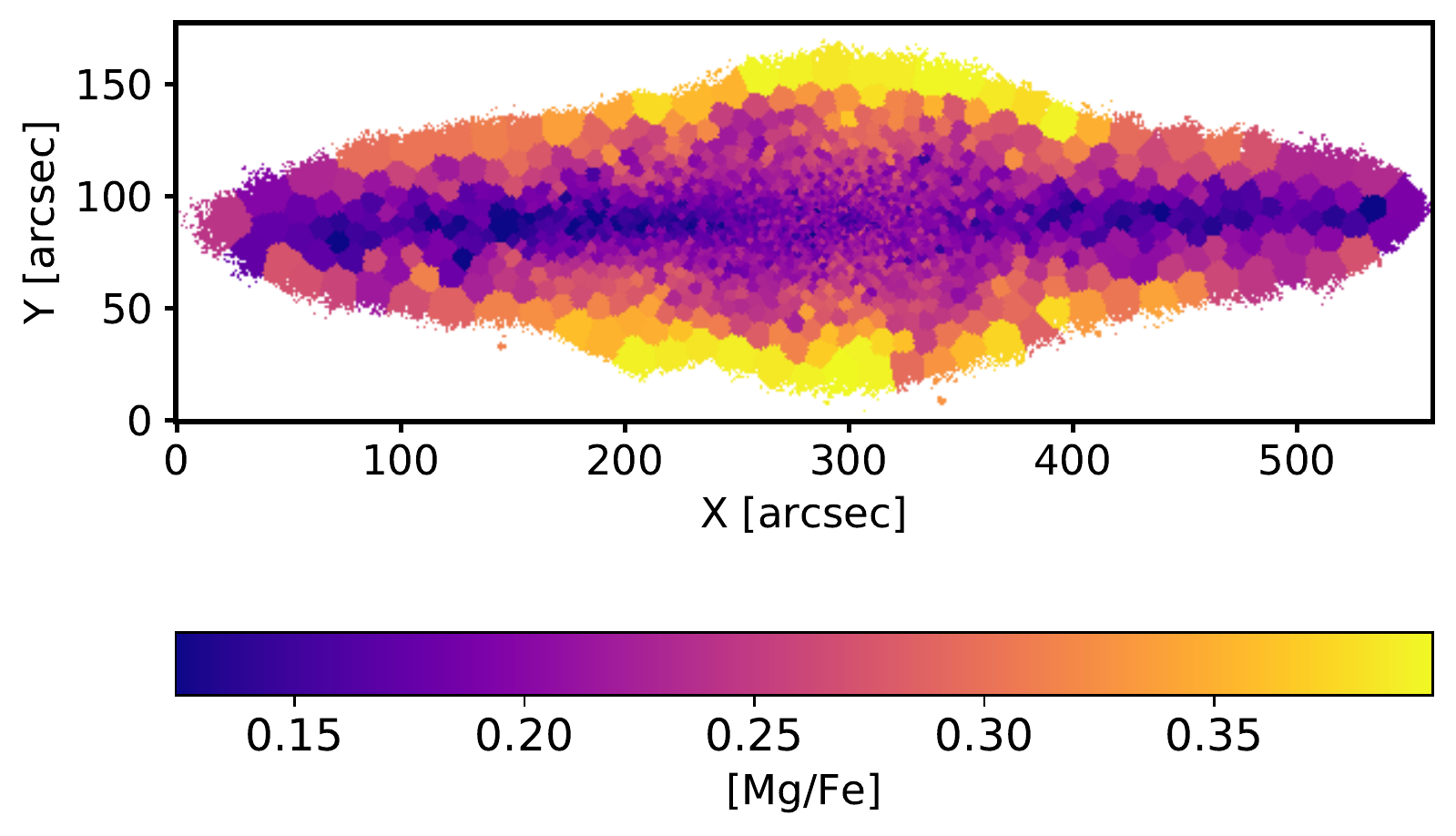}
      \includegraphics[width=7.9cm]{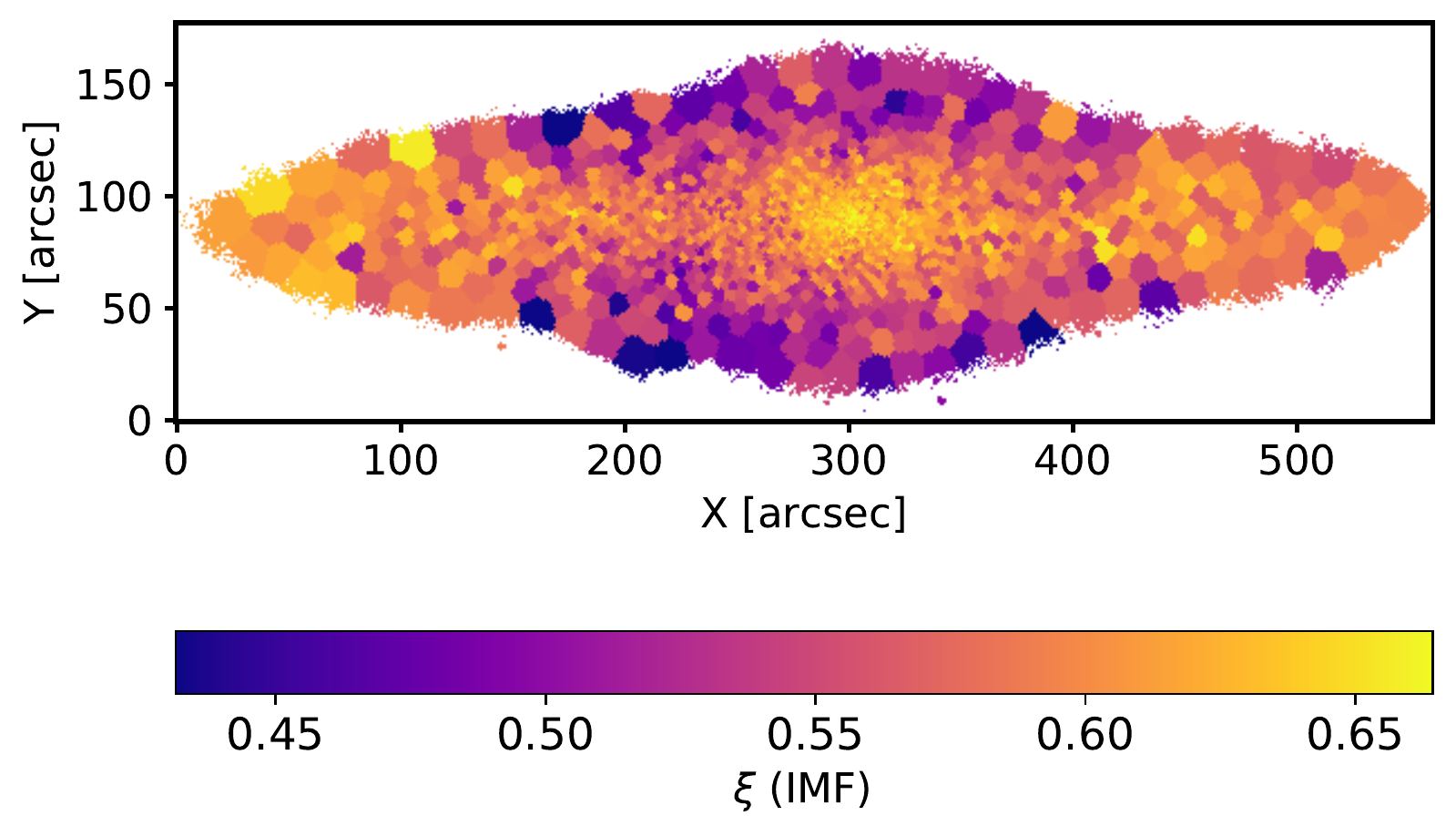}
      \caption{F3D stellar population maps of FCC\,170. From left to right and top to bottom: age, metallicity, [Mg/Fe], and IMF slope maps} 
   \end{figure*}

   \begin{figure*}
      \centering
      \includegraphics[width=6.1cm]{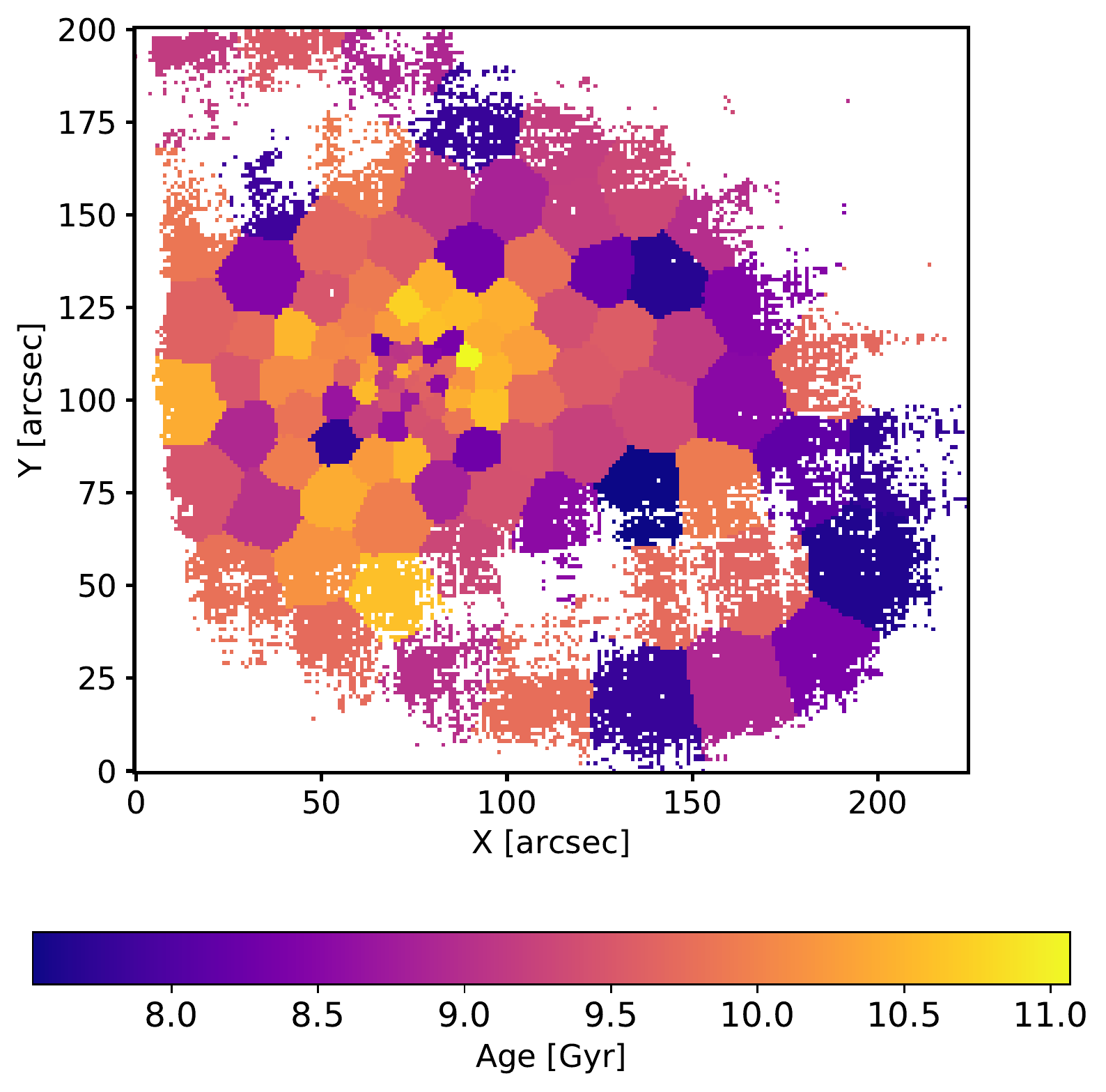}
      \includegraphics[width=6.1cm]{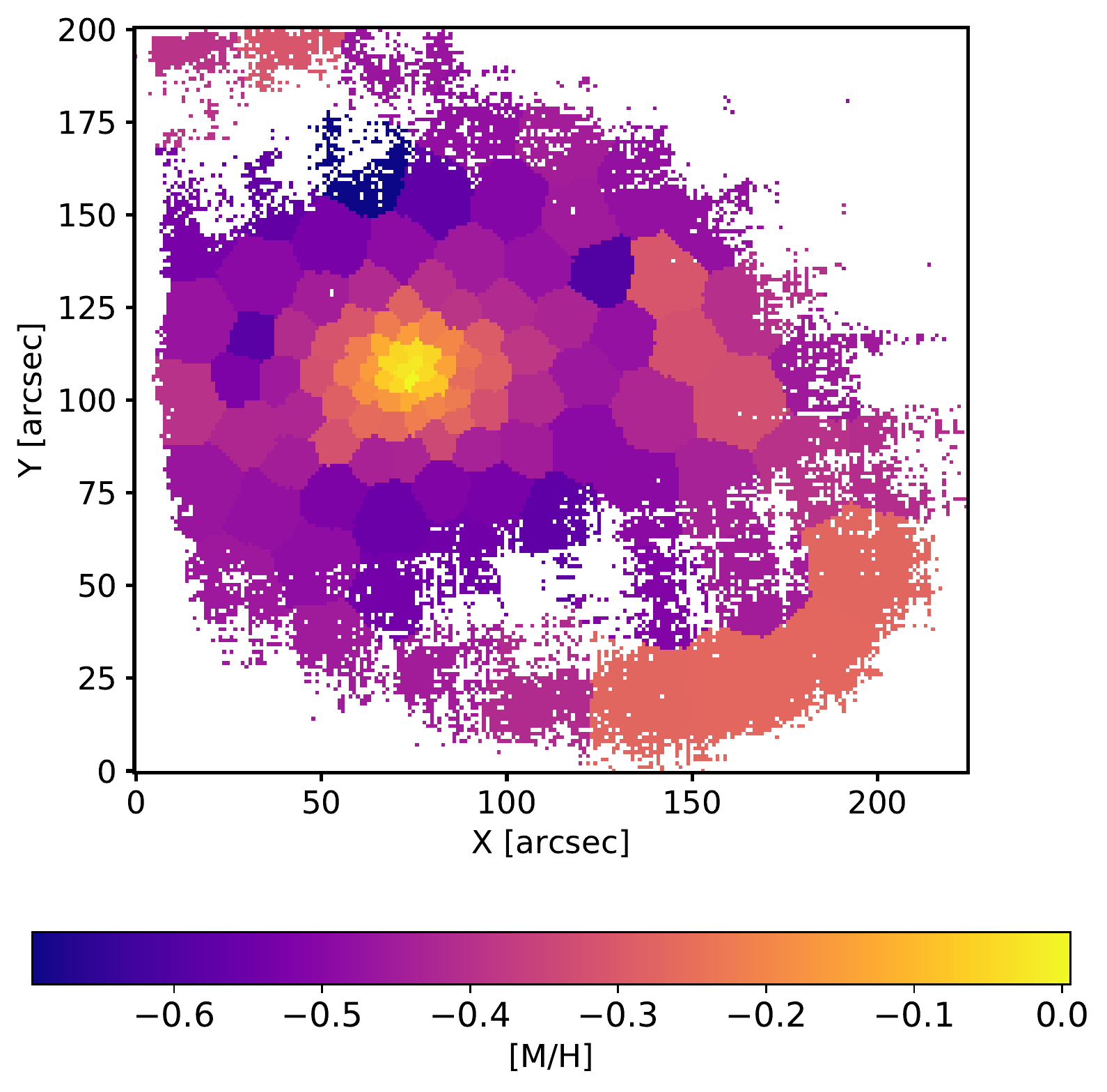}
      \includegraphics[width=6.1cm]{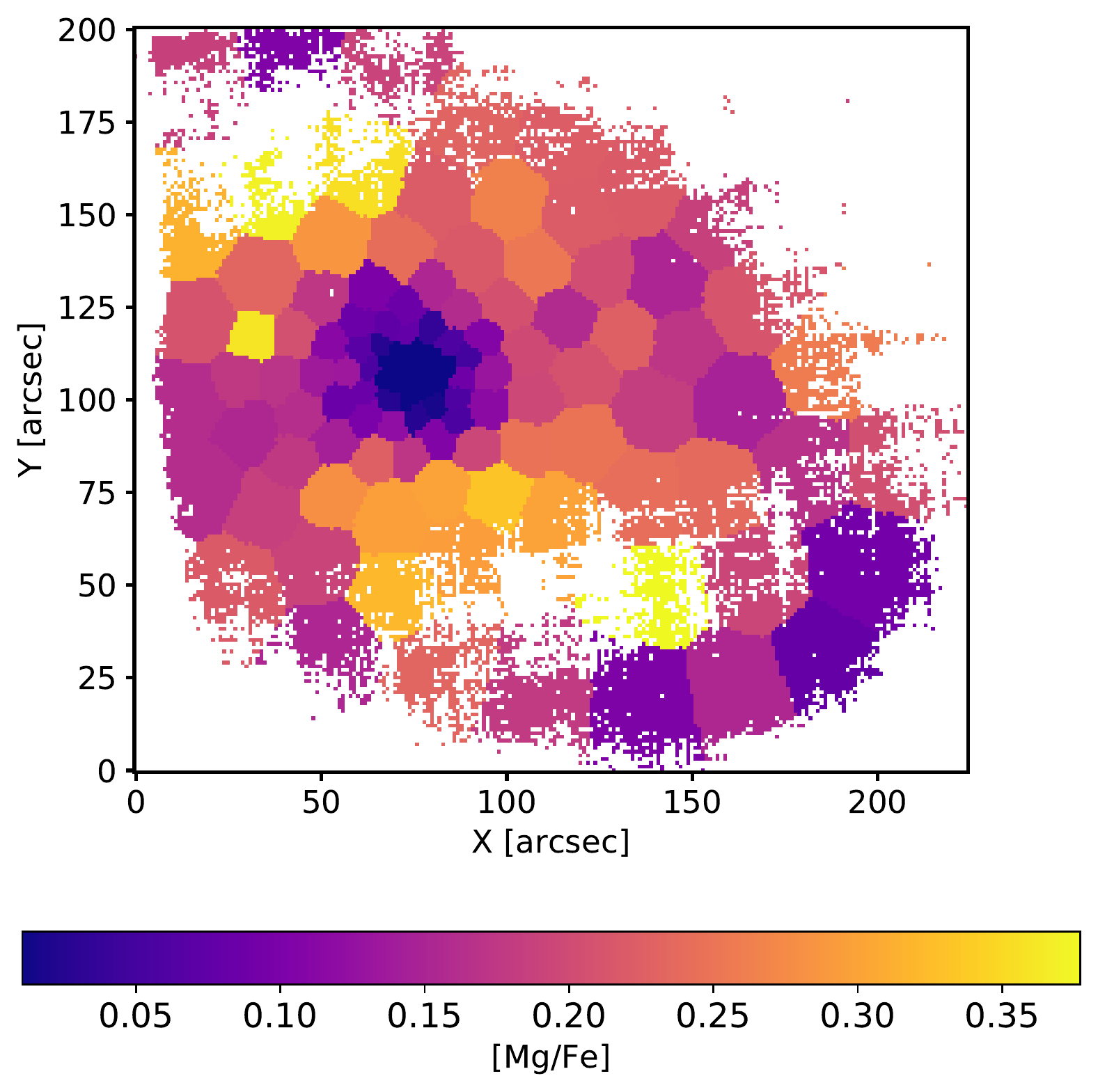}
      \includegraphics[width=6.1cm]{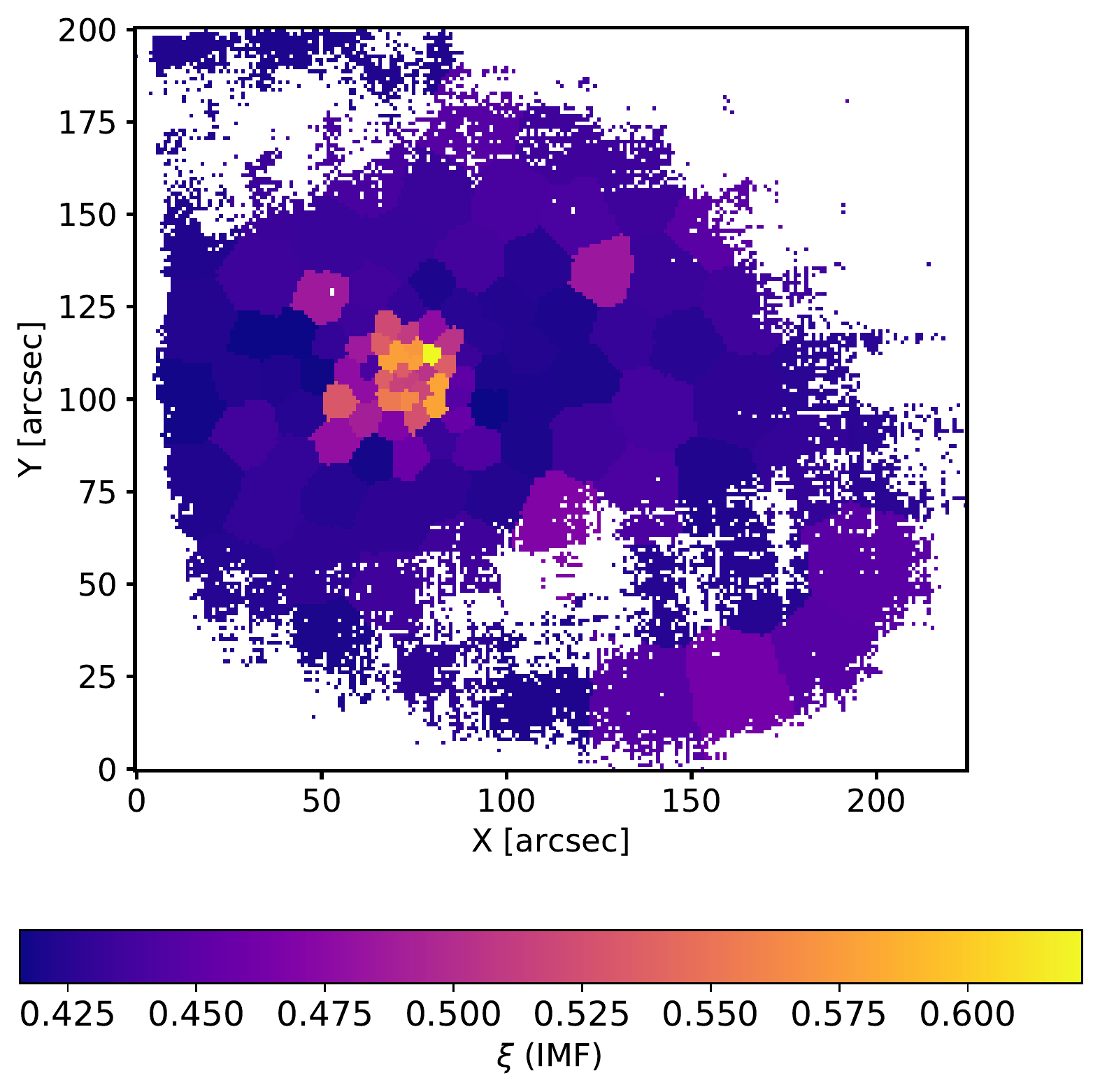}
      \caption{F3D stellar population maps of FCC\,176.} 
   \end{figure*}

   \begin{figure*}
      \centering
      \includegraphics[width=8.5cm]{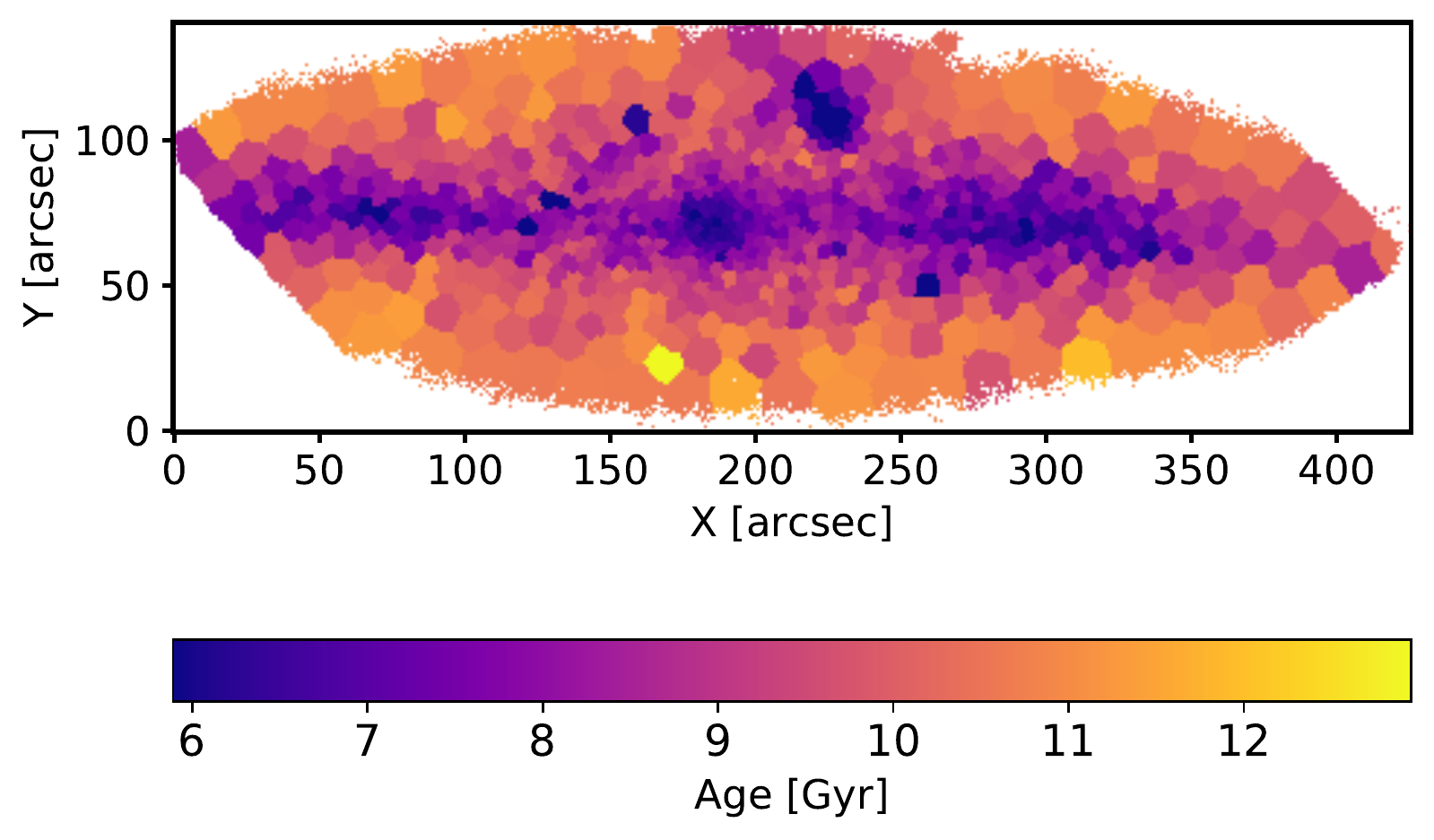}
      \includegraphics[width=8.5cm]{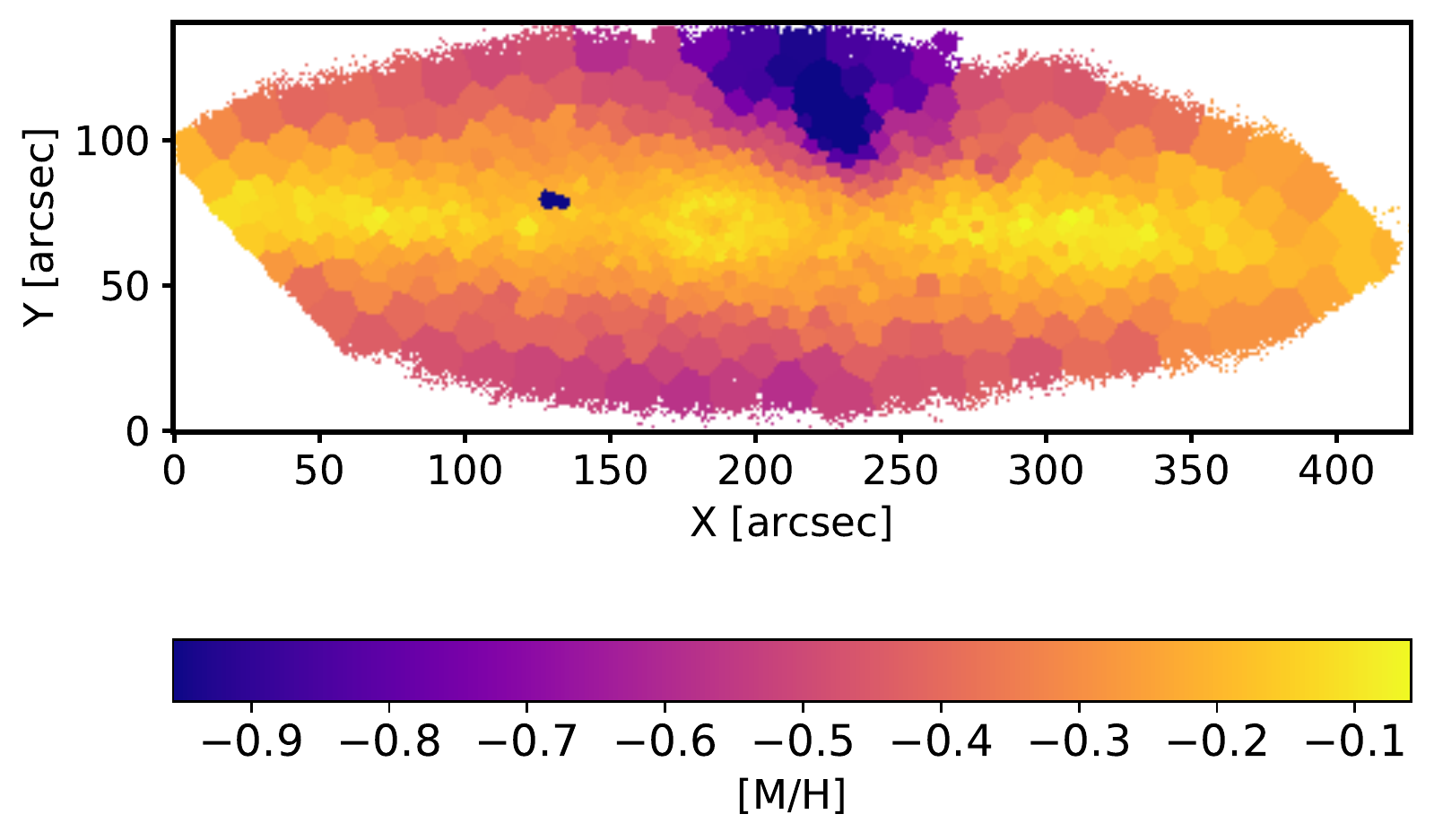}
      \includegraphics[width=8.5cm]{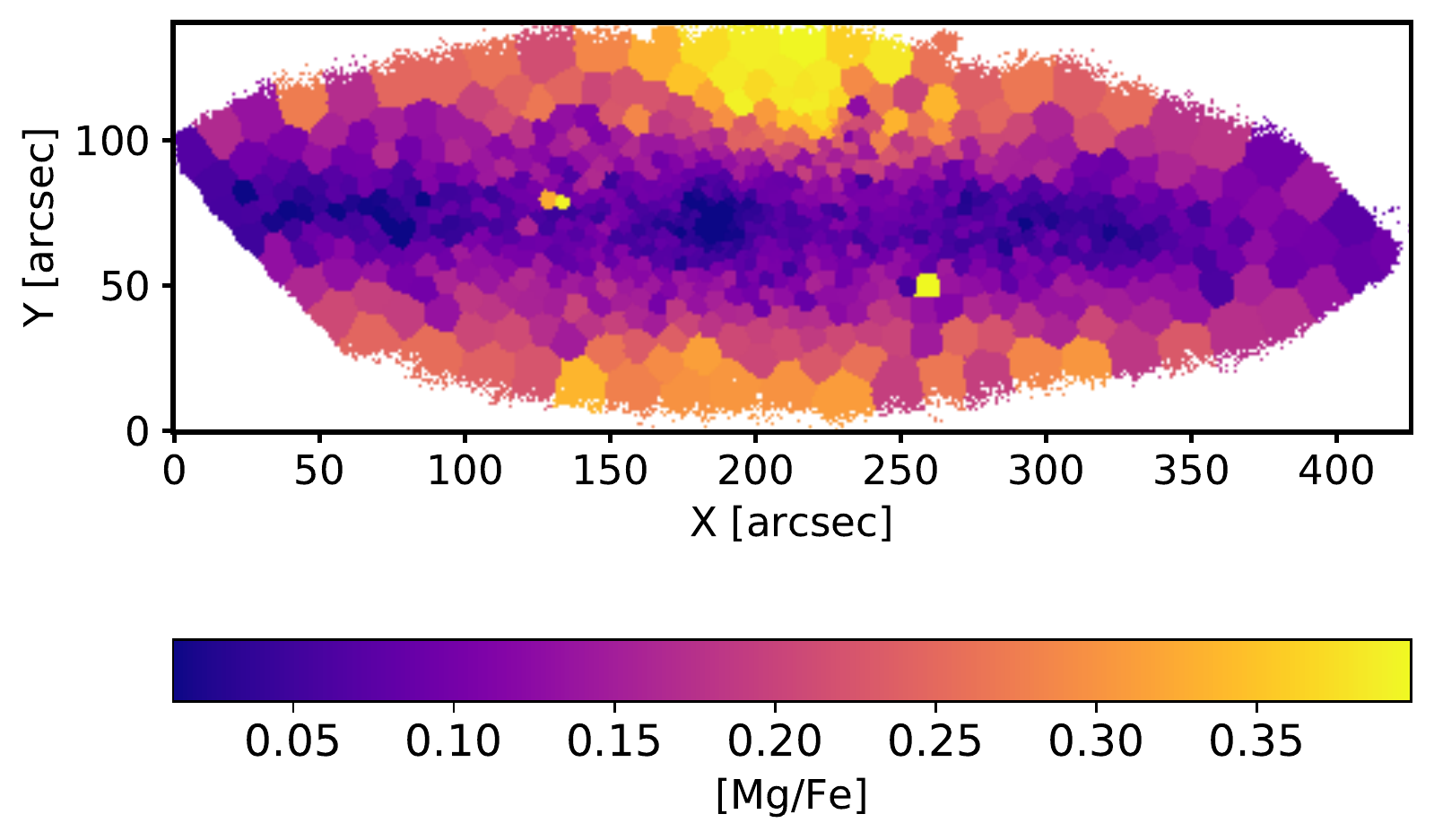}
      \includegraphics[width=8.5cm]{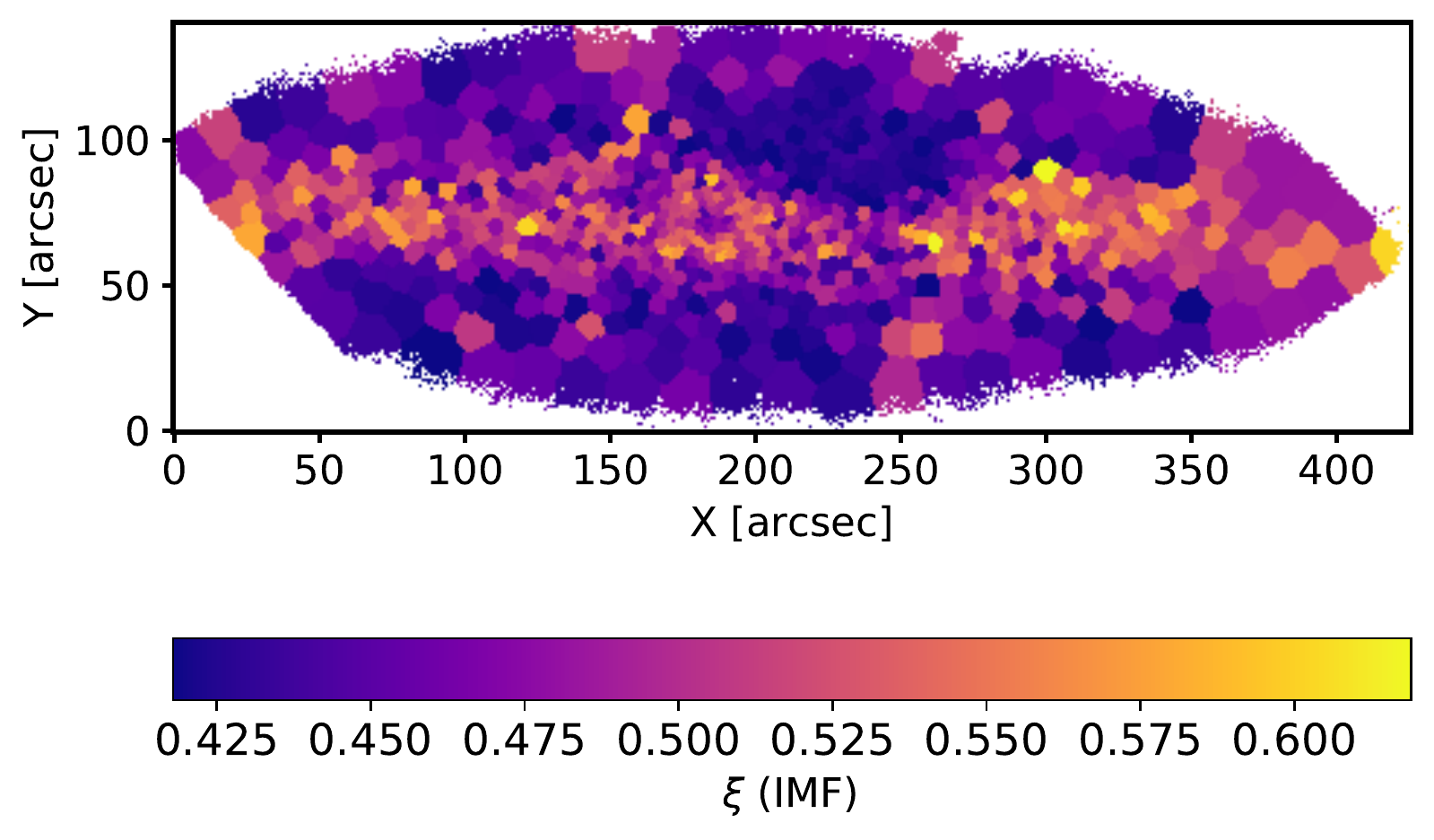}
      \caption{F3D stellar population maps of FCC\,177. From left to right and top to bottom: age, metallicity, [Mg/Fe], and IMF slope maps} 
   \end{figure*}

   \begin{figure*}
      \centering
      \includegraphics[width=6.1cm]{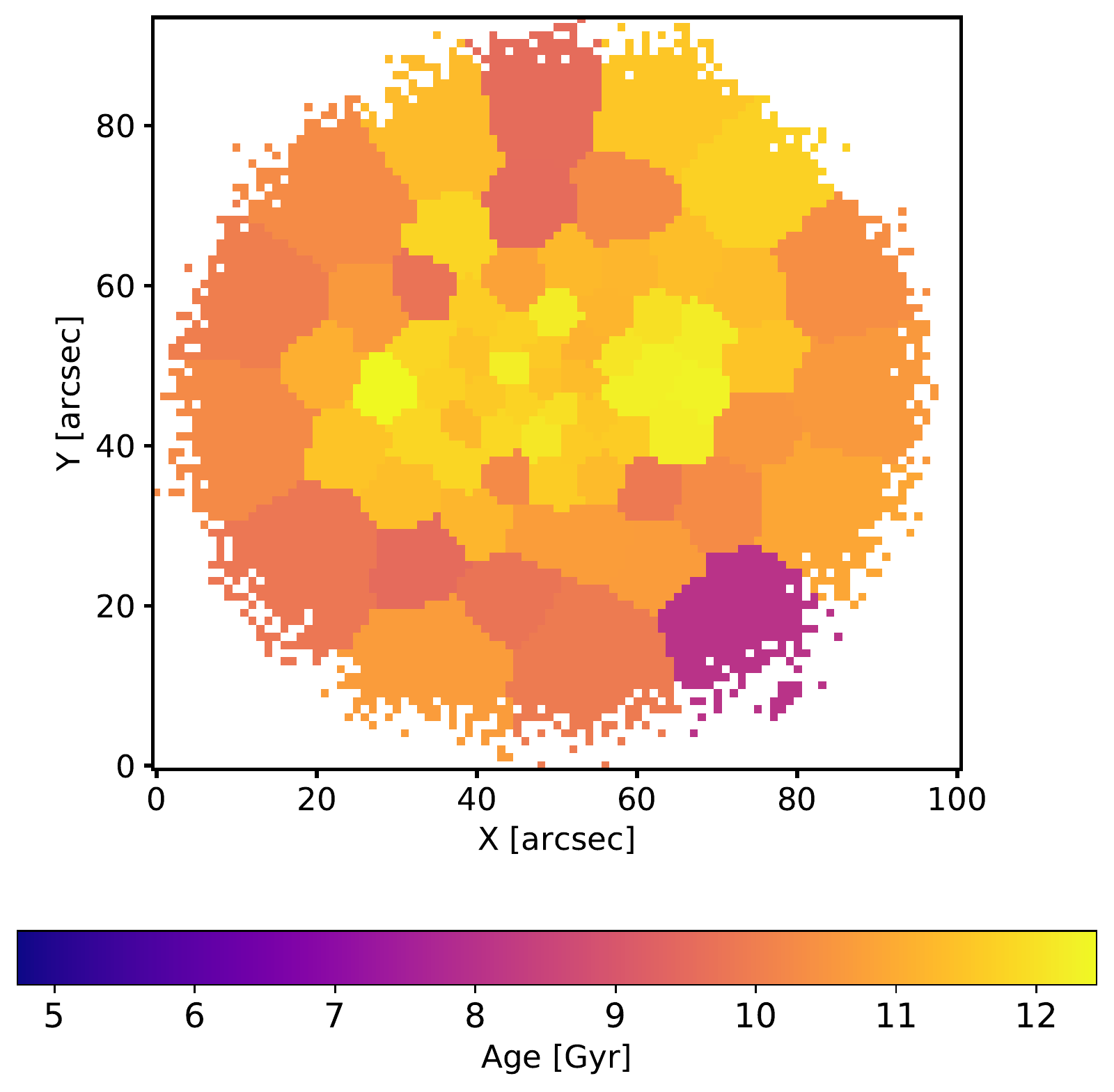}
      \includegraphics[width=6.1cm]{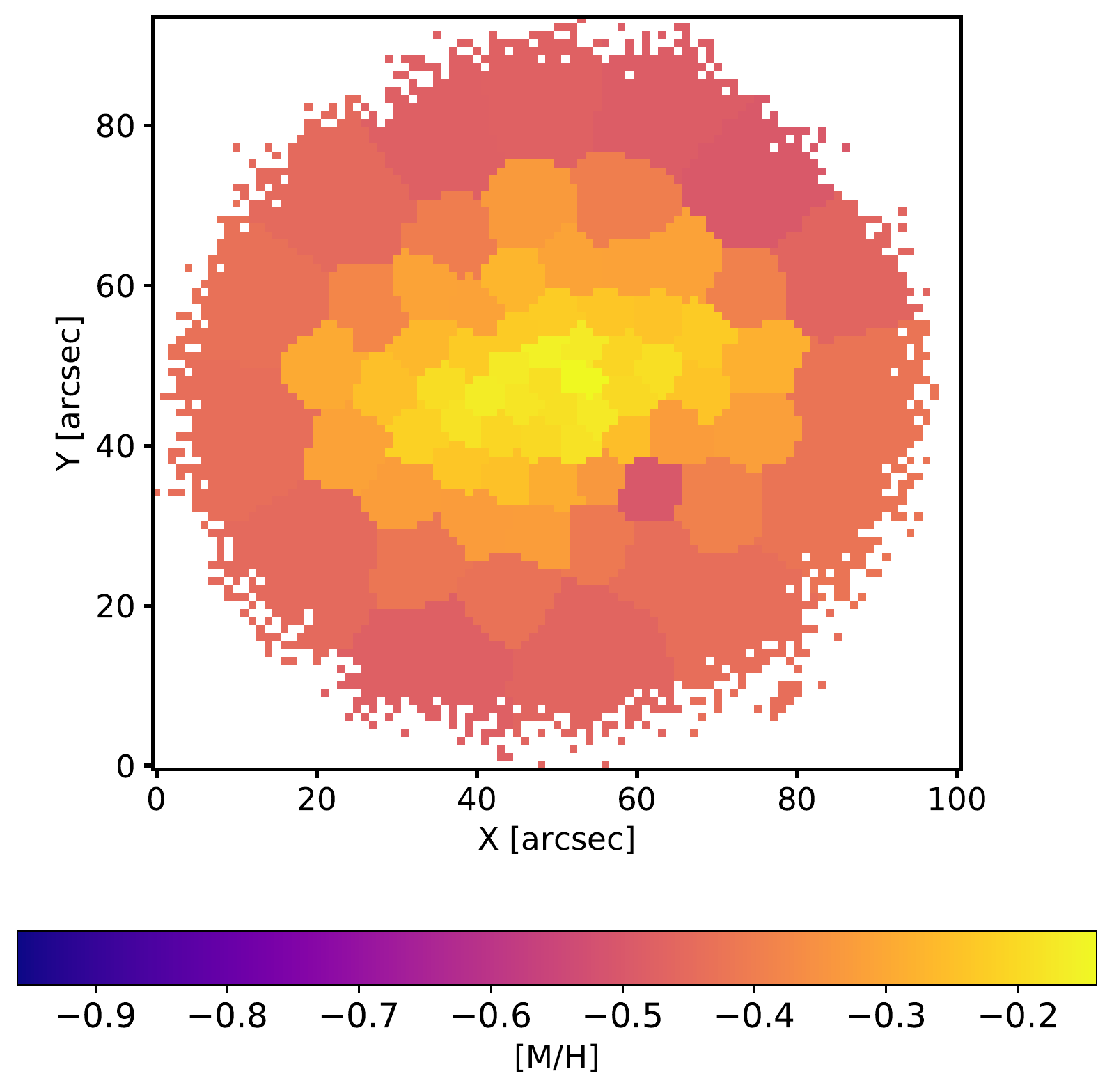}
      \includegraphics[width=6.1cm]{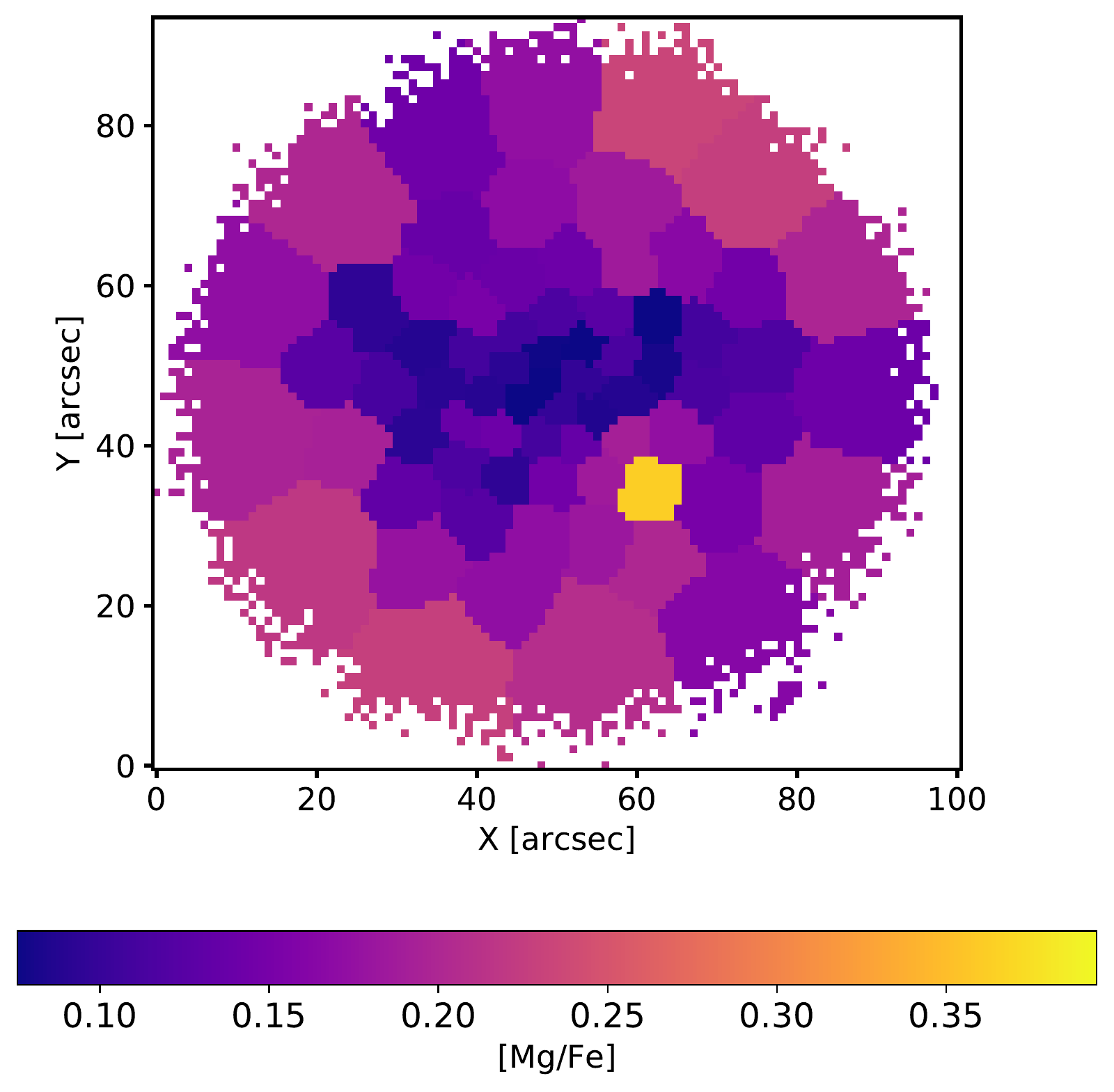}
      \includegraphics[width=6.1cm]{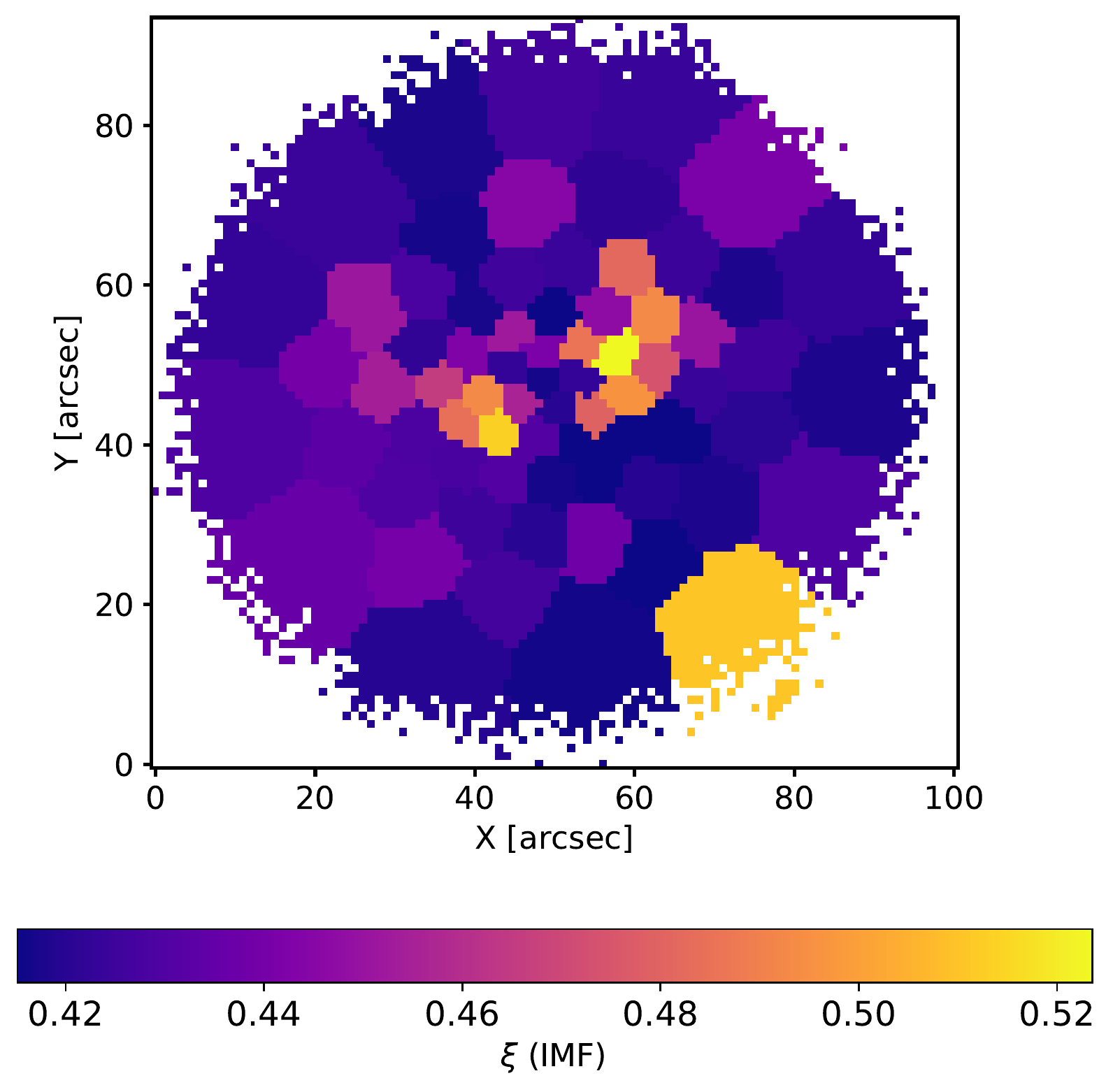}
      \caption{F3D stellar population maps of FCC\,182. From left to right and top to bottom: age, metallicity, [Mg/Fe], and IMF slope maps} 
   \end{figure*}

   \begin{figure*}
      \centering
      \includegraphics[width=8.cm]{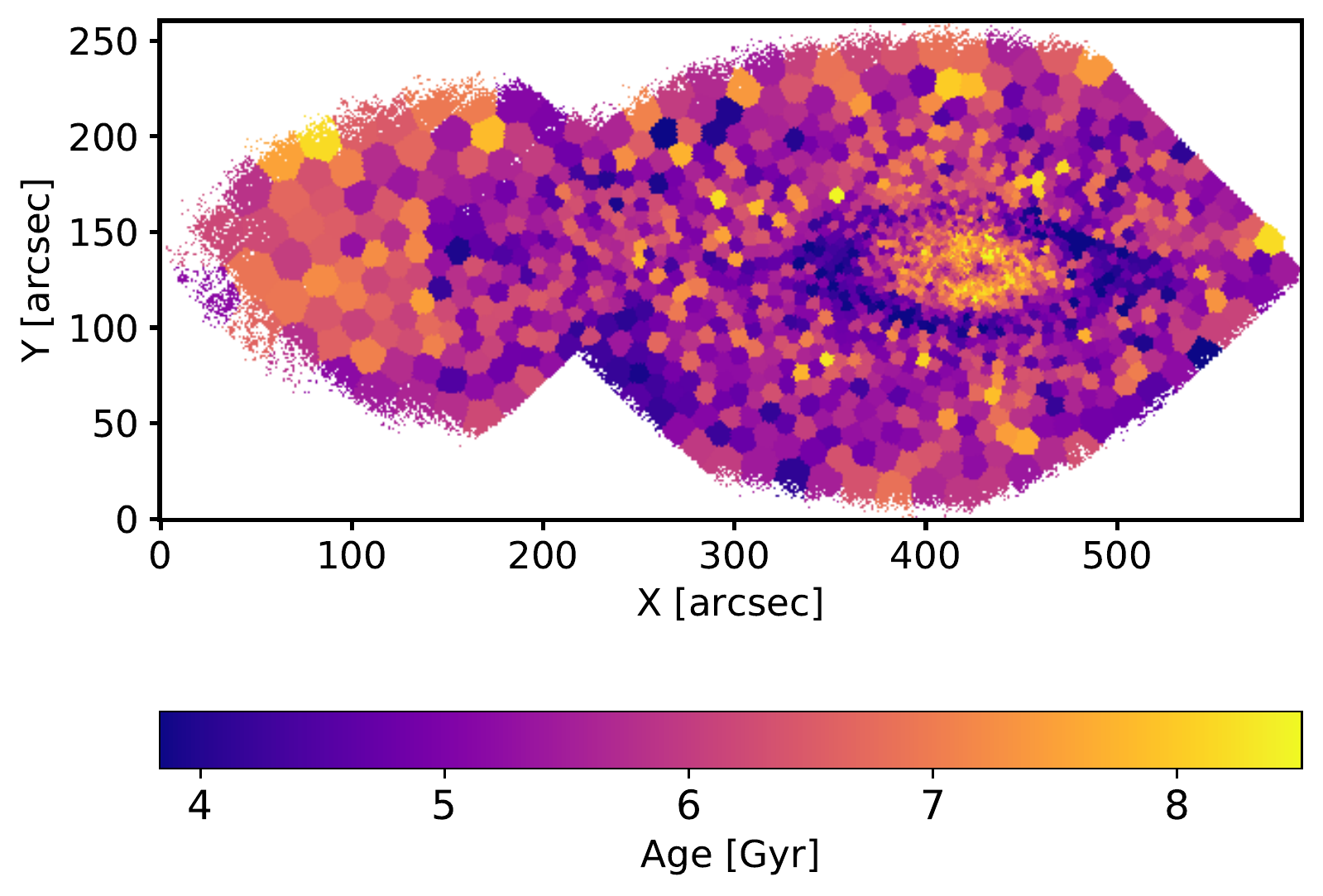}
      \includegraphics[width=8.cm]{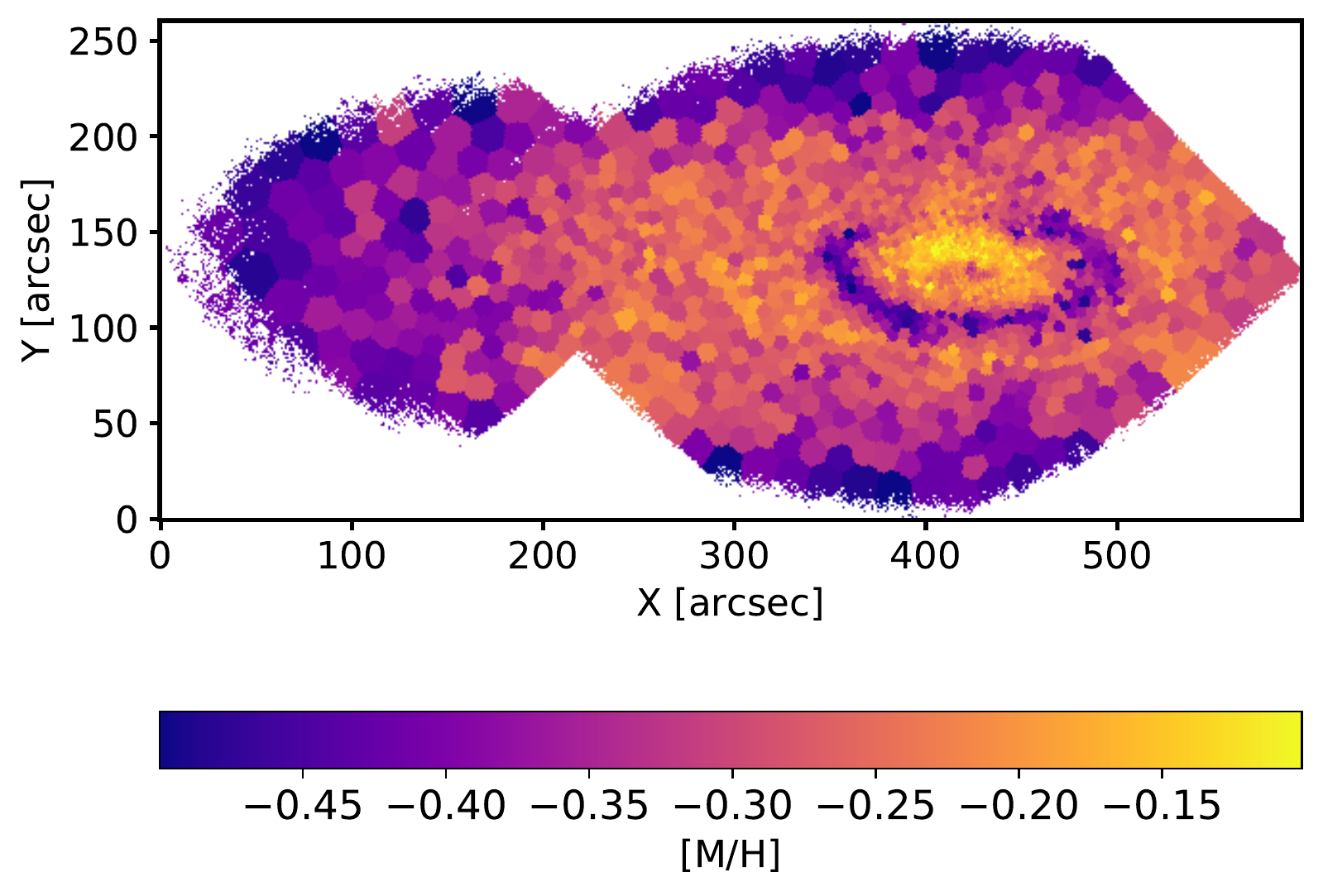}
      \includegraphics[width=8.cm]{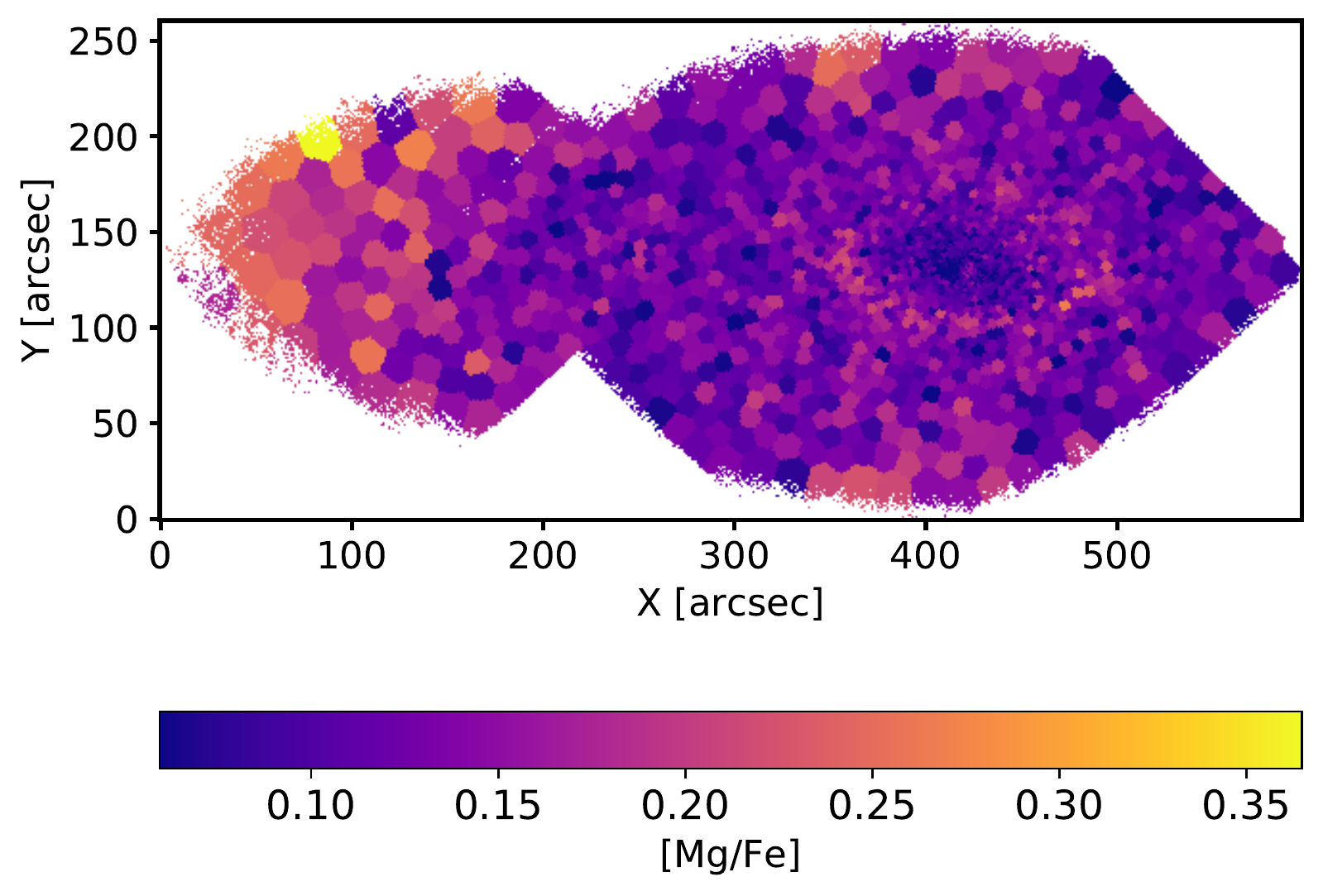}
      \includegraphics[width=8.cm]{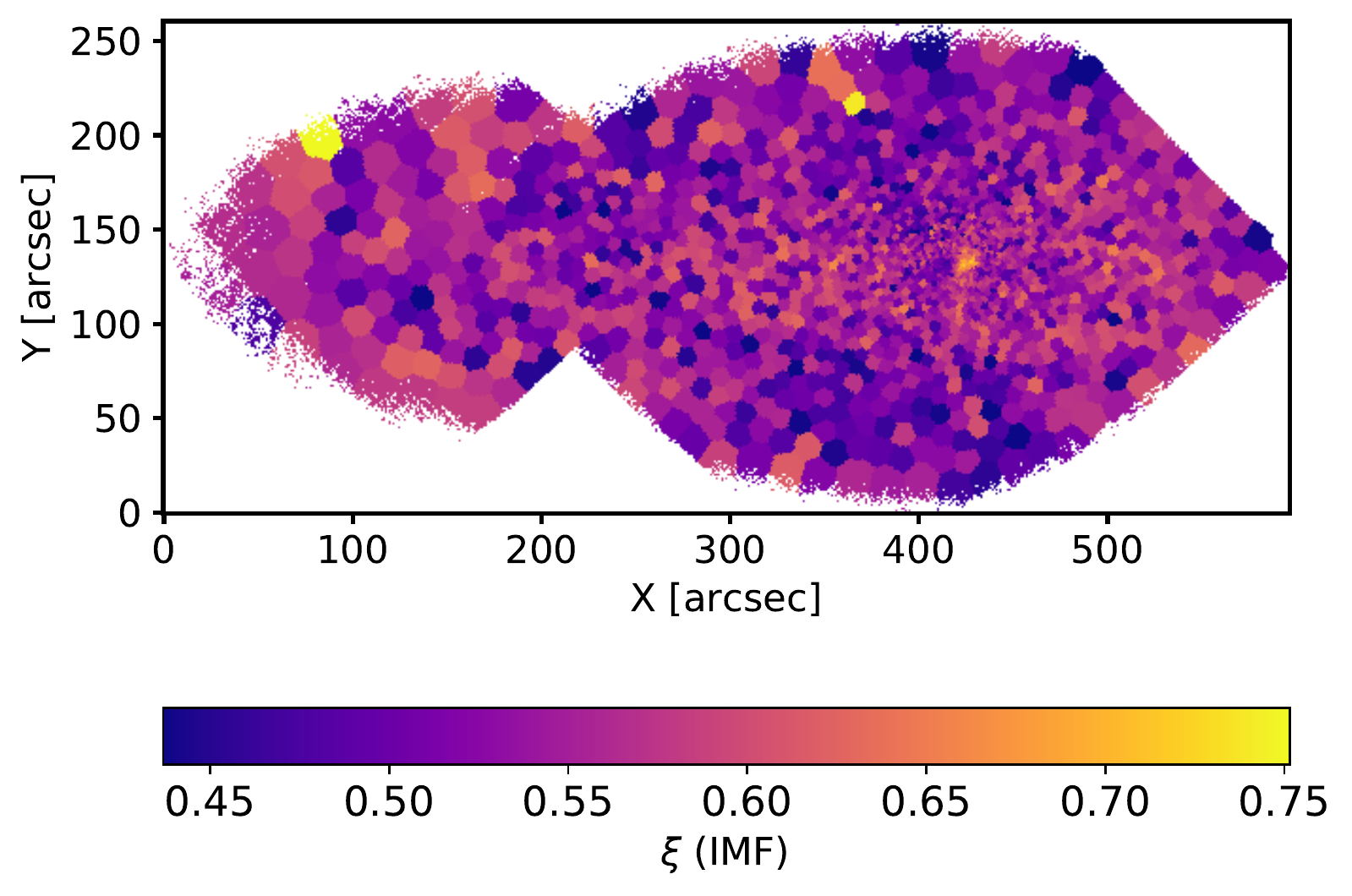}
      \caption{F3D stellar population maps of FCC\,179.} 
   \end{figure*}

   \begin{figure*}
      \centering
      \includegraphics[width=6.1cm]{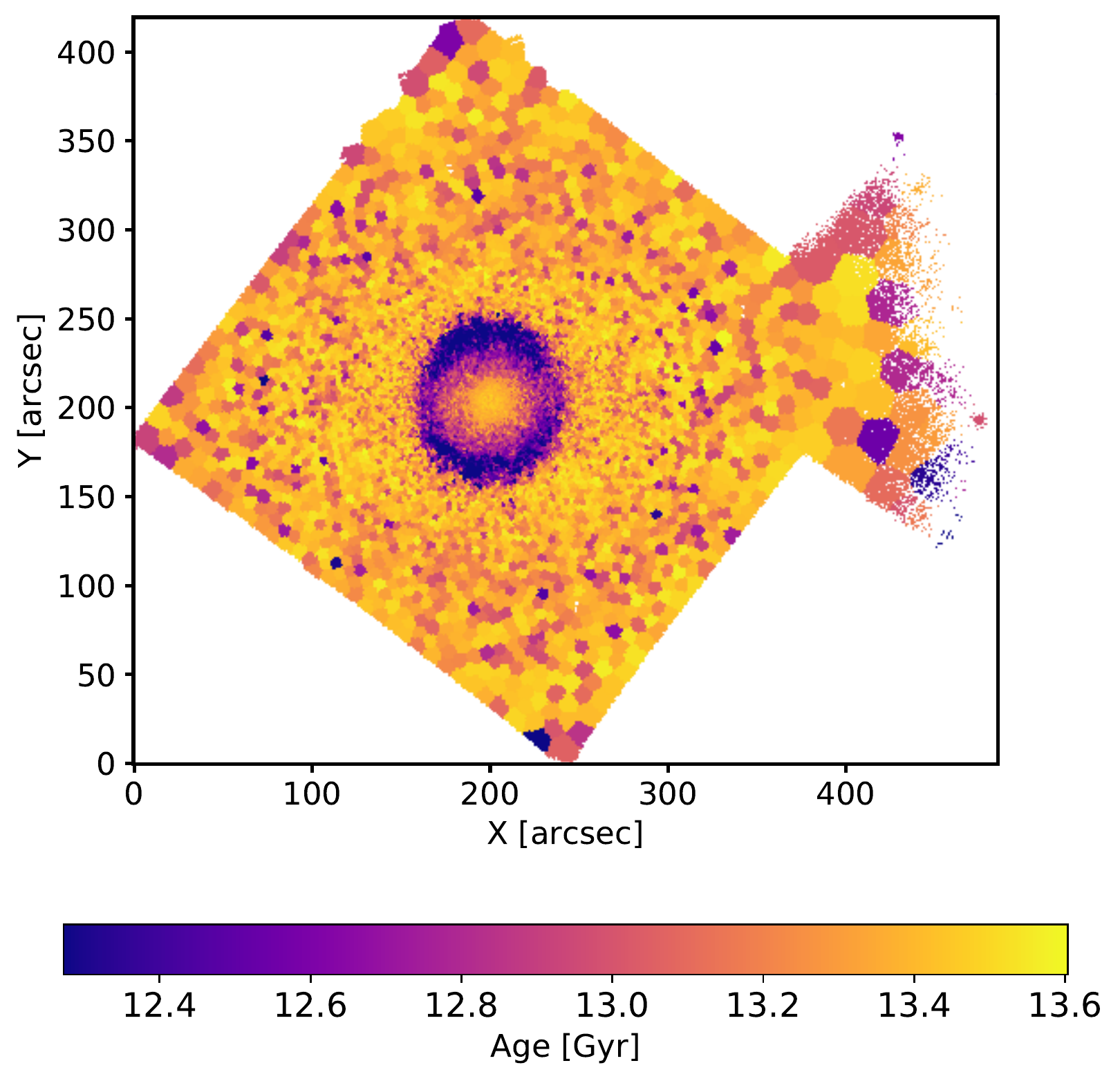}
      \includegraphics[width=6.1cm]{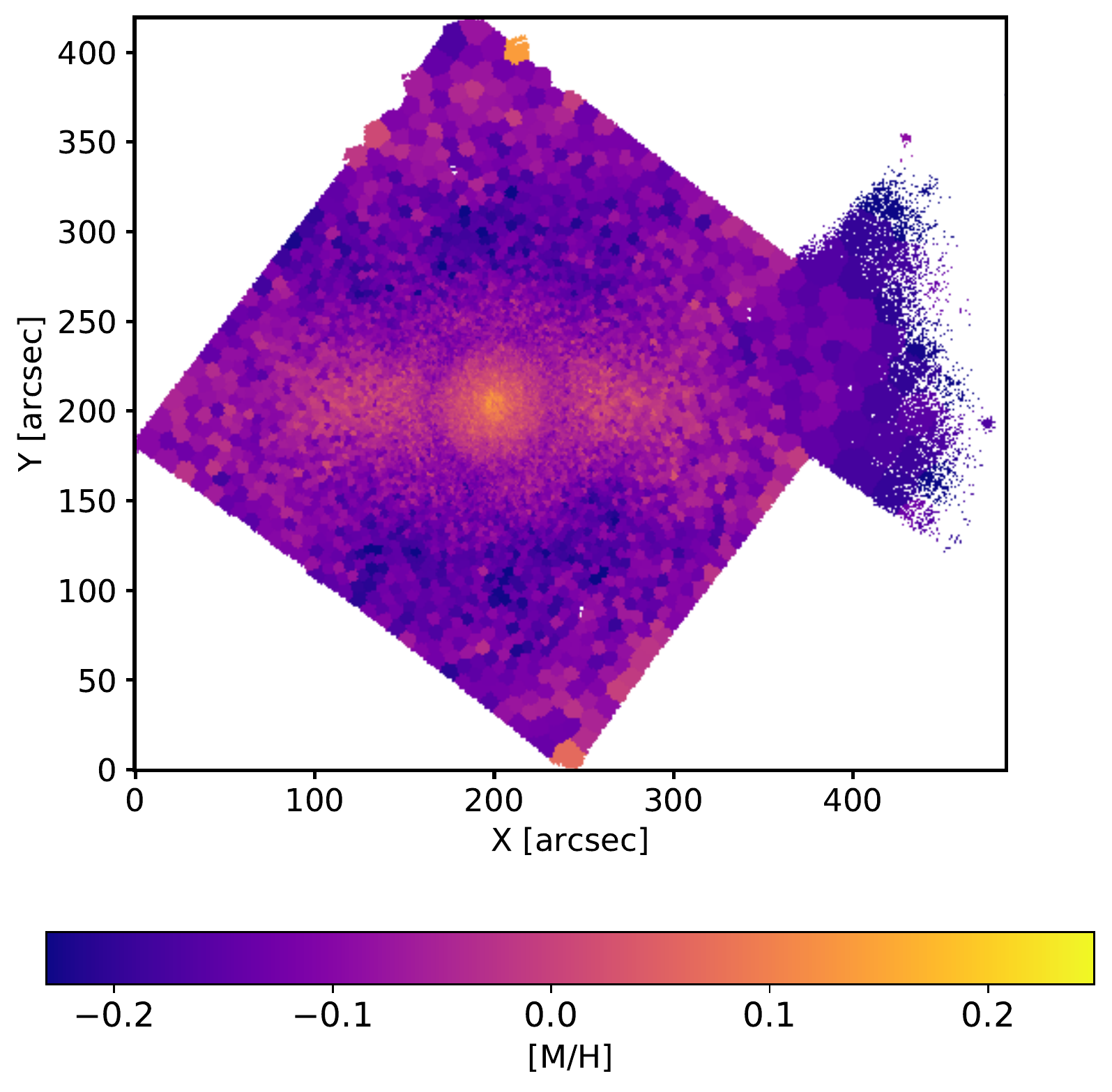}
      \includegraphics[width=6.1cm]{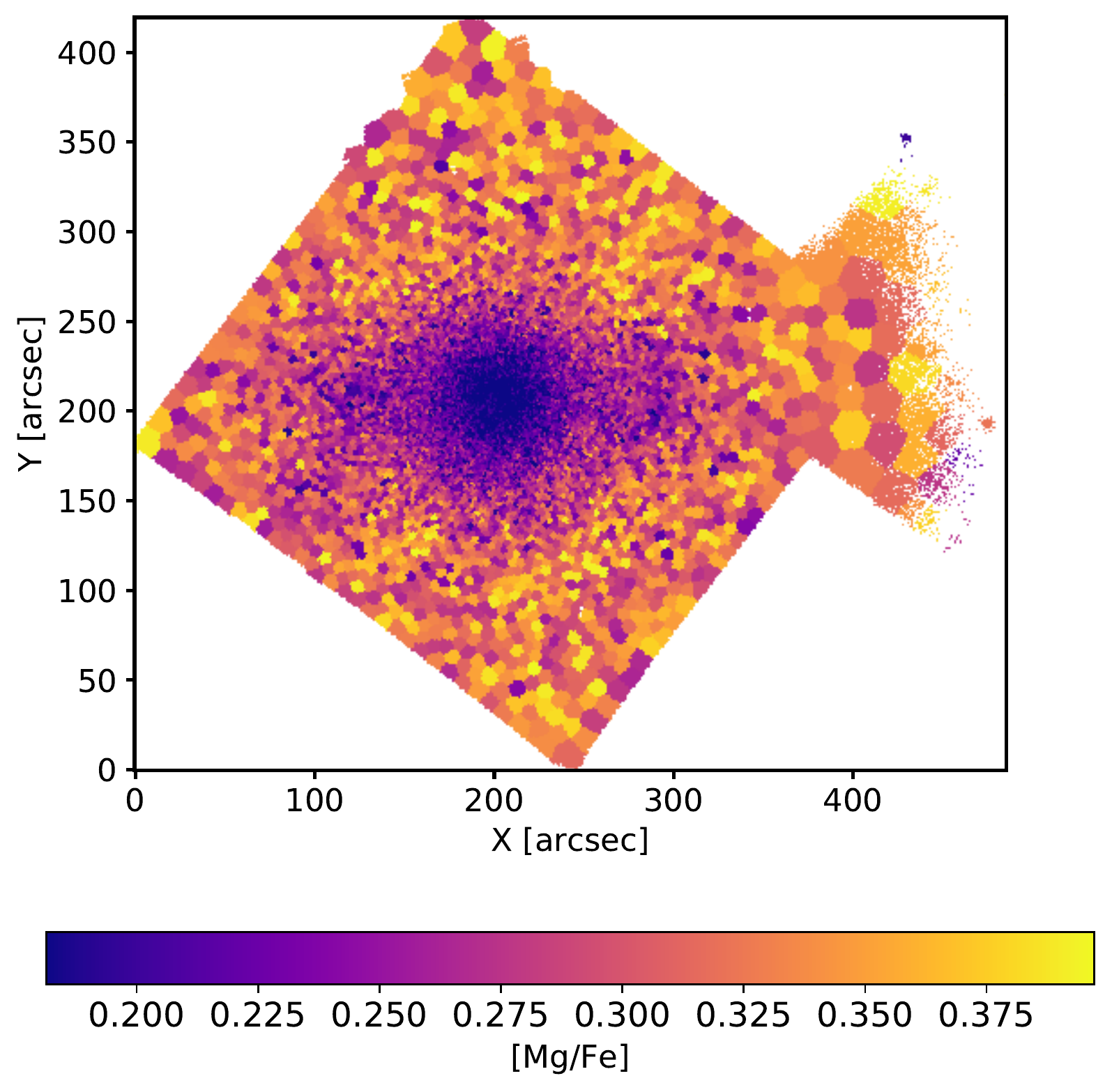}
      \includegraphics[width=6.1cm]{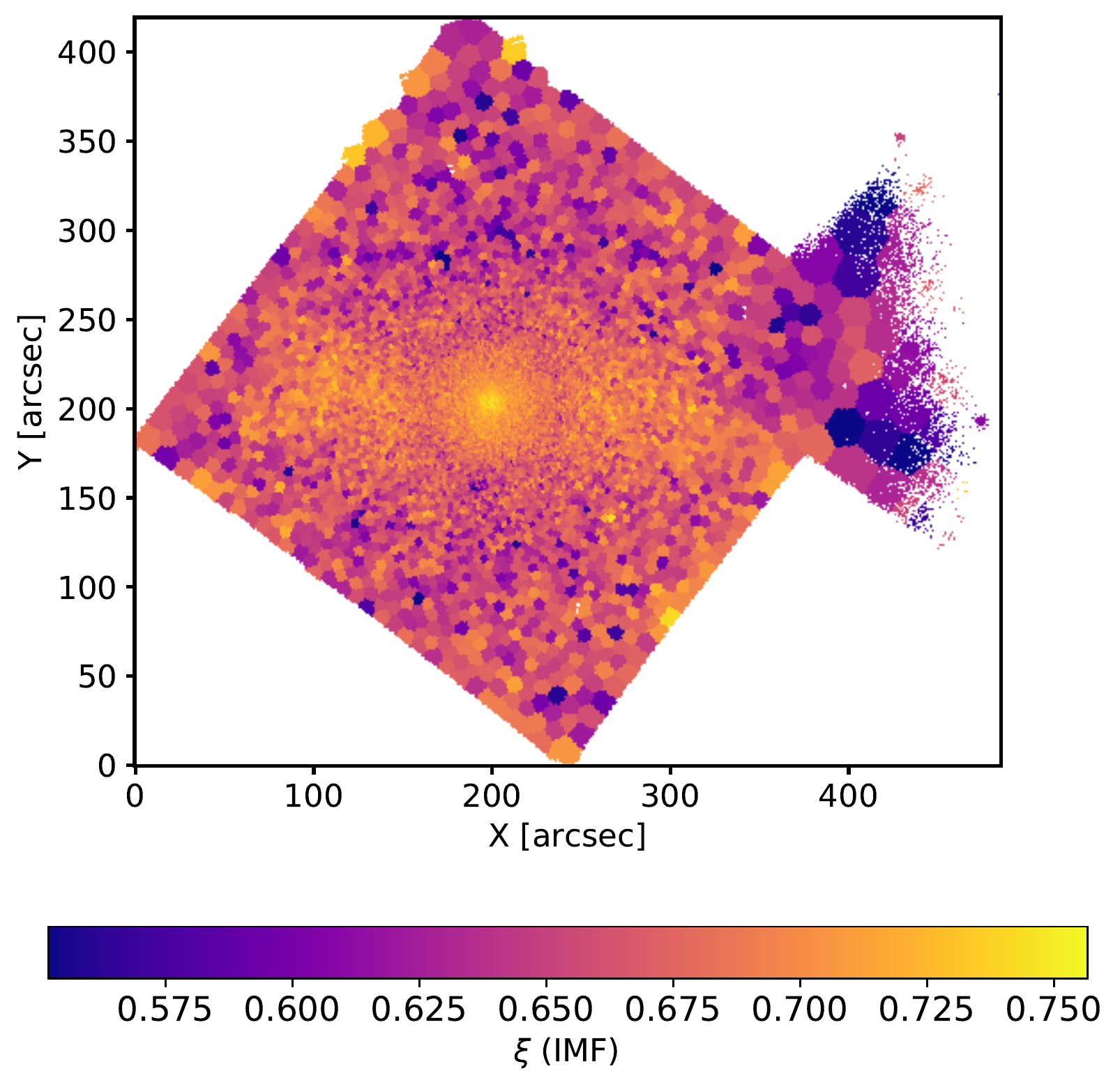}
      \caption{F3D stellar population maps of FCC\,184.} 
      \label{fig:fcc184}
   \end{figure*}

   \begin{figure*}
      \centering
      \includegraphics[width=6.1cm]{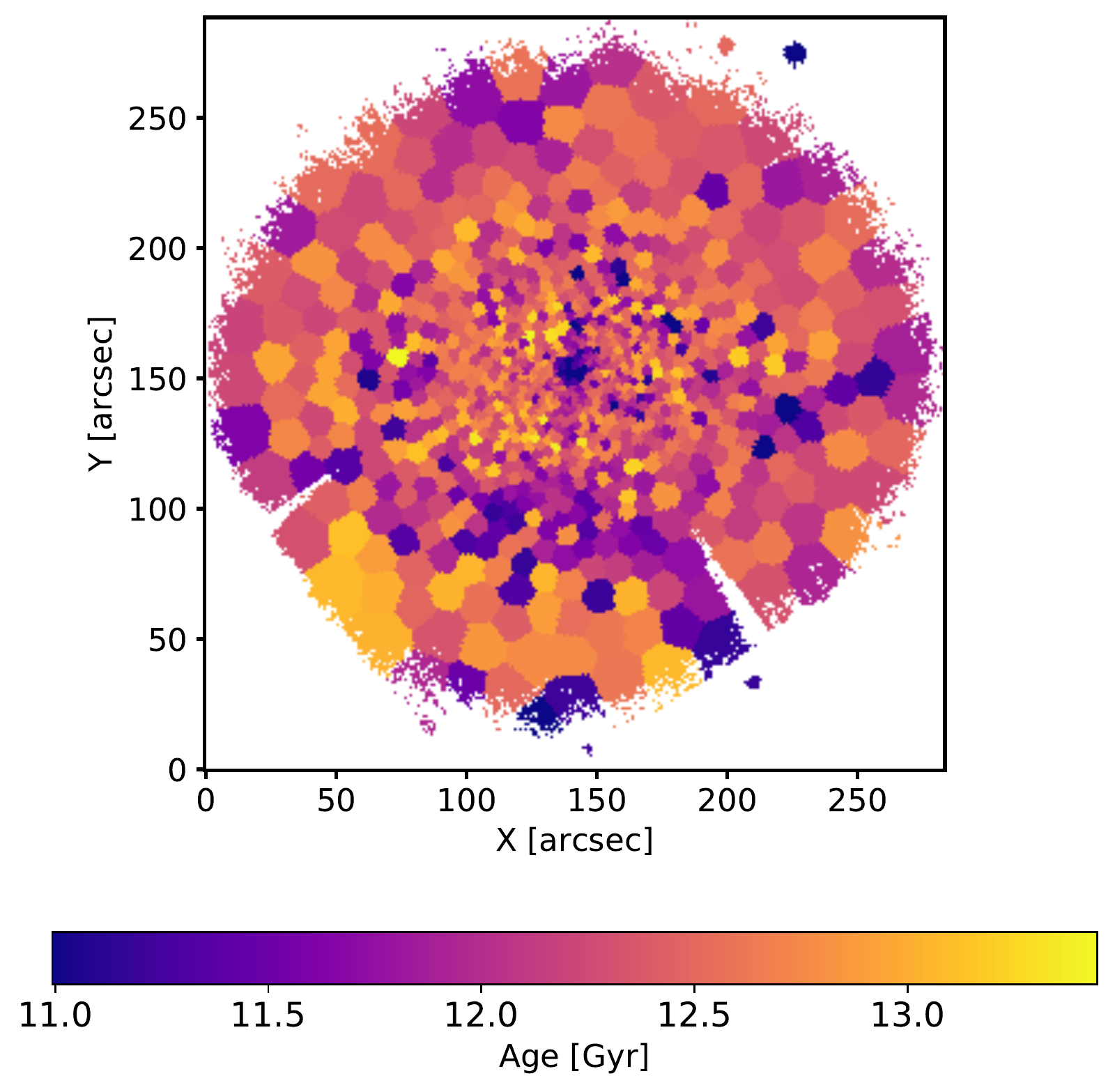}
      \includegraphics[width=6.1cm]{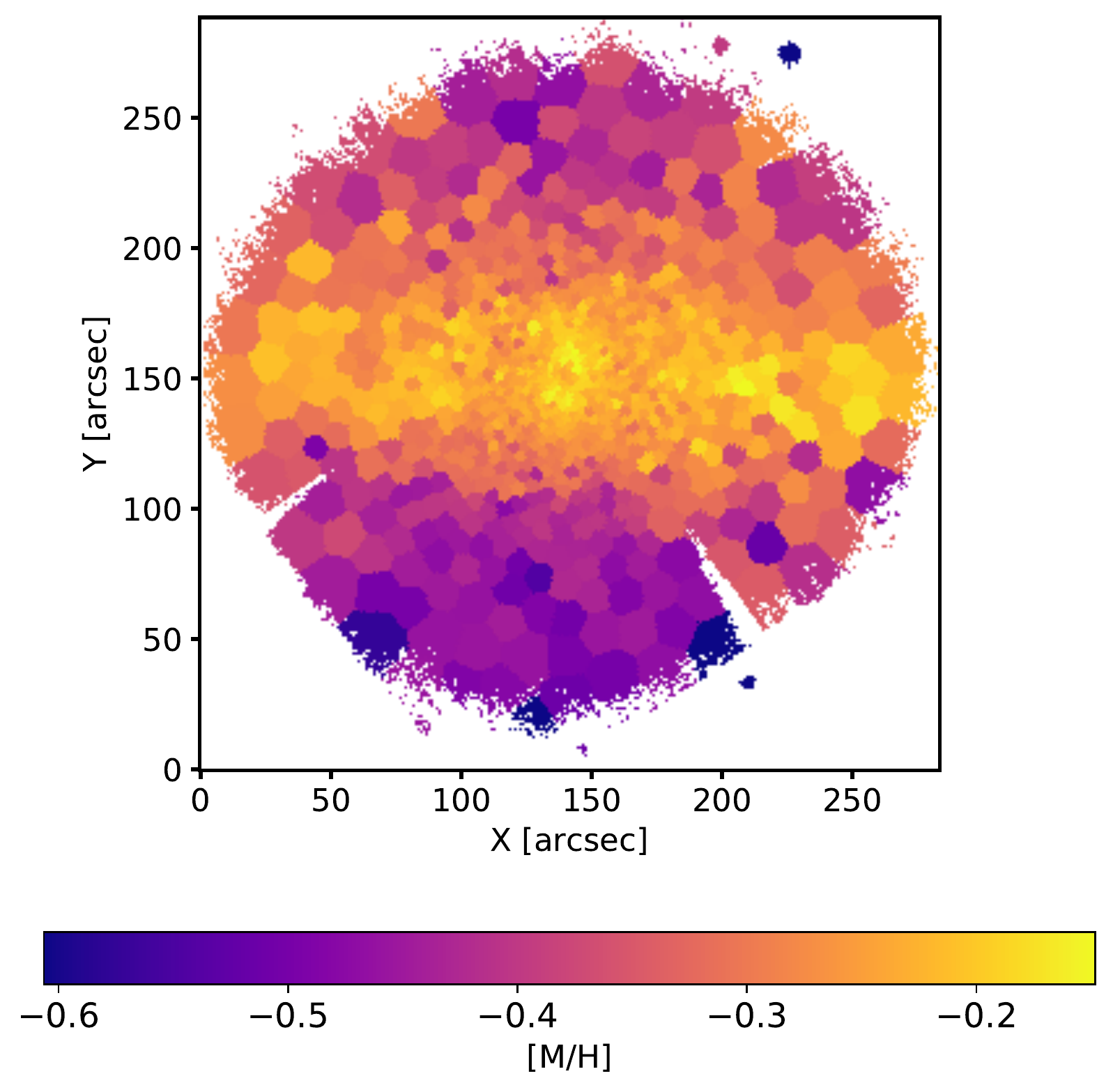}
      \includegraphics[width=6.1cm]{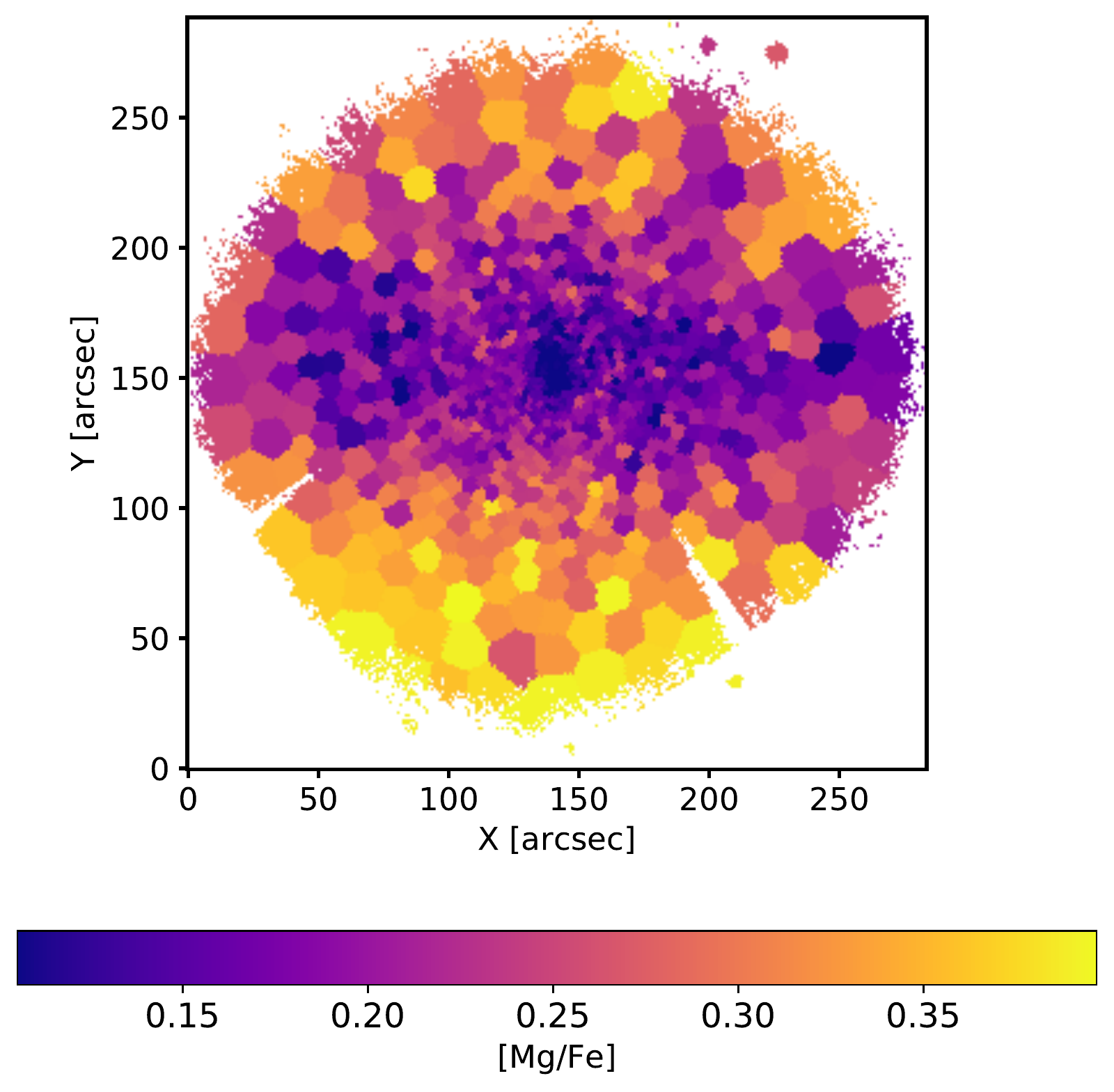}
      \includegraphics[width=6.1cm]{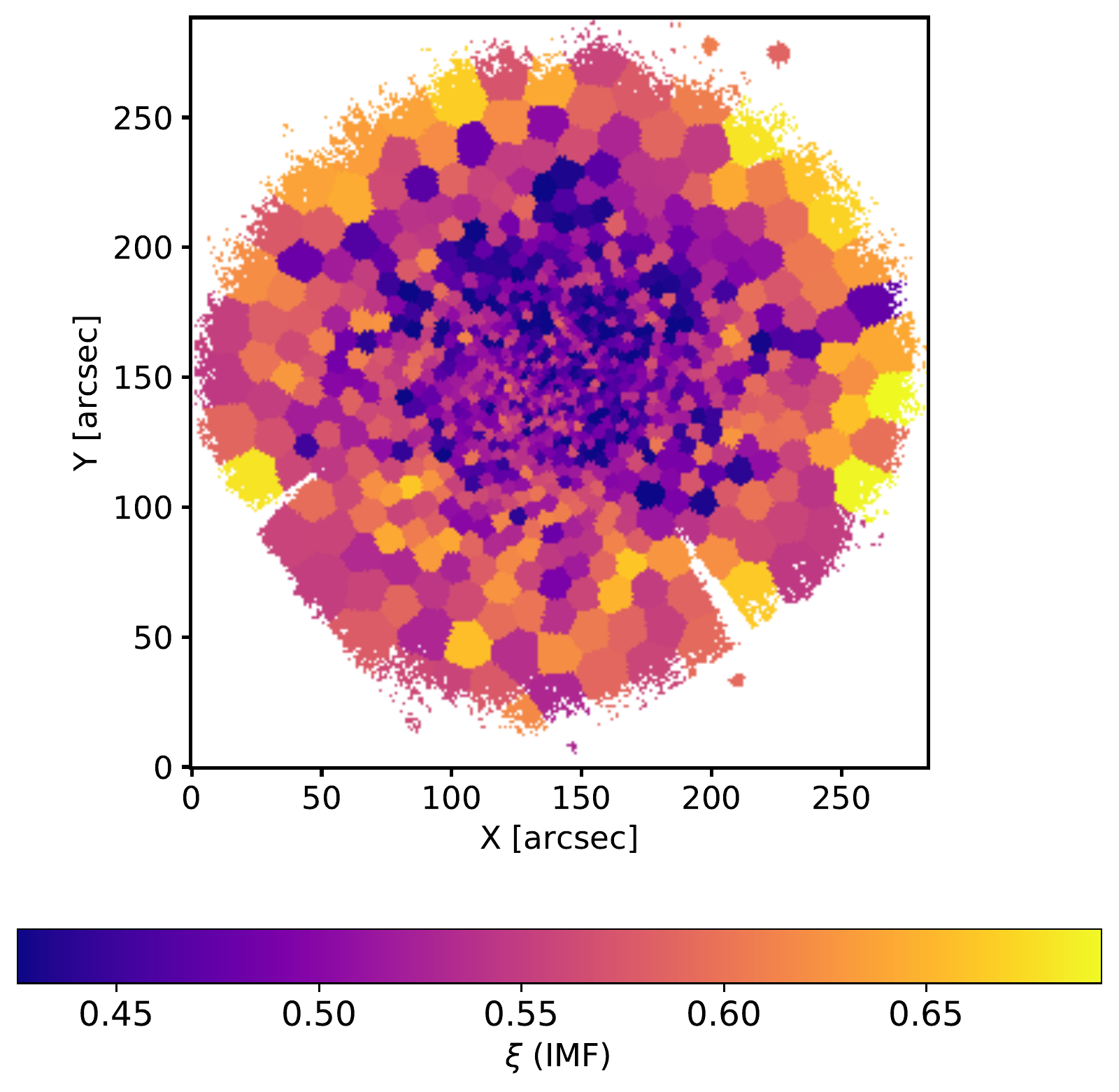}
      \caption{F3D stellar population maps of FCC\,190.} 
   \end{figure*}

   \begin{figure*}
      \centering
      \includegraphics[width=6.1cm]{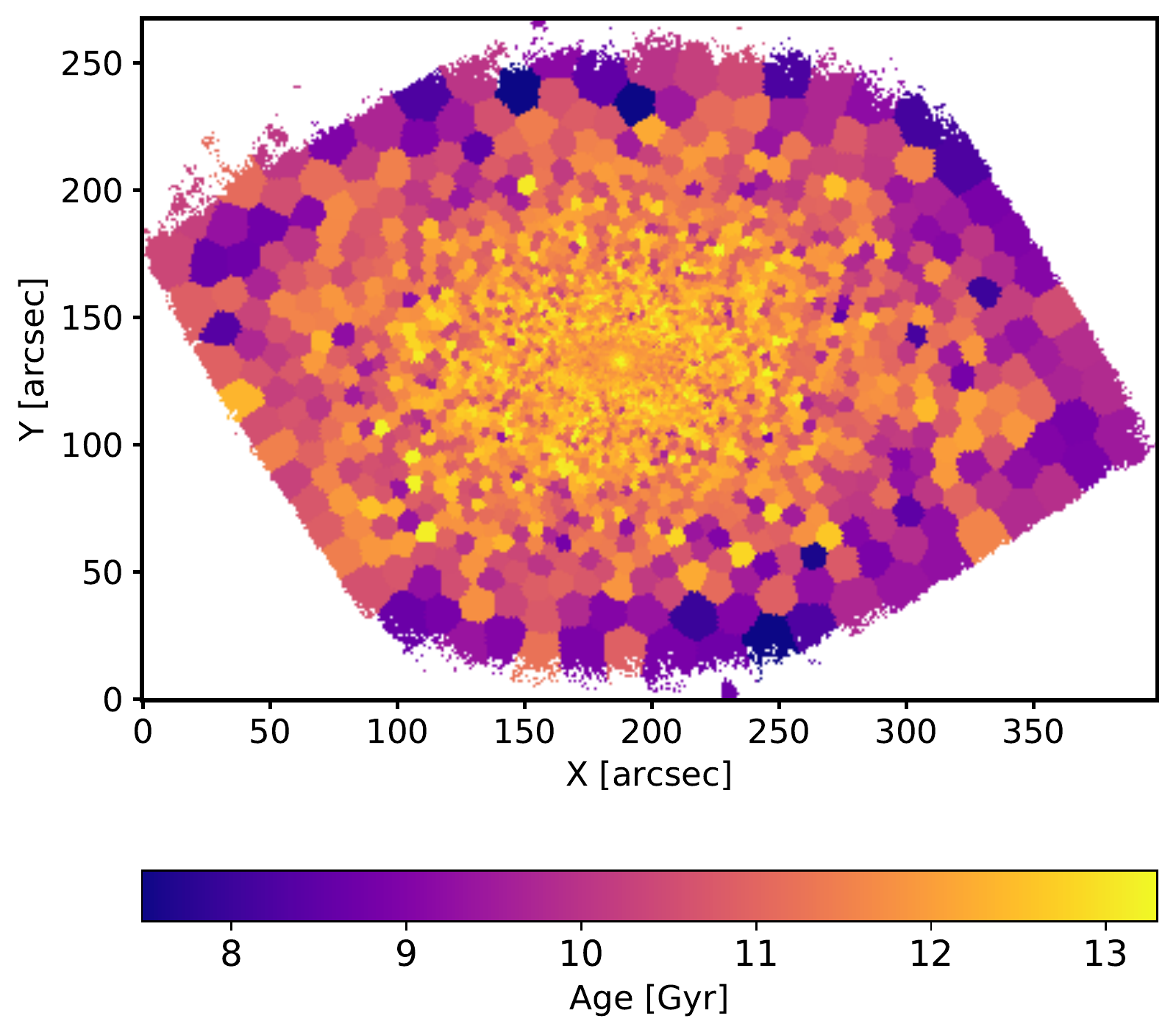}
      \includegraphics[width=6.1cm]{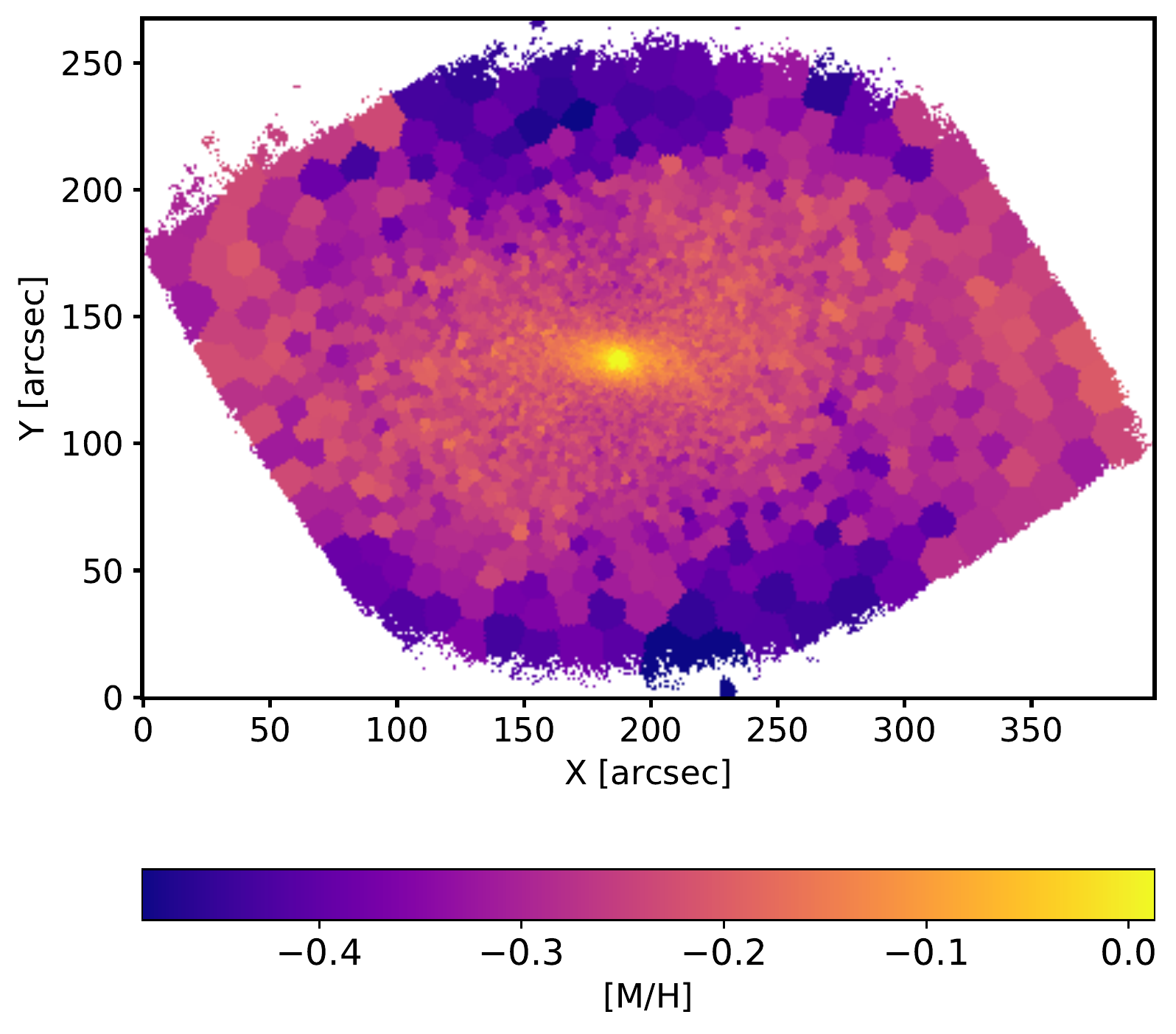}
      \includegraphics[width=6.1cm]{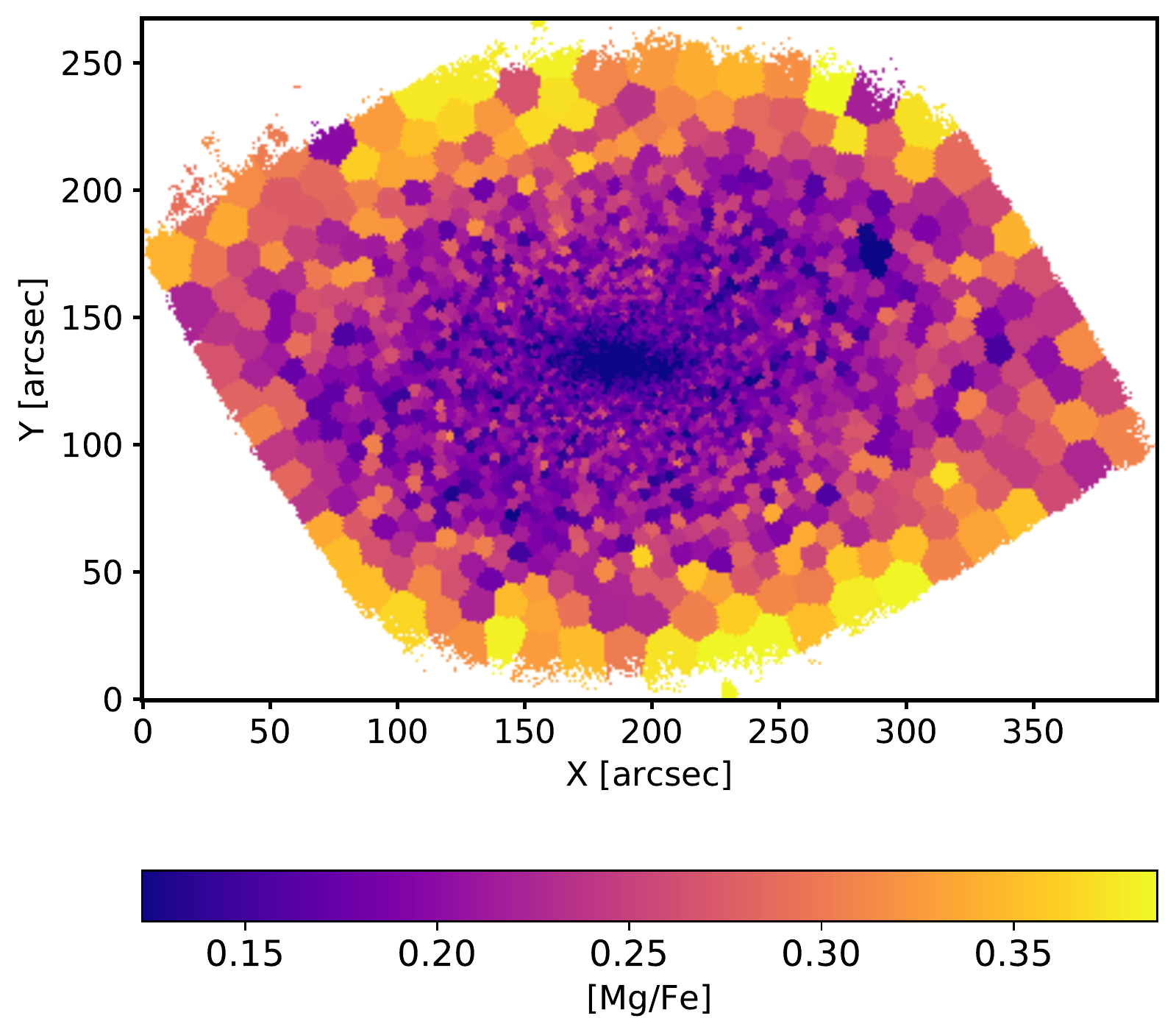}
      \includegraphics[width=6.1cm]{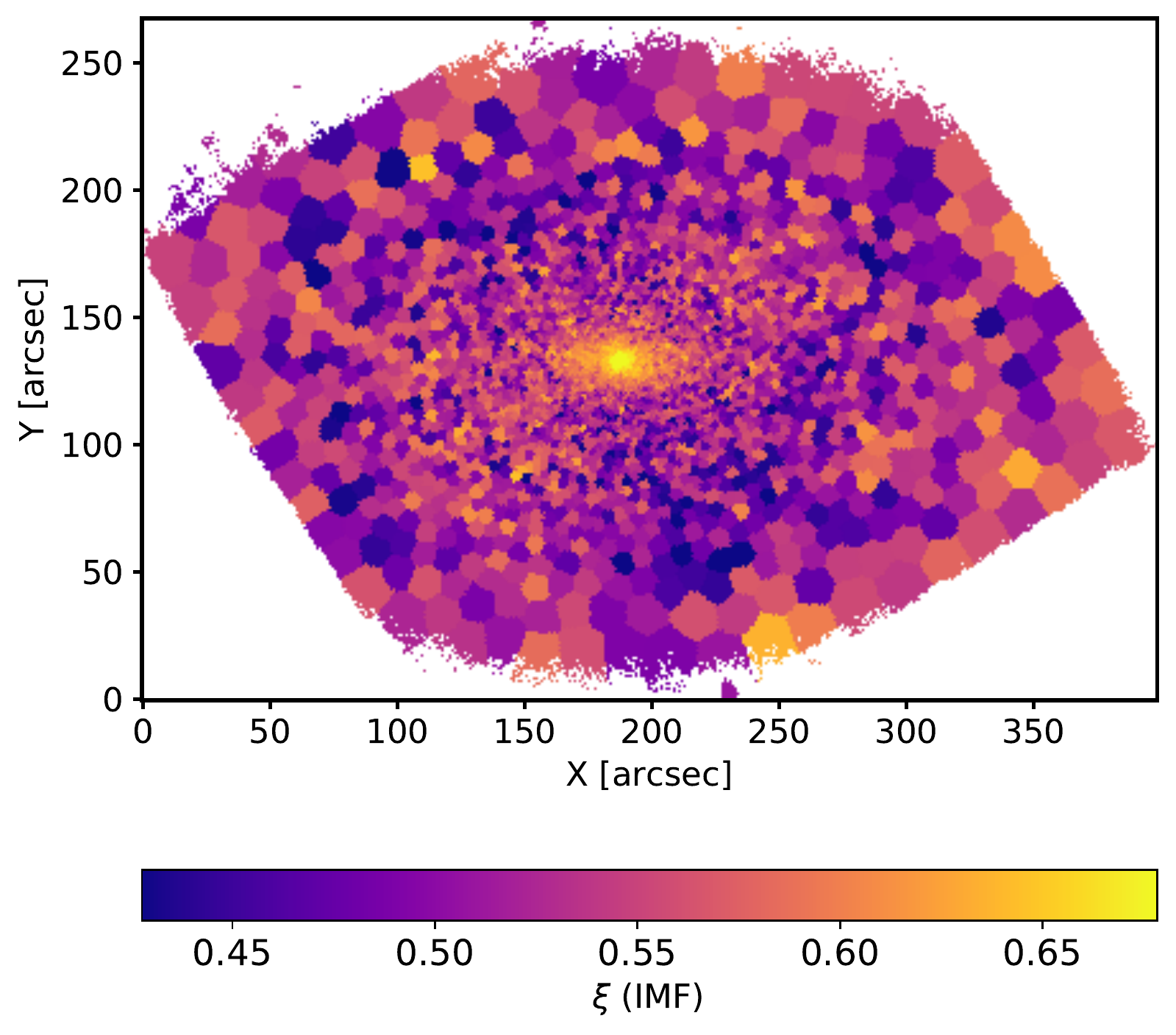}
      \caption{F3D stellar population maps of FCC\,193.} 
   \end{figure*}

   \begin{figure*}
      \centering
      \includegraphics[width=6.1cm]{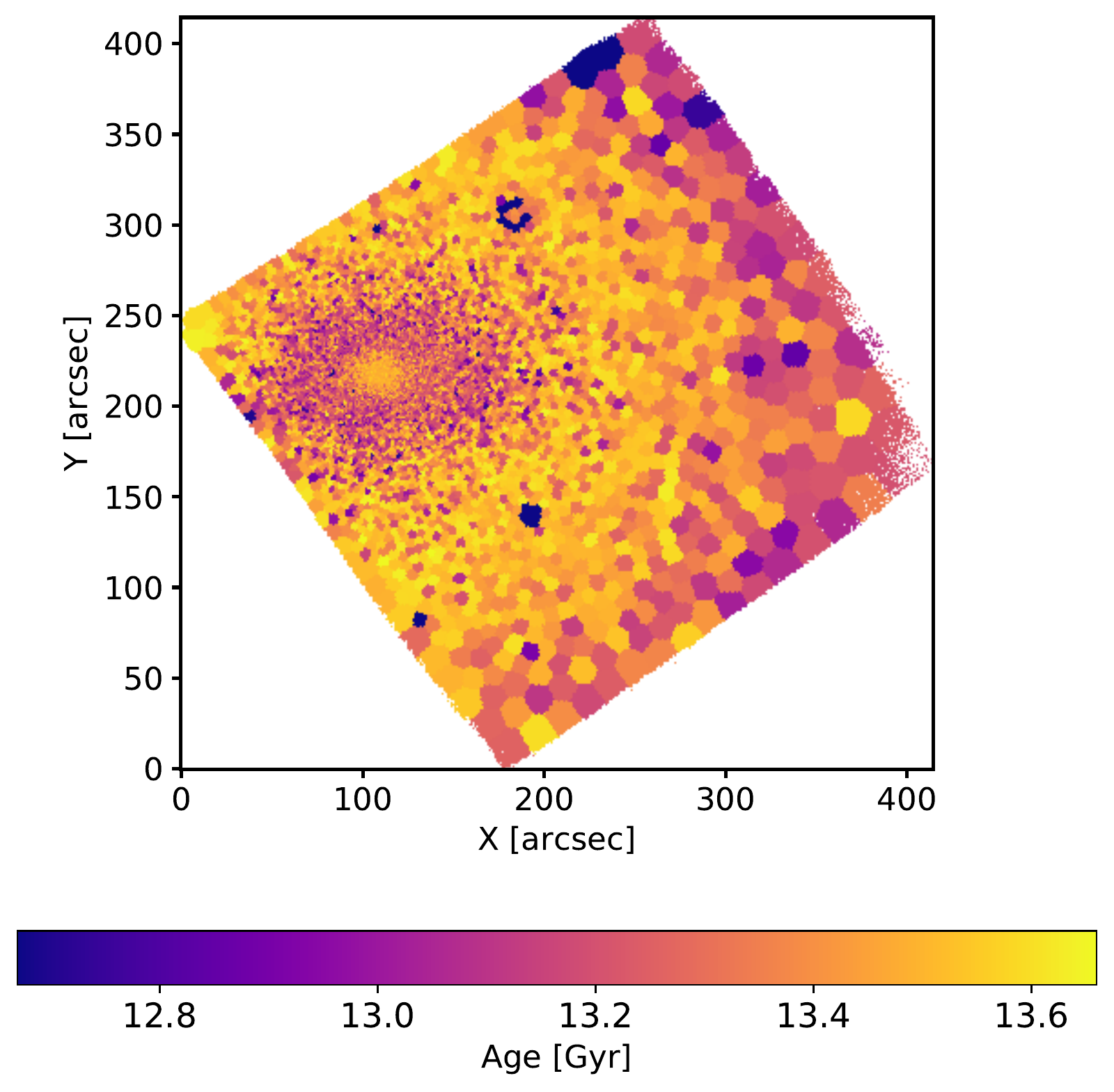}
      \includegraphics[width=6.1cm]{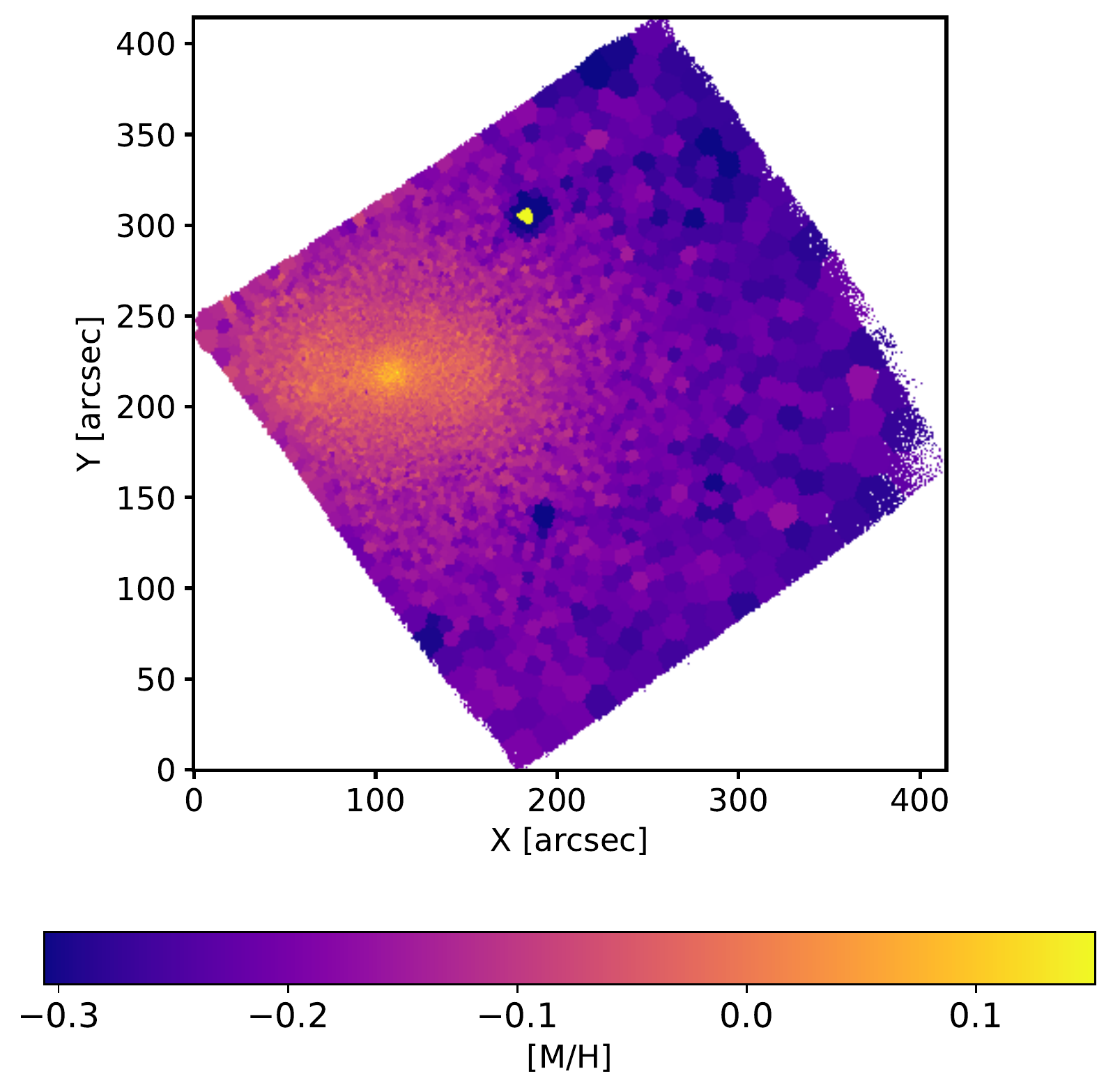}
      \includegraphics[width=6.1cm]{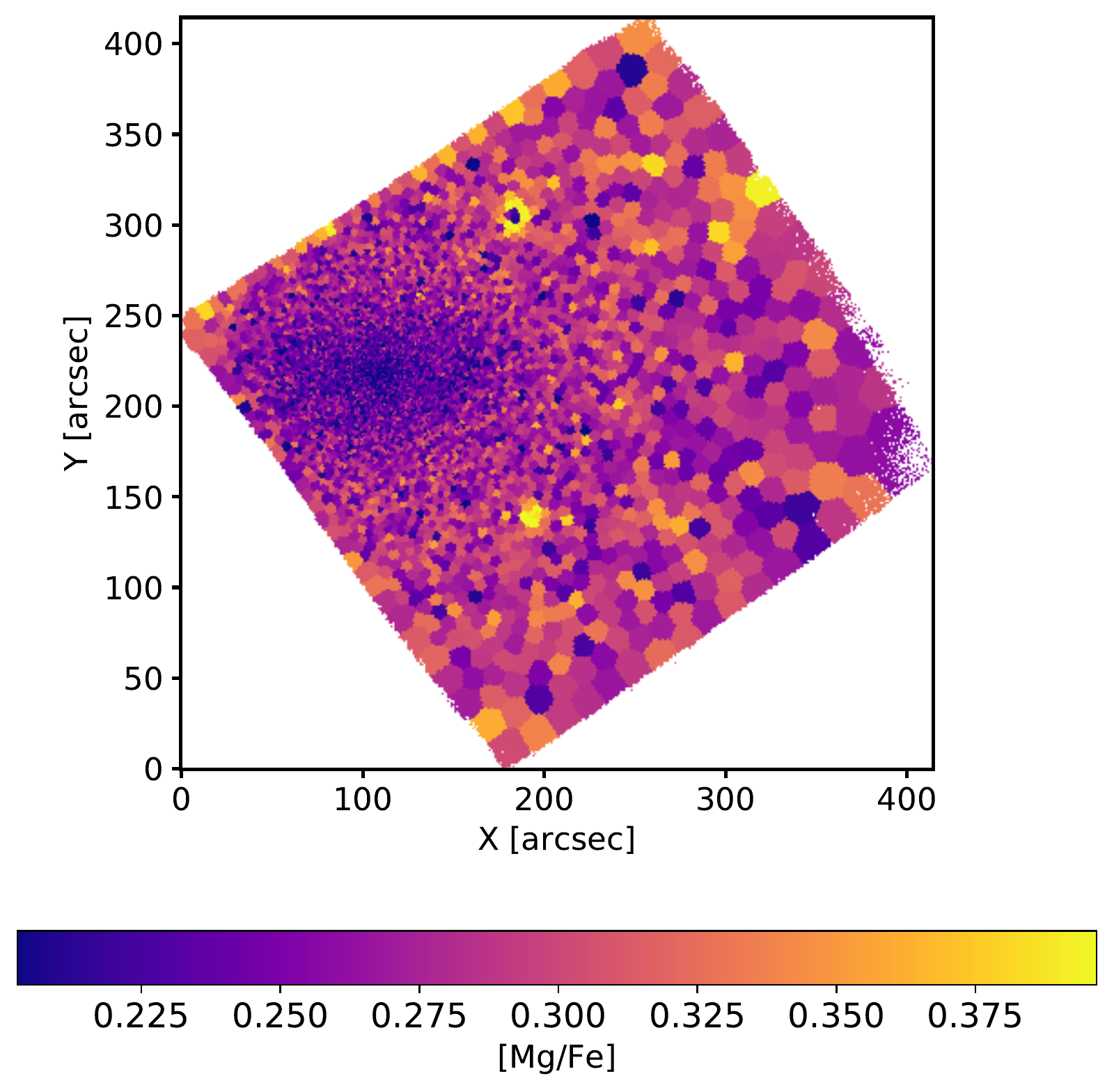}
      \includegraphics[width=6.1cm]{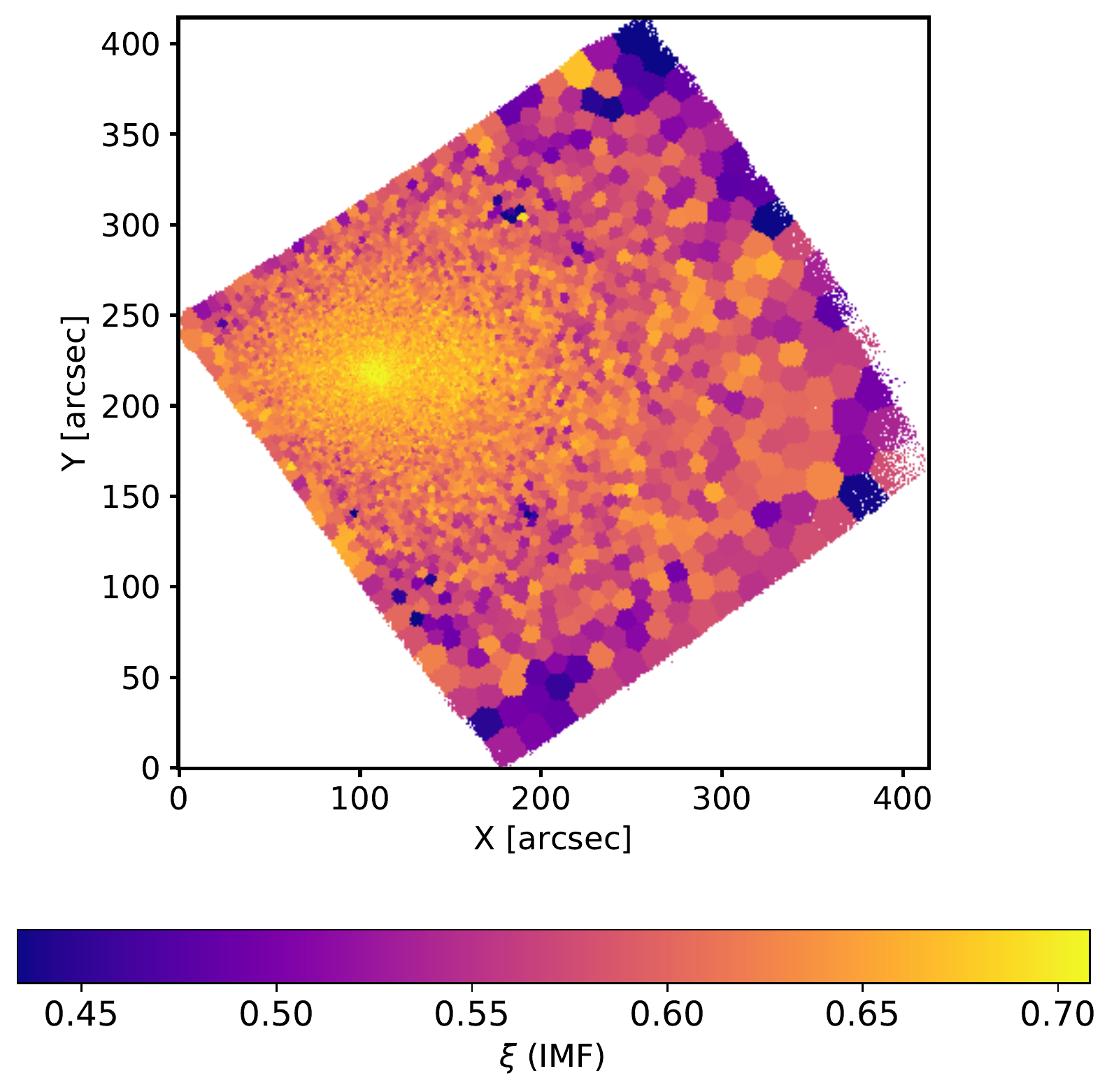}
      \caption{F3D stellar population maps of FCC\,219.} 
   \end{figure*}

   \begin{figure*}
      \centering
      \includegraphics[width=6.1cm]{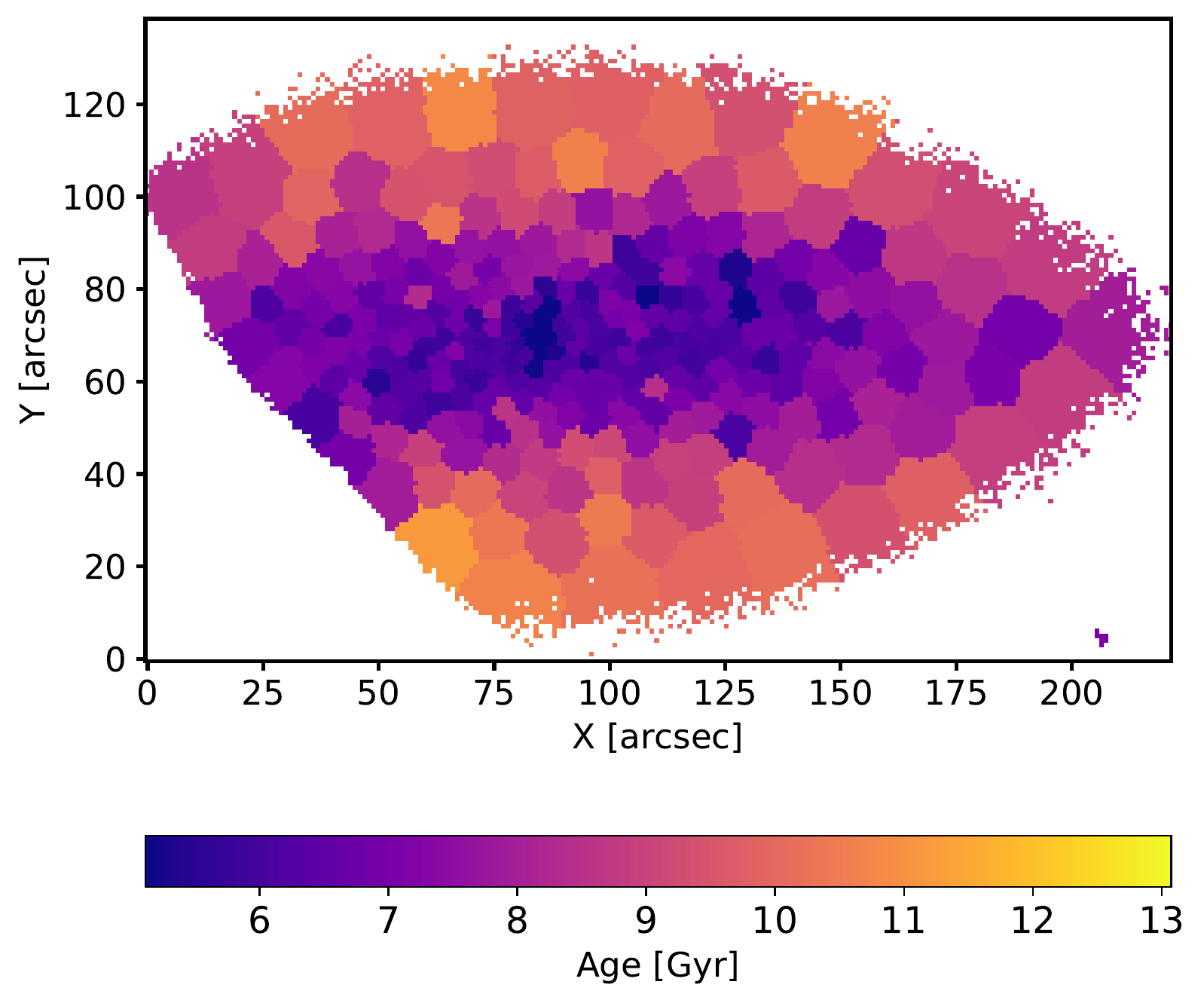}
      \includegraphics[width=6.1cm]{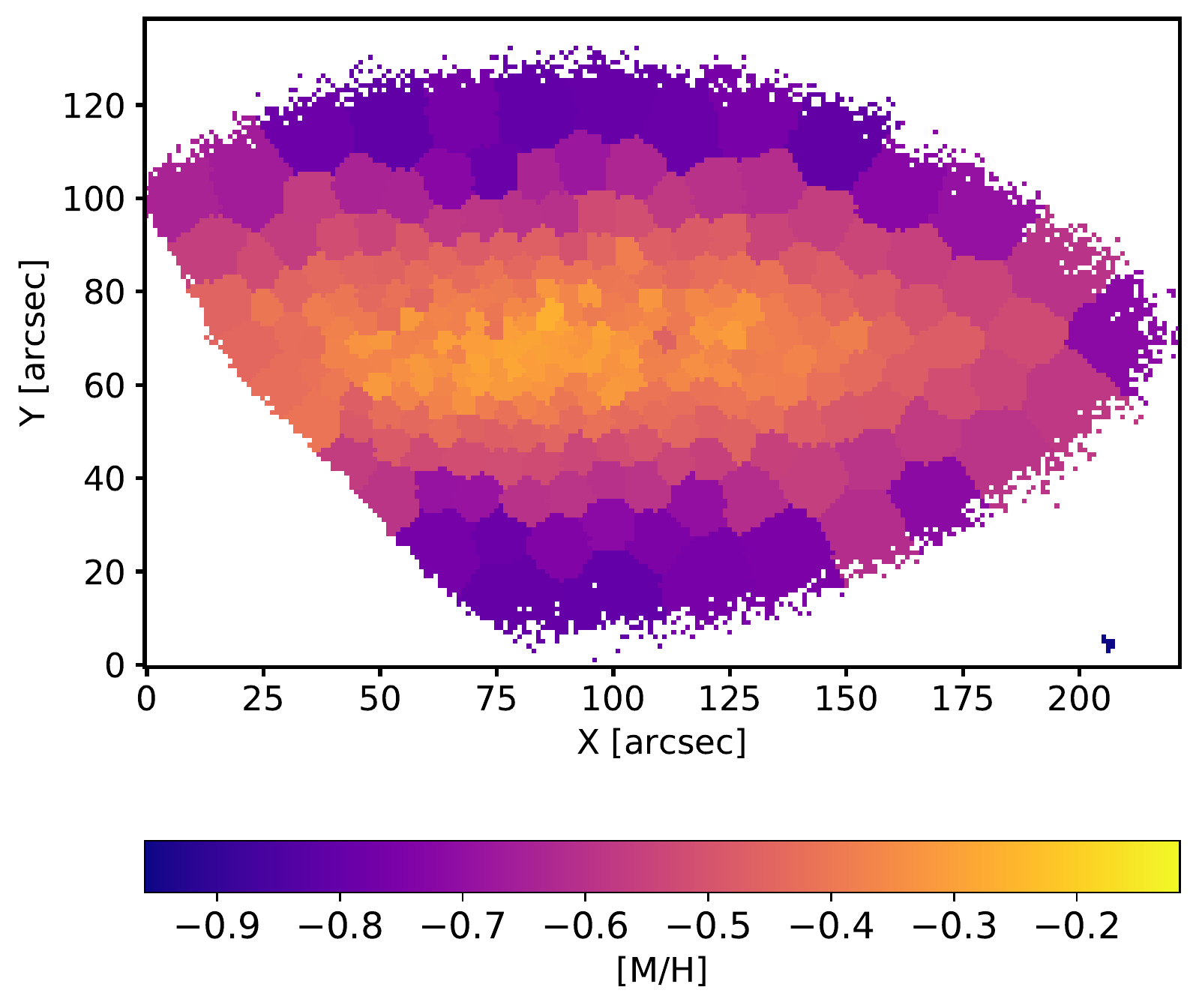}
      \includegraphics[width=6.1cm]{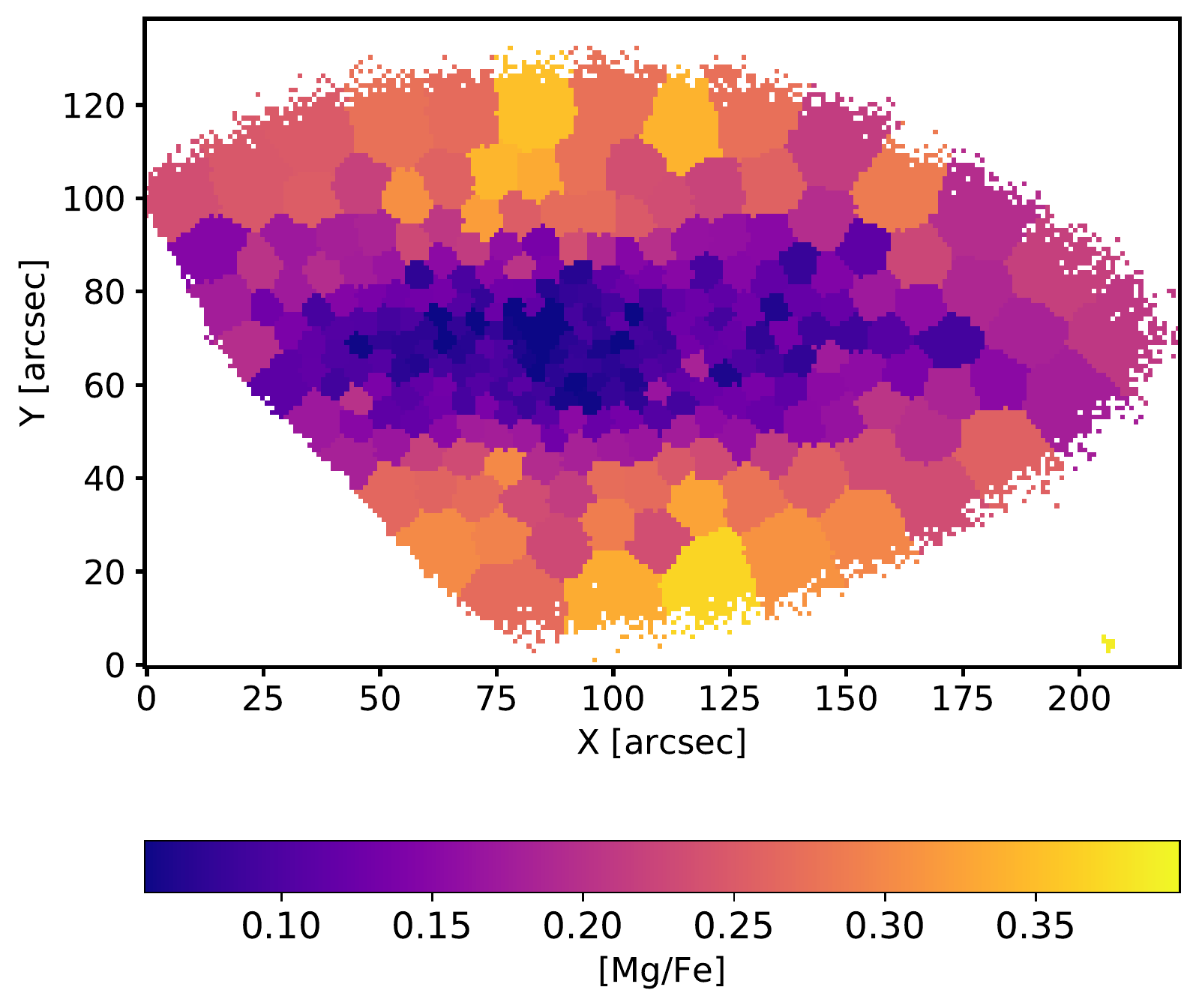}
      \includegraphics[width=6.1cm]{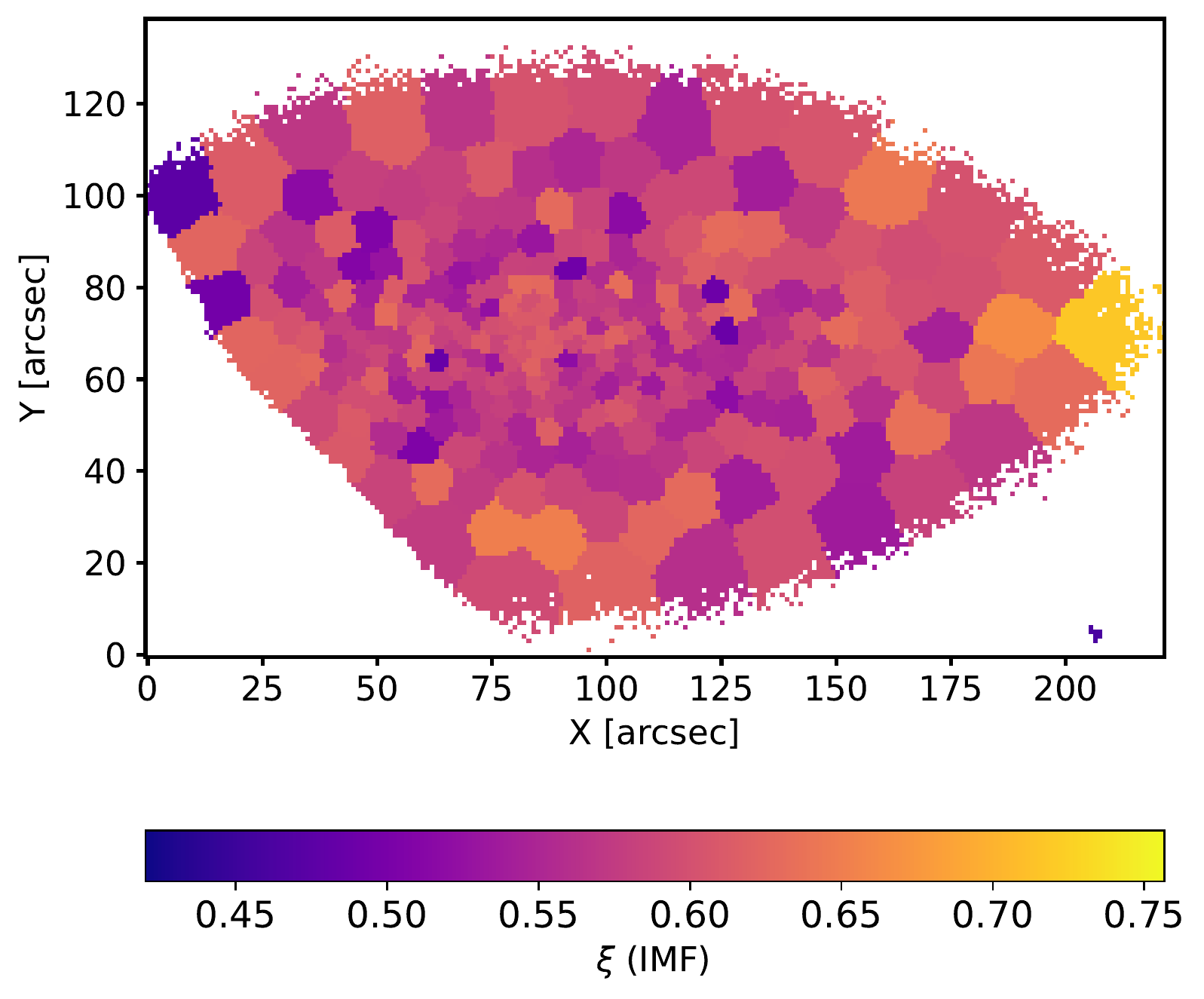}
      \caption{F3D stellar population maps of FCC\,255.} 
   \end{figure*}

   \begin{figure*}
      \centering
      \includegraphics[width=6.1cm]{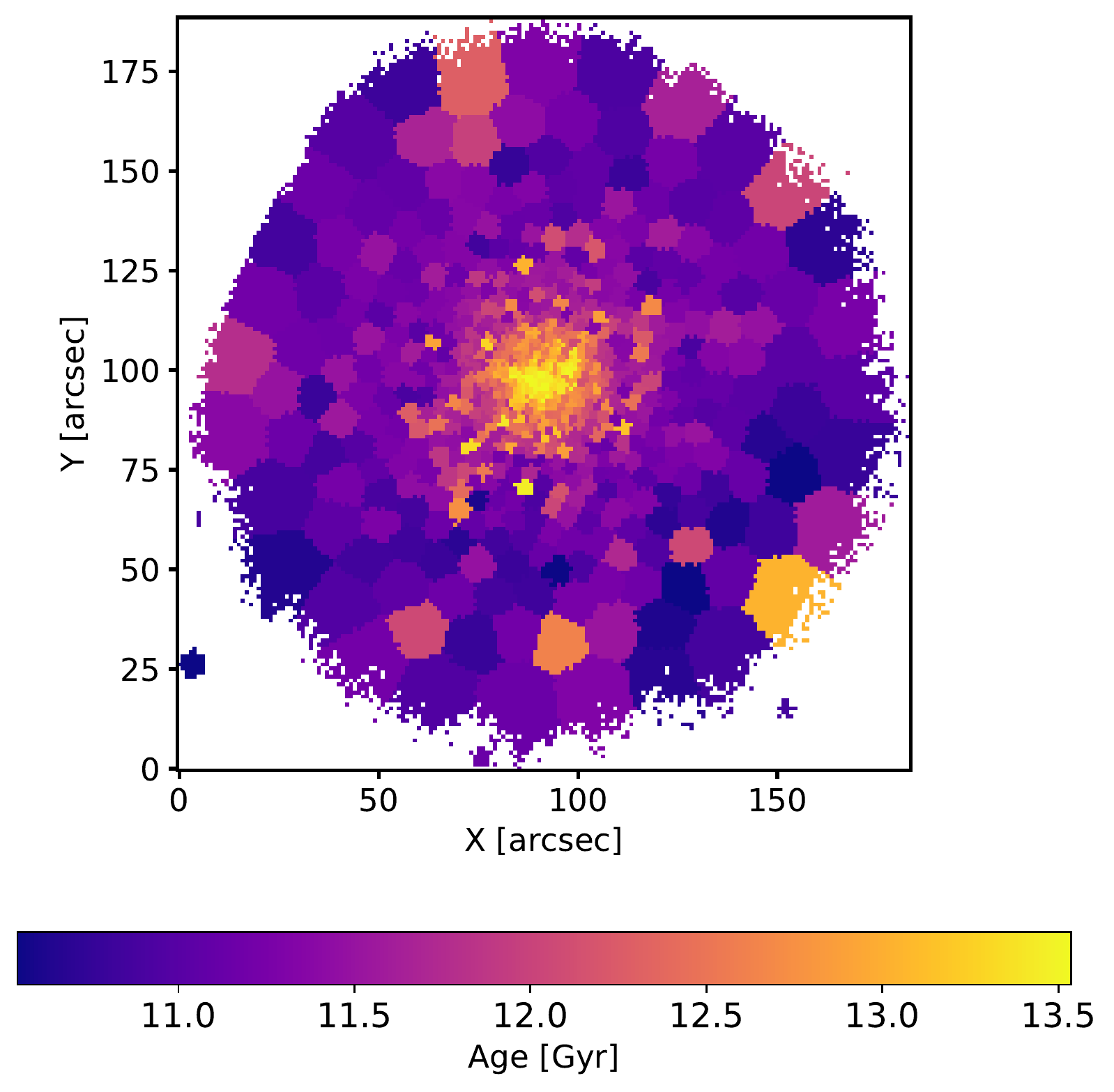}
      \includegraphics[width=6.1cm]{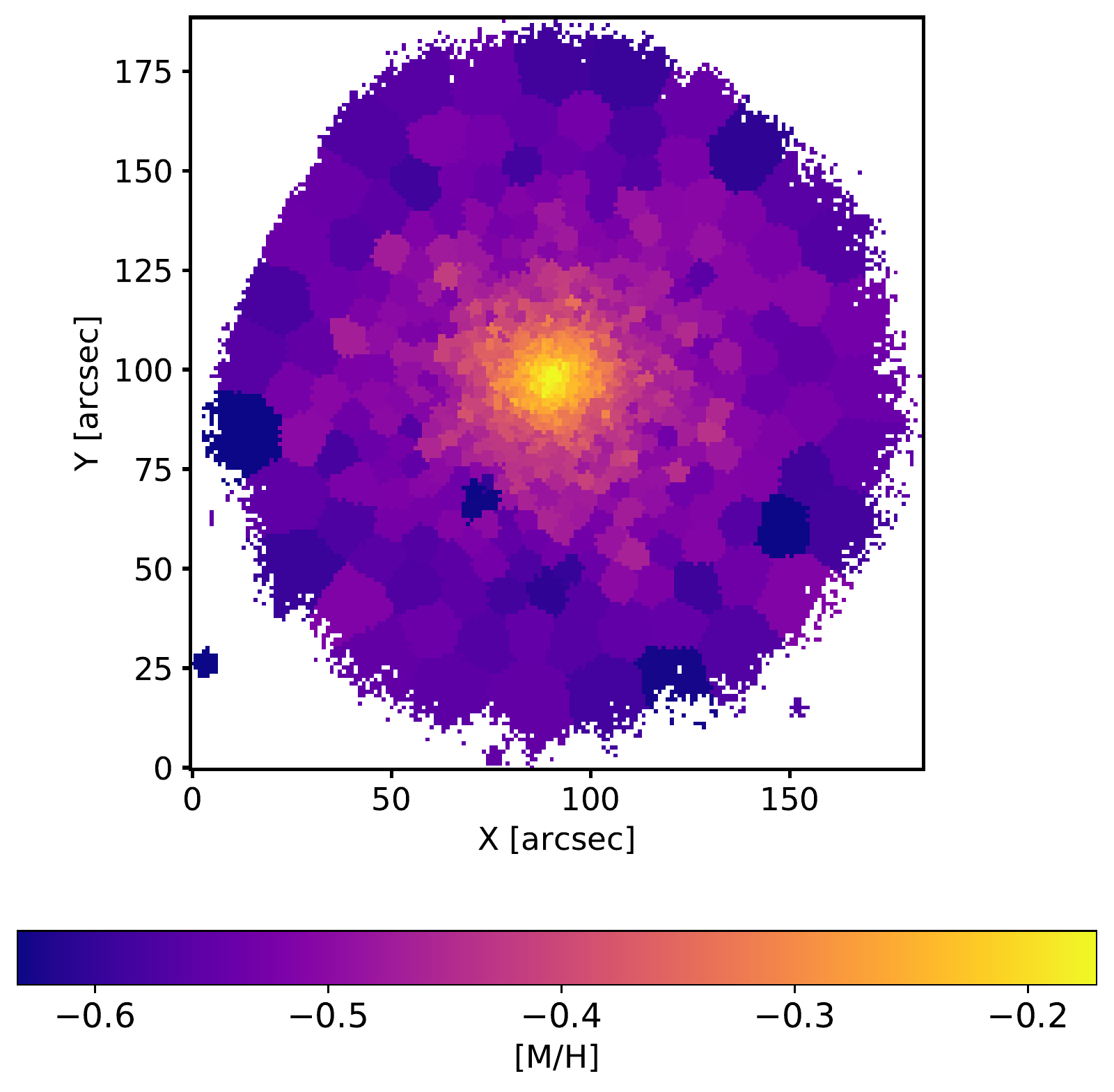}
      \includegraphics[width=6.1cm]{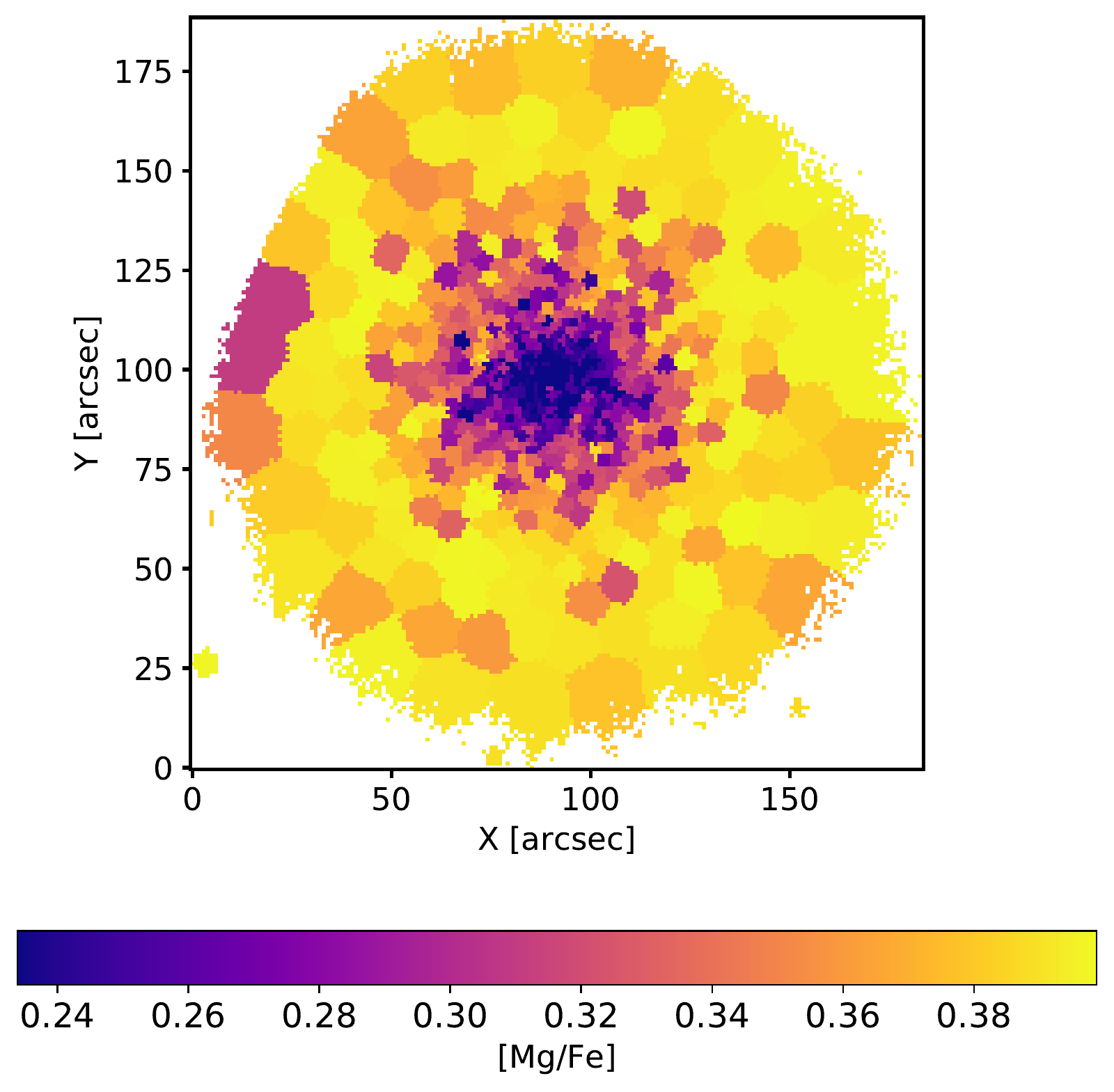}
      \includegraphics[width=6.1cm]{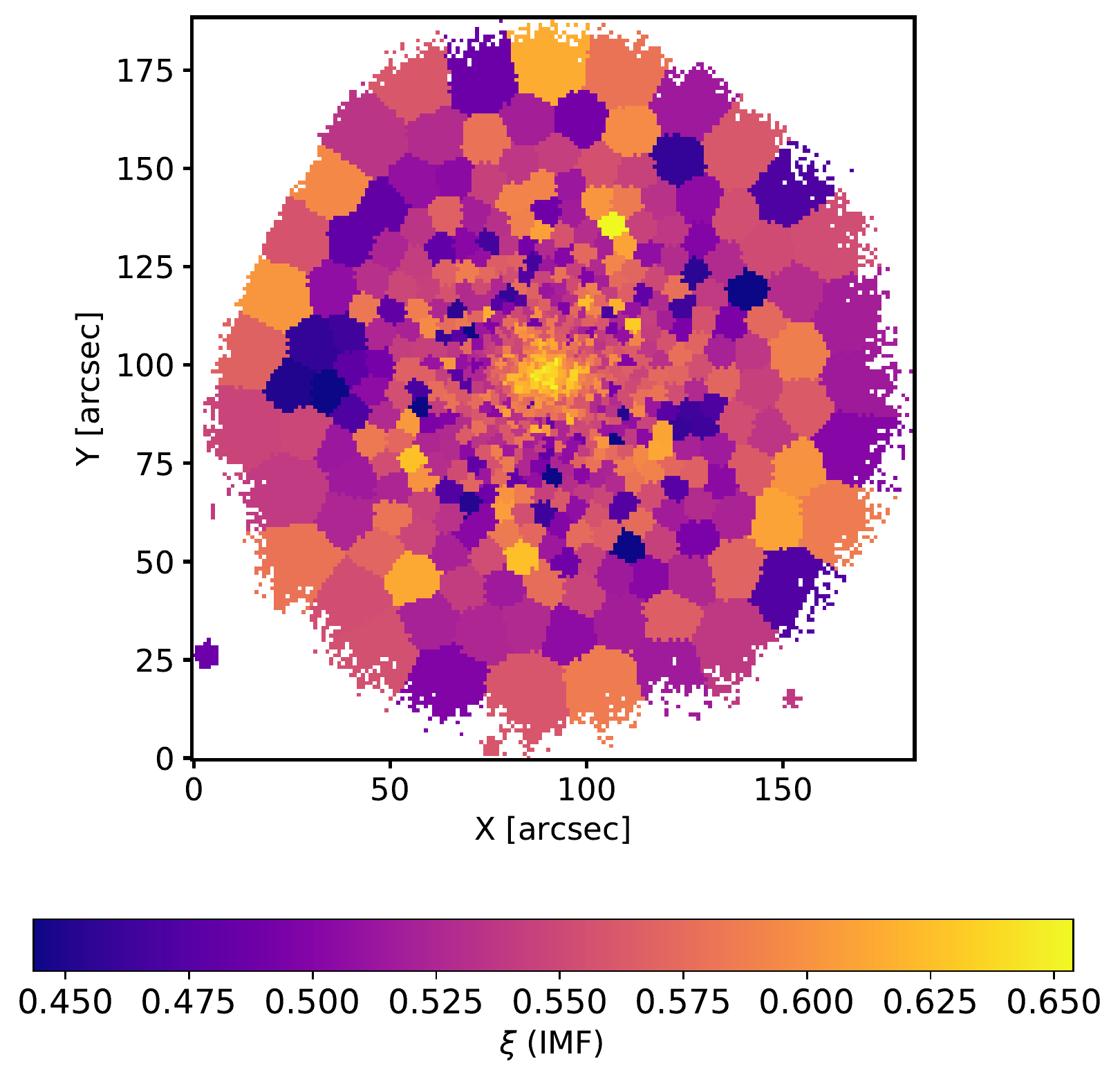}
      \caption{F3D stellar population maps of FCC\,249.} 
   \end{figure*}

   \begin{figure*}
      \centering
      \includegraphics[width=6.1cm]{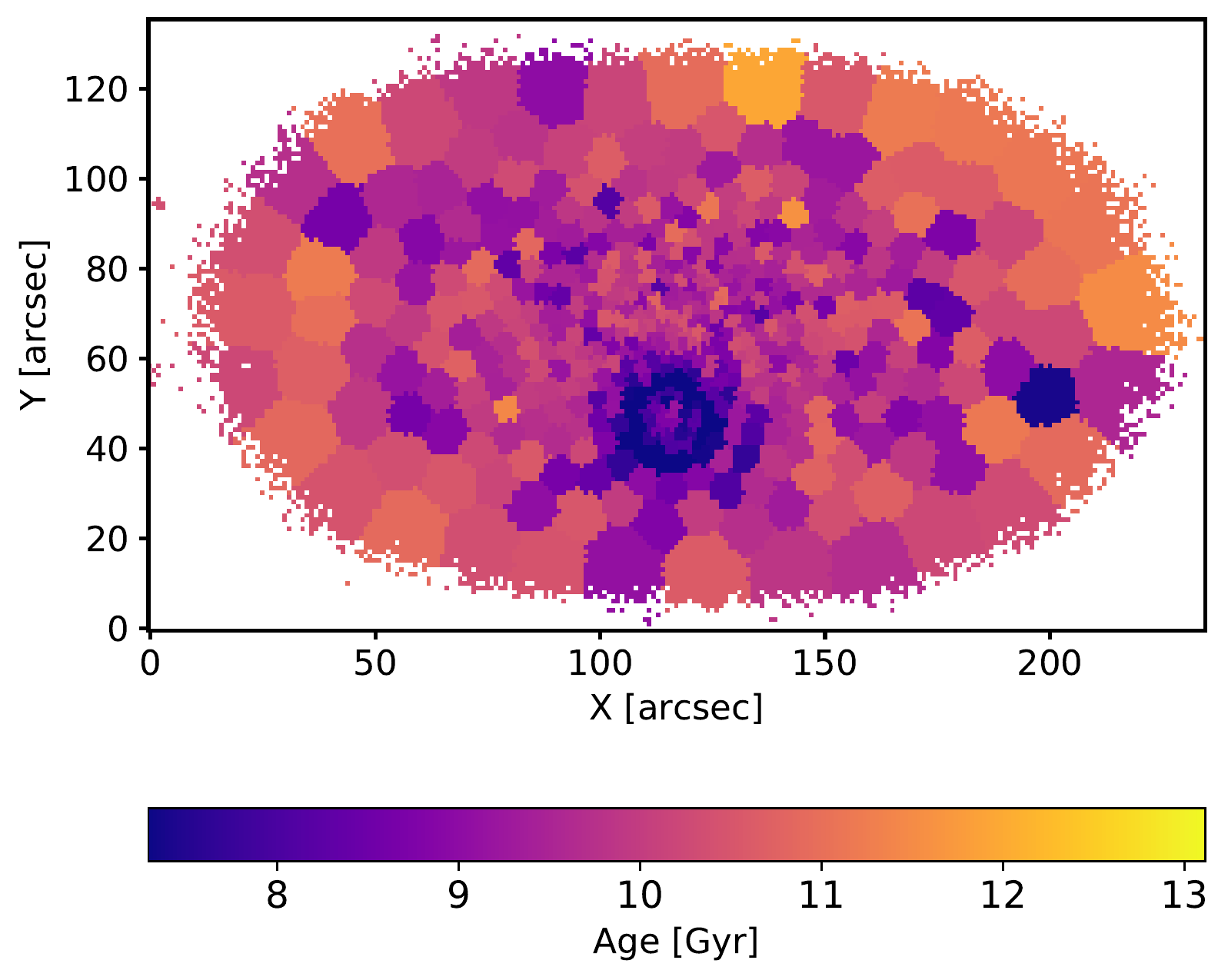}
      \includegraphics[width=6.1cm]{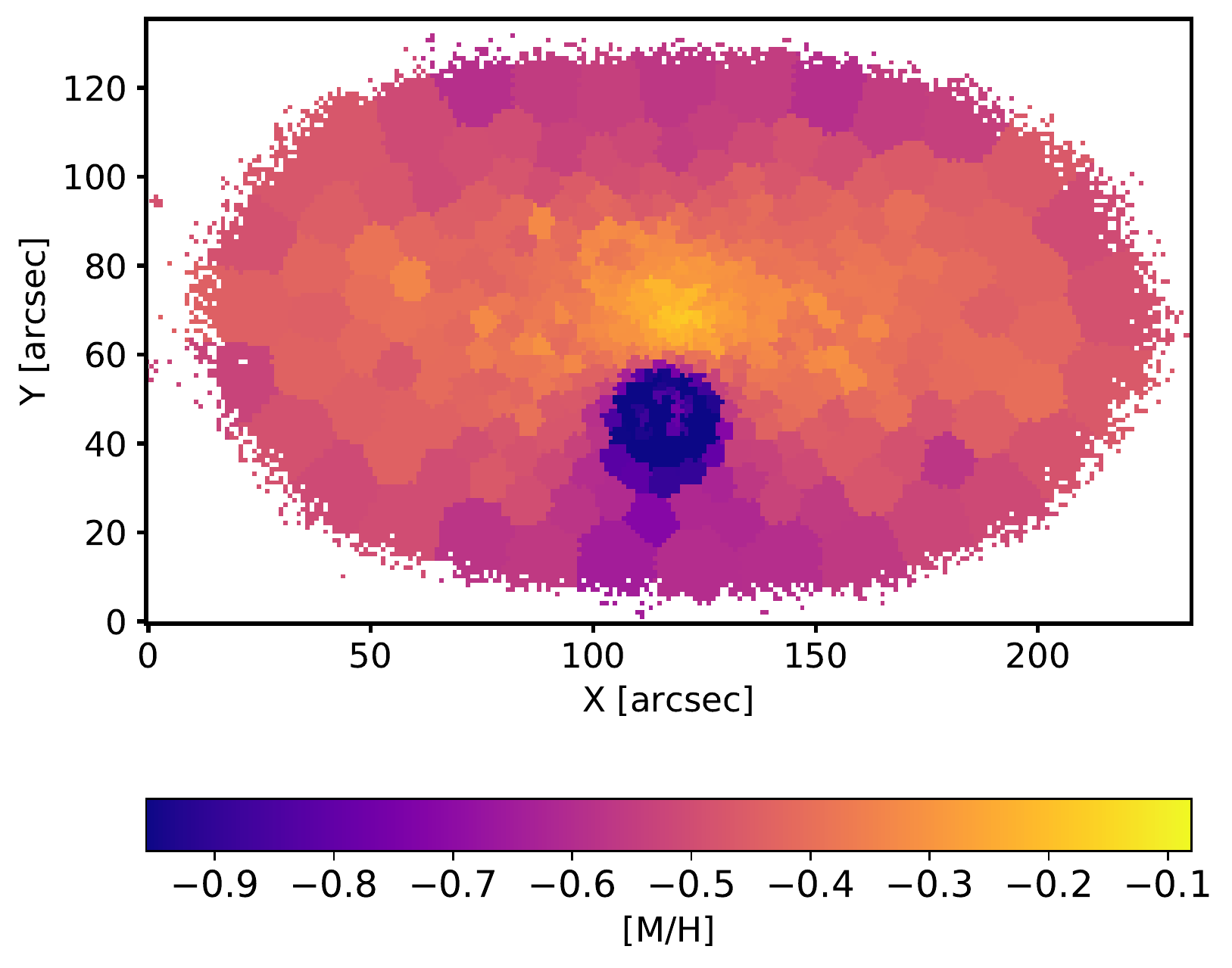}
      \includegraphics[width=6.1cm]{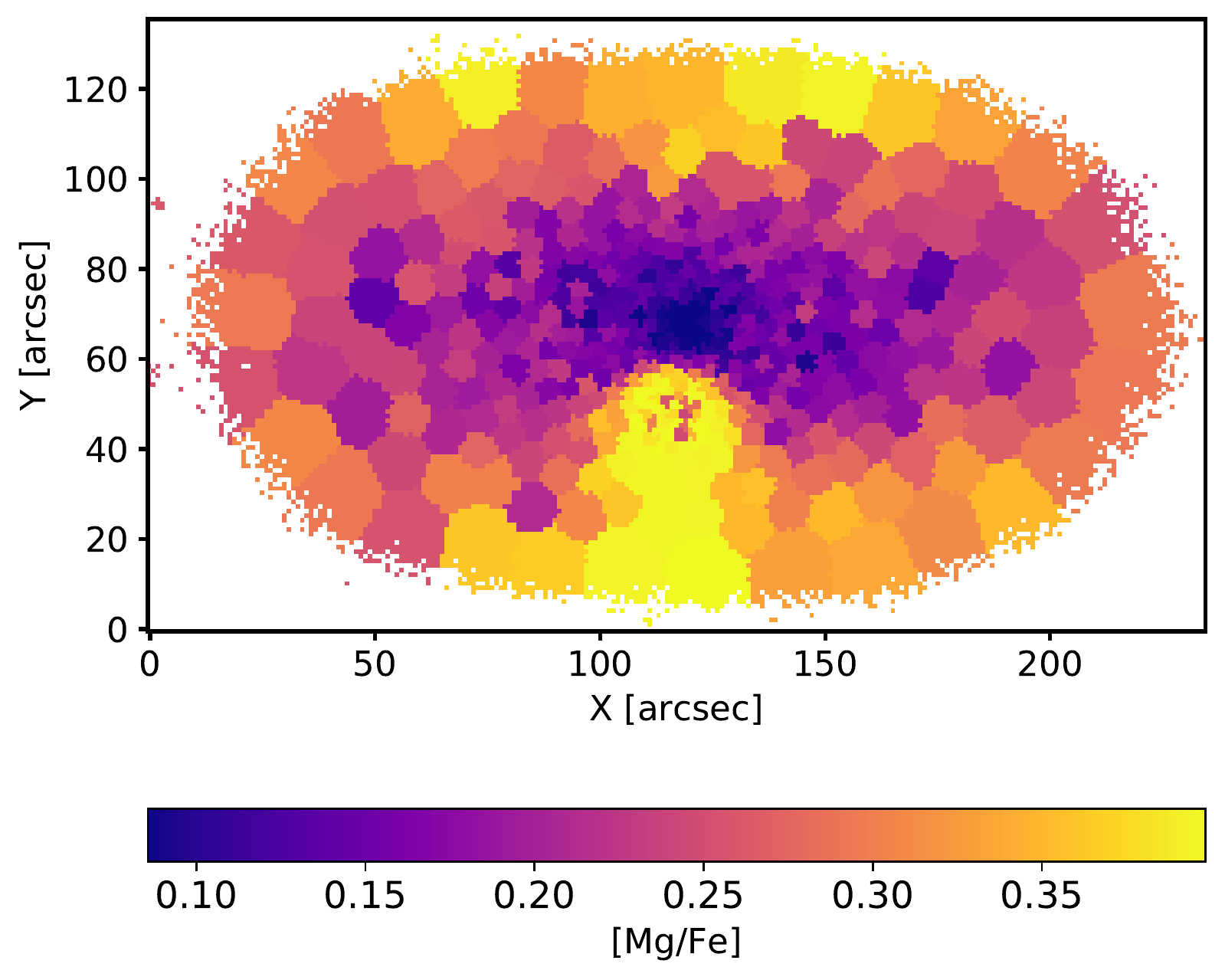}
      \includegraphics[width=6.1cm]{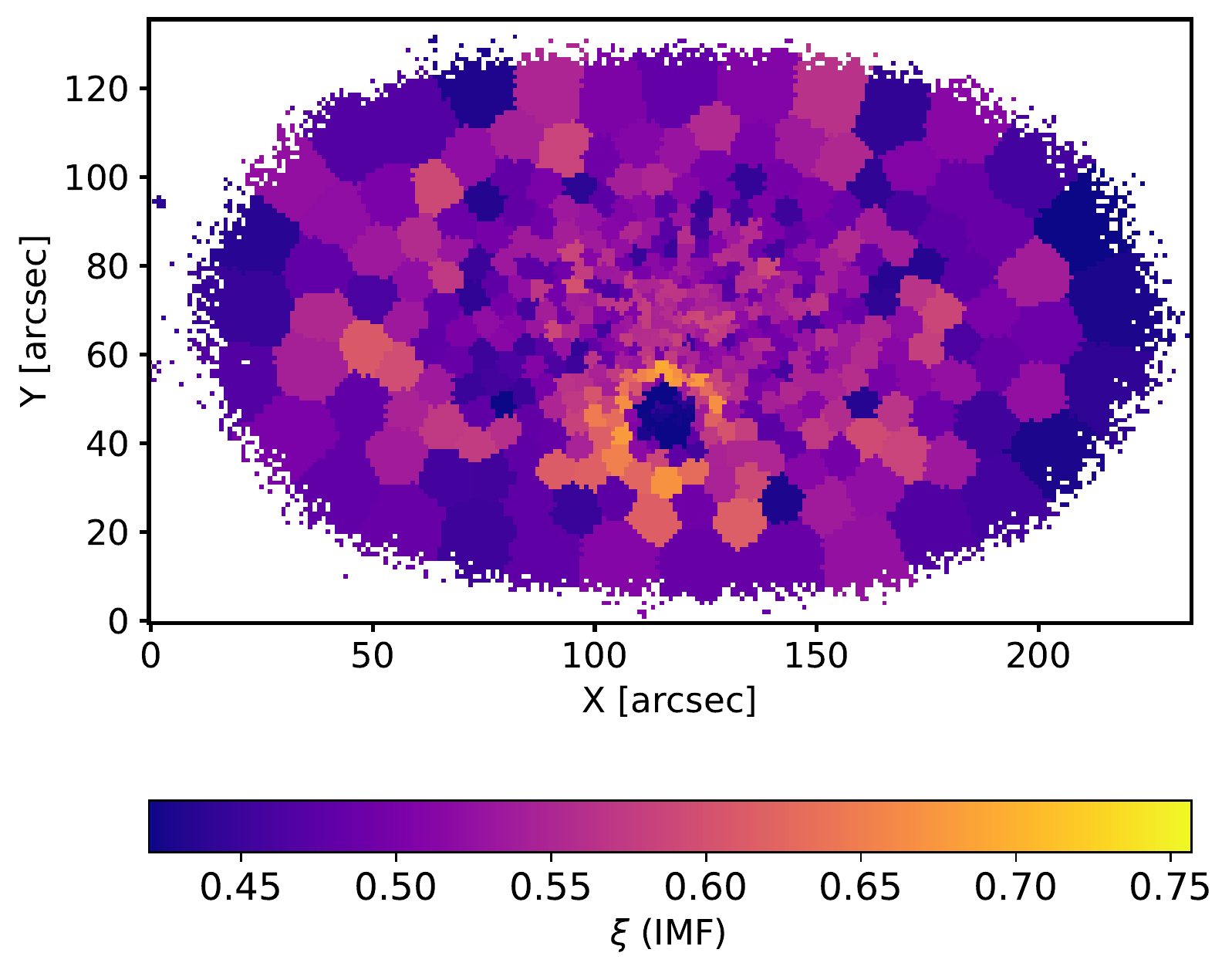}
      \caption{F3D stellar population maps of FCC\,277.} 
   \end{figure*}

   \begin{figure*}
      \centering
      \includegraphics[width=6.1cm]{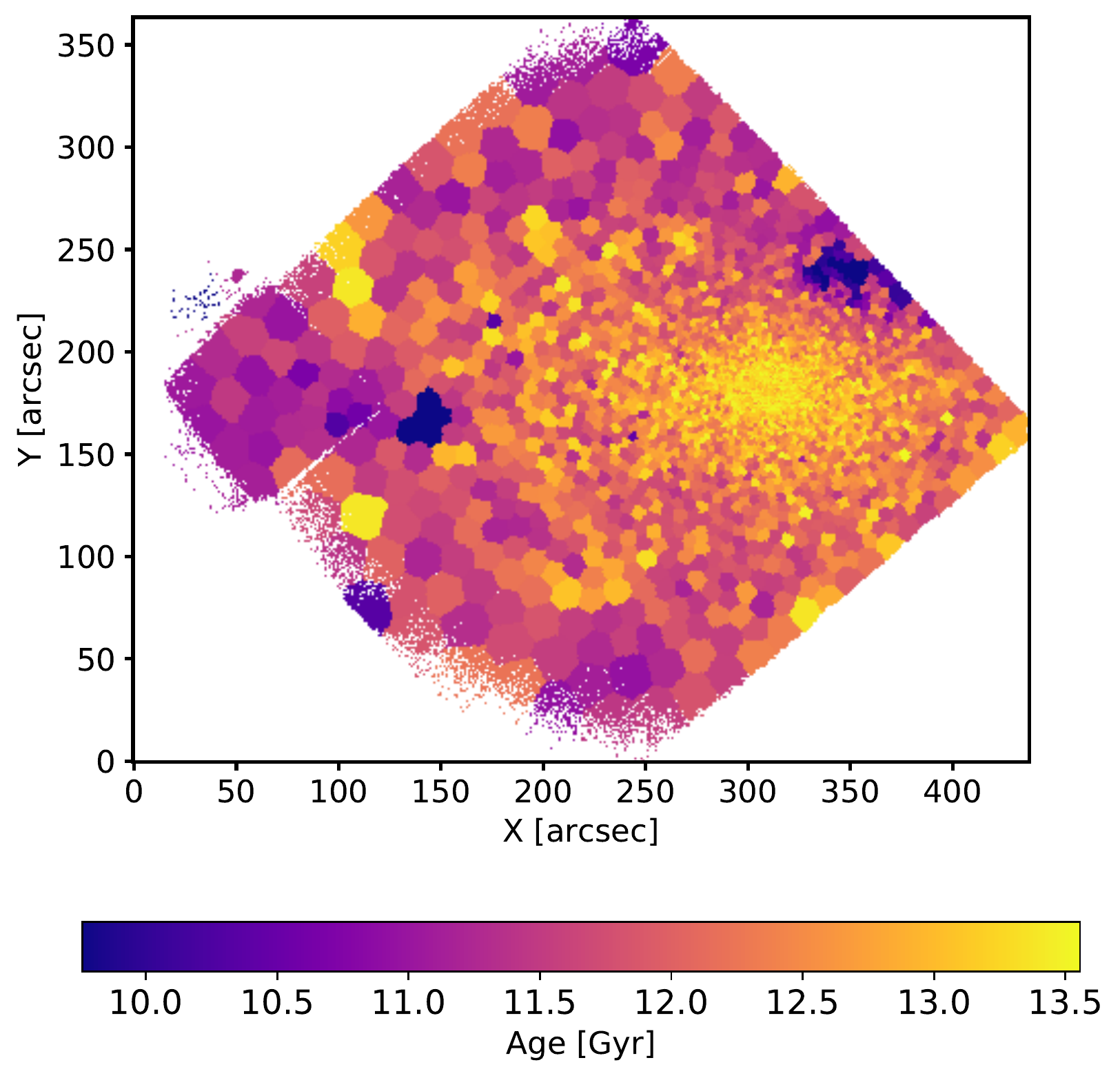}
      \includegraphics[width=6.1cm]{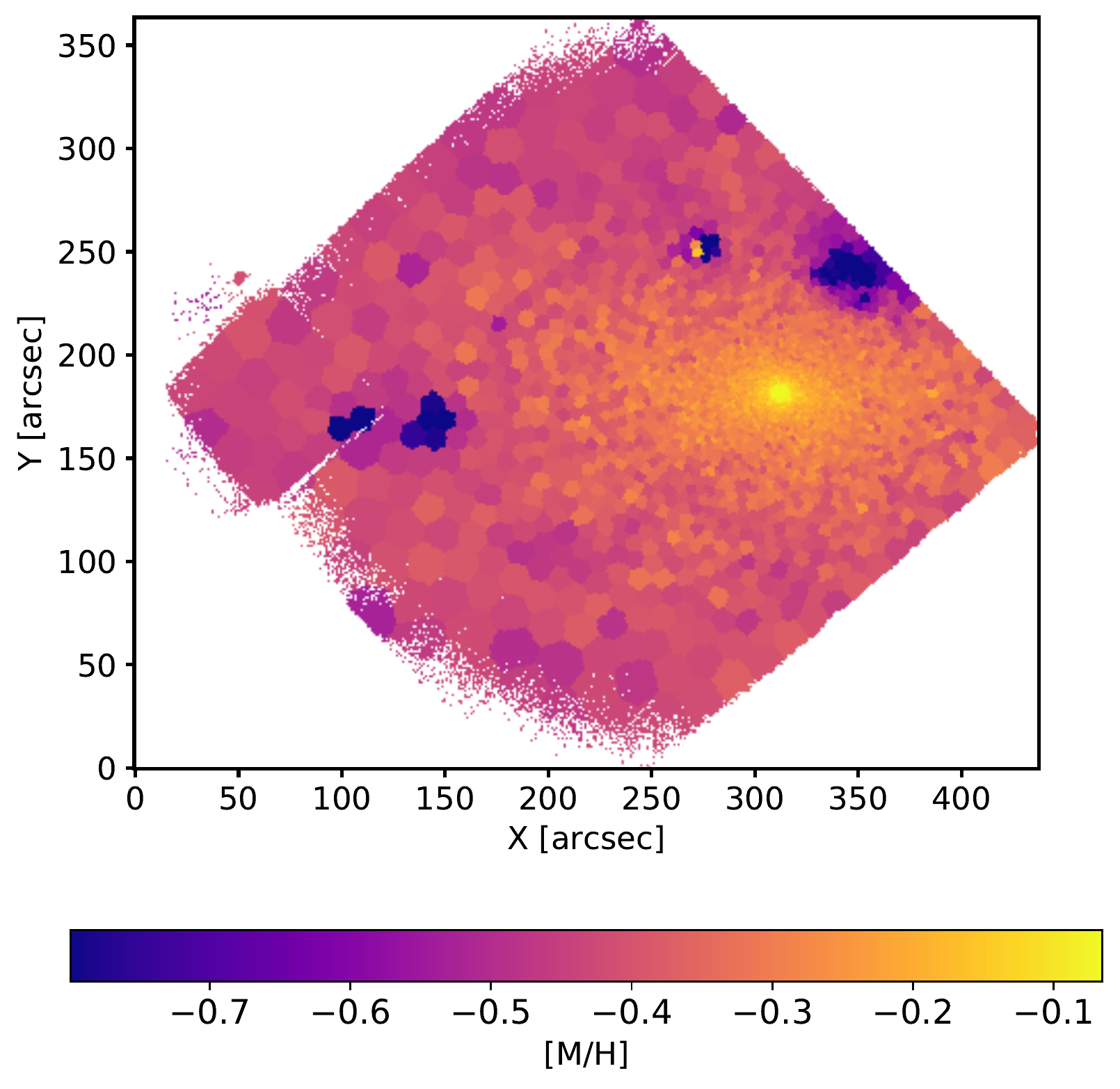}
      \includegraphics[width=6.1cm]{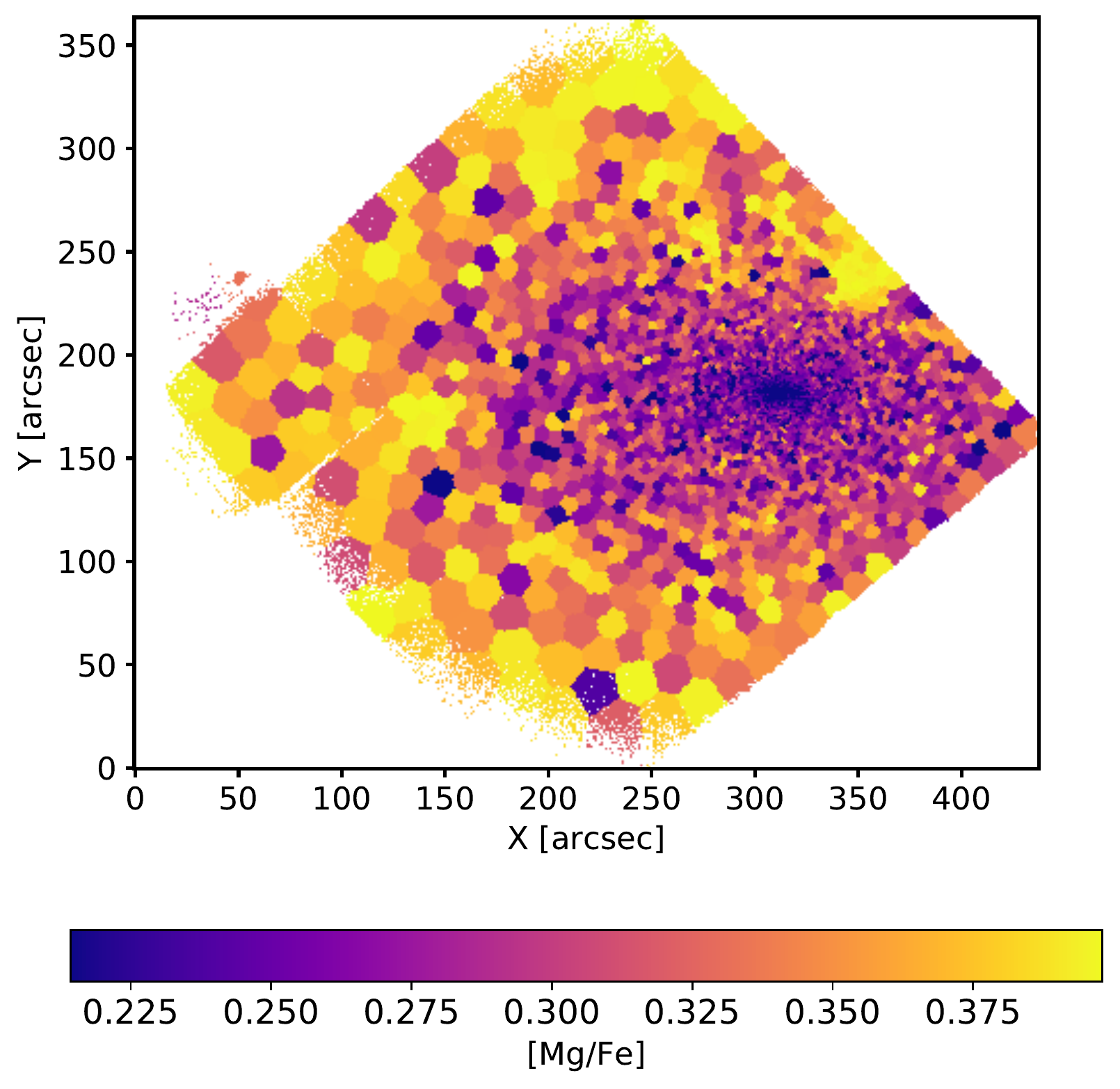}
      \includegraphics[width=6.1cm]{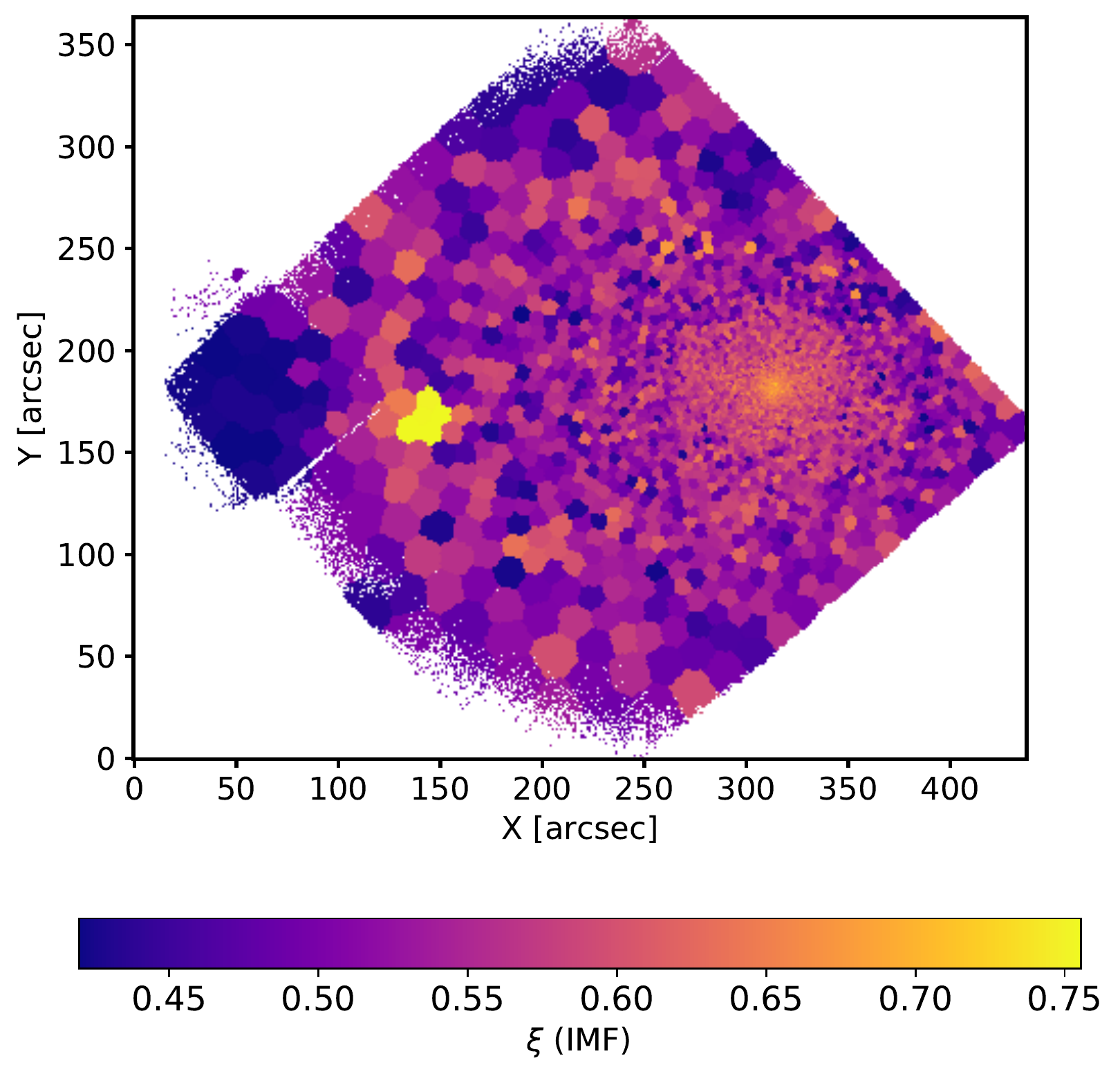}
      \caption{F3D stellar population maps of FCC\,276.} 
   \end{figure*}

   \begin{figure*}
      \centering
      \includegraphics[width=6.1cm]{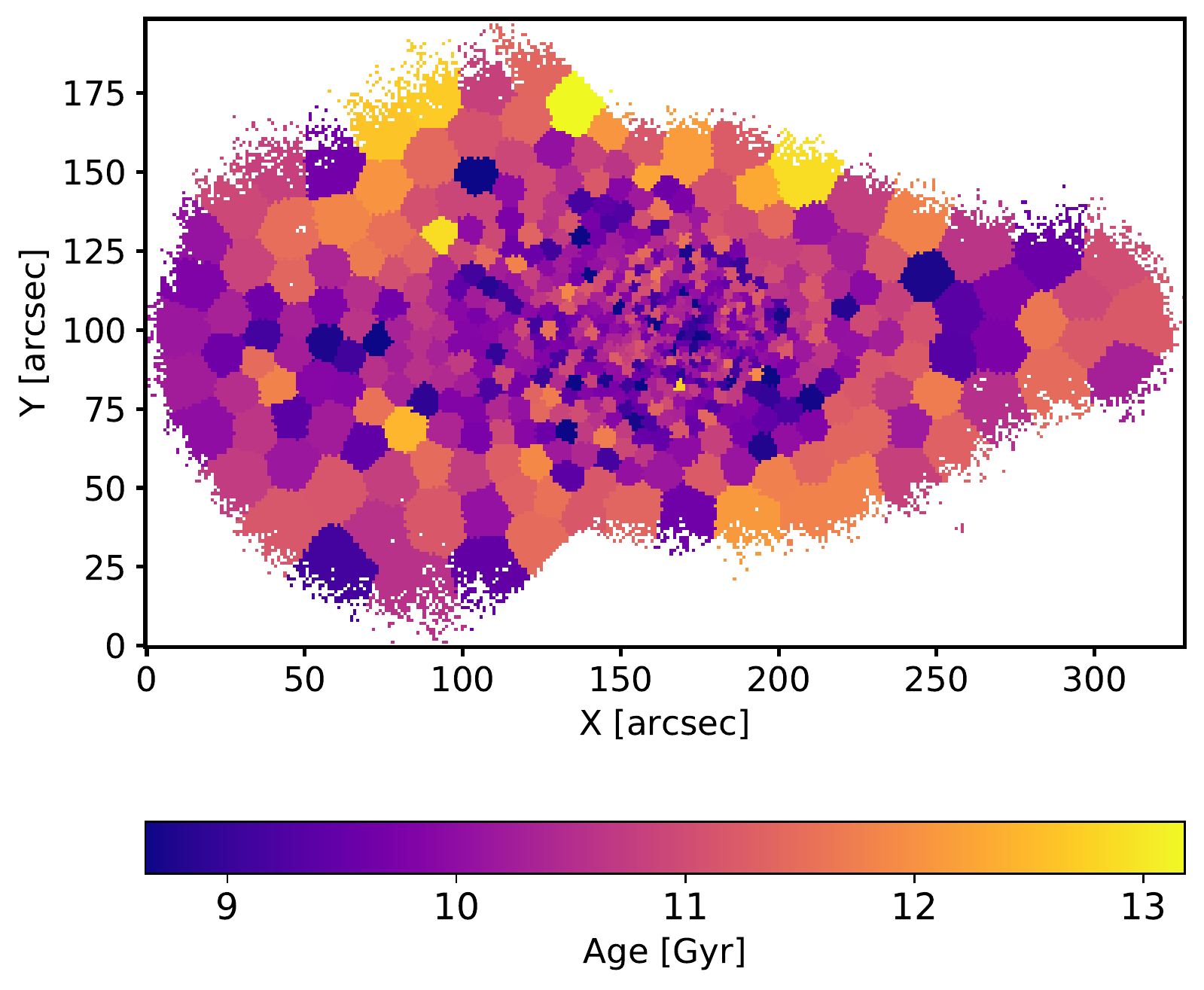}
      \includegraphics[width=6.1cm]{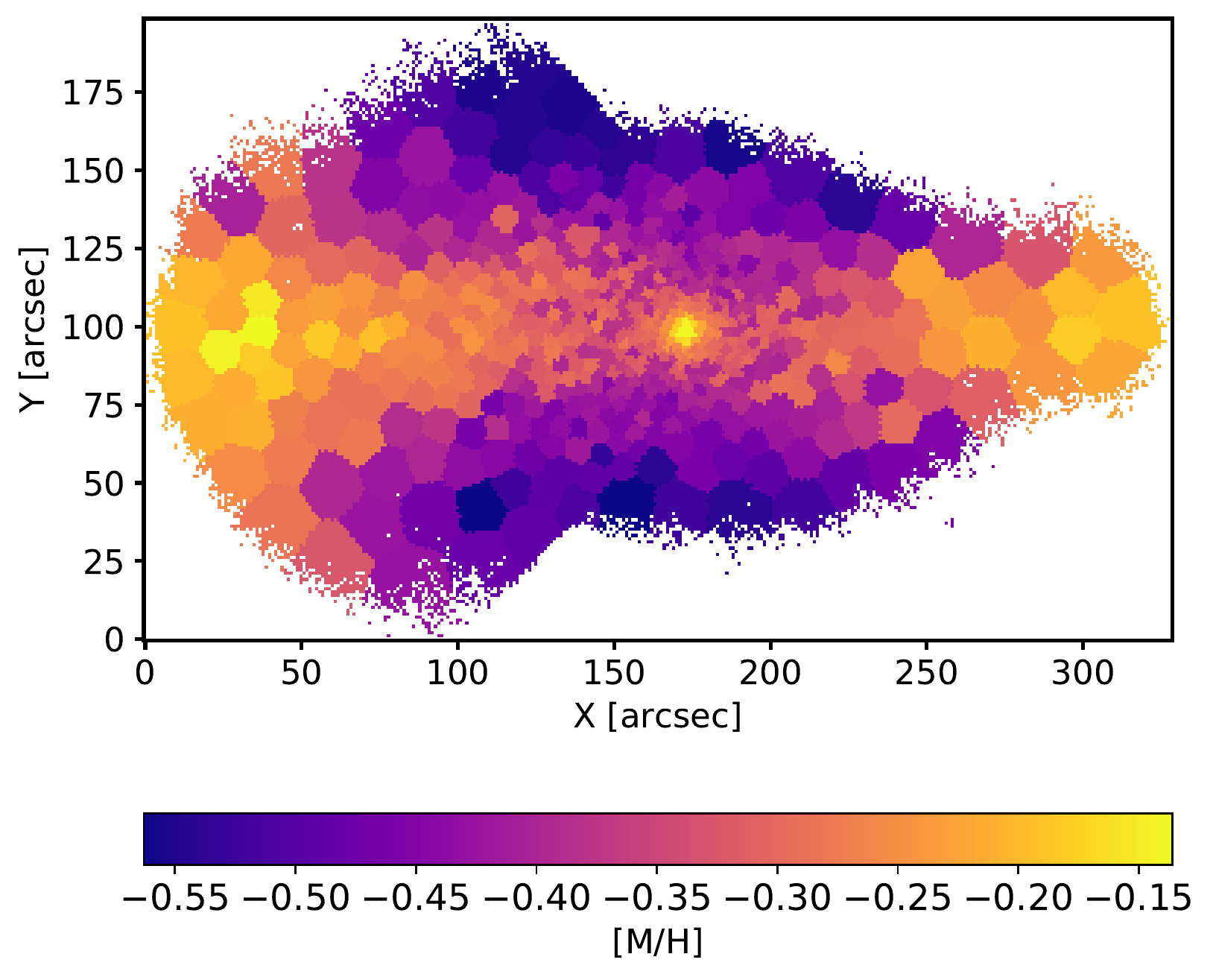}
      \includegraphics[width=6.1cm]{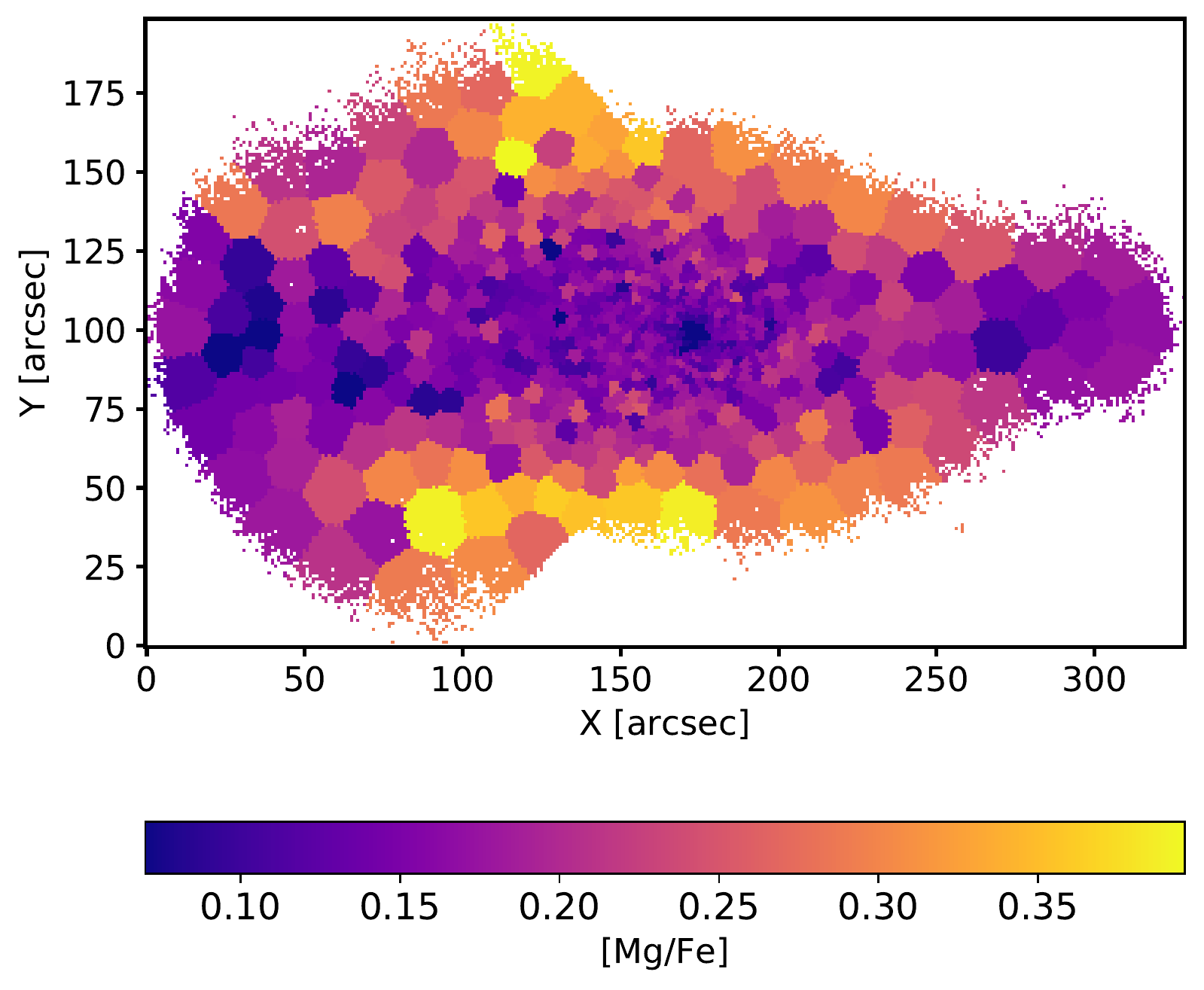}
      \includegraphics[width=6.1cm]{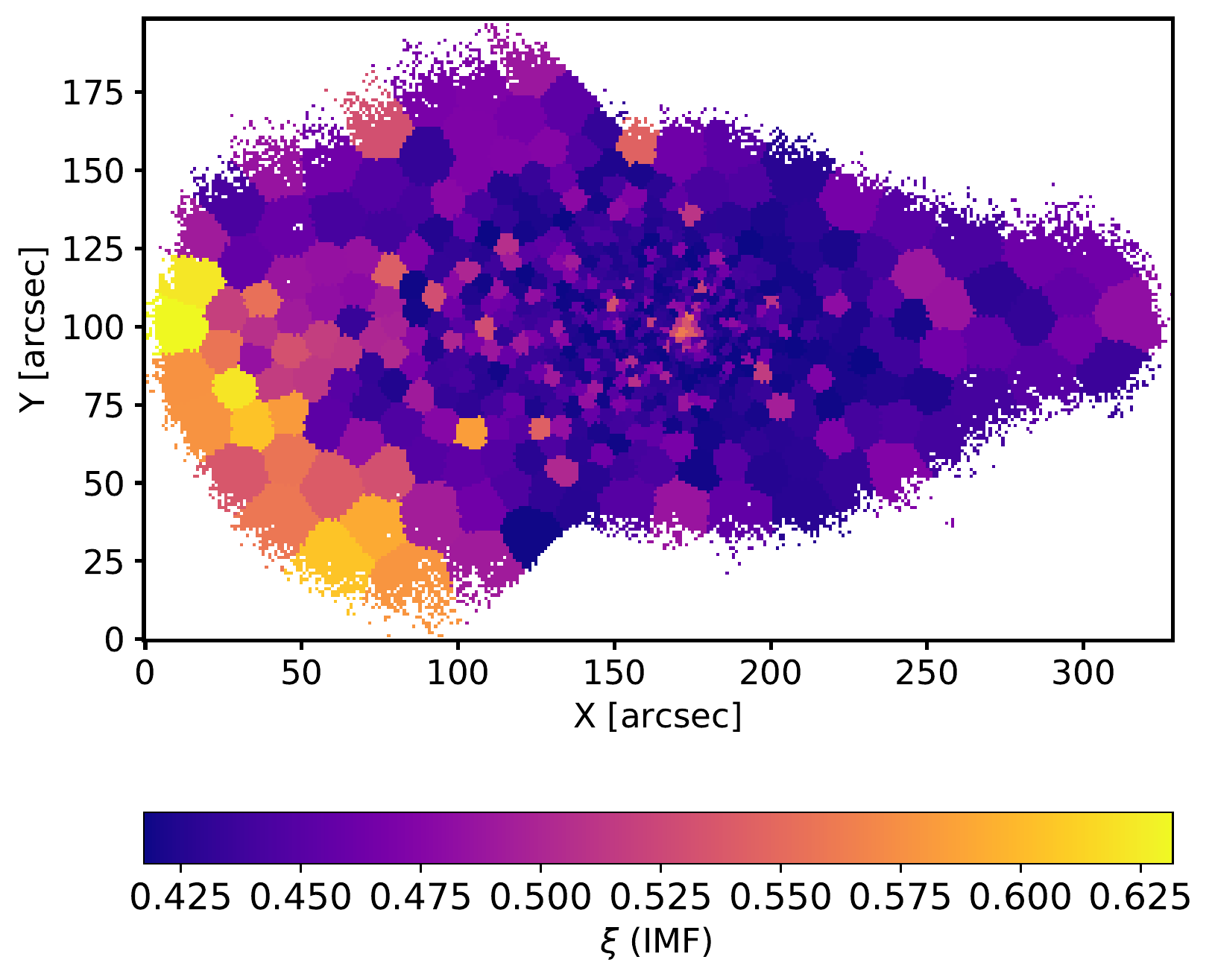}
      \caption{F3D stellar population maps of FCC\,310.} 
   \end{figure*}

   \begin{figure*}
      \centering
      \includegraphics[width=6.1cm]{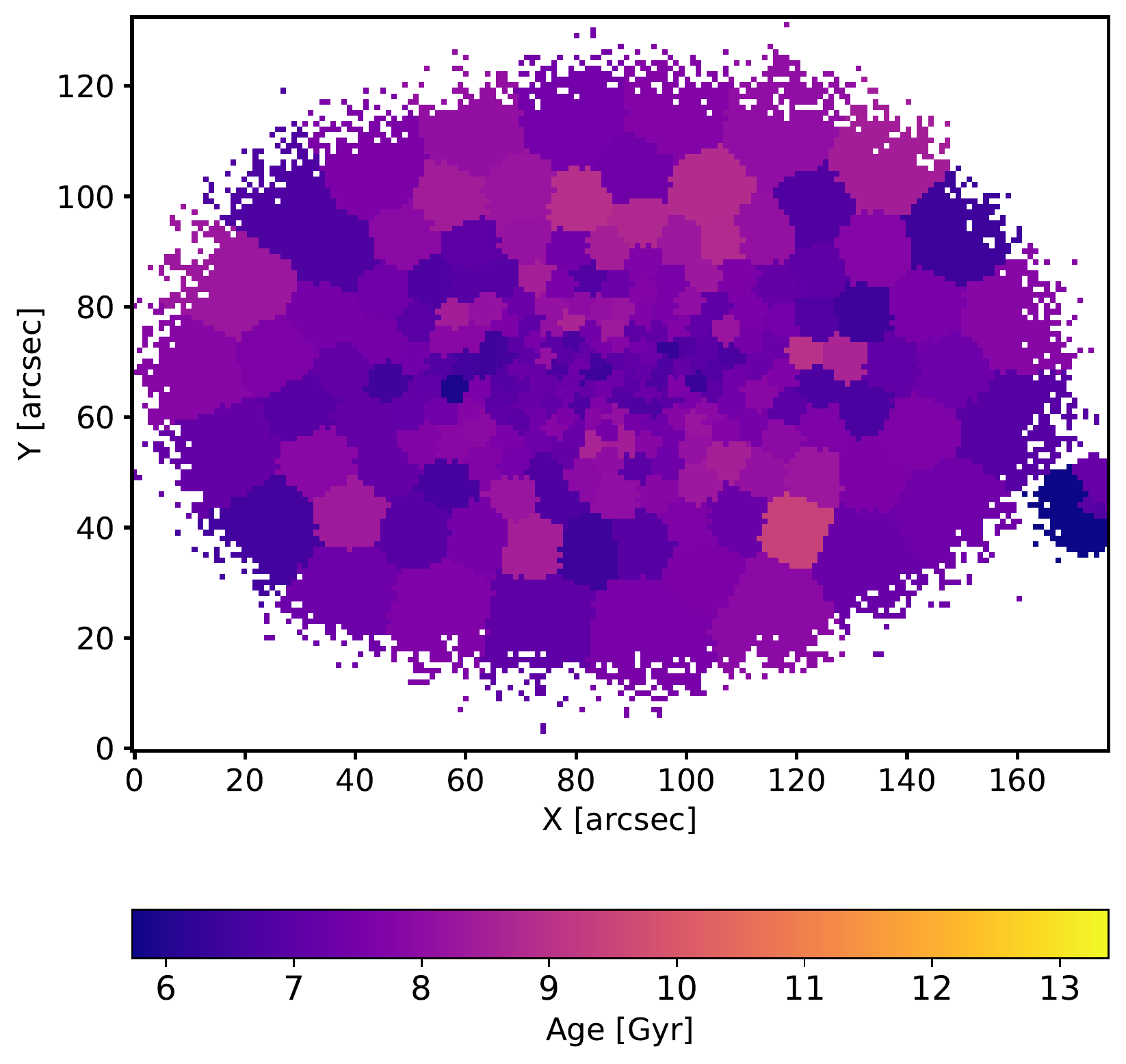}
      \includegraphics[width=6.1cm]{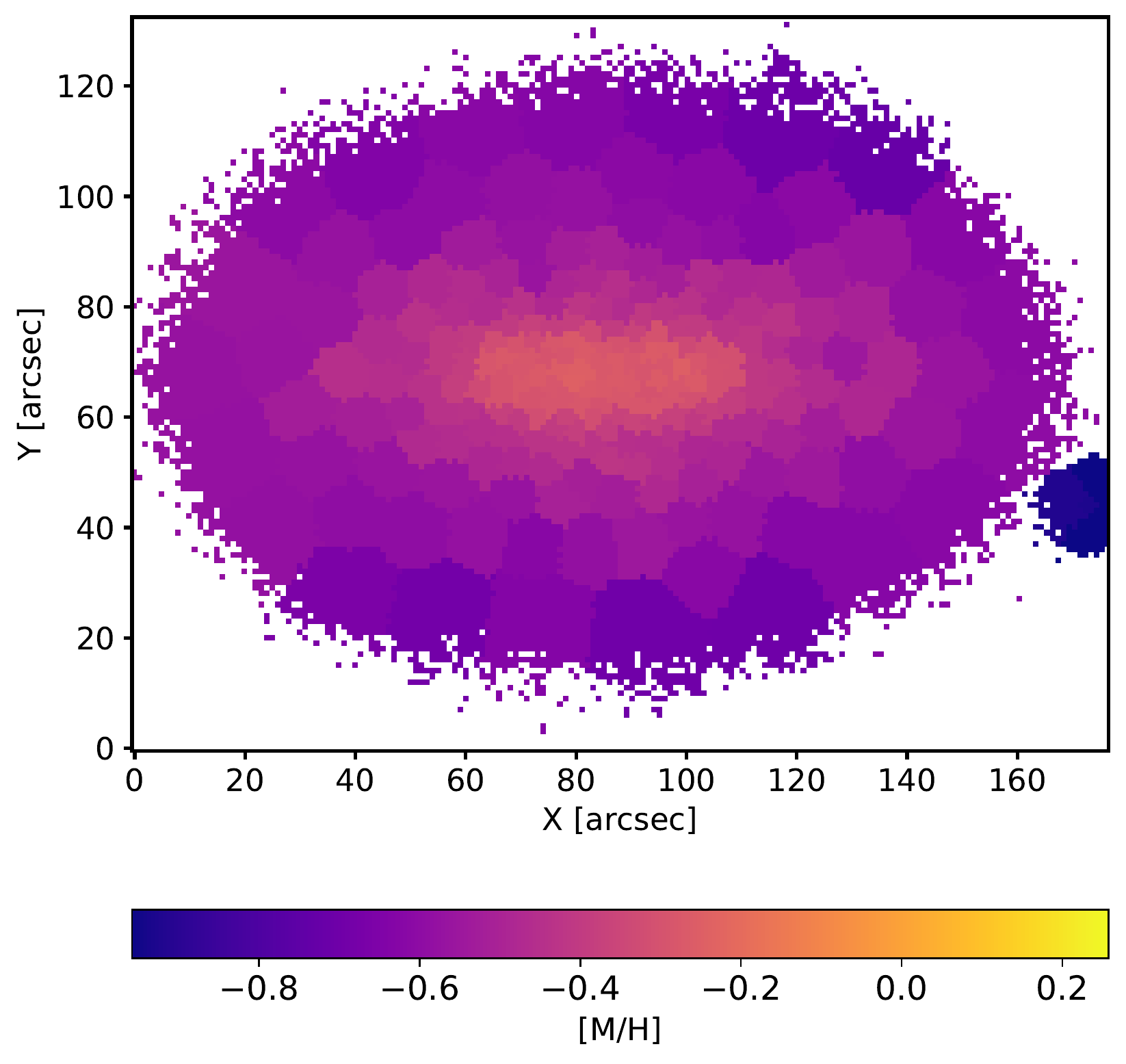}
      \includegraphics[width=6.1cm]{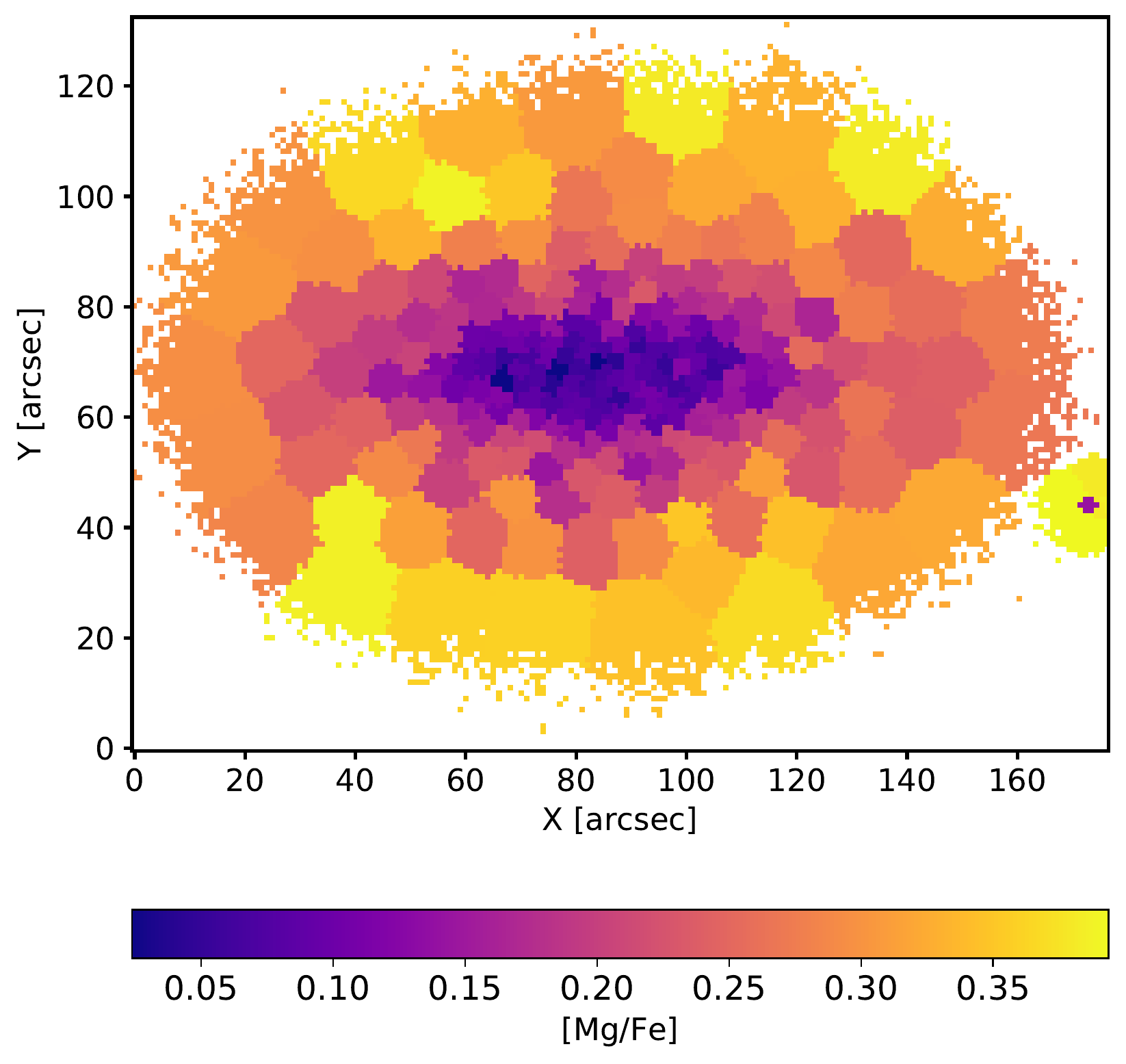}
      \includegraphics[width=6.1cm]{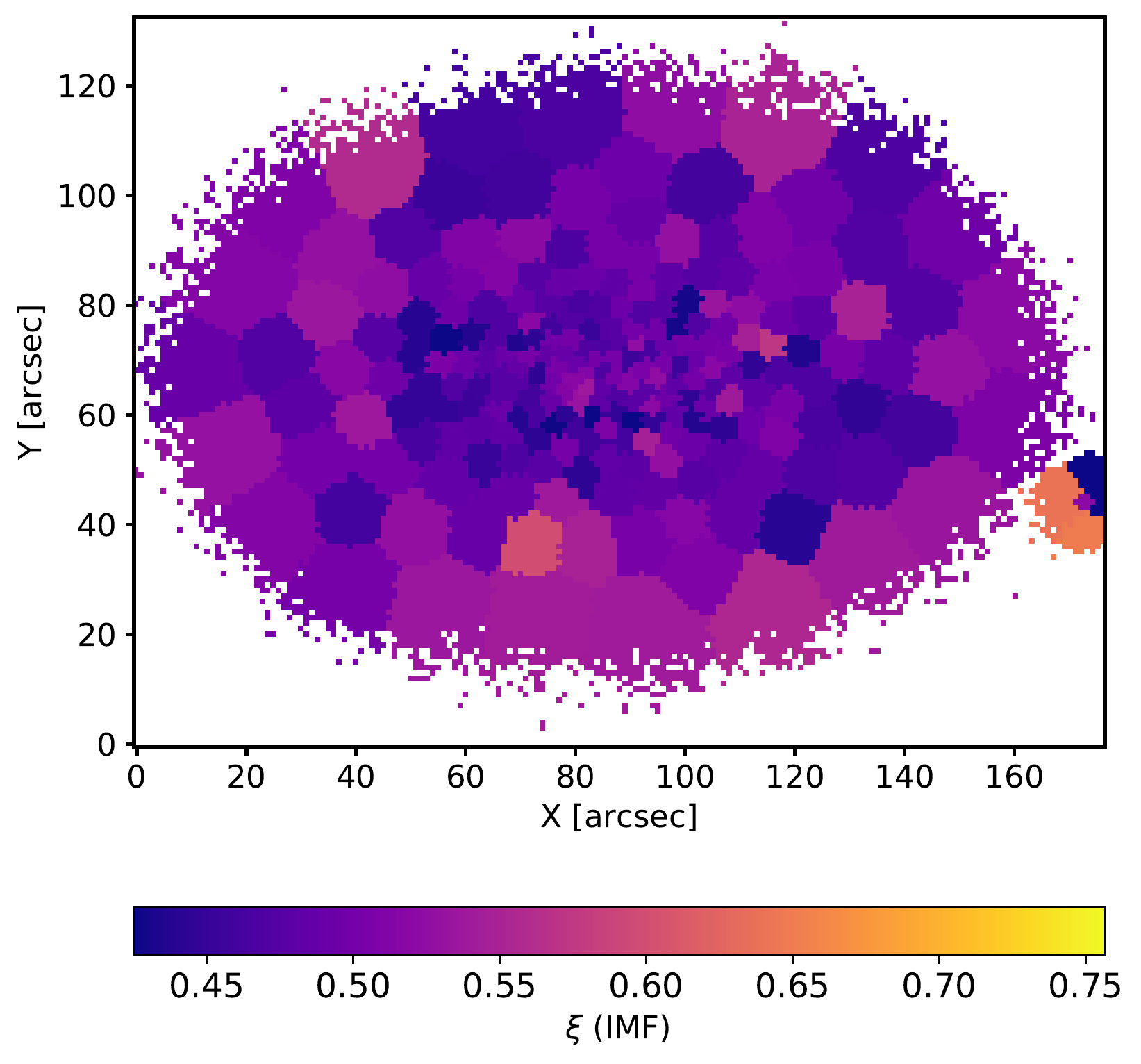}
      \caption{F3D stellar population maps of FCC\,301.} 
   \end{figure*}

\end{appendix}

\end{document}